\documentclass[10pt]{amsart}
\usepackage{amsmath,amssymb,amsthm,amscd,mathrsfs}
\usepackage{graphics,graphicx,epsfig}
\usepackage[all]{xy}
\usepackage{color}
\CompileMatrices
\numberwithin{equation}{section}
\newtheorem{prop}{Proposition}[section]

\newtheorem{coro}[prop]{Corollary}

\newtheorem{defi}[prop]{Definition}
\newtheorem{conj}[prop]{Conjecture}
\numberwithin{equation}{section}
\newcommand{\be}{\begin{equation}}
\newcommand{\ee}{\end{equation}}

\newcommand{\IP}{\mathbb{P}}

\newcommand\IZ{\mathbb {Z}}

\newcommand\IQ{\mathbb {Q}}
\newcommand{\IC}{\mathbb{C}}

\newcommand{\IR}{\mathbb{R}}

\newcommand{\ba}{\begin{array}}
\newcommand{\ea}{\end{array}}

\newcommand{\CV}{{\mathcal V}}

\newcommand{\IF}{{\mathbb F}}

\newcommand{\wF}{{\widetilde F}}

\newcommand{\CS}{{\mathcal S}}
\newcommand{\CB}{{\mathcal B}}
\newcommand{\CK}{{\mathcal K}}

\newcommand{\IH}{{\mathbb H}}
\newcommand{\bal}{\begin{aligned}}
\newcommand{\eal}{\end{aligned}}

\newcommand{\CZ}{{\mathcal Z}}

\newcommand{\half}{{1\over 2}}

\newcommand{\longto}{\longrightarrow}

\newcommand{\ch}{{\mathrm{ch}}}

\newcommand{\CO}{{\mathcal O}}

\newcommand{\CA}{{\mathcal A}}
\newcommand{\CH}{{\mathcal H}}
\newcommand{\CF}{{\mathcal F}}
\newcommand{\CG}{{\mathcal G}}

\newcommand{\CM}{{\mathcal M}}
\newcommand{\CW}{{\mathcal W}}
\newcommand{\CY}{{\mathcal Y}}
\newcommand{\CL}{{\mathcal L}}
\newcommand{\CJ}{{J}}
\newcommand{\CR}{{\mathcal R}}
\newcommand{\CN}{{\mathcal N}}
\newcommand{\CC}{{\mathcal C}}

\newcommand{\RHom}{{\mathrm{RHom}}}
\newcommand{\CQ}{{\mathcal Q}}

\newcommand{\CT}{{\mathcal T}}

\newcommand{\calP}{{\mathcal P}}

\newcommand{\IL}{{\mathbb L}}
\newcommand\fro{{\overline{\underline{\Omega}}}}

\title{Geometric engineering of (framed) BPS states}
\author[W.-y. Chuang, D.-E. Diaconescu, J. Manschot, G. W. Moore, Y. Soibelman]{Wu-yen Chuang${}^1$, Duiliu-Emanuel Diaconescu${}^2$, 
Jan Manschot${}^{3,4}$, Gregory W. Moore${}^5$, 
Yan Soibelman${}^6$}
\address{${}^1$ Department of Mathematics, National Taiwan University, Taipei, Taiwan}
\email{wychuang@gmail.com}
\address{${}^2$ NHETC, Rutgers University\\
Piscataway, NJ 08854-0849 USA}
\email{duiliu@physics.rutgers.edu}
\address{${}^{3}$ Max Planck Institute for Mathematics, Vivatsgasse 7, 53111 Bonn, Germany}
\address{${}^{4}$Bethe Center for Theoretical Physics, Bonn University, Nu\ss allee 12, 53115 Bonn, Germany}
\email{manschot@uni-bonn.de}
\address{${}^5$ NHETC, Rutgers University\\
Piscataway, NJ 08854-0849 USA}
\email{gmoore@physics.rutgers.edu}
\address{${}^6$ Department of Mathematics, Kansas State
University, Manhattan, KS 66506-2602 USA}
\email{soibel@math.ksu.edu}
\date{}
\begin{document}
\begin{abstract}
BPS quivers for $\CN=2$ $SU(N)$ gauge theories are derived via geometric engineering from derived categories of toric Calabi-Yau threefolds. While the outcome is in agreement of previous low energy constructions, the geometric approach leads to several new results. An absence of walls conjecture is formulated for all values of 
$N$, relating the field theory BPS spectrum to large radius 
D-brane bound states. Supporting evidence is presented 
as explicit computations of BPS degeneracies in some
examples. These computations also prove the existence of BPS states of arbitrarily high spin and infinitely many marginal stability 
walls at weak coupling. 
Moreover, framed quiver models for framed BPS states are naturally derived from this formalism, as well as a mathematical formulation of framed and unframed BPS degeneracies in terms of motivic and cohomological
Donaldson-Thomas invariants.  
We verify the conjectured absence of BPS states with ``exotic''
$SU(2)_R$ quantum numbers using motivic DT invariants. 
This application
is based in particular on a complete recursive algorithm which determines
the unframed BPS spectrum at any point on the Coulomb branch in terms
of noncommutative Donaldson-Thomas invariants for framed quiver
representations.

\end{abstract}
\maketitle

\tableofcontents

\section{Introduction}\label{sectionone}
The BPS spectrum of four dimensional $\CN=2$ gauge theories has been a constant
subject of research since the discovery of the Seiberg-Witten solution. An incomplete sampling of references includes 
\cite{Seiberg:1994rs,Seiberg:1994aj,Ferrari:1996sv,Bilal:1996sk, Klemm:1996bj,Bilal:1997st,Mikhailov:1997jv,Mikhailov:1998bx, Shapere:1999xr,Fraser:1996pw,Hollowood:1997pp,Taylor:2001hg}.
Very recent intense activity in this field was motivated  by the connection \cite{GMN}
between wallcrossing on the Coulomb 
branch and the Kontsevich-Soibelman formula
\cite{wallcrossing}.
An incomplete sampling of references includes 
 \cite{GMN,GMNII,CV-I,framedBPS,CNV,Alim:2011ae,
 Alim:2011kw,
Chen:2011gk,spectral_networks,Cecotti:2012va,Cecotti:2012sf,Cecotti:2012gh,Xie:2012gd}.
 For recent reviews see \cite{review,Cecotti:2012se}.
 
 On
 the other hand, it has been known for a while that many
$\CN=2$ gauge theories are obtained in geometric engineering as a low energy
limit of string theory dynamics in the presence of Calabi-Yau
singularities \cite{Aspinwall:1995xy,Kachru:1995wm,Klemm:1996bj,geom_eng,KMV}. This leads immediately to a close
connection between the gauge theory BPS spectrum and the BPS
spectrum of string theory in the presence of such singularities.
The latter consists of supersymmetric D-brane bound states wrapping
exceptional cycles, and hence can in principle be analyzed using
derived category methods \cite{homological-mirror, Dbranes_categories, Pistability,Dbranes_monodromy,Sharpe:1999qz,Aspinwall:2001pu,Dbranes-mirror}.
In principle geometric engineering is expected to provide
a microscopic string
theory derivation for the BPS quivers found in 
\cite{Denef:2002ru,Denef:2007vg,CNV,Alim:2011kw}
by low energy methods. Indeed the BPS quivers constructed in loc. cit.
for $SU(N)$ gauge theories were first derived by Fiol in \cite{Fiol:2000pd}
using fractional branes on quotient singularities. It is quite remarkable
that this construction was confirmed ten years later by completely different low energy methods. A similar approach, employing a
more geometric point of view  has been subsequently employed in \cite{SW-derived,DGS}
for $SU(2)$ gauge theories. Their results are again in agreement
with the low energy constructions.

The goal of the present work is to proceed to a systematic 
study of the gauge theory BPS spectrum via categorical and geometric
methods. Special emphasis is placed on higher rank gauge theories, where the BPS spectrum is not completely known on the entire Coulomb
branch, many problems being at the moment open. In order to keep the
paper to be of reasonable length, only pure $SU(N)$ gauge theories will
be considered in this paper. In this case the local toric threefolds are 
resolved $A_{N-1}$ quotient singularities fibered over $\IP^1$, such that 
the singularity type does not jump at any points on the base. Their
derived categories are equivalent by tilting to derived categories of 
modules over the path algebra of a quiver with potential determined by an exceptional collection of line bundles. Physically, these quivers encode 
the quantum mechanical effective action of a collection of fractional 
branes on the toric threefold. 
Taking the field theory limit amounts to a truncation of the 
fractional brane quiver, omitting the branes which become very heavy 
in this limit together with  the adjacent arrows. The resulting quiver for pure $SU(N)$ gauge theory 
is of the form
\[
\xymatrix{
{q}_{N-1} \bullet  \ar@<.5ex>[rrr]|{c_{N-1}} \ar@<-.5ex>[rrr]|{d_{N-1}}
& & & \bullet p_{N-1}\ar@/_1pc/[ddlll]|{r_{N-2}}
\ar@/^1pc/[ddlll]|{s_{N-2}}  \\
& &  & \\
q_{N-2}\bullet \ar@<.5ex>[rrr]|{c_{N-2}} \ar@<-.5ex>[rrr]|{d_{N-2}}
\ar[uu]|{b_{N-2}}
& & &\bullet p_{N-1}\ar[uu]|{a_{N-2}}  \\
\vdots & & & \vdots \\
q_{i+1} \bullet\ar@<.5ex>[rrr]|{c_{i+1}} \ar@<-.5ex>[rrr]|{d_{i+1}}
& &  &\bullet p_{i+1}\ar@/_1pc/[ddlll]|{r_i}
\ar@/^1pc/[ddlll]|{s_i} \\
& & & \\
q_{i}\bullet \ar@<.5ex>[rrr]|{c_i} \ar@<-.5ex>[rrr]|{d_i}
\ar[uu]|{b_i}
& & & \bullet p_{i}\ar[uu]|{a_i}  \\
\vdots && & \vdots \\
q_{2}\bullet \ar@<.5ex>[rrr]|{c_2} \ar@<-.5ex>[rrr]|{d_2}
& & & \bullet p_2 \ar@/_1pc/[ddlll]|{r_1}
\ar@/^1pc/[ddlll]|{s_1} \\
& & & \\
q_{1}\bullet \ar@<.5ex>[rrr]|{c_1} \ar@<-.5ex>[rrr]|{d_1} \ar[uu]|{b_1}
& & & \bullet p_{1}\ar[uu]|{a_1}
\\
}
\]
with a potential 
\[
\CW=\sum_{i=1}^{N-2} \left[r_i(a_ic_i-c_{i+1}b_i)+s_i(a_id_i-d_{i+1}b_i)\right].
\]
This is the same as the quiver found in \cite{Fiol:2000pd}, 
and is mutation equivalent to the quivers found in \cite{CNV,Alim:2011kw} by different methods.
This approach can be extended to   
gauge theories with flavors allowing 
the $A_{N-1}$ singularity to jump at special points on the base.

In order to set the stage, geometric engineering and the field theory limit of Calabi-Yau compactifications is carefully reviewed in
Section \ref{sectiontwo}. Special emphasis is placed on categorical
constructions, in particular exceptional collections of line bundles on
toric Calabi-Yau threefolds. In particular an explicit construction of such collections is provided for toric Calabi-Yau threefolds $X_N$ engineering pure $SU(N)$ gauge theory.
Not surprisingly, it is then shown that the associated fractional
brane quiver is the same as the one obtained in \cite{Fiol:2000pd}
by orbifold methods. As opposed to the construction in loc. cit.,
the geometric approach provides a large radius limit presentation
of fractional branes in terms of derived objects on $X_N$.
The main outcome of Section \ref{sectiontwo} is a conjectural
categorical description of gauge theory BPS states in terms of
a triangulated subcategory $\CG\subset D^b(X_N)$. As shown
by detailed {\bf A}-model computations in 
 Section \ref{twotwo}, $\CG$ is a
truncation of $D^b(X_N)$ generated by fractional branes with  finite central charges 
in the field theory limit. It is perhaps worth noting that this conclusion involves certain delicate cancellations between tree level
and world-sheet instanton contributions which were never spelled out in the literature. 

According to \cite{Dbranes_categories,Pistability,
Dbranes_monodromy} supersymmetric D-brane configurations on $X_N$
are identified with
$\Pi$-stable objects in the derived category $D^b(X_N)$,
or in rigorous mathematical formulation, Bridgeland stable objects
\cite{stabtriangulated}. Therefore one is naturally led to conjecture that
gauge theory BPS states will be constructed in terms of Bridgeland stable objects in $D^b(X_N)$ which belong to $\CG$. However it is
important to note that agreement of 
the low energy constructions
with \cite{CNV,Alim:2011ae,Alim:2011kw} requires a
stronger statement. Namely, that gauge theory BPS states must be
constructed in terms of an intrinsic stability condition on $\CG$.
Mathematically, these two statements are not equivalent since in general
a stability condition on the ambient derived category does not automatically induce one on the subcategory $\CG$. It is however
shown in Section \ref{twothree} that this does hold for {\it quivery}
or {\it algebraic} stability conditions, analogous to those constructed in
\cite{stabnoncompact,locptwo}.
The above statement {\it fails} for geometric large radius
limit stability conditions, such as $(\omega,B)$-stability, which is analyzed in Section \ref{sectionthree}.
Section \ref{sectiontwo} concludes with a detailed comparison of
gauge theoretic BPS indices and the motivic Donaldson-Thomas invariants constructed in \cite{wallcrossing}. In particular it is shown
that the  protected spin characters defined in \cite{framedBPS}
correspond mathematically to a $\chi_y$-genus type specialization  of the motivic
invariants. In contrast, the unprotected spin characters introduced in
\cite{DM-crossing, DGS} are related to virtual Poincar\'e or Hodge
polynomials associated to the motivic invariants. This is explained in
Section \ref{twofour}, together with a summary of positivity
conjectures for gauge theory BPS states states formulated in \cite{framedBPS}. 

We note here that different mathematical 
constructions of categories and stability conditions for BPS 
states is carried out by Bridgeland and Smith in \cite{quadratic,quivers-Fukaya}, and, as part of a more general 
framework, by Kontsevich and Soibelman in \cite{KoSo}.
The connection between their 
work and this paper will be explained in Section 
\ref{BPSmirror}.

Section \ref{localmirror} consists of a detailed analysis of the field theory limit 
in terms of the local mirror geometry for $SU(2)$ gauge theory. 
The results confirm the conclusions of Section \ref{twotwo} 
and also set the stage for the {\it absence of walls conjecture} 
formulated in the next section.

Section \ref{sectionthree} is focused on large radius supersymmetric D-brane configurations on $X_N$
and their relation to gauge theory BPS states. 
Motivated by the $SU(2)$ example in Section \ref{SUtwo},
we are led to conjecture a precise relation between large radius
and gauge theory BPS states, called the {\it absence of walls conjecture}. 
As explained in the beginning of Section 
\ref{localmirror}, for general $N$ the complex structure moduli 
space of the local mirror to $X_N$ is parameterized by $N$ 
complex coordinates $z_i$, $0 \leq i\leq N-1$. The large 
complex structure limit point (LCS) 
lies at the intersection of the $N$ boundary 
divisors $z_i=0$, $0 \leq i\leq N-1$. On the other hand, the 
scaling region defining the field theory limit is centered at the 
intersection between the divisor 
$z_0=0$ and the discriminant $\Delta_N$, 
as sketched below. 
 \bigskip

 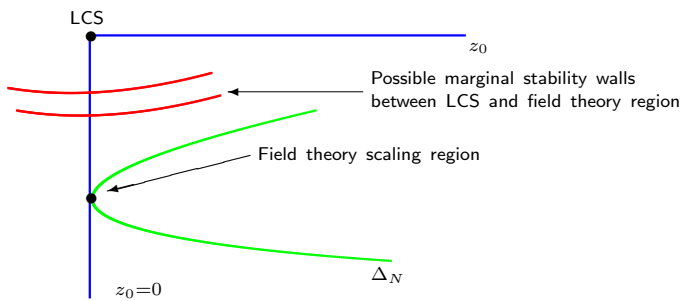
\begin{figure}[h]
  \setlength{\unitlength}{1cm}
\hspace{20pt}
\begin{picture}(5,3.5)
\linethickness{0.3mm}
\put(0,3.5){\color{blue}\line(1,0){5}}
\put(0,0){\color{blue}\line(0,1){3.5}}
\put(-0.07,3.4){$\bullet$} 
{\color{green} \qbezier(3,2.5)(-3.4,1)(4,0.5)}
{\color{red} \qbezier(-1.2,2.8)(0,2.6)(1.5,3)}
{\color{red} \qbezier(-1.2,2.5)(0,2.3)(1.5,2.7)}
\put(-0.3,1.25){$\bullet$}
\put(-0.5,3.7){${}_{\sf LCS}$}
\put(2,1.9){${}_{\sf Field\ theory\ scaling\ region}$}
\linethickness{0.1mm}
\put(1.9,1.9){\vector(-4,-1){1.9}}
\put(0.1,0.1){${}_{z_0=0}$}
\put(4.8,3.35){${}_{z_0}$}
\put(3.5,0.3){${}_{\Delta_N}$}
\put(3.5,2.9){${}_{\sf Possible\ marginal\ stability\ walls}$}
\put(3.5,2.6){${}_{\sf between\ LCS\ and\ field\ 
theory\ region}$}
\put(3.4,2.75){\vector(-1,0){1.8}}
\end{picture}
\caption{Schematic representation of the complex structure moduli 
space for general $N\geq 2$.}
\label{fig:moduli}
\end{figure}
\bigskip

\noindent
In principle there could exist marginal stability walls between 
the LCS limit point and the field theory 
scaling region as sketched in Figure \ref{fig:moduli}. 
Therefore a correspondence between large radius BPS states
and gauge theory BPS states is not expected on general grounds. 
We conjecture that for all charges 
$\gamma \in \Gamma$ which support BPS states of finite mass 
in the field theory limit it is possible to choose a path connecting 
the two regions in the moduli space which avoids all such walls. 
This implies a one-to-one correspondence between BPS states 
in these two limits, which was first observed for $SU(2)$ gauge theory 
in \cite{DGS}.

Section \ref{threetwo} contains a precise mathematical formulation 
of this conjecture employing the notion of {\it limit weak coupling} BPS spectrum. Intuitively,  the limit spectrum should be thought of as an
extreme weak coupling limit of the BPS spectrum where all instanton
and subleading polynomial corrections to the $\CN=2$ prepotential are turned off. Then the absence of walls conjecture implies that 
the limit weak coupling spectrum is identified with a certain limit of 
the large radius BPS spectrum. 
As a first test of this conjecture we next show that all large
radius supersymmetric D-branes in this limit, with charges in the gauge
theory lattice $\Gamma\simeq K^0(\CG)$,
actually belong to the triangulated subcategory $\CG$.
This is a nontrivial result, and an important categorical test of the
field theory limit of Calabi-Yau compactifications. 

In order to carry out further tests, the 
large radius BPS spectrum of $SU(3)$ theory 
is then investigated in Section \ref{threefive}. The geometrical setup 
 determines a Cartan subalgebra of $SU(3)$ together with a set of 
simple roots $\{\alpha_1,\alpha_2\}$. 
We determine the degeneracy of states with magnetic charge 
$\alpha_1+\alpha_2$. The results 
show that one can find BPS states with arbitrarily high spin at weak coupling. 

Section \ref{sectionfour} presents some exact weak coupling results
for BPS degeneracies in  $SU(3)$ gauge theories with magnetic
changes $(1,m)$ with $m\geq 1$. Explicit formulas are derived both
for $m=1$ by a direct
analysis of the moduli spaces of stable quiver representations.
 It is also shown
that for any $m\geq 1$ the BPS degeneracies vanish in a specific chamber in the moduli space of stability conditions. This yields exact
results by wallcrossing, explicit formulas being written only for $m=2$.
It should be noted at this point that the above results are not in 
agreement  with those
obtained in \cite{Fraser:1996pw} by monodromy arguments. 
The weak coupling spectrum found in \cite{Fraser:1996pw} is only a 
subset of the BPS states found here by quiver computations. In addition,
 it is explicitly
shown that there exist BPS
states of arbitrarily high spin and infinitely many marginal stability walls
at weak coupling. This is also in agreement with the semiclassical analysis 
of \cite{Gauntlett:1999vc,semiclassical} based on counting zero
modes of a Dirac operator on the  monopole moduli space.
Finally, these results are shown to be in agreement with their 
large radius counterparts in  Section \ref{nowallsSUthree}, confirming the
predictions of the absence of walls conjecture. 

Section \ref{strongsection} exhibits a strong coupling chamber
for $SU(N)$ gauge theories where
the BPS spectrum is in agreement with previous results
\cite{Alim:2011kw,spectral_networks}. In contrast with loc. cit., here
this chamber is obtained by a direct analysis of the spectrum of stable
quiver representations. As a corollary, a {\it deceptive} adjacent
chamber is found in Section \ref{deceptivesect} where the BPS spectrum
coincides with the one generated in \cite{Fraser:1996pw} by
monodromy transformations. However, the disposition of the
central charges in the complex plane shows that the deceptive
chamber cannot be a weak coupling chamber, hence justifying its name.

Building on the geometric methods developed so far, framed
quiver models are constructed in
Section \ref{sectionfive} for framed BPS states corresponding to
simple magnetic line defects. From a geometric point of view,
such line defects are engineered by D4-branes wrapping
smooth noncompact divisors in the toric threefold $X_N$.
This framework leads to a rigorous
mathematical construction of such states in terms of weak stability
conditions\footnote{The meaning of ``weak stability conditions '' is explained in \cite{generating}.}  for framed quiver representations depending on an
extra real parameter $\delta$ related to the phase of the line defect
\cite{JM,framedBPS,NH,N-derived}. The wallcrossing
theory of \cite{wallcrossing} is shown to be applicable to such
situations, resulting in a mathematical derivation of the framed wallcrossing formula of \cite{framedBPS}. Moreover, in Section
\ref{recursionsect}, a detailed
analysis of the chamber structure on the $\delta$-line leads to a
complete recursive algorithm, determining the BPS spectrum at
any point on the Coulomb branch in terms of the noncommutative
Donaldson-Thomas invariants defined in \cite{szendroi-noncomm}.
It should be emphasized that this argument solely relies on
wallcrossing on the $\delta$-line, and is therefore valid at any fixed point on the Coulomb branch where this particular quiver description is
valid. As an application, we show 
in Section \ref{exotics} that 
the recursion formula 
implies the absence of exotics
conjecture for framed and unframed BPS states
first articulated in \cite{framedBPS}. 
%The argument assumes 
%without proof 
%the existence of spin $SL(2,\IC)$ 
%action on spaces of gauge theory BPS states, 
%as well as all foundational aspects of motivic Donaldson-Thomas %invariants stated in \cite{wallcrossing}. 

Note that rigorous positivity results are obtained in a similar context 
in \cite{DMSS} by proving a purity result for the cohomology of 
the sheaf of vanishing cycles. It is interesting to note that the 
the technical conditions used in \cite{DMSS} are not in general 
satisfied in gauge theory examples. Hence we are led to conjecture that such positivity results will hold under more 
general conditions, not yet understood from a mathematical 
point of view.

Finally, Section \ref{sectionsix} addresses the same issues from the
perspective of cohomological Hall algebras, introduced by Kontsevich and
Soibelman in \cite{COHA} as well as their framed stability conditions
introduced in \cite{berkeley}. A geometric construction is outlined
in this context for the action of the spin $SL(2,\IC)$
group on the space of BPS states. Moreover, absence of exotics is conjectured to follow in this
formalism from a hypothetical
Atiyah-Bott fixed point theorem for
the cohomology with rapid decay at infinity defined in \cite{COHA}.

\subsection{A (short) summary for
 mathematicians}\label{summary}
In this section we summarize the main results of this work 
for a mathematical audience. 
Recent physics results on BPS states \cite{GMN,GMNII,CV-I,framedBPS,CNV,Alim:2011ae,
 Alim:2011kw,
Chen:2011gk,spectral_networks,Cecotti:2012va,Cecotti:2012sf,Cecotti:2012gh,Xie:2012gd} point towards a general 
conjectural correspondence assigning to an $\CN=2$ supersymmetric gauge theory 
\begin{itemize}
\item[$(i)$] a triangulated CY3 category $\CG$, and 
\item[$(ii)$] a map $\varrho: \CC \to {\rm Stab}(\CG)$ from 
the universal cover $\CC$ of the gauge theory Coulomb branch 
to the moduli space of Bridgeland stability conditions 
on $\CG$.
\end{itemize}
The central claim is then:
\bigskip

{\it $(G.1)$ The BPS spectrum of the gauge theory at any point 
$a\in \CC$ is determined by the motivic Donaldson-Thomas invariants 
\cite{wallcrossing} of $\varrho(a)$-semistable objects of $\CC$.}
\bigskip

Since supersymmetric quantum field theories do not admit a 
rigorous mathematical construction, a natural question is 
whether the above correspondence can be converted into a 
rigorous mathematical statement. One answer to this question 
is presented in \cite{quadratic, quivers-Fukaya,KoSo} (building on the main ideas of 
\cite{GMNII}.) 
The present paper proposes a different approach 
to this problem based instead 
on geometric engineering of gauge theories 
\cite{Aspinwall:1995xy,Kachru:1995wm,Klemm:1996bj,geom_eng,KMV}. As explained in Subsection \ref{BPSmirror} below, geometric engineering 
and the construction of \cite{quadratic,quivers-Fukaya,KoSo} 
are related 
by mirror symmetry, modulo certain subtle issues concerning 
the field theory limit. 

Very briefly, geometric engineering 
is a physics construction assigning an $\CN=2$ gauge theory 
to a certain toric  Calabi-Yau threefold with singularities. 
It is not known  whether any gauge theory 
can be obtained this way, but  a large class of such theories 
admit such a geometric construction. 
For example $SU(N)$
 gauge theories with $N_f \leq 2N$ 
fundamental hypermultiplets and quiver gauge theories with 
gauge group $\prod_i SU(N_i)$ belong to this class, as shown in 
\cite{KMV}.

 Accepting geometric engineering as a black box, 
the  present paper identifies the category $\CG$ with a triangulated 
subcategory of the derived category $D^b(X)$. This identification 
is based on a presentation of $D^b(X)$ in terms of an exceptional 
collection of line bundles $\{\CL_\alpha\}$
\cite{branes_toric,dimer_special, tiltingII}. Any such 
collection determines a dual collection of objects $\{P_\alpha\}$ 
of $D^b(X)$ such that ${\rm RHom}_X(\CL_\alpha,P_\beta) 
={\underline \IC} \delta_{\alpha, \beta}$. These are usually called 
fractional branes in the physics literature. Then the 
conjecture proposed in this paper is:
\bigskip

{\it $(G.2)$ There exists a subset $\{\CL_{\alpha'}\}\subset \{\CL_\alpha\}$ 
such that the gauge theory category 
$\CG$ is the triangulated subcategory 
of $D^b(X)$ generated by the fractional branes $\{P_{\beta'}\}$ 
satisfying ${\rm RHom}_X(\CL_{\alpha'},P_{\beta'})=0$. }
\bigskip

For illustration, this is explicitly shown in Sections \ref{twoone} and \ref{twotwo} 
for pure $SU(N)$ gauge theory of arbitrary rank. More general  models can be treated analogously, explicit statements being 
left for future work. 

Granting the above statement, the results of 
\cite{tilting,rep_assoc_coh,morita_derived} further 
identify $\CG$ with a category of twisted complexes 
of modules over the path algebra of a quiver with potential 
$(Q,W)$. Moreover, a detailed analysis of geometric engineering 
as in Section \ref{twotwo} further yields an assignment of 
central charges $z_{\beta'}:\CC\to \IC$ to the objects 
$\{P_{\beta'}\}$. 
 Therefore one obtains a well defined stability condition in 
 ${\rm Stab}(\CG)$ for any point $a\in \CC$ where 
 the images $z_{\beta'}(a)$ belong to a half-plane $\IH_\phi$. 
 This defines a map $\varrho_{(Q,W)}:\CC_{(Q,W)} 
 \to {\rm Stab}(\CG)$ over a certain subspace 
 $\CC_{(Q,W)}\subset \CC$. We further conjecture that, using mutations,  
 one can extend this map 
 to a map $\varrho:\CC \to 
 {\rm Stab}(\CG)$, and moreover the image of $\varrho$ is 
 contained in the subspace of algebraic (or quivery) stability 
 conditions in the terminology of \cite{t_local, stabnoncompact, locptwo}. 

The above construction also leads to a mathematical model for 
framed BPS states of simple magnetic line defects \cite{framedBPS} in terms of moduli spaces of {\it weakly 
stable} framed 
quiver representations. This is explained in Section \ref{sectionfive}. 

In this framework, one is naturally led to a series of 
conjectures, or at least questions of mathematical interest.
First note that four dimensional Lorentz invariance predicts 
the existence of a Lefschetz type $SL(2,\IC)_{spin}$-action 
on the cohomology of the sheaf of vanishing cycles 
of the potential $W$ on moduli spaces of 
stable quiver representations. In addition there is a 
second $SL(2,\IC)_R$-action, encoding the
$R$-symmetry of the gauge theory. The action 
of the maximal torus $\IC^\times_R\subset SL(2,\IC)_R$
is determined by 
the Hodge structure on the above cohomology groups, 
as explained in Section \ref{twofour}. 

Assuming the existence of the above actions 
a series of positivity conjectures are 
formulated in  \cite{framedBPS}, and reviewed 
in Section \ref{twofour}. 
The strongest of these conjectures
claims that the $\IC^\times_R$-action is 
trivial, and the virtual Poincar\'e polynomial of the vanishing 
cycle cohomology 
decomposes into a sum of irreducible 
$SL(2,\IC)_{spin}$ {\it integral spin} 
characters with positive integral coefficients. 
This is called the {\it no exotics conjecture}. 

Granting the existence of the $SL(2,\IC)_{spin}$-action, 
in order to prove the no exotics conjecture 
it suffices to prove that all refined DT invariants 
belong to the subring generated by $(xy)^{1/2}, (xy)^{-1/2}$. 
This follows from the integrality result proven in \cite{COHA}.
Here we provide an alternative proof for pure $SU(N)$ 
gauge theory in Section 
\ref{exotics} using a framed wallcrossing argument.
Furthermore, as explained in the last paragraph 
of Section \ref{exotics}, physical arguments suggest
that the no exotics conjecture should hold for 
refined DT invariants of toric Calabi-Yau threefolds in 
general. Again four dimensional 
Lorentz invariance predicts a Lefschetz type action 
on the moduli space of stable quiver representations. 
Moreover, there is also a $\IC_R^\times$-action 
\cite{DM-crossing} corresponding to an $U(1)_R$-symmetry. 
Combining all these statements, one is led to claim that 
a no exotics result will hold for toric Calabi-Yau threefolds, 
if one can prove that the motivic DT invariants belong 
to the subring generated by $\IL^{1/2}, \IL^{-1/2}$, 
as conjectured in \cite{wallcrossing}. 
For DT invariants defined in terms of 
algebraic stability conditions, this follows from the 
results of \cite{COHA}. For geometric stability conditions, 
this follows from the results of \cite{COHA}
and the motivic wallcrossing formula 
\cite{wallcrossing,COHA}. Explicit computations in some 
examples have been carried out in 
\cite{motivic-conifold,motivic-crepant,CKK}.

It is important to note that some cases of the no 
exotics conjecture 
are proven in \cite{quant-cluster,DMSS} via purity results for the 
vanishing cycle cohomology. However, the proof relies 
on certain technical assumptions -- such as compactness of 
the moduli space in \cite{DMSS} -- which are not generically satisfied for gauge theory quivers. Physics arguments
predict that similar results should hold in a much larger class 
of examples of quivers with potential, although the mathematical 
reason for that is rather mysterious.

Finally, note that the above conjectures are formulated 
in the language of cohomological Hall algebras \cite{COHA} 
in Section \ref{sectionsix}. In particular a series of 
conjectures of \cite{berkeley} are generalized to 
moduli spaces of weakly stable
framed quiver representations.

In addition, geometric engineering also suggests an
{\it absence of walls} conjecture
stating an equivalence between 
refined DT invariants of 
large radius limit stable objects of $D^b(X)$ 
and refined DT invariants of gauge theory 
quiver representations. The precise statement 
requires some preparation and is given in Section 
\ref{threetwo}. As explained there it claims 
the existence of special paths in the complex 
K\"ahler moduli space of $X$ avoiding certain 
marginal stability walls. 
\bigskip

\subsection{BPS categories and mirror 
symmetry}\label{BPSmirror}

For completeness, we explain here
a general framework emerging from string theory dualities, which ties together  geometric engineering, $\CN = 2$ theories of 
class S, and the constructions
of \cite{quadratic,quivers-Fukaya,KoSo}. Our treatment will be rather sketchy with the details and is highly
conjectural. Our purpose here is merely to give a bird's eye framework for relating several different approaches
to the BPS spectrum of $\CN = 2$ theories. 

We will restrict ourselves to the gauge theories of class S
introduced in \cite{Witten:1997sc,Gaiotto:2009we,GMNII}.
These are in one-to-one correspondence with the following data 
\begin{itemize}
\item a compact Riemann surface $C$ with a collection of marked 
points $\{p_i\}$ 
\item a Hitchin system with gauge group $G$ on $C$ with prescribed singularities at the marked points $\{p_i\}$. 
\end{itemize} 
Let $\CH$ denote the total space of the Hitchin system 
and $\pi: \CH\to \CB$ the Hitchin map. The target $\CB$ 
of the Hitchin map is an affine linear space and the 
fibers of $\pi$ are Prym varieties. We will denote by $\Delta 
\subset \CB$ the discriminant of the map $\pi$.

The connection with M-theory is based on the 
spectral cover construction of the Hitchin system. 
Let 
$D=\sum_i {p_i}$ denote the divisor of marked points 
on $C$, and $S_D$ the total space of the line bundle
$K_C(D)$ on $C$. Let also $S=S_D \setminus 
\cup_{i}K_C(D)_{p_i}$ be the complement  
of the union of fibers  of $K_C(D)$ at the marked 
points. Note that $S$ is isomorphic to the complement 
of the union of fibers $T_{p_i}^*C$ in the total space 
of the cotangent bundle $T^*C$. In particular $S$ is 
naturally a holomorphic symplectic surface. 

If the Hitchin system has simple regular singularities at 
the marked points, the total space $\CH$ 
is identified with a moduli space of 
pairs $({\bar \Sigma}, {\bar F})$ where ${\bar \Sigma} \subset S_D$ is
a compact effective 
divisor in $S_D$ and ${\bar F}$ a torsion free
sheaf on $\Sigma$. 
At generic points in the moduli space ${\bar \Sigma}$ is reduced 
and irreducible and ${\bar F}$ is a rank one torsion free sheaf. 
For physics reasons, it is more 
convenient to think of the data $({\bar \Sigma},{\bar F})$ as a 
non-compact curve  $\Sigma \subset S$ 
and a torsion free sheaf $F$ on $\Sigma$  
with prescribed behavior at ``infinity'' i.e. at the 
points of intersection with the fibers $K_C(D)_{p_i}\subset 
S_D$. In the following we will assume such a spectral cover 
construction to hold even if the Hitchin system has irregular singularities. 

The holomorphic symplectic surface $S$ can be used 
to construct an M-theory background $\IR^{3,1}\times S
 \times \IR^3$. The data $(\Sigma, F)$ determines a supersymmetric 
M five-brane configuration with world-volume 
of the form $\IR^{3,1}\times \Sigma$.
Now the connection with \cite{quadratic,quivers-Fukaya,KoSo} can be explained 
employing M-theory/IIB duality. Suppose two out of the 
three transverse directions are compactified on a rectangular 
torus such that the M-theory background becomes 
$\IR^{3,1} \times S \times S^1_M \times S^1_A \times \IR$. 
Then a standard chain of string dualities shows that such a 
configuration is dual to a IIB background on 
a Calabi-Yau threefold $Y$. 

The construction of $Y$ for Hitchin 
systems with no singularities, i.e. no marked points $p_i$ 
has been carried out in \cite{jacobians}. More precisely,
according to \cite{jacobians},
 any Hitchin system $\CH\to \CB$ 
of ADE type determines naturally a family $\CY\to \CB$ 
of Calabi-Yau threefolds such that 
\begin{itemize}
\item For any $b\in \CB\setminus \Delta$, 
$Y_b$ is smooth and isomorphic to the total space of a conic bundle over 
the holomorphic symplectic surface $S$ with discriminant $\Sigma$. 
\item For any point $b\in \CB \setminus \Delta$ 
the intermediate Jacobian $J(Y_b)$ is isogenous to the 
Prym $\pi^{-1}(b)$. 
\end{itemize}
The family is defined over the entire base $\CB$, and $Y_b$ 
is isomorphic to the total space of a singular conic bundle 
over $S$ at points 
$b\in \Delta$. Furthermore note that by construction 
all fibers $Y_b$, $b\in \CB\setminus \Delta$ are equipped 
with a natural symplectic structure. 

The duality argument sketched above leads to the conjecture 
that one can construct a family $\CY\to \CB$ with analogous properties for Hitchin systems $\CH\to \CB$ with prescribed 
singularities at marked points. Since string duality preserves the spectrum of BPS 
states, one is further led to the following conjecture, 
which provides a string theoretic framework for the constructions 
of \cite{quadratic,quivers-Fukaya,KoSo}.
\bigskip 

{\it $(F.1)$ For any $b\in \CB \setminus \Delta$, let 
$\CF(Y_b)$ be the Fukaya category of $Y_b$ 
generated by compact lagrangian cycles. Let 
${\widetilde \CB}$  denote the universal cover 
of $ \CB \setminus \Delta$. Then for any point  
${\tilde b} \in {\widetilde \CB}$ over  
$b\in \CB \setminus \Delta$ 
there is a unique point $\sigma_{\tilde b} \in 
{\rm Stab}(\CF(Y_b))$ in the moduli space of 
Bridgeland stability conditions on $\CF(Y_b)$ such that 
the gauge theory BPS spectrum at the point ${\tilde b}$ 
is determined by the motivic DT invariants of moduli 
spaces of $\sigma_{\tilde b}$-semistable objects in 
$\CF(Y_b)$. 

Furthermore, there is a natural equivalence of triangulated 
$A_\infty$-categories of all 
categories $\CF(Y_b)$, $b\in \CB\setminus \Delta$ 
with a fixed triangulated $A_\infty$-category $\CF$. Hence one obtains 
a map $\varrho: {\widetilde \CB} \to {\rm {Stab}}(\CF)$ 
as predicted in the first paragraph of Section 
\ref{summary}, with $\CG\simeq \CF$. }
\bigskip

The construction of the family $\CY\to \CB$ 
was carried out in \cite{KoSo},
where the case of arbitrary irregular singularities was considered.
Loc. cit. generalizes the results of \cite{jacobians} 
to a wide class of non-compact Calabi-Yau threefolds. 
It also gives a mathematically precise meaning to Conjecture 
$(F.1)$ above and relates the DT-invariants of Fukaya categories 
from Conjecture $(F.1)$ to the geometry of the corresponding 
Hitchin integrable system.

In order to explain the relation with the geometric engineering
of the present paper, 
recall that any toric Calabi-Yau threefold $X$ is related by local 
mirror symmetry 
\cite{HV,Morrison:1995yh} to a family $\CZ$
of non-compact Calabi-Yau threefolds. As explained in more detail in Section 
\ref{localmirror}, 
the mirror family is a family of  hypersurfaces 
of the form 
\[
P_\alpha(v,w)=xy,
\]
where $(v,w, x,y)\in (\IC^\times)^2 \times \IC^2$, and 
$P_\alpha$ is a polynomial function depending on some complex 
parameters $\alpha$. 
Each such hypersurface is a 
conic bundle $Z_\alpha$ over $  (\IC^\times)^2 $ with 
discriminant $P_\alpha(z,w)=0$. Homological mirror symmetry 
predicts an equivalence of triangulated $A_\infty$-categories 
\be\label{eq:hommirror}
D^b(X) \simeq {\rm Fuk}(Z_\alpha)
\ee
for any smooth $Z_\alpha$ in the family, 
where ${\rm Fuk}(Z_\alpha)$ is the Fukaya category of 
$Z_\alpha$.

In local mirror variables, the field theory limit is presented as a degeneration of the family $\CZ$. Referring the reader to 
Section \ref{localmirror} for more details, the parameters 
$\alpha$ are written in the form $\alpha=\alpha(u, \epsilon)$ for another set 
of parameters $u$ to be identified with the Coulomb branch 
variables of the field theory, $u\in \CB$.
 Then one takes the limit $\epsilon \to 0$ 
obtaining a family of threefolds $\CZ_0$ over a parameter 
space $\CB$. Note that this degeneration has been studied 
explicitly in the physics literature \cite{geom_eng,KMV},
but some geometric aspects would deserve a more detailed 
analysis. To conclude, string duality arguments predict the following conjecture: 

{\it $(F.2)$ 
The limit family $\CZ_0\to \CB$ is the same as the family 
of threefolds $\CY\to \CB$ in $(F.1)$.  
Moreover the equivalence \eqref{eq:hommirror} 
restricts to an equivalence 
\be\label{eq:subequiv}
\CG\simeq \CF,\ee
 where $\CG\subset D^b(X)$ is the category 
 defined in $(G.2)$}

Note that this conjecture predicts an interesting class of 
examples of homological mirror symmetry. 
The category 
$\CG$ is defined algebraically as the subcategory of $D^b(X)$
spanned by a subset of fractional branes, while 
$\CF$ is obtained from ${\rm Fuk}(Z_\alpha)$ by degeneration. 
Hence it is natural to ask whether the category 
$\CG$ can be obtained directly by constructing the mirror 
of the threefold  family $\CY\to \CB$. 
\bigskip

{\it Acknowledgements.} 
We are very grateful to Paul Aspinwall for collaboration 
and very helpful discussions 
at an early stage of the project. 
We thank Arend Bayer, Tom Bridgeland, Clay Cordova, Davide 
Gaiotto, Dmitry Galakhov, Zheng Hua, 
Amir Kashani-Poor, Albrecht Klemm, Maxim Kontsevich,
 Pietro Longi, Davesh Maulik,
Andy Neitzke, Andy Royston, and  Balazs Szendr{\"o}i
 for very helpful discussions and
correspondence. 
W.Y.C. was supported by NSC grant 101-2628-M-002-003-MY4 and 
a fellowship from the Kenda Foundation.
D.E.D was partially supported by NSF grant PHY-0854757-2009.
J.M. thanks the Junior Program of the Hausdorff Research Institute
for hospitality.
G.M. was partially supported by DOE grant DE-FG02-96ER40959 and by a grant from the Simons Foundation 
($\sharp$ 227381). D.E.D also thanks Max Planck Institute, Bonn, 
the Simons Center for Geometry and Physics, and 
the Mathematics Department of National Taiwan University for hospitality 
during the completion of this work.
GM also gratefully acknowledges
partial support from the Institute for Advanced Study and the Ambrose Monell Foundation. The research of Y.S. was partially supported by NSF
grant. He thanks IHES for excellent research conditions.

\section{Geometric engineering, exceptional collections,
and quivers}\label{sectiontwo}

This section contains a detailed construction of a discrete family
$X_N$, $N\geq 2$ of toric Calabi-Yau threefolds employed in geometric
engineering \cite{Aspinwall:1995xy,Kachru:1995wm,Klemm:1996bj,geom_eng,KMV}
of pure $SU(N)$ gauge theories with eight supercharges.
Physical aspects of this correspondence will be discussed in
Section \ref{twotwo}.

Let $Y_a$ be the total space of the rank two bundle
$\CO_{\IP^1}(a)\oplus \CO_{\IP^1}(-2-a)$
over $\IP^1$, where $a\in \IZ$. For any $N\in \IZ$, $N\geq 2$,
there is a fiberwise $\IZ_N$-action on $Y_a$ with weights
$\pm 1$ on the two summands. The quotient $Y_a/\IZ_N$
is a singular toric threefold with a line of
quotient $A_N$ singularities which admits a smooth Calabi-Yau toric
resolution $X_N$.
 For concreteness, let $a=0$ in the following
 \footnote{Different values of $a$ will lead to different Calabi-Yau
threefolds, and the category of branes on these 3-folds will depend
nontrivially on $a$. It is expected, however, that the field theoretic
subcategories of interest in this paper will in fact be $a$-independent.
Whether this is really so is left to future investigation.}. 
Then $X_N$ is
defined by the toric data
\be\label{eq:tordataA}
\begin{array}{llllllllll}
 & x_0 & x_1 & x_2 & x_3 & \ldots & x_{N-1} & x_{N}
 & y_1 & y_2 \\
% &  &  &  &  &  &  &  &  &  \\
\IC^\times_{(1)}& 1 & -2 & 1 & 0 & \ldots & 0 & 0 & 0 & 0 \\
\IC^\times_{(2)} & 0 & 1 & -2 & 1 & \ldots & 0 & 0 & 0 & 0 \\
&  &  &  &  &  &  &  &  &  \\
 &    &     &    &    & \vdots &   &    &   &   \\
 \IC^\times_{(N-1)}&  0&0  & 0 &0   &\ldots   &  -2 & 1 &  0& 0 \\
\IC^\times_{(N)} & -2 & 0 & 0 & 0 & \ldots & 0 & 0 & 1 & 1 \\
 \end{array}
\ee
with disallowed locus
\be\label{eq:forbiddenA}
\bigcup_{\substack{0\leq i,j\leq N\\ |i-j|\geq 2}} \{x_i=x_j=0\}\cup \{y_1=y_2=0\}.
\ee
The toric fan of $X_N$ is the cone in $\IR^3$ over the
planar polytope in Fig. (2.a) embedded in the coordinate hyperplane
${\vec r}\cdot {\vec e}_3=1$.
\bigskip
  \begin{figure}[h]
  \setlength{\unitlength}{1mm}
\hspace{-100pt}
\begin{picture}(80,40)
\put(35,5){\line(0,1){30}}%\put(7.5,7.7){$_{D_4^-}$}
\put(20,5){\line(1,0){30}}%\put(31,31.5){$_{D_2^-}$}
\put(20,5){\line(1,2){15}}%\put(31,19){$_{D_1^-}$}
\put(50,5){\line(-1,2){15}}%\put(15,31.5){$_{D_3^-}$}
\put(20,5){\line(3,2){15}}
\put(20,5){\line(3,4){15}}
\put(50,5){\line(-3,2){15}}
\put(50,5){\line(-3,4){15}}
\put(34,4){$\bullet$}
\put(34,14){$\bullet$}
\put(34,24){$\bullet$}
\put(19,4){$\bullet$}
\put(49,4){$\bullet$}
\put(34,34){$\bullet$}
\put(32,0){(2.a)}
\put(85,5){\line(0,1){30}}%\put(7.5,7.7){$_{D_4^-}$}
\put(70,5){\line(1,0){30}}%\put(31,31.5){$_{D_2^-}$}
\put(70,5){\line(1,2){15}}%\put(31,19){$_{D_1^-}$}
\put(100,5){\line(-1,2){15}}%\put(15,31.5){$_{D_3^-}$}
%\put(70,5){\line(3,2){15}}
%\put(70,5){\line(3,4){15}}
%\put(100,5){\line(-3,2){15}}
%\put(100,5){\line(-3,4){15}}
\put(84,4){$\bullet$}
%\put(84,14){$\bullet$}
%\put(84,24){$\bullet$}
\put(69,4){$\bullet$}
\put(99,4){$\bullet$}
\put(84,34){$\bullet$}
\put(82,0){$(2.b)$}
\end{picture}
\caption{The toric polytope for $X_3$ and the singular threefold
$Y_0/\IZ_3$. The polytope (2.a) for $X_N$ is similar,
except it will contain $N-1$ inner
points on the vertical axis. }
%\label{torictrans}
\end{figure}
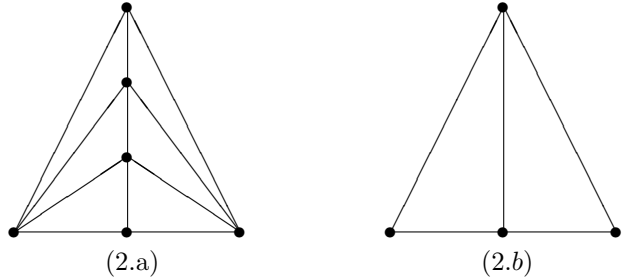

\bigskip

\noindent
Note that the toric data of the singular
threefold $Y_0/\IZ_N$ is the same, the disallowed locus being
\[
\{y_1=y_2\}=0.
\]
The toric fan of the singular threefold is represented
in Fig (1.b).

As expected, there is a natural toric projection $\pi:X_N \to \IP^1$,
its fibers being isomorphic to the canonical resolution of the two dimensional $A_N$ singularity.
The divisor class of the fiber is $H=(y_1)=(y_2)$.
The inner points of the polyhedron correspond to the $N-1$ irreducible
compact toric divisors $S_i\subset X_N$
determined by $x_i=0$, $i=1,\ldots,N-1$. Each of them is isomorphic to  a Hirzebruch surface,
$S_i\simeq \IF_{2i}$, $i=1,\ldots, N-1$. 

For completeness, we recall that a Hirzebruch surface $\IF_m$, 
$m\in \IZ$ is a holomorphic $\IP^1$-bundle over $\IP^1$. 
It has two canonical sections $\Sigma_-, \Sigma_+$ and the 
homology $H_2(\IF_m)$ is generated by $\Sigma_-, \Sigma_+,C$, where $C$ is the fiber class. The intersection form is 
\[ 
\Sigma_-^2 = -m, \qquad \Sigma_+^2 = m, \qquad 
\Sigma_\pm \cdot C =1, \qquad C^2 =0
\]
and there is a relation
\[
\Sigma_+=\Sigma_- + m C.
\]
The canonical bundle is 
\[ 
K_{\IF_m} = - c_1(\IF_m) = -\Sigma_- -\Sigma_+ -2C,
\]
and 
\[
\int_{\IF_m} c_2(\IF_m) =4. 
\] 

In addition $X_N$ contains two noncompact toric 
divisors $S_0,S_N$ determined by $x_0=0$ and
$x_N=0$ respectively. The first,
$S_0$ is isomorphic to $\IC\times \IP^1$
and the second, $S_N$, is isomorphic to the total space of the line bundle $\CO_{\IP^1}(-2N-2)$.
Note that
$S_{i}$ and $S_{i+1}$ intersect transversely along a $(2i, -2i-2)$ rational curve $\Sigma_{i}$,
$i=0,\ldots, N-1$,
which is a common section of both surfaces over $\IP^1$.
All other intersections
are empty.
Note also that the  equations
\[
y_1=0,\qquad x_i=0, \qquad i=1,\ldots, N
\]
determine a fiber $C_i$ in each divisor $S_i$,
 a compact rational curve for $i=1,\ldots, N-1$,
and a complex line for  $i=0,N$.
These curve classes satisfy the relations
\be\label{eq:curverelA}
\Sigma_{i}=\Sigma_{i-1}+2iC_i, \qquad i=1,\ldots, N-1,
\ee
which follow for example from \cite[Prop. 2.9. Ch. V]{har}. 

The rational Picard group of $X_N$ is generated by $N$
divisors classes $D_1,\ldots, D_{N-1},H$, one for each 
factor of the torus $(\IC^\times)^{N}$. This is so because for 
each $\IC^\times$ factor we can associate a canonical associated line bundle to the principal torus bundle over the quotient. From the weights of the action on homogeneous coordinates in 
\eqref{eq:tordataA} we see  that a section of $D_i$ can be taken to be $x_0^ix_1^{i-1} \cdots x_{i-1}$, $1\leq i\leq N-1$. 
The canonical  toric divisors are equivalent to a linear combination 
of the generators $D_1,\ldots, D_{N-1},H$ with coefficients 
determined by the columns of the charge matrix in \eqref{eq:tordataA}. In particular 
\be\label{eq:SandD}
S_i = -\sum_{j=1}^{N-1}{\sf C}_{i,j}D_j, \qquad i=1,\ldots, N-1, 
\qquad S_N = D_{N-1},
\ee
where ${\sf C}_{i,j}$ is the Cartan matrix of $SU(N)$
normalized to have $+2$ on the diagonal. 
One can obviously invert these relations, obtaining
$D_i = -\sum_{i=1}^{N-1}{\sf C}_{ij}^{-1}S_j$, $i=1,\ldots, N-1$, where
the coefficients ${\sf C}_{ij}^{-1}$ are fractional.
 Alternatively, relations \eqref{eq:SandD} can be 
 recursively inverted starting with $D_{N-1}=S_N$. 
 This yields the integral linear relations 
 \be\label{eq:linebundlesB}
D_i= \sum_{j=1}^{N-i} jS_{i+j}, \qquad i=1,\ldots,N-1,
\ee
which will be used in the construction of an exceptional
collection on $X_N$. Note that this equation involves 
$S_N$, hence is compatible with $D_i = -\sum_{i=1}^{N-1}{\sf C}_{ij}^{-1}S_j$.
Moreover note the following intersection numbers
\be\label{eq:intersmatrix}
(C_i\cdot D_j)_{X_N} = \delta_{ij},\qquad
 (C_i\cdot S_j)_{X_N} = -{\sf C}_{ij}\qquad i,j=1,\ldots, N-1.
\ee

For the construction of line defects in  Section \ref{fiveone} 
it is important to note  that each class $D_i$ contains
a smooth irreducible surface given by 
\be\label{eq:diveqA}
\big[a_{1,i} y_1^{2i} + a_{2,i}  y_2^{2i}\big]
\big(x_0^{i} x_1^{i-1}\cdots x_{i-1}\big) + b_i
x_{i+1}x_{i+2}^2\cdots
x_{N}^{N-i} =0,
\ee
with $a_{1,i}, a_{2,i}, b_i\in \IC$, $i=1,\ldots, N-1$, generic coefficients. This follows from the fact that the global holomorphic 
sections of the line bundles $\CO_X(S_i)$ are homogeneous 
polynomials in the toric coordinates $(x_0,\ldots, x_N, y_1,y_2)$ 
with $(\IC^\times)^N$ charge vector equal to the $x_i$ column 
of the charge matrix \eqref{eq:tordataA}. Then using 
equations \eqref{eq:linebundlesB} one computes the 
charges of the sections of $\CO_X(D_i)$, $1\leq i \leq N-1$. 
Smoothness follows from the observation that
the homogeneous toric coordinates
in equation \eqref{eq:diveqA} are naturally divided
into two groups, $(x_{k})_{1\leq k\leq i-1}$ and
$(x_l)_{i+1\leq l \leq N}$. According to equation
\eqref{eq:forbiddenA},
no two variables $x_k$, $x_l$ with
$1\leq k \leq i$ and $i+2\leq x_l\leq N$
are allowed to vanish simultaneously. Since $y_1,y_2$ are also not allowed to vanish simultaneously, a straightforward computation
shows that the divisors \eqref{eq:diveqA}
are smooth and irreducible for  generic coefficients
$a_{1,i}, a_{2,i}, b_i\in \IC$. Abusing notation, the same notation
will be used for the divisor classes $D_i$ and a generic smooth irreducible representative in each class. The distinction will be clear
from the context.

\subsection{Exceptional collections and
fractional branes}\label{twoone}

Adopting the definition of \cite{branes_toric}, a
 full strong exceptional collection of line bundles on a
toric threefold $X$ is
%\rem\ I am still bothered that this does not quite coincide with
%other definitions I have seen of this term \rem\
 a finite set $\{\CL_\alpha\}$ of line bundles
which generate $D^b(X)$ and satisfy
\[
{\rm Ext}^k_X(\CL_\alpha, \CL_\beta) =0
\]
for all $k>0$, and all $\alpha, \beta$.
Given such a collection the direct sum
$\CT =\oplus_\alpha \CL_\alpha$ is a tilting object in
the derived category $D^b(X)$ as defined in  \cite{tilting,rep_assoc_coh,morita_derived}.
Then the results of loc. cit.
imply that the functor
${\rm RHom}(\CT, \ \bullet\ )$ determines
an equivalence of the derived category
$D^b(X)$ with the derived category of modules over
the finitely generated algebra ${\rm End}_X(\CT)^{op}$.
%When $X$ is a toric Calabi-Yau
%threefold, this algebra is isomorphic to the path algebra of a quiver
%with potential.
%The
%quiver in question is furthermore identified with the ${\rm Ext}^1$-%quiver
%of a collection of compact objects of $D^b(X)$, called fractional %branes.
%\rem\ The above sentences are highly nontrivial statements! They %require
%a ref. Or a proof. Or at least an acknowledgement that even if they %are
%folklore, they are highly nontrivial. \rem\

Full strong exceptional collections of line bundles
on toric Calabi-Yau threefolds can be constructed
\cite{dimer_special, tiltingII}
using the dimer models introduced in  \cite{Franco:2005sm,Franco:2006gc,
Hanany:2006nm,
Hanany:2005ve}.
A different construction
for the threefolds $X_N$, $N\geq 2$,
exploiting the fibration structure $X_N\to \IP^1$
is presented in  Appendix
\ref{appA}. The resulting exceptional collection
consists of the line bundles
\be\label{eq:linebundlesA}
L_i = \CO_{X_N}(D_i), \qquad
M_i = \CO_{X_N}(D_i+H),\qquad
i=1,\ldots,N,
\ee
where $D_i$, $i=1,\ldots, N-1$ are the divisor classes
given in \eqref{eq:linebundlesB} and $D_N=0$. So 
$L_N=\CO_{X_N}$. 
Therefore there is an equivalence of derived categories
\be\label{eq:dercatequiv}
D^b(X_N) \simeq D^b({\rm End}(T)^{op}-{\rm mod}),
\qquad E\mapsto {\rm RHom}_{X_N}(T, E),
\ee
where
$T=\big(\oplus_{i=1}^N L_i\big) \oplus
\big(\oplus_{i=1}^N M_i\big)$,
and ${\rm End}(T)$ is the endomorphism algebra of $T$.
According to Appendix \ref{appA},
this algebra is isomorphic to the path algebra of the quiver
\eqref{eq:endalgB} with the quadratic relations given in equation\eqref{eq:Endrelations}. Reversing the arrows yields the periodic quiver
$\CQ$ below
\be\label{eq:fractbranesB}
\xymatrix{
& & & \vdots & \ar@/_0.2pc/[dl]|{r_N} & {}\quad 
\ar@/^0.2pc/[dll]|{s_N} & \vdots &  & \\
& & & q_N\bullet \ar@<.5ex>[rrr]|{c_N} \ar@<-.5ex>[rrr]|{d_N}
\ar@{-}[u]|{b_N} 
& & & \bullet p_N \ar@/_1pc/[ddlll]|{r_{N-1}}
\ar@/^1pc/[ddlll]|{s_{N-1}}  
\ar@{-}[u]|{a_N} & &
\\
& & & & &  &  & & \\
& & & q_{N-1} \bullet \ar@<.5ex>[rrr]|{c_{N-1}} 
\ar@<-.5ex>[rrr]|{d_{N-1}}
\ar[uu]|{b_{N-1}}
& & & \bullet p_{N-1} \ar[uu]|{a_{N-1}}  & & \\
& & & \vdots & & & \vdots & & \\
& & & q_{i+1}\bullet
 \ar@<.5ex>[rrr]|{c_{i+1}} \ar@<-.5ex>[rrr]|{d_{i+1}}
& &  & \bullet p_i \ar@/_1pc/[ddlll]|{r_i}
\ar@/^1pc/[ddlll]|{s_i} & & \\
& & & & & & & & \\
& & & q_i\bullet \ar@<.5ex>[rrr]|{c_i} 
\ar@<-.5ex>[rrr]|{d_i}
\ar[uu]|{b_i}
& & &  \bullet p_i \ar[uu]|{a_i}  & & \\
& & & \vdots && & \vdots & & \\
& & & q_2 \bullet \ar@<.5ex>[rrr]|{c_2} \ar@<-.5ex>[rrr]|{d_2}
& & & \bullet p_2 \ar@/_1pc/[ddlll]|{r_1}
\ar@/^1pc/[ddlll]|{s_1} & & \\
& & & & & & & & \\
& & & q_1\bullet \ar@<.5ex>[rrr]|{c_1} \ar@<-.5ex>[rrr]|{d_1} \ar[uu]|{b_1}
& & & \bullet p_1\ar[uu]|{a_1} 
\ar@/^0.2pc/@{-}[dl]|{s_N} \ar@/_0.2pc/@{-}[dll]|{r_N}
& & 
\\
& & &\vdots \ar[u]|{b_N} & & &\vdots \ar[u]|{a_N} & & \\
}
\ee
where the vertices $p_i,q_i$ correspond to the line bundles $L_i,M_i$, $i=1,\ldots, N$ respectively.
At the same time, the relations \eqref{eq:Endrelations} are
derived from the cubic potential
\be\label{eq:potentialA}
\bal
\CW= & \sum_{i=1}^{N-1} 
\left[r_i(a_ic_i-c_{i+1}b_i)+s_i(a_id_i-d_{i+1}b_i)
\right]\\
& \ \ \ \ \ \ + r_N(a_Nc_N- c_1b_N)+
s_N(a_Nd_N-d_1b_N).
\\
\eal
\ee
The resulting quiver with potential $(\CQ,\CW)$ has a dual interpretation \cite{quiv_superpotential, Herzog:2005sy, Herzog:2006bu},
 as the ${\rm Ext}^1$ quiver of a collection of
fractional branes $(P_i,Q_i)_{1\leq i\leq N}$. The latter are objects of $D^b(X_N)$ corresponding to the simple $(\CQ,\CW)$-modules associated to the vertices $(u_i,v_i)_{1\leq i\leq N}$
under the equivalence \eqref{eq:dercatequiv}. The simple 
module associated to a particular node is the representation 
consisting of a dimension 1 vector space assigned to the given 
node and trivial vector spaces otherwise.
They are  uniquely determined by the orthogonality conditions\footnote{Here ${\rm RHom}_{X_N}(\ ,\ )$ denotes the right derived functor of global ${\rm Hom}_{X_N}(\ ,\ )$, which assigns to a pair of sheaves $E,F$ 
the linear space of global sheaf morphisms $E\to F$. 
For any pair $(E,F)$, 
${\rm RHom}_{X_N}(E,F)$ is a finite complex of vector spaces whose cohomology groups are isomorphic to the global extension groups 
${\rm Ext}^k_{X_N}(E,F)$. See \cite{hom_alg} for abstract definition and properties.} 
\be\label{eq:onconditions}
\bal
& {\RHom}_{X_N}(L_i,P_j)=\delta_{i,j}{\underline \IC}, \qquad
& {\RHom}_{X_N}(L_i,Q_j)=0\\
& {\RHom}_{X_N}(M_i,P_j)=0, \qquad
& {\RHom}_{X_N}(M_i,Q_j)=\delta_{i,j}{\underline \IC},\\
\eal
\ee
$1\leq i,j\leq N$,
where ${\underline \IC}$ denotes the one term complex
of vector spaces with   $\IC$ in degree zero.
As shown in Appendix \ref{appA}, the following collection 
of objects satisfy conditions \eqref{eq:onconditions}.
\be\label{eq:fractbranesA}
\bal
& P_i=F_i[1], \qquad Q_i=F_i(-H)[2],\qquad i=1,\ldots, N-1, \\
& P_N = F_N, \qquad Q_N=F_N(-H)[1]
\eal
\ee
where
\be\label{eq:fractbranesAB}
\bal
& F_i=\CO_{S_i}(-\Sigma_{i-1}), \quad i=1,\ldots, N-1,\\
&
F_N = \CO_{S},\qquad S=\sum_{i=1}^{N-1}S_i.\eal
\ee
For future reference we note here that 
\be\label{eq:chPQA}
\bal 
& \ch_0(P_i) =0, \qquad \ch_1(P_i) = -S_i, \qquad
\ch_2(P_i) = - (i+1)C_i, \qquad 
\chi(P_i) = 0\\
& \ch_0(Q_i) =0, \qquad \ch_1(Q_i) = S_i, \qquad
\ch_2(Q_i) = i C_i, \qquad 
\chi(Q_i) = 0\\
\eal
\ee
for $1\leq i\leq N-1$, respectively
\be\label{eq:chPQB}
\bal 
& \ch_0(P_N) = 0,\qquad \ch_1(P_N) = S, \\
& \ch_2(P_N) = \Sigma_0 + \sum_{i=1}^N (i+1)C_i, 
\qquad 
\chi(P_N) = 1\\
& \ch_0(Q_N) =0, \qquad \ch_1(Q_N) = -S \\
& \ch_2(Q_N) = -\Sigma_0 -\sum_{i=1}^N i C_i, 
\qquad 
 \chi(Q_N) =-1,\\
\eal
\ee
where $S_i$, $1\leq i\leq N-1$, 
will also stand for a degree $2$ cohomology class via pushforward, and similarly for $C_i, \Sigma_0$. 
For completeness recall that the holomorphic Euler character 
$\chi(E)$ of an object $E$ of $D^b(X_N)$ with compact 
support is defined as 
\be\label{eq:chiderivedobj}
\chi(E) = \sum_{k\in \IZ} (-1)^k{\rm dim}\,
{\rm Hom}_{D^b(X_N)}(\CO_{X_N},E[k]).
\ee
Since $E$ is compactly supported and bounded, this is a 
finite sum and all vector spaces ${\rm Hom}_{D^b(X_N)}(\CO_{X_N},E[k])$, $k\in \IZ$, are finite dimensional.
For $E=F[p]$, with $F$ a sheaf with compact support and $p\in \IZ$, this definition agrees with the standard definition
of the holomorphic Euler character of $F$ up to sign, 
\be\label{eq:chisheaf}
\chi(E) = (-1)^p\chi(F) = (-1)^p \sum_{k\in \IZ} H^k(X_N,F).
\ee
Here $H^k(X_N,F)$ are the $\check{\rm C}$ech cohomology groups 
of $F$. Since $F$ has compact support on $X_N$, the 
$\check{\rm C}$ech cohomology groups 
are finite dimensional and vanish for $k<0$ and 
$k> {\rm dim}\, {\rm supp}(F)$. Furthermore note 
the Riemann-Roch formula 
\be\label{eq:chiRR}
\chi(F) = \int_{X_N} \ch(F) {\rm Td}(X_N).
\ee

The quiver $\CQ$ is then identified with the ${\rm Ext}^1$-quiver of the collection of fractional branes $(P_i,Q_i)_{1\leq i\leq N}$.
The nodes $p_i,q_i$ correspond to the objects $P_i,Q_i$,
$i=1,\ldots,N$ respectively while the arrows between any two
nodes are in one-to-one correspondence with basis elements 
of the ${\rm Ext}^1$-space between the associated objects. 
Moreover note that the equivalence \eqref{eq:dercatequiv} 
relates the objects $(P_i,Q_i)$ to the simple
quiver representations supported respectively at each of the nodes
$(p_i,q_i)$, $1\leq i\leq N$. In contrast, the line bundles 
$L_i,M_i$
are related to the projective modules canonically associated to the
nodes $p_i,q_i$ respectively.

The potential \eqref{eq:potentialA} is related to the $A_\infty$-structure on the triangulated subcategory $\CF\subset D^b(X_N)$ generated
by the fractional branes $(P_i,Q_i)_{1\leq i\leq N}$, as explained below.
Consider an object in this category
of the form
\[
\bigoplus_{i=1}^N (V_i\otimes P_i)\oplus \bigoplus_{i=1}^N (W_i \otimes Q_i),
\]
where $V_i,W_i$, $i=1,\ldots, N$
are finite dimensional vector spaces.
This object is identified by the equivalence \eqref{eq:dercatequiv} to a representation $\rho$ of $(\CQ,\CW)$
assigning the vector spaces $V_i,W_i$ to the nodes $p_i,q_i$,
$i=1,\ldots, N$ respectively, and the zero map to all arrows.
Physically, this is a collection of fractional branes on $X_N$.
The  space of open string zero modes between such a collection of fractional branes is isomorphic to the extension group
${\rm Ext}^1_{\CF}(\rho,\rho)$. The latter is in turn  isomorphic
to the linear space
\be\label{eq:linspace}
\bal
V_\rho= & \bigoplus_{i=1}^N {\rm Hom}(W_i,V_i)^{\oplus 2} \oplus 
\\ & \bigoplus_{i=1}^{N-1} ({\rm Hom}(V_i,V_{i+1})\oplus 
{\rm Hom}(W_i,W_{i+1}) \oplus {\rm Hom}(V_{i+1},W_i)^{\oplus 2}).\\
\eal
\ee
%where the sum is over all arrows of $\CQ$.
%Here $t({\sf a})$ denotes the tail of ${\sf a}$ and
%$h({\sf a})$ the head of ${\sf a}$ for each arrow ${\sf a}$.
%Then $V_{t({\sf a})}$ and $V_{h({\sf a})}$ are the vector
%spaces associated to the tail and the head of ${\sf a}$.

Using canonical projective
resolutions for simple modules as in Appendix \ref{extensions},
one can construct
a cyclic $A_\infty$ structure on $\CF$. The cyclic $A_\infty$ structure
determines in particular a holomorphic superpotential $\CW_\rho$
on the above extension space as explained in detail in \cite{AinftyCFT, Dbrane_superpotential,quiv_superpotential,algCS}. This is the tree level superpotential
in the effective gauge theory on the fractional D-brane configuration
with multiplicities ${\rm dim}(V_i),{\rm dim}(W_i)$ at the vertices 
of $\CQ$.
By analogy with \cite{quiv_superpotential, algCS}, it is
conjectured here that the superpotential $\CW_\rho$
is identified with the cubic function on $V_\rho$
determined by $\CW$.
This statement was proven in \cite{quiv_superpotential, algCS}
for local toric Fano surfaces, and it was explained to us 
by Zheng Hua
that the proof of \cite{algCS} based on projective 
resolutions will go through in our case as well. 
In the following it will be assumed that this is the case for the
fractional branes $(P_i,Q_i)_{1\leq i\leq N}$, omitting a 
rigorous proof. An independent physical argument 
will be 
given in Section \ref{orbifoldsection} below, which 
provides an orbifold construction of the exceptional 
collection $(L_i,M_i)$, $1\leq i \leq N$.

%A second conjectural statement will be assumed in this paper,
%asserting an equivalence of triangulated categories
%\be\label{eq:compactequiv}
%D^b_{cpt}(X_N) \simeq \CF
%\ee
%where $D^b_{cpt}(X_N)$ is the derived category of
%$X_N$ with compact support. Again for local Fano surfaces, the %analogous statement is a rigorous result due to
%\cite{sheaves_local}.

\subsection{Orbifold quivers}\label{orbifoldsection}
By construction $X_N$ is the resolution of the quotient 
$Y_2/ \IZ_N$, where $Y_2$ is the total space 
of the rank two bundle $\CO_{\IP^1}\oplus \CO_{\IP^1}(-2)$. 
Note that $Y_2\simeq \IC \times Z$, where 
$Z$ is the canonical resolution of the $\IC^2/\IZ_2$ 
quotient singularity. The $\IC^2/\IZ_2$ orbifold 
contains two fractional 
branes corresponding to the objects  
\[
G_1 = \CO_{C}, \qquad G_2 = \CO_{C}(-1)[1]
\] 
supported on the exceptional cycle $C\simeq \IP^1\subset Z$ 
of the resolution \cite{DM,fractbr}. 
According to \cite{DM}, the effective action  
of a configuration of fractional branes $G_1^{\oplus n_1}\oplus 
G_2^{\oplus n_2}$ is obtained by dimensional reduction
of a $\CN=2$ quiver gauge theory in four dimensions. 
The $\CN=1$ chiral multiplet content of this theory is 
encoded in the quiver diagram 
\bigskip 

\be\label{eq:Yfract}
\xymatrix{ 
\bullet \ar@(ul,dl)|{B} \ar@/^1pc/[rr]|{C} \ar@/^2pc/[rr]|{D} 
& & \bullet \ar@(ur,dr)|{A}
\ar@/^1pc/[ll]|{R} \ar@/^2pc/[ll]|{S}
}
\ee
and the superpotential is given by
\be\label{eq:Ysuperpot}
W = {\rm Tr}\, A (CR+DS) - {\rm Tr}\, B(RC +SD).
\ee

Now consider the orbifold $Y_2/\IZ_N$. Using the rules 
of \cite{DM}, for each D-brane $G_1$, $G_2$ in the 
covering theory, one obtains a collection of fractional 
branes  $(G_1, \rho_i)$, respectively $(G_2, \rho_i)$, 
$0\leq i\leq N-1$, where $\rho_i$ is the $i$-th canonical 
irreducible representation of $\IZ_N$. The representations 
$\rho_i$ encode the action of the orbifold group on the 
Chan-Paton line bundles of the fractional branes. 
At the same time, the orbifold group acts on the 
$N=1$ chiral superfields as 
\[ 
\bal 
& A \to e^{-2i\pi/N}A,  
\qquad C \to C, \qquad 
R\to e^{2i\pi/N} R\\
& B \to e^{-2i\pi/N}B,  
\qquad D\to D, \qquad 
S\to e^{2i\pi/N} S\\
\eal
\]
for $j=1,2$. 

The effective action for any collection 
\[
\bigoplus_{i=0}^{N-1} (G_1,\rho_i)^{\oplus d_i} 
\oplus 
\bigoplus_{i=0}^{N-1} (G_1,\rho_i)^{\oplus e_i}
\]
is obtained by projecting the quiver \eqref{eq:Yfract} 
and the superpotential \eqref{eq:Ysuperpot} 
onto orbifold invariant fields. This yields precisely 
a quiver of the form \eqref{eq:fractbranesB}, 
with a cubic superpotential of the form \eqref{eq:potentialA}. 
The fields $a_i,b_i,c_i,d_i,r_i,s_i$ in \eqref{eq:fractbranesB}
are the invariant components of $A,B,C,D,R,S$ respectively. 

The above construction can be set on firmer mathematical grounds using the results of \cite{BKR}. 
According to loc. cit., there is an equivalence of 
derived categories 
\be\label{eq:XYequiv}
D^b(X_N) \simeq D^b_{\IZ_N}(Y_2)
\ee
where $D^b_{\IZ_N}(Y_2)$ is the $\IZ_N$-equivariant derived 
category of $Y_2$. This equivalence is determined by a 
Fourier-Mukai functor given explicitly in \cite{BKR}. 
Since the objects $G_1,G_2$ are scheme theoretically supported 
on the exceptional cycle $C$, which is fixed by the 
$\IZ_N$ action, the pairs $(G_1,\rho_i)$, 
$(G_2,\rho_i)$ are naturally objects of $D^b_{\IZ_N}(Y_2)$. 
Therefore mathematically, one is led to the claim that the equivalence \eqref{eq:XYequiv} maps the fractional 
branes $(P_i,Q_i)$ to the objects $(G_1,\rho_i)$, 
$(G_2,\rho_i)$, $1\leq i\leq N$. In principle, one 
can employ the methods of \cite{KV} in a relative setting, 
but we will leave the details for future work. 

\subsection{Field theory limit {\bf A}}\label{twotwo}
This section is focused on physical aspects of geometric engineering,
explaining the relation between the toric Calabi-Yau threefolds
$X_N$ and pure $SU(N)$ gauge theory with eight supercharges.
More specifically, it will be explained in detail how the
the rigid special geometry of the Coulomb branch is obtained as a
scaling limit
 of the special geometry of the complex K\"ahler moduli space
of $X_N$. This limit is usually referred to as the field theory limit,
and can be formulated either in terms of the  local mirror
{\bf B}-model \cite{geom_eng,KMV}, or directly in terms
of the large radius prepotential of $X_N$
\cite{Iqbal:2003ix, Iqbal:2003zz,Eguchi:2003sj,Konishi:2003qq,LLZ}.
In the first case one obtains
the family of Seiberg-Witten curves as a scaling limit of a family of curves encoding the  mirror {\bf B}-models.
In the second, the semiclassical gauge theory prepotential is
obtained by taking a similar scaling limit of the large radius
limit prepotential, including genus zero world-sheet instanton corrections. The second approach will be employed below to derive
the central charges of the fractional branes
$(P_i,Q_i)_{1\leq i\leq N}$ in the field theory limit.

A convenient parameterization of the
K\"ahler cone is obtained observing that
the  Mori cone of $X_N$ is generated by the
curve classes $\Sigma_{0}, C_i$, $i=1,\ldots, N-1$.
Moreover, the vertical divisor class $H$ has intersection
numbers
\[
H\cdot \Sigma_{0}=1, \qquad H\cdot C_i =0, \qquad i=1,\ldots, N-1.
\]
Given the intersection numbers \eqref{eq:intersmatrix}, it follows that
the K\"ahler class of $X_N$ can be naturally written as
\be\label{eq:kahlerclass}
\omega =t_0H+\sum_{i=1}^{N-1} t_iD_i,
\ee
with $t_i>0$, $i=0,\ldots, N-1$. Obviously,
\[
\int_{\Sigma_{0}} \omega = t_0, \qquad
\int_{C_i}\omega
=t_i, \quad 1\leq i\leq N-1.
\]
The complexified K\"ahler class will be written similarly as
\[
B+\sqrt{-1}\omega = s_0H+\sum_{i=1}^{N-1} s_iD_i
\]
with $s_i=b_i+\sqrt{-1}t_i$, $i=0,\ldots, N-1$.

In the large radius limit $t_0,t_i>>1$, $i=1,\ldots, N-1$,
the special coordinates ${\widetilde s}_i$, $i=0,\ldots,N-1$.
 are related to $s_i$ by the mirror map,
\[
{\widetilde s}_i = s_i + {\rm exponentially\ small\  corrections\ at\ large\ radius}.
\]
They are also identified via
homological mirror symmetry with the central charges of a collection of 
$K$-theory classes
\be\label{eq:D2branes}
\Upsilon_0= [\CO_{\Sigma_{0}}(-1)],\qquad
\Upsilon_i = [\CO_{C_i}(-1)], \qquad i=1,\ldots, N-1
\ee
representing $D2$-branes supported by
the Mori cone generators. Note that we have chosen the Chan-Paton
bundles to have degree $(-1)$ in order for the total D0-charge, including
gravitational contributions, to be trivial.
The precise relation is
\[
{\widetilde s}_i = {1\over M_s}Z(\Upsilon_i)
\]
where $M_s$ is the string mass scale.  

The next task is to construct the effective action 
for normalizable IIA modes on $X_N$ and show that it 
reduces to known gauge theory results in the field theory limit.
A conceptual problem is that the $N=2$ prepotential is not 
intrinsically defined for local Calabi-Yau models. In principle, 
one has to 
find a suitable realization of the local model as a degeneration 
of a compact Calabi-Yau threefold, and obtain the $N=2$ prepotential
as a limit of the $N=2$ prepotential of the compact model. 
On general grounds the prepotential of the compact model 
has the form 
\[
\CF=\CF^{pert} + {\zeta(3)\chi\over 
(2\pi\sqrt{-1})^3} + \CF^{inst}
\]
where $\CF^{pert}$ is a perturbative polynomial part deduced from 
\eqref{eq:largeRZ}
and $\CF^{inst}$ encodes genus zero world-sheet instanton
effects. In contrast, the periods of the compact cycles associated 
to the $A_{N-1}$ degeneration are intrinsically defined in the local limit, 
as shown in detail below. So our strategy will be to analyze their 
behavior in the field theory limit and show that the 
finite periods in this limit are consistent with the Seiberg-Witten 
$SU(N)$ prepotential. 

The lattice of compact D-brane charges on $X_N$ is 
isomorphic to the compactly supported $K$-theory lattice of $X_N$,
$K^0_{cpt}(X_N)$. It is equipped with an antisymmetric pairing 
$$\langle\ ,\ \rangle : K^0_{cpt}(X_N)\times K^0_{cpt}(X_N)\to 
\IZ,$$ the restriction of the natural pairing 
\[
K^0(X_N) \times K^0_{cpt}(X_N) \to \IZ
\]
where $K^0(X_N)$ is the $K$-theory lattice of $X_N$ with 
no support condition. Note that $K^0(X_N)$ is generated 
as a ring by
the line bundles $[\CO_{X_N}(S_i)]$, $0\leq i\leq N$ and
 $\CO_{X_N}(H)$. Given a line bundle $\CL$ on $X_N$ and 
a sheaf $F$ with compact support, 
\[ 
\langle [\CL], [F]\rangle = \chi(\CL^{-1}\otimes_{X_N} F).
\]
where 
$\otimes_{X_N}$ denotes the tensor product of $\CO_{X_N}$-modules. This notation will be frequently 
 used throughout this paper. 
Moreover, note the relation 
\[ 
[\CO_{X_N}(D)] = 1- [\CO_D]
\]
for any effective divisor $D$ on $X_N$, where 
$1=[\CO_{X_N}]\in K^0(X_N)$. 
Therefore $K^0(X_N)$ is also generated as a ring by $1$ and 
the divisor classes $\CO_{S_i}$, $0\leq i\leq N$, $\CO_H$. 
Then the $K$-theory with compact support will be generated 
as a $\IZ$-module by $K$-theory classes of the form 
\[
\big[\otimes_{i=0}^N \CO_{S_i}^{\otimes k_i}
 \otimes \CO_H^{\otimes l}\big], \qquad k_i,l \in \IZ_{\geq 0}
\] 
which do not involve the generator $1$. This expression 
can be simplified using the defining equations 
$x_i=0$ for the $S_i$, $0\leq i\leq N$, respectively $y_1=0$
for $H$. For example it follows that 
$\CO_{S_i} \otimes \CO_{S_j} =0 $, for $\vert i-j\vert >1$. 
Using such identities  it follows that $K^0_{cpt}(X_N)$-theory 
is generated as a $\IZ$-module by 
\[
\bal
& [\CO_p], \qquad  \Upsilon_0=[\CO_{\Sigma_0}(-1)], \qquad
\Upsilon_i = [\CO_{C_i}(-1)], \qquad 1\leq i\leq N-1\\
& \qquad \qquad \ \Lambda^i = [\CO_{S_i}(-\Sigma_{i-1}-(i+1)C_i)],
\qquad 
1\leq i\leq N-1.\\
\eal
\]
Here $\CO_p$ is the class of the skyscraper sheaf associated to
a point $p\in X_N$. 
Again we have chosen to twist the structure sheaves of the 
divisors $\CO_{S_i}$, $1\leq i\leq N-1$ by appropriate 
line bundles such that the D2-brane charge is zero. The nontrivial inner  products of the 
generators are
\[
\langle \Upsilon_i , \Lambda^j \rangle = -{\sf C}_{ij}, \qquad 1\leq i\leq N-1, 
\]
all other products being zero. In particular the pairing $\langle\ ,\ \rangle$ 
has a nontrivial annihilator generated by $[\CO_p], \Upsilon_0$. 

In order to use special geometry relations, note that there is an 
alternative set of rational generators
where the classes $\Lambda^i$ are 
replaced by the rational linear combinations 
\be\label{eq:DfourbranesA}
\Upsilon^i = -\sum_{j=1}^{N-1} {\sf C}^{-1}_{ij} \Lambda^j, \qquad 
1\leq i\leq N-1.
\ee
These generators satisfy the orthogonality relations
\be\label{eq:onconditionsG}
\langle \Upsilon_i,\Upsilon^j\rangle =\delta_i^j, \qquad i,j=1,\ldots, N-1.
\ee
The next goal is to study the behavior of the $\CN=2$ central charges 
of $\Upsilon^i$, $1\leq i\leq N-1$, in the field theory limit. 

The central charge $Z(\Upsilon)$ for a $K$-theory class $\Upsilon$ with compact support has the large radius
expansion \cite{anomaly_inflow}
\be\label{eq:largeRZ}
Z(\Upsilon) = -M_s\int_{X_N} e^{-({\widetilde s}_0H +
\sum_{i=1}^{N-1}{\widetilde s}_kD_k)} \ch(\Upsilon) \sqrt{{\rm Td}(X)}+Z^{inst}(\Upsilon)
\ee
where $Z^{inst}(\Upsilon)$ are exponentially small  genus zero 
world-sheet instanton 
corrections\footnote{This formula is often attributed to 
\cite{anomaly_inflow,Minasian:1997mm} and it is certainly closely related to the (correct!) results 
of those papers. However, when writing the central charge one should not 
forget (as some authors do) to include the correction to the prepotential 
proportional to $\zeta(3) \chi(X_N)$. This term affects only the D6-brane central charge not D4 and D2. Hence it is irrelevant here since the D6-brane 
is infinitely heavy in the local limit, and has no effect on the field theory dynamics.}.
Special geometry constraints imply that the instanton corrections 
to the central charges $Z(\Upsilon^i)$, $1\leq i\leq N-1$, 
are given by 
\[
Z^{inst}(\Upsilon^i) = M_s {\partial \CF_{X_N}^{inst}\over 
\partial {\widetilde s}_i}
\]
where $\CF^{inst}_{X_N}$ is the sum of the genus zero
world-sheet instanton 
corrections 
\be\label{eq:GWpotA}
\CF^{inst}_{X_N} = -{1\over (2\pi \sqrt{-1})^3}\sum_{\substack{
n_i\in \IZ_{\geq 0}\\
0\leq i\leq N-1\\}} \sum_{k\geq 1} {1\over k^3}
N(n_i) \prod_{i=0}^{N-1} {\rm exp}\big(
2\pi k\sqrt{-1} \sum_{i=0}^{N-1} n_i {\widetilde s}_i\big).
\ee
The coefficients $N(n_i)\in \IZ$ are virtual numbers of genus zero curves
in the homology class $n_0\Sigma_{0}+\sum_{i=1}^{N-1} n_iC_i$.
Although $X_N$ is noncompact, these numbers are intrinsically 
defined via counting curves
preserved by a torus action on $X_N$ which leaves the global holomorphic three-form invariant. Hence $N(n_i)$ are equivariant
genus zero
Gromov-Witten invariants which can be exactly computed using  local mirror symmetry \cite{geom_eng, KMV,HV}. For the purpose of geometric 
engineering note that there is a decomposition 
\[
\CF^{inst}_{X_N} = \CF^{vert}_{X_N} + \CF^{hv}_{X_N}
\]
where $\CF^{vert}_{X_N}$ is the contribution of the vertical 
curve classes i.e. terms with $n_0=0$ while $\CF^{hv}$ is the sum over 
mixed horizontal and vertical classes i.e. all terms with $n_0>0$. 

There is an explicit expression for the 
 vertical part of the instanton prepotential \cite{geom_eng,KMV},
written as a sum over the positive roots $\alpha\in \Delta_N^+$
 of $SU(N)$. Each positive root  
 \[
 \alpha =\sum_{i=1}^{N-1} n_i(\alpha) \alpha_i, \qquad n_i(\alpha)\in \IZ_{\geq 0} 
 \]
 determines a vertical curve class 
 \[ 
 C_\alpha = \sum_{i=1}^{N-1} n_i(\alpha) C_i
 \]
 where $\{\alpha_i\}$, $1\leq i\leq N-1$ is a set of simple roots. 
 The Gromov-Witten invariant of each curve class $C_\alpha$ is 
 \cite{geom_eng}
 \[
 N(C_\alpha) = -2 
 \]
 and there are no other vertical contributions except 
 for multicovers. Therefore  
 \be\label{eq:GWpotB}
\CF_{X_N}^{vert}=-{1\over 4\pi^3 \sqrt{-1}}
\sum_{\alpha \in \Delta_N^+}\sum_{k\geq 1}
{1\over k^3} {\rm exp}\big(2\pi
k\sqrt{-1}  \langle{\widetilde s},\alpha\rangle\big).
\ee
where 
\[
\langle {\widetilde s},\alpha\rangle =\sum_{i=1}^{N-1}
n_i(\alpha) {\widetilde s}_i.
\]
An exact expression for the  second term, $\CF_{X_N}^{hv}$ 
can be obtained either using local mirror symmetry or the topological vertex 
\cite{topvert}.

In order to compute the central charges 
$Z(\Upsilon^i)$, $1\leq i\leq N-1$
note that  
\be\label{eq:D4branesC}
\ch_0(\Upsilon^i)=0,\qquad
\ch_1(\Upsilon^i) =-\sum_{j=1}^{N-1}{\sf C}_{ij}^{-1}S_j,
\qquad \ch_2(\Upsilon^i) = 0,\qquad 
\chi(\Upsilon^i) = 0\ee
for $1\leq i\leq N-1$. 
Moreover the toric data \eqref{eq:tordataA} and
relations \eqref{eq:SandD} yield the following relations 
\be\label{eq:intrelations}
(H\cdot S_i\cdot S_j)_{X_N}= -{\sf C}_{ij},\qquad
(H\cdot D_i\cdot S_j)_{X_N}= \delta_{ij}, \qquad 
(H\cdot H \cdot S_i)=0
\ee
in the intersection ring of $X_N$. 
Finally, by adjunction,
\be\label{eq:adjunction}
\int_{S_i} c_2(X_N)|_{S_i} =\int_{S_i} (c_2(S_i)-c_1({S_i})^2) = 
12 \chi(\CO_{S_i})- 2 \int_{S_i} c_1(S_i)^2 = -4
\ee
for all compact divisors $S_i\simeq \IF_{2i}$, $1\leq i\leq N-1$.
Then using equations \eqref{eq:D4branesC}, \eqref{eq:intrelations}, 
\eqref{eq:adjunction} in \eqref{eq:largeRZ}, a
straightforward computation yields
\be\label{eq:DfourZ}
\bal
Z(\Upsilon^i) = & M_s\bigg[
\sum_{j=1}^{N-1}{\sf C}_{ij}^{-1} {\widetilde s}_0 {\widetilde s}_j
+{1\over 2} \sum_{j,k,l=1}^{N-1} {\widetilde s}_j {\widetilde s}_k
{\sf C}_{il}^{-1}(S_l\cdot D_j\cdot D_k)_{X_N} +{1\over 6}\sum_{j=1}^{N-1} {\sf C}^{-1}_{ij}\bigg] \\
& \ + Z^{inst}(\Upsilon^i) \\ \eal
\ee
for $1\leq i\leq N-1$. 

Following \cite{Iqbal:2003zz,Eguchi:2003sj,Konishi:2003qq,Iqbal:2003ix,LLZ} the field theory limit is the $\epsilon\to 0$ 
limit of the string theory, where 
\be\label{eq:fieldlimitA}
{\widetilde s}_0 = -{N\sqrt{-1}\over \pi}
\left(c_0+\ln\epsilon\right), \qquad {\widetilde s}_i =
{a_i\over M_0}\epsilon,\qquad M_s={M_0\over \epsilon}.
\ee
Here $M_0$ is an arbitrary scale, 
$c_0\in \IR_{>0}$ a fixed constant term, and $\epsilon\in \IR$, 
 $0<\epsilon < e^{-c_0}$, the scaling parameter. 
 A priori the large radius
instanton expansion \eqref{eq:GWpotA}
 might be divergent in this limit since the complex K\"ahler parameters 
 ${\widetilde s}_i$ are very small. 
It was however shown in \cite{Iqbal:2003zz,Eguchi:2003sj,Konishi:2003qq,Iqbal:2003ix,LLZ} that $\CF^{hv}_{X_N}$ 
has a finite limit 
as $\epsilon \to 0$, which agrees with the 
semiclassical instanton expansion of the gauge theory  with a 
QCD scale given by 
\be\label{eq:azero}
\Lambda^{2N-2} = 2^{2N}\bigg({M_0\over 2\pi}\bigg)^{2N-2} e^{2Nc_0}.
\ee
According to \cite{geom_eng,KMV}, the vertical instanton contributions
$\CF^{vert}_{X_N}$
are expected to yield the one loop correction 
to the gauge theory prepotential in the 
$\epsilon\to 0$ limit.  
This will be confirmed below by a detailed analysis 
of the $\epsilon \to 0$ limit of the central charges 
$Z(\Upsilon^i)$, 
$1\leq i\leq N-1$.
In particular,  it will be shown that they 
have a finite limit as $\epsilon\to 0$
as a result of a fairly delicate cancellations between the polynomial
terms and the vertical part of the instanton prepotential.
In \cite{Iqbal:2003ix, Iqbal:2003zz,Eguchi:2003sj,Konishi:2003qq,LLZ} the $\epsilon \to 0$ limit of 
$\CF_{X_N}^{hv}$ has been shown to be well-defined and in fact given by the instanton contribution to the field theory 
prepotential 
\be\label{eq:instlimit} 
\lim_{\epsilon\to 0} \CF_{X_N}^{hv}
= \CF_{SU(N)}^{inst},
\ee
but the perturbative and vertical contributions were not 
discussed in detail. Here we focus on the truncation 
$Z^{pv}(\Upsilon^i)$ 
of the central charges to polynomial 
and vertical instanton terms.
Equations  
\eqref{eq:GWpotB} and \eqref{eq:fieldlimitA} yield
\[ 
\bal 
{\partial \CF_{X_N}^{vert}\over \partial  {\widetilde s_i}} 
= -{1\over 2\pi^2}\sum_{\alpha\in \Delta^+_N}\sum_{k\geq 1}
{n_i(\alpha)\over k^2} e^{2\pi k\epsilon \sqrt{-1}
 \langle a,\alpha\rangle/M_0},
\eal
\] 
where 
\[ 
\langle a,\alpha\rangle = \sum_{i=1}^{N-1} n_i(\alpha) a_i,
\]
and 
\[
\bal
{\partial^2 \CF_{X_N}^{vert}\over \partial {\widetilde s}_i
\partial {\widetilde s}_j } = {\sqrt{-1}\over \pi}  
\sum_{\alpha\in \Delta^+_N}
n_i(\alpha)n_j(\alpha)
\ln\big(1- e^{2\pi \epsilon \sqrt{-1}\langle a, \alpha\rangle/M_0}\big).
\eal
\]
The second derivative has a small $\epsilon$ expansion of the form 
\[
\bal
{\partial^2 \CF_{X_N}^{vert}\over \partial {\widetilde s}_i
\partial {\widetilde s}_j } = {\sqrt{-1}\over \pi}  
\sum_{\alpha\in \Delta^+_N}
n_i(\alpha)n_j(\alpha)
\ln\left(-{2\pi \epsilon\sqrt{-1}\over M_0} \langle a, \alpha\rangle\right) + O(\epsilon). 
\eal
\]
This implies that the first derivative will have an expansion of the form
\[ 
\bal 
{\partial \CF_{X_N}^{vert}\over \partial  {\widetilde s_i}} 
= &\ 
c +{\epsilon\sqrt{-1}\over \pi M_0}
\sum_{\alpha \in \Delta_N^+}
n_i(\alpha) \langle a,\alpha\rangle \bigg[\ln \left(-{2\pi\epsilon\sqrt{-1}\over M_0}
\langle a,\alpha\rangle\right) -1\bigg]+O(\epsilon^2)
\eal
\]
where $c$ is a constant. Since all terms in the above expression 
except $c$ vanish in the $\epsilon \to 0$ limit, $c$ is
the value of the first derivative at $\epsilon=0$, 
\[
c= -{1\over 2\pi^2} 
\bigg(\sum_{\alpha\in \Delta^+_N}n_i(\alpha)\bigg) \bigg(\sum_{k\geq 1} {1\over k^2}\bigg) =- {1\over 12} \sum_{\alpha\in \Delta^+_N}n_i(\alpha).
\]
Then
 the leading terms of the central charges $Z^{pv}(\Upsilon^i)$, $1\leq i\leq N-1$ in the $\epsilon \to 0$ limit are 
\be\label{eq:DfourZB}
\bal
Z^{pv}(\Upsilon^i)  \sim {M_0\over \epsilon}  \bigg[ &
-{\epsilon N\sqrt{-1}\over \pi M_0}\sum_{j=1}^{N-1} 
{\sf C}_{ij}^{-1} (c_0+ \ln \epsilon) a_j \\
& +{1\over 12} \bigg(2\sum_{j=1}^{N-1} {\sf C}^{-1}_{ij}
-\sum_{\alpha\in \Delta^+_N} n_i(\alpha)\bigg)\\
& 
+{\epsilon \sqrt{-1}\over \pi M_0} 
\sum_{j=1}^{N-1}\sum_{\alpha\in \Delta^+_N} 
n_i(\alpha) n_j(\alpha) a_j 
\ln \epsilon \\
& + {\epsilon\sqrt{-1}\over \pi M_0} \sum_{\alpha\in \Delta^+_N} n_i(\alpha)
\langle a, \alpha\rangle 
\bigg[\ln \left(-2\pi{\sqrt{-1}\over M_0}
\langle a,\alpha\rangle\right) -1\bigg]
\bigg].\\
\eal
\ee
Now note that the term proportional to $1/\epsilon$ in the 
expression of $Z^{pv}(\Upsilon^i)$ 
cancels because of the following identity 
\be\label{eq:cartanidA}
\sum_{\alpha\in \Delta_N^+} n_i(\alpha) = 2 \sum_{j=1}^{N-1} 
{\sf C}_{ij}^{-1}.
\ee
This is equivalent to the known identity 
\[ 
\rho = {1\over 2}\sum_{\alpha\in\Delta_N^+} \alpha = \sum_{i=1}^{N-1}\lambda_i 
\] 
where $\rho$ is the Weyl vector and $\lambda_i$, $1\leq i\leq N-1$ 
the fundamental weights of $SU(N)$. 
Moreover, the terms proportional to $\ln\epsilon$, 
\[
{\sqrt{-1}\over \pi} \bigg(\sum_{j=1}^{N-1} a_j\bigg(
-N{\sf C}_{ij}^{-1} +\sum_{\alpha \in \Delta^+_N} 
n_i(\alpha) n_j(\alpha) \bigg) \bigg)
\]
also cancel because of a second identity, 
\be\label{eq:cartanidB}
\sum_{\alpha \in \Delta^+_N} 
n_i(\alpha) n_j(\alpha) = N {\sf C}_{ij}^{-1},
\ee 
which is proven below. 

Define the Cartan-Killing form with its natural normalization  
${\rm Tr} (Ad(X) Ad(Y)) = (X,Y)_{CK}$. 
Then the usual decomposition of the Lie algebra into root
spaces implies that on the dual space we have
\be\label{eq:idone}
(\alpha, \beta)_{CK} = \sum_{\gamma \in \Delta_N} 
(\alpha, \gamma)_{CK} (\beta, \gamma)_{CK}
\ee
for any roots $\alpha, \beta$, where 
 $\Delta_N$ is the set of roots of $SU(N)$. 
 Let $(X,Y)'$ be the Killing form normalized such that the roots have 
 length two. Then 
 \[ 
 (X,Y)_{CK} = {(X,Y)'\over 2N},
 \]
and \eqref{eq:idone} yields 
\be\label{eq:idtwo}
\sum_{\gamma\in \Delta_N^+} 
(\alpha, \gamma)'(\beta,\gamma)' = N (\alpha, \beta)'.
\ee
Of course, this can be extended linearly so it is also
true if we replace $\alpha, \beta$ by any linear combination of roots.  In particular,
we may replace them by fundamental weights $\lambda_i, \lambda_j$.
Since 
\[
(\lambda_i, \lambda_j)' = C_{ij}^{-1}, 
\]
equation \eqref{eq:idtwo} becomes \eqref{eq:cartanidB}. 

In conclusion, collecting all terms, it follows that the 
perturbative and vertical parts of the central charges 
$Z(\Upsilon^i)$, $1\leq i\leq N-1$, have a finite $\epsilon \to 0$ 
limit:
\be\label{eq:dualperlim}
\lim_{\epsilon\to 0} 
Z^{pv}(\Upsilon^i) = -{N\sqrt{-1}\over \pi} \sum_{j=1}^{N-1} 
{\sf C}_{ij}^{-1}
c_0a_j + {\sqrt{-1}\over \pi} 
\sum_{\alpha\in \Delta_N^+} n_i(\alpha)
\langle a,\alpha \rangle
\bigg[\ln \left(-{2\pi\sqrt{-1}\over M_0}
\langle a,\alpha\rangle\right) -1\bigg].
\ee
Using again identity \eqref{eq:cartanidB} and equation \eqref{eq:azero}, 
this can be written  
\be\label{eq:DfourZE}
\bal 
\lim_{\epsilon\to 0} 
Z^{pv}(\Upsilon^i)  
 = & {N\over \pi}\bigg[{\pi\over 2}+\sqrt{-1}
\left({c_0+N\ln(2)\over N-1}-1\right)\bigg] \sum_{j=1}^{N-1} 
{\sf C}_{ij}^{-1}a_j \\
& + {\sqrt{-1}\over \pi} 
\sum_{\alpha\in \Delta_N^+} n_i(\alpha)
\langle a,\alpha \rangle
\ln
{\langle a,\alpha\rangle\over \Lambda}.\\
\eal
\ee
If we identify 
\[ 
\lim_{\epsilon\to 0} Z^{pv}(\Upsilon^i) 
= {\partial \CF_{SU(N)}^{pert}\over \partial a_i}
\]
then we find, up to an additive constant 
\be\label{eq:gaugeprep}
\CF_{SU(N)}^{pert} 
= {N\over 2}\tau_0 \sum_{i,j=1}^{N-1} {\sf C}^{-1}_{ij} a_ia_j 
+ {\sqrt{-1}\over 2\pi} \sum_{\alpha\in \Delta_N^+} 
\langle a,\alpha\rangle^2 \ln {\langle a,\alpha\rangle \over \Lambda}
\ee
with 
\[
\tau_0 = {1\over 2} + {\sqrt{-1}\over \pi} 
\left({c_0+N\ln(2)\over N-1}-{3\over 2}\right).
\]
Thus we find 
\[
\lim_{\epsilon \to 0} \CF_{X_N}^{pv} = \CF_{SU(N)}^{pert}
\]
and together with equation \eqref{eq:instlimit} this implies 
\[
\lim_{\epsilon\to 0} \CF_{X_N} = \CF_{SU(N)}.
\]

Finally, note that the above results also determine the behavior of 
the central charges of the fractional branes $(P_i,Q_i)$, 
$1\leq i\leq N$ in the field theory 
limit. 
The $K$-theory classes of the sheaves  $F_i$, $i=1,\ldots,N$ in \eqref{eq:fractbranesAB} are given by
 \[
\bal 
& [F_i]  = -{\sf C}_{ij} \Upsilon^j +(i+1)\Upsilon_i, \qquad 1\leq i\leq N-1, 
\\
& [F_N] = -\sum_{i,j=1}^{N-1}{\sf C}_{ij} \Upsilon^j+ \Upsilon_0 + 
\sum_{i=1}^{N-1}(i+1)\Upsilon_i +[\CO_p].\\
\eal
\]
Therefore $Z(F_N)$ diverges as $Z(\Upsilon_0)\sim M_0\epsilon^{-1}$, while 
 \[
  {Z(F_i)} = -{\sf C}_{ij} Z(\Upsilon^j) + (i+1) a_i, \qquad 1\leq i\leq N-1 
  \] 
are finite in the $\epsilon \to 0$ limit.   
  This shows that the fractional branes $(P_N,Q_N)$
are very heavy and decouple from the low energy dynamics in this
limit while $(P_i,Q_i)$, $1\leq i\leq N-1$ 
are dynamical BPS particles with central charges
\be\label{eq:ZfractbranesB}
Z_{gauge}(P_i) = {\sf C}_{ij} a^D_j - (i+1)a_i,
\qquad
Z_{gauge}(Q_i) = -{\sf C}_{ij} a^D_j + ia_i,
\ee
with $i=1,\ldots, N-1$.

This result allows us to employ geometric engineering
methods in the study of the gauge theory BPS spectrum.
Using the detailed discussion of the field theory limit
one can construct a dictionary between
 D-brane bound states and gauge theory BPS
particles. First note that the
abelian gauge fields $A^{(i)}$, $i=1,\ldots, N-1$ in the
low energy effective action are obtained by KK reduction
of the three-form field,
\[
C^{(3)}=\sum_{i=1}^{N-1} A^{(i)}\wedge \eta_{S_i},
\]
 on a set of harmonic two-forms
$\eta_{S_i}$ related by Poincar\'e duality to $S_i$.

D-branes wrapping
the compact holomorphic cycles in $X_N$
yield massive BPS particles in the low energy theory
whose electric and magnetic charges are determined by the
standard couplings to background RR fields using relations
\eqref{eq:SandD}, \eqref{eq:intersmatrix}.
A D2-brane with $K$-theory class
$\Upsilon_i=[\CO_{C_i}(-1)]\in K^0_{cpt}(X_N)$, $i=1,\ldots, N-1$,
yields a massive BPS particle whose world-line  coupling  to the
abelian gauge fields $A^{(i)}$ is given by
\[
{\rm exp}\bigg[\sqrt{-1} \int_{C_i\times \IR}
C^{(3)}\bigg]=
{\rm exp}\bigg[-\sqrt{-1} \sum_{j=1}^{N-1}{\sf C}_{ij}
\int_{\IR} A^{(j)}\bigg]
\]
Therefore it has electric
electric charge vector
$(-{\sf C}_{ij})_{1\leq j \leq N-1}$ and trivial magnetic
charges.
These particles will be identified with the
massive $W$-bosons in field theory.
A D4-brane
with $K$-theory class $\Lambda^i=[\CO_{S_i}(-\Sigma_{i-1}-(i+1)C_i)]$, $i=1,\ldots, N-1$ yields a magnetic monopole with magnetic charge
$(\delta_{ij})_{1\leq i\leq j}$. This can be checked by a similar
argument. Note that the integral homology cycle
\[
{\widetilde C}_i = -N\sum_{j=1}^{N-1}{\sf C}^{-1}_{ij}C_j
\]
has intersection numbers
\[
({\widetilde C}_i \cdot S_j)_{X_N} =N\delta_{ij}
\]
with the compact four-cycles $S_j$.
Then pick a smooth representative
${\widetilde C}_i$ and let $S^2_r$
be a two-sphere of very large radius in $\IR^3$
in the rest frame of the BPS particle, centered at the
origin. Note that
the four-cycle $S^2_r\times {\widetilde C}_i\subset \IR^3
\times X_N$ is a linking cycle for $\{0\}\times S_i\subset
\IR^3\times X_N$ with linking number $N$, and has linking number 0 with the cycles $\{0\}\times S_j$, $j\neq i$.
Since a D4-brane wrapped on $S_i$ carries one unit of magnetic charge with respect to the three-form field $C^{(3)}$, it follows that
\[
N\delta_{ij}= \int_{S^2_r\times {\widetilde C}_j} dC^{(3)} =
\int_{S^2_r\times {\widetilde C}_j} \sum_{k=1}^{N-1}
dA^{(k)} \wedge \eta_{S_k} =
N\sum_{k=1}^{N-1}\delta_{jk} \int_{S^2_r} F^{(k)}
= N  \int_{S^2_r} F^{(j)}.
\]

As expected, the $K$-theory classes $\Upsilon_i$,
$\Lambda^i$, $i=1,\ldots, N-1$ belong to the
sublattice generated by the $K$-theory classes of the fractional
branes $[P_i], [Q_i]\in K^0_{cpt}(X_N)$, $i=1,\ldots, N-1$, which have finite mass
in the field theory limit. In fact one can easily check by a Chern
class computation that the sublattice generated by
$\Upsilon_i$,
$\Lambda^i$, $i=1,\ldots, N-1$
is identical to the one generated by
$[P_i], [Q_i]$, $i=1,\ldots, N-1$.
Moreover there is an orthogonal direct sum decomposition 
\be\label{eq:Ktheorydecomp} 
K^0_{cpt}(X_N) = {\rm Span}\{ \CO_p, \CO_{\Sigma_0}(-1)\} 
\oplus {\rm Span}\{ [P_i],[Q_i]\}
\ee 
with respect to the pairing $\langle \ ,\ \rangle$ such that the induced 
pairing on the second term is nondegenerate. 

The above arguments lead to the conclusion that  the symplectic 
infrared charge lattice $\Gamma$
of the gauge theory is identified with 
\be\label{eq:gaugelattice} 
\Gamma \simeq {\rm Span}\{ [P_i],[Q_i]\}_{1\leq i\leq N-1}.
\ee
On general grounds, the infrared lattice of electric and magnetic
charges $\Gamma$ does not admit a canonical splitting
into
electric and magnetic complementary sublattices,
$\Gamma\simeq \Gamma^{e}\oplus \Gamma^{m}$.
However there is a canonical splitting
in the semiclassical limit,
where $\Gamma^e$ is generated by the charges
of massive $W$-bosons and $\Gamma^m$ by the charges of magnetic
monopoles.
More precisely $\Gamma^e$ is the root lattice of the gauge group
$G$ while
$\Gamma^m$ is the coroot lattice.
Note that these are not dual lattices. The dual of the coroot
lattice is the weight lattice. In field theory the quotient of $\Gamma$
by the annihilator is symplectic.
Geometrically, $\Gamma^e$ is identified with the
sublattice of $K^0_{cpt}(X_N)$ generated by the vertical curve classes
$\Upsilon_i$, $i=1,\ldots, N-1$,
while $\Gamma^m$ is identified with the sublattice
generated by $\Lambda^i$, $i=1,\ldots, N-1$.

In addition one can also obtain
line defects by wrapping D4 and D2-branes on
noncompact cycles in $X_N$. A D4-brane supported on
a noncompact divisor
$D_i$ of the form \eqref{eq:diveqA} flows in the infrared to a simple
line defect which has in the present conventions magnetic
charge vector $(-{\sf C}_{ij}^{-1})_{1\leq j\leq N-1}$.
The electric charges of the line defect are determined by the
Chan-Paton line bundle on $D_i$. It will be shown in
Section \ref{sectionfive} that there is a simple choice
of Chan-Paton line bundles which yields trivial electric charges.
With this choice the charge vectors of the simple line defects are 
precisely identified with the projections ${\overline \Upsilon}^i$ of the rational $K$-theory generators 
\eqref{eq:DfourbranesA} to the lattice $\Gamma$. 

Note that the magnetic charge vector of a line defect engineered
by a D4-brane wrapping a divisor $D_i$ does not belong to
$\Gamma^m$, since it has fractional entries. This is in fact in agreement with the gauge theory classification of line defects
\cite{framedBPS}.
According to loc. cit. the magnetic
charges of a line defect sits in  the magnetic weight lattice
$\Gamma^{mwt}$ as a  $\Gamma^m$-torsor.

In conclusion, geometric engineering predicts that
gauge theory BPS states are identified with
bound states of the fractional branes $(P_i,Q_i)_{1\leq i\leq N-1}$.
On physical grounds the low energy dynamics of such bound states
will be determined by
a truncation of the quiver $(\CQ,\CW)$ where the vertices
$u_N,v_N$ and all adjacent arrows are removed. This yields a
smaller quiver with potential $(Q,W)$, where $W$ is obtained
by truncating \eqref{eq:potentialA} accordingly.
A precise mathematical study of such bound states requires the
notion of
$\Pi$-stability introduced in
\cite{Dbranes_categories,Pistability,Dbranes_monodromy}, which was mathematically formulated by
Bridgeland in \cite{stabtriangulated}.

\subsection{Stability conditions}\label{twothree}
According to \cite{Dbranes_categories,Pistability,
Dbranes_monodromy} supersymmetric D-brane bound states
must be Bridgeland
stable objects \cite{stabtriangulated}
in the derived category $D^b(X_N)$.
At the same time the finite mass bound states in the field
theory limit must belong to the subcategory $\CG$
spanned by the subset of fractional branes $(P_i,Q_i)_{1\leq i\leq N-1}$.
A natural question is whether such objects can be intrinsically described
as stable objects with respect to a stability condition on $\CG$, as
suggested by RG flow decoupling arguments. In general this is not the
case on mathematical grounds, as explained in more detail below.
However it will also be shown that this condition is satisfied for
a natural class of stability conditions from the quiver point
of view. In contrast, this property fails for large radius limit stability
conditions, as discussed in detail in Section \ref{sectionthree}.

Recall that a Bridgeland stability condition at a point in the complex
K\"ahler moduli spaces is specified by a t-structure on
$D^b(X_N)$ satisfying a compatibility condition with the
central charge function. The precise definition of a t-structure will
not be
needed in the following. It suffices to note that any t-structure
determines an abelian
subcategory  $\CA\subset D^b(X_N)$, the heart of the
$t$-structure,
such that exact sequences in $\CA$ are exact triangles in the
ambient derived category.
The compatibility condition requires the central charges of all objects of $\CA$ to  lie in a complex half-plane
of the form
 \[
\IH_\phi=\{{\rm Im}(e^{i\phi}z)>0\}\cup \{{\rm Im}(e^{i\phi} z)=0,
\ {\rm Re}(e^{i\phi}z)\leq 0\},
\]
for some $\phi\in \IR$. Therefore for any nontrivial
object $F$ of $\CA$
one can define a phase $\varphi(F) \in [\phi,\pi+\phi)$ of $Z(F)$.
All stable objects must belong to $\CA$ up to shift, and
an object
$F$ of $\CA$ is (semi)stable if
$\varphi(F)\ (\geq)\ \varphi(F')$ for any proper nontrivial subobject
$0\subset F'\subset F$ in $\CA$.

Now let $\CG$ be the smallest triangulated subcategory of the
derived category $D^b(X_N)$
generated by the fractional branes $(P_i,Q_i)_{1\leq i\leq N-1}$.
For a given point in the K\"ahler moduli space,
supersymmetric D-brane bound states in $\CG$ are
stable objects of $\CA$ which belong to $\CG$.
Note that $\CG$ satisfies the conditions of
\cite[Lemma 1.3.19]{BBD},
therefore the given $t$-structure on $D^b(X_N)$ induces a t-structure on $\CG$. Therefore
the intersection $\CA\cap \CG$ is an abelian subcategory
$\CA^\CG\subset \CA$, the heart of the induced t-structure. However, the test subobjects
$0\subset F'\subset F$ in the definition of stability do not necessarily
belong to $\CA^\CG$. Therefore in general the D-brane bound states
will not be defined intrinsically by a stability condition on $\CG$. In the present case, there is however a natural class of stability conditions where this potential complication does not arise.

Since $D^b(X_N)\simeq D^b(\CQ,\CW)$, there is a canonical
bounded t-structure whose heart $\CA$ is the abelian category of $(\CQ,\CW)$-modules. The heart $\CA^\CG$
of the induced t-structure on $\CG$ is the
abelian category of modules over the path algebra of the truncated
quiver $(Q,W)$ defined at the end of Section \ref{twotwo}.

It is clear that all subobjects and
all quotients of an object $\rho$ of $\CA^\CG$ also belong to
$\CA^G$. Therefore in this case the stable objects of $\CA$
belonging to $\CG$ are defined by an intrinsic stability condition
on $\CA^\CG$.
By analogy with the local $\IP^2$ model treated in  \cite{stabnoncompact,locptwo},
such stability conditions on $\CG$
are obtained by assigning complex numbers $(z,w)=(z_i,w_i)$
 to the nodes $(p_i,q_i)$, $i=1,\ldots, N-1$, of $Q$, all lying in the half-plane $\IH_\phi$. In order to fix notation, the dimension vector
 of a representation $\rho$ with underlying vector spaces 
 $(V_i,W_i)_{1\leq i\leq N-1}$ will be denoted by $(d_i,e_i)_{1\leq 
 i\leq N-1}$. 
Then the  $(z,w)$-slope of a representation $\rho$ of dimension vector
$(d_i,e_i)$ at the nodes $(p_i,q_i)$, $i=1,\ldots, N-1$, respectively
is defined by
\[
\mu_{(z,w)}(\rho) = - {{\rm Re}(\sum_{i=1}^{N-1} e^{i\phi}(d_iz_i+e_iw_i))
\over {\rm Im}(\sum_{i=1}^{N-1} e^{i\phi}(d_iz_i+e_iw_i))}.
\]
A representation $\rho$ of dimension vector 
$(d_i,e_i)_{1\leq i\leq N-1}$
 is $(z,w)$-(semi)stable if
\[
\mu_{(z,w)}(\rho')\ (\leq)\ \mu_{(z,w)}(\rho)
\]
for all subrepresentations $0\subset \rho'\subset \rho$.
For simplicity, it is often convenient
 to consider stability parameters of the form
\be\label{eq:kingparameters}
z_i= r(-\theta_i+\sqrt{-1}),\qquad w_i=r(-\eta_i+\sqrt{-1}),\qquad i=1,\ldots, N-1
\ee
where $r, \theta_i,\eta_i \in \IR$, $r\in \IR_{>0}$ such that
 $\phi$ may be taken trivial.
In this case the slope reduces to
\[
\mu_{(\theta,\eta)}(\rho) = {\sum_{i=1}^{N-1}(d_i\theta_i+e_i\eta_i)
\over \sum_{i=1}^{N-1} (d_i+e_i)},
\]
where $\theta=(\theta_i)_{1\leq i\leq N-1}$, $\eta=(\eta_i)_{1\leq i\leq N-1}$.
One can further reduce to the GIT stability conditions constructed by King
in \cite{King} observing that a representation $\rho$ is
$(\theta,\eta)$-(semi)stable if and only if it is
$({\bar \theta}, {\bar \eta})$-(semi)stable
where
\[
{\bar \theta}_i = \theta_i - {\sum_{i=1}^{N-1}
 (d_i\theta_i+e_i\eta_i)\over
\sum_{i=1}^{N-1}(d_i+e_i)}, \qquad
{\bar \eta}_i = \eta_i - {\sum_{i=1}^{N-1} (d_i\theta_i+e_i\eta_i)\over
\sum_{i=1}^{N-1}(d_i+e_i)}, \qquad i=1,\ldots,N-1.
\]
Note that $({\bar \theta}, {\bar \eta})$ satisfy
\be\label{eq:gitparam}
\sum_{i=1}^{N-1}(d_i{\bar \theta}_i+e_i{\bar \eta}_i) =0.
\ee
Stability parameters satisfying equation \eqref{eq:gitparam}
will be referred to as King stability parameters. In some situations
working with such parameters leads to significant simplifications.

For physical stability conditions  the stability parameters
$(z_i,w_i)$, $1\leq i\leq N-1$, are determined by the central
charges \eqref{eq:ZfractbranesB} assigned to the corresponding fractional branes:
\be\label{eq:stabparametersA} 
z_i ={1\over \Lambda} \big({\sf C}_{ij}a_j^D -(i+1)a_i\big),\qquad 
w_i={1\over \Lambda} \big(-{\sf C}_{ij}a_j^D +ia_i\big), 
\qquad 1\leq i \leq N-1.
\ee
More precisely there exists a subset $\CC_{(Q,W)}$
of the universal cover of (the smooth locus of) the
 Coulomb branch
where the central charges \eqref{eq:ZfractbranesB}
belong to a half-plane $\IH_\phi$, for some $\phi\in \IR$.
Then the above construction yields a map
$\CC_{(Q,W)}\to {\rm Stab}(\CG)$ to the moduli
space of Bridgeland stability conditions on $\CG$.

This map can be extended to a larger subset
using quiver mutations
to change the $t$-structure on $\CG$ as in  \cite{stabnoncompact}. For all stability conditions
obtained this way,
 the heart of the underlying $t$-structure
 is an abelian category of
modules over the path algebra of a quiver with potential 
$(Q',W')$  related by a mutation to $(Q,W)$.
Such stability conditions will be called algebraic, following the
terminology of \cite{locptwo}. The
subset of ${\rm Stab}(\CG)$ parameterizing such stability
conditions will be denoted by
${\rm Stab}^{\sf alg}(\CG)\subset {\rm Stab}(\CG)$.
 In conclusion one obtains a map
\be\label{eq:BPSmap}
\varrho: \CC^{\sf alg}_\CG \to {\rm Stab}^{\sf alg}(\CG)
\ee
defined on some subset $\CC^{\sf alg}_\CG$ of the universal cover of the Coulomb branch.
The field theory limit
leads to the conjecture that the gauge theory BPS spectrum at a point $u\in \CC^{\sf alg}_\CG$ is determined by the spectrum of Bridgeland stable
representations at the point $\varrho(u)$. Numerically, the BPS
degeneracies are identified with counting invariants of
stable objects in $\CG$ as explained in the next subsection.
This is in agreement with the quivers found in \cite{Fiol:2000pd,CNV,Alim:2011kw}.
Furthermore, it is also natural to conjecture that
in fact the domain of definition of $\varrho$ covers the
whole universal cover of the Coulomb branch of the field theory.
 That is, for any point in the Coulomb branch one can find an algebraic stability condition
on $\CG$ encoding the complete BPS spectrum at that
point.

For completeness, note that the derived category $D^b(X_N)$
is expected to admit
a different class of stability conditions, analogous to the
geometric stability conditions constructed in \cite{locptwo}.
In fact such stability conditions must be used if one
is interested in the spectrum of supersymmetric D-brane
bound states in a neighborhood of the large radius limit.
A rigorous construction of geometric Bridgeland stability conditions
is beyond the scope of the present paper.
More physical insight can be gained assuming their existence and  examining its consequences for the gauge theory BPS spectrum. This is the goal of  Section \ref{sectionthree}.

It might be useful to some readers to have an informal summary
of the main point of this section, expressed in more physical terms.
In this paper we are
viewing gauge theory BPS states as string theory BPS states which
remain ``light'' (i.e. of finite energy) in a certain
``field theoretic limit.'' In the type IIA string picture, the field theoretic
limit is a limit in which there is also a hierarchy of
scales within the Calabi-Yau manifold (see equation \eqref{eq:fieldlimitA}.) Some D-brane BPS states have
infinite energy in this limit (simply because they have nonzero tension and
wrap cycles which have infinite volume), but some D-brane BPS states
have a finite energy in this limit. Thus we use interchangeably the
terms ``light BPS states,'' ``finite energy BPS states,"  and ``field-theoretic
BPS states."

Now, both in field theory and in string theory the BPS states are expected
to be objects in a category. When the field theory is viewed as a limit of
string theory, evidently the gauge theory BPS states should be objects in
a subcategory of the string theory category.

In general two (or more) BPS states can interact and form a BPS boundstate, but that bound state only exists for certain vacuum 
parameters -- because the vacuum parameters
 determine the strength of the force between
constituents. 
The interaction energy is strictly negative away from walls of marginal
stability. The stability conditions on a category tell us
 when BPS states can be considered to be boundstates of
collections of other BPS states.  If the field-theoretic BPS
states are objects in a subcategory of a string-theoretic category
containing all BPS states then there are two possible notions of
boundstates:  We could consider only boundstates made of
field-theoretic BPS constituents or we could consider boundstates
of all possible string-theoretic BPS constituents. These notions
are, in principle, different because it is   quite possible that a light, field-theoretic
BPS state is (in the string theory) a boundstate of heavy string-theoretic
D-brane states. These heavy states might interact with a large negative
binding energy,  producing light states. Such a phenomenon produces an
obstruction to formulating a good stability condition on the field-theoretic
subcategory:  We might have ``spurious''  decays of BPS states in the field theory
in the sense that they are not made of honest field-theoretic BPS states.
Therefore, we would like a criterion
whereby we can determine if a BPS state is a boundstate purely of
field-theoretic BPS states. This is the physical interpretation of
an ``intrinsic stability
condition on $\CA^\CG$.''

   In fact, spurious decays do not happen in
our examples, but it is not easy to see that this is so 
in the large radius picture
based on $(\omega,B)$-stability. On the other hand, at string scale distances
there is an alternative picture of the BPS states in terms of quiver quantum
mechanics. In the quiver quantum mechanics picture it turns out that
there actually is a natural criterion (i.e. a t-structure on the derived category of
$(\CQ,\CW)$ modules) in which case it is easy to see that states which
are light in the field-theoretic limit can only be boundstates of BPS
particles which are also themselves light in the field theoretic limit.

 \subsection{BPS degeneracies and Donaldson-Thomas
  invariants}\label{twofour}

In this section we discuss the relation between various flavors of BPS
degeneracies used by physicists and various flavors of Donaldson-Thomas
invariants used by mathematicians. The proper identification of these
quantities will be a crucial working hypothesis in this paper.

Let us begin with the physical BPS degeneracies.
We recall the definition of protected spin characters from
\cite{framedBPS}.
 The Hilbert space of gauge theory BPS states
 carries an action of
$SU(2)_{spin}\times SU(2)_R$ where the first factor
$SU(2)_{spin}\subset Spin(1,3)$ is 
the little group of a massive particle in four dimensions
and the second is the $R$-symmetry group of the gauge
theory.
\footnote{The R-symmetry group of a theory is the group of
global symmetries which commutes with the Poincar\'e group
but does not commute with the supersymmetries. In our
case we normalize the $R$ symmetry generators so that
$2J_R$ has weights $\pm 1$ on the supercharges.}
 The irreducible representations of this group
will be denoted by
$(j_{spin},j_R)\in {1\over 2}\IZ_{\geq 0}
\times {1\over 2}\IZ_{\geq 0}$. Moreover,
as a representation of $SU(2)_{spin}\times SU(2)_R$
  the Hilbert space has   the form
\[
{\CH}_{BPS} \simeq \CH_{HH} \otimes \CH_{int}
\]
where $\CH_{HH}$ is the center-of-mass half-hypermultiplet
and $\CH_{int}$ is
the Hilbert space of internal quantum states of the BPS particles.
As a representation of $SU(2)_{spin}\times SU(2)_R$
$\CH_{HH}$ is   $(1/2,0)\oplus (0,1/2)$. The low energy
gauge group is abelian and global gauge transformations
act on $\CH_{BPS}$. The decompositions into isotypical
components defines the grading by the electromagnetic
charge lattice $\Gamma$.
The space $\CH_{HH}$ is neutral under global gauge transformations
so there is an induced grading
\be
\CH_{int}\simeq \oplus_{\gamma\in \Gamma}
\CH_{int}(\gamma).
\ee
The spaces $\CH_{int}(\gamma)$   depend in a piecewise constant manner\footnote{More formally,
 there is a piecewise defined flat connection on the piecewise-defined
 bundle of BPS states
 over the moduli space.}  on the
 order parameters $u$  of the Coulomb branch. The notation
 $\CH_{BPS}(\gamma; u)$, $\CH_{int}(\gamma;u)$
 will be used whenever this dependence needs to be emphasized.

Let $J_{spin}, J_R$ be Cartan generators of
$SU(2)_{spin}, SU(2)_R$ normalized to have  half-integral weights. The protected spin character for unframed
BPS states is defined in \cite{framedBPS} as
\be
{\rm Tr}_{\CH_{BPS}(\gamma; u)}(2J_{spin})(-1)^{2J_{spin}}(-y)^{2(J_{spin}+J_R)}.
\ee
The key property of the protected spin character is that it is an
\emph{index}, a result easily obtained from the
representation theory of the $\CN=2$ $d=4$ supersymmetry algebra:
 Massive, i.e. non-BPS representations, do not contribute
to this character.  Now, the protected spin character
 can be written as $(y-y^{-1}){\Omega(\gamma; u;y)}$,
where
\be\label{eq:spinchar}
{\Omega(\gamma; u;y)} ={\rm Tr}_{\CH_{int}(\gamma; u)}y^{2J_{spin}} (-y)^{2J_R}.
\ee
Note that in situations where the $SU(2)_R$ symmetry is broken
down to a $U(1)_R$  R-symmetry we can still define the RHS of \eqref{eq:spinchar},
although there is no longer a good reason for it to be an index, in general.

%In fact a doubly refined spin character
%\be\label{eq:doublyrefined}
%\Omega(\gamma; x,y,u) = {\rm Tr}_{\CH_{int}(\gamma,u)}
%(-x)^{J_{spin}+J_R} (-y)^{J_{spin}-J_R},
%\ee
%was defined
%in \cite{DM-wall} employing only the $U(1)_R$ subgroup.

Reference \cite{framedBPS} stated a pair of conjectures
concerning the protected spin character, known as
the \emph{positivity conjecture} and the \emph{no-exotics
conjecture}. These are meant to apply only to field-theoretic
(and not string-theoretic)  BPS states.
The positivity conjecture asserts that
${\Omega(\gamma; u,y)}$, regarded as a function of
$y$, can be written as a positive integral
linear combination of $SU(2)$ characters. That is:
\be
{\Omega(\gamma; u,y)}=\sum_{n \geq 1}
d(\gamma;u;n) \chi_{n}(y)
\ee
where
\be
\chi_{n}(y) := {\rm Tr}_{n} y^{2J } = \frac{y^n - y^{-n}}{y-y^{-1}}
\ee
is the character in the $n$-dimensional representation of $SU(2)$
and the $d(\gamma;u;n)$ are piecewise constant functions of $u$.
The positivity conjecture states that $d(\gamma;u;n)\geq 0$ for all $\gamma$ and
all points $u$ on the Coulomb branch.  It would follow if the center
of $SU(2)_R$ acts trivially on $\CH_{int}$, i.e., that $\CH_{int}$
contains only integral spins. We will call this
the \emph{integral spin property}. It is stronger than the
positivity conjecture.   The even stronger \emph{no-exotics} conjecture
posits that in fact only states with trivial $SU(2)_R$ quantum
numbers contribute to the protected spin character. When there
are no exotics the naive spin character coincides with the
protected spin character. In Section \ref{sectionsix} and 
also below we will discuss
criteria for the absence of exotics, and also string-theory
examples where exotics are present.

Turning now to the mathematical perspective, one can define \cite{wallcrossing,COHA}
motivic Donaldson-Thomas invariants for moduli spaces of
stable objects in the triangulated category $\CG$.
Employing an algebraic stability condition $\varrho(u)$ at a point
$u\in \CC_\CG^{alg}$,  the invariant
$DT^{mot}(\gamma, z(u),w(u))$ is the virtual
motive of the moduli space
of $(z,w)$-stable $(Q,W)$-modules with
dimension vector $\gamma$, taking values in an appropriate ring of motives.
See Appendix \ref{motives}  for the minimal material on motives needed to follow
the following discussion.

\def\fh{\mathfrak{h}}

As explained in Appendix \ref{motives}, the Hodge type
Donaldson-Thomas invariant
$$DT(\gamma;z,w;x,y)\in \IQ(x^{1/2}, y^{1/2})$$
is the image of $DT^{mot}(\gamma;z,w)$ under a homomorphism from the
ring of motives to the ring of Laurent polynomials
$ \IQ(x^{1/2},y^{1/2})$. It can therefore be written in   the form
\be\label{eq:hodgeDT}
DT(\gamma,z,w;x,y)= \sum_{r,s\in {1\over 2}\IZ}
\fh^{r,s}(\gamma;z,w)
x^{r}y^{s}
\ee
The coefficients $\fh^{r,s}(\gamma,z,w)$ are
by construction
non-negative integers. Moreover, as explained below,
physics arguments \cite{DM-crossing,DGS} lead to
the conjecture that they satisfy a duality relation,
$\fh^{r,s}(\gamma;z,w)=
\fh^{-r,-s}(\gamma;z,w)$.  As observed in Appendix
\ref{motives}, if the moduli space
of $(z,w)$-stable quiver representations is a smooth
projective variety $\CM(\gamma,z,w)$
of complex dimension $m$,
\be
\fh^{r,s}(\gamma;z,w) = h^{r+m/2,s+m/2}(\CM(\gamma;z,w))
\ee
where the latter are the standard Hodge numbers. (In particular,  $\fh^{r,s}$
is only nonzero for integral $r,s$ when $m$ is even and half-integral
$r,s$ when $m$ is odd. )
In what follows we will be particularly concerned with
the specialization:
\be\label{eq:Special-1}
DT(\gamma;z,w;y,y)=\sum_{r,s\in \frac{1}{2}\IZ} \fh^{r,s}(\gamma;z,w) y^{r+s}
\ee
and it will also be useful to define
\be\label{eq:Special-2}
DT^{ref}(\gamma;z,w;y) :=\sum_{r,s\in \frac{1}{2}\IZ} (-1)^{r-s}\fh^{r,s}(\gamma;z,w) y^{2r}
\ee
We will refer to \eqref{eq:Special-2} as the \emph{refined Donaldson-Thomas invariants}.
Note that $\fh^{r,s}$ is nonzero only when $r-s$ is integral,
as observed at the end of Appendix \ref{motives}.

Now let us turn to the relation between the physical and mathematical
counting functions. Our working hypothesis is, that when the
moduli space of BPS states is smooth we can identify
\be\label{eq:BPScoh}
\CH_{int}(\gamma;u) \cong \oplus_{p,q} H^{p,q}(\CM(\gamma;z(u),w(u)))
\ee
Moreover, under this isomorphism  the action of the spin group $SU(2)_{spin}$
should be identified with the standard Lefschetz action on cohomology. Thus,
$2J_{spin}$ acts on the $(p,q)$-graded piece   as $p+q-m$.
Furthermore,  $2J_R$ acts with weight
$p-q$ on the $(p,q)$-graded piece. Granting these identifications
 the protected spin character \eqref{eq:spinchar}
becomes
\be\label{eq:PCS-Hodge-1}
\Omega(\gamma;u; y)=\sum_{p,q\in\IZ }(-1)^{p-q}y^{2p-m} h^{p,q}({\CM}(\gamma; z(u),w(u)))
\ee
for compact and smooth moduli spaces.

A historical remark might be clarifying to some readers at this
point. The identification of spin $SU(2)$ with Lefshetz $SU(2)$
acting on cohomology of BPS spaces goes back to Witten
\cite{Witten:1996qb}.   The specialized Hodge-polynomials \eqref{eq:Special-1}
were alleged in  \cite{DM-crossing, DG,DGS} to coincide with  the
{\it un-protected} spin character ${\rm Tr}_{\CH_{int}(\gamma;z,w)}
y^{2J_{spin}}$, even though the un-protected spin character
is not an index. Moreover, it was also proposed in \cite{DM-crossing}
that $2J_R$ acts as $p-q$, at least when the moduli space of BPS states is smooth.
In general we do not expect to be able to compute unprotected quantities
exactly. At special loci there could be, for example, massive BPS multiplets saturating
the BPS bound, thus invalidating the identification
in \eqref{eq:Special-1}. As we discuss in Sections \ref{exotics}, 
\ref{sectionsix}
below the surprising successes of many computations based on the spin character can be explained
in some examples where the absence of exotic BPS representations can be
proven. However a notable exception has been found in
\cite{wall-pairs}, where convincing evidence has been found
for the isomorphism \eqref{eq:BPScoh} in the presence of exotic
BPS states.

What is the mathematical import of \eqref{eq:PCS-Hodge-1} ?
Recall that the $\chi_{\tilde y}$-genus of a smooth projective
variety $V$ is defined by
\be
\chi_{\tilde y} (V)  :=
\sum_{p,q\in \IZ } (-1)^{p+q} {\tilde y}^p h^{p,q}(V).
\ee
Therefore
\be\label{eq:chigenusconjA}
\Omega(\gamma; u; y)= y^{-m}
\chi_{\tilde y}({\CM}(\gamma, z(u),w(u)))\big|_{\tilde y=y^2}
\ee
A natural extension of this claim is that 
\be\label{eq:chigenusconjB}
\Omega(\gamma; u; y)= \sum_{r,s\in \frac{1}{2}\IZ} (-1)^{r-s} y^{2r} \fh^{r,s}(\gamma;z(u),w(u))
\ee
for \emph{any} charge $\gamma$ and point $u$ on the Coulomb branch,
even when the moduli spaces of BPS states are singular.

Comparing with \eqref{eq:Special-2} our extended conjecture \eqref{eq:chigenusconjB}
identifies the protected spin character $\Omega(\gamma; u; y)$
with a refined DT invariant.
Granting this identification,  the  absence of exotics conjecture
translates into the condition
${\mathfrak h}^{r,s}(\gamma; z,w)=0$ for all $r \neq s$.
If this holds,
\be
\Omega(\gamma;u ; y)= \sum_{r\in \frac{1}{2}\IZ }  y^{2r} \fh^{r,r}(\gamma;z(u),w(u))
\ee
If the moduli space is smooth we can further write:
\be
\Omega(\gamma;u ; y) = y^{-m} P(\CM(\gamma; z(u),w(u)); y^2)
\ee
where $P$ is the Poincar\'e polynomial.

Finally, note that the
specialization of $DT^{ref}(\gamma, z,w;y)$ at $y=(-1)$ coincides
with the specialization of $DT(\gamma,z,w;y)$ at $y=(-1)$,
and equals the numerical Donaldson-Thomas invariants
$DT(\gamma,z,w)$. Relation \eqref{eq:chigenusconjB} then implies
that the numerical invariants $DT(\gamma, z,w)$ are identified with
the BPS indices $\Omega(\gamma;u)$.

\section{Field theory limit {\bf B}}\label{localmirror}
This section reviews the B-model formulation of the field theory
limit for SU(2) gauge theory, following the earlier geometric engineering literature
\cite{Kachru:1995wm,Klemm:1995wp,Klemm:1996bj,geom_eng,KMV}. Our main point here
is to establish some results on periods and their analytic continuation
from the large complex structure point to the field theory point
so that we can check our no-walls conjecture (in section 4.2 below)
in the B-model. 
Similar results for $SU(2)$ were obtained in  \cite{SW-derived, DGS} employing slightly different local Calabi-Yau models. 
Here we will employ the local $\IF_2$ model and focus on analytic 
continuation of BPS central charges between LCS limit 
point and the field theory scaling region.

According to \cite{geom_eng,KMV,HV}, 
the local mirror of the toric Calabi-Yau threefolds $X_N$, $N\geq 2$,
is a family of conic bundles over $(v,w)\in (\IC^\times)^2$ 
given by 
\be\label{eq:locmirrorA}
P(v,w)=xy,
\ee
where $(x,y)\in \IC^2$.
 In terms of homogeneous
coordinates $(\alpha_i,\beta_1,\beta_2)\in (\IC^{\times})^{N+3}$
on the moduli space,
the polynomial $P(v,w)$ is given by
\be\label{eq:locmirrorB}
P(v,w) = \beta_1v+{\beta_2\over v} + \sum_{i=0}^{N}
\alpha_i w^i.
\ee
The homogeneous parameters $\alpha_i, \beta_1,\beta_2$, 
$0\leq i \leq N$, are subject to a scaling gauge symmetry 
\[
\alpha_i \to  \lambda_s^{k_s^i}\alpha_i, \qquad 
\beta_1 \to \lambda_s^{k_s^{N+1}}\beta_1,\qquad 
\beta_2\to \lambda_s^{k_s^{N+2}}\beta_2,\qquad 
0\leq i\leq N+2,
\]
where $\{k_s=(k_s^j)\}$, $1\leq s\leq 3$, $0\leq j\leq N+2$,
is an integral basis of the kernel of the charge matrix 
$(Q^a_{j})$, $a\leq 1\leq N$, $0\leq j\leq N+2$, in \eqref{eq:tordataA}. 
The gauge invariant algebraic coordinates $z_i$, $i=0,\ldots, N-1$
on the moduli space are
given by
\be\label{eq:algcoordinates}
z_i=\alpha_{i-1}\alpha_i^{-2}\alpha_{i+1}, \qquad 1\leq i \leq N-1,
\qquad
z_0 = \beta_1\beta_2\alpha_0^{-2}
\ee
since $(\alpha_i,\beta_1,\beta_2)$ all have weight one under the 
scaling gauge symmetry.
There is a unique (up to scaling)
holomorphic three-form 
\[
\Omega = {1\over y} dx dw dv
\]
on the conic bundles
\eqref{eq:locmirrorA}
whose periods $\Pi$  satisfy the GKZ system
\be\label{eq:GKZ}
{\partial \over \partial \alpha_{i-1}} {\partial \over
\partial \alpha_{i+1}} \Pi = {\partial^2\over \partial \alpha_{i}^2}\Pi,
\qquad 1\leq i \leq N-1,
\qquad
{\partial \over \partial \beta_1} {\partial \over
\partial \beta_2} \Pi = {\partial^2\over \partial \alpha_{0}^2}\Pi.
\ee
Note that the mirror map
is of the form
\[
z_i=e^{2\pi \sqrt{-1}(b_i+\sqrt{-1}t_i)}
\]
near the LCS limit point, $z_i\to 0$, $i=0,\ldots, N-1$.
The field theory limit is a scaling limit of the form \cite{geom_eng,KMV}
\[
z_0 \sim \epsilon^{2N},\qquad z_i\sim \epsilon^0, \qquad i=1,\ldots, N-1
\]
which identifies the curve $P(v,w)=0$ with the Seiberg-Witten curve
of pure $SU(N)$ gauge theory. 
As shown below for $N=2$, this is the {\bf B} model counterpart of the 
scaling limit studied in Section \ref{twotwo} 
in terms of ${\bf A}$-model variables. 
%limit makes the quantum volume of
%the horizontal curve class $\Sigma_{0}$ very large keeping the
%quantum volumes of the vertical curve classes $C_i$, $i=1,\ldots, %N-1$
%finite. The former is then identified with the QCD scale of the %gauge theory,
%a non-dynamical parameter, while the latter are identified with the order
%parameters on the Coulomb branch.

In the case $N=2$, corresponding to $SU(2)$ gauge theory, the toric threefold $X_2$ constructed in section \ref{sectiontwo} 
is isomorphic to the total space of the anticanonical bundle
of the Hirzebruch surface $S_1=\IF_2$. The Mori cone of $X_2$ is generated by the fiber class $C_1$ and the section class 
$\Sigma_0$. 
The Mori vectors are given by (\ref{eq:tordataA}): 
$$\ell^{(0)}=(-2,0,0,1,1), \qquad 
\ell^{(1)}=(1,-2,1,0,0).$$
Equation  (\ref{eq:algcoordinates}) gives us the two coordinates on the
moduli space: $z_0=\beta_1\beta_2/\alpha_0^{2}$ and
$z_1=\alpha_0\alpha^{-2}_1\alpha_2$. The mirror map relates
$\ln(z_i)\sim 2\pi i \tilde s_i$ where $\tilde s_0$, $\tilde
s_1$  are the special flat coordinates on the complex K\"ahler moduli 
space associated to the generators $\Sigma_0$, $C_1$ respectively.

The Picard-Fuchs operators follow from (\ref{eq:GKZ}) and are equal to: 
\begin{eqnarray}
\label{eq:PFlocalF2}
&\CL_0&=\theta_{0}^2-z_0(\theta_{1}-2\theta_{0})(\theta_{1}-2\theta_{0}-1), \nonumber \\
&\CL_1&=\theta_{1}(\theta_1-2\theta_0)-z_12\theta_{1}(2\theta_{1}+1). \nonumber
\end{eqnarray}
with $\theta_{i}=z_i\frac{\partial}{\partial z_i}$. In the vicinity of the
large complex structure limit $|z_0|, |z_1|\ll 1$, the periods can be obtained by
introducing 
\be
\Pi(z_0,z_1;r_0,r_1)=\sum_{n_0,n_1=0}^\infty
\frac{z_0^{n_0+r_0}z_1^{n_1+r_1}}{\prod_{i=1}^5\Gamma(\sum_{\alpha=0,1} \ell^{(\alpha)}_i(n_\alpha+r_\alpha)+1)}.
\ee
and 
evaluating its derivatives with respect to
 $r_i$ at $r_i=0$, $i=0,1$, of (see e.g.
\cite{Hosono:1993qy, Chiang:1999tz}).
The action of the Picard-Fuchs generator on 
$\Pi(z_1,z_2;r_0,r_1)$ gives a simpler function 
which vanishes upon taking derivatives with respect to 
$r_0,r_1$ and setting them to $0$. 
Using Euler's reflection formula
$\Gamma(1-x)\,\Gamma(x)=\frac{\pi}{\sin \pi x}$, 
the resulting expressions for the periods are:
\be\label{eq:tD}
\bal
\Pi_c&=\left. \Pi(z_0,z_1;r_0,r_1) \right|_{r_0=r_1=0}=1,\\
2\pi \sqrt{-1}\,  \Pi_0&=\frac{\partial}{\partial r_0}\left. \Pi(z_0,z_1;r_0,r_1)
\right|_{r_0=r_1=0}\\
&=\ln(z_0)+2 \sum_{m=1}^\infty\frac{\Gamma(2m)}{\Gamma(m+1)^2}z_0^m,\\
2\pi \sqrt{-1}\, \Pi_1&=
\frac{\partial}{\partial r_1}\left. \Pi(z_0,z_1;r_0,r_1)
\right|_{r_0=r_1=0}\\
&=
\ln(z_1)-\sum_{m=1}^\infty\frac{\Gamma(2m)}{\Gamma(m+1)^2}z_0^m
\\
&\quad +2\sum_{m=0,n=1}^\infty \frac{\Gamma(2n)}{\Gamma(-2m+n+1)\,\Gamma(n+1)\,\Gamma^2(m+1)}z_0^mz_1^n,\\
\eal 
\ee
\[
\bal
(2\pi \sqrt{-1})^2\, \Pi_D&=\left( \frac{\partial^2}{\partial r_1^2}+\frac{\partial^2}{\partial r_0 \partial r_1}\right)\left. w_0(z_0,z_1;r_0,r_1)\right|_{r_1=r_2=0}\\
&=\ln(z_1)^2+\ln(z_0)\ln(z_1)-\ln(z_0)\,
\sum_{m=1}^\infty\frac{\Gamma(2m)}{\Gamma(m+1)^2}z_0^m\\
&\quad +2\ln(z_0z_1^2)\sum_{m=0,n=1}^\infty
\frac{\Gamma(2n)}{\Gamma(-2m+n+1)\,\Gamma(n+1)\,\Gamma(m+1)^2}z_0^mz_1^n\\
&\quad
-\frac{2\pi^2}{3}-2\sum_{m=1}^\infty\frac{\Gamma(2m)}{\Gamma(m+1)^2}\left(\psi(2m)-\psi(m+1)
\right)z_0^m\\
&\quad +4\sum_{m=0,n=1}^\infty\frac{\Gamma(2n)\,\left(2\psi(2n)-\psi(m+1)-\psi(n+1) \right)}{\Gamma(-2m+n+1)\,\Gamma(n+1)\,\Gamma(m+1)^2}z_0^mz_1^n,
\eal 
\]
where $\Gamma(x)$ and $\psi(x)=\frac{d}{dx}\ln\,\Gamma(x)$ are the
usual gamma and digamma function. Physically, 
$\Pi_c$ is identified by mirror symmetry with the central charge of a 
D0-brane on $X_2$, while  $\Pi_0$ and $\Pi_1$ 
are identified with the central charges of the D2-branes 
$\CO_{C_1}(-1)$, $\CO_{\Sigma_0}(-1)$ 
wrapping the fiber $C_1$, and the section $\Sigma_0$ respectively. 
In section \ref{twotwo} their $K$-theory charges were denoted by 
$\Upsilon_{0},\Upsilon_1$. Since $\Pi_c=1$, the flat 
coordinates $\tilde s_i$ are given by 
\[ 
{\widetilde s}_i =\Pi_i, \qquad i=0,1.
\]
The fourth period 
 ${\tilde s}_D=\Pi_D$ will be associated similarly to a D4-brane on $X_2$, 
 which will be identified once $\Pi_D$ is expanded in terms
 of  flat coordinates near the LCS limit point. 
 
To determine the {\bf A}-model instanton corrections one inverts the
relations for $\tilde s_0$ and $\tilde s_1$. The series in $z_0$
appearing in $\tilde s_0$ can be summed up to an elementary function: 
\begin{eqnarray}
\label{eq:F1}
F_1(z)&=&\sum_{m=1}^\infty 
\frac{\Gamma(2m)}{\Gamma(m+1)^2}\,z^m\\
&=&\sum_{m=1}^\infty
\frac{\Gamma(m+\textstyle\half)}{2\sqrt{\pi}\,\Gamma(m+1)}\,\frac{(4z)^m}{m}\nonumber
\\
&=&2\int^z dt \left(-\frac{1}{4t}+\frac{1}{4t}\frac{1}{\sqrt{1-4t}}
\right)\nonumber \\ 
&=&-\ln(\textstyle \half+\half\sqrt{1-4z}),\nonumber
\end{eqnarray}
where for the second equal sign we used the duplication formula
$\Gamma(x)\,\Gamma(x+\textstyle{\half})=2^{1-2x}\sqrt{\pi}\,\Gamma(2x)$,
and for the third
$\sum_{n=0}^\infty\frac{\Gamma(\alpha+n)}{\Gamma(\alpha)\Gamma(n+1)}t^n=(1-t)^{-\alpha}$. Now
one can easily verify the inverse relation $q_0=\exp(2\pi \sqrt{-1}\,\tilde s_0 )$:
$$z_0=\frac{q_0}{(1+q_0)^2}.$$
Inverting the third equation 
in \eqref{eq:tD} iteratively, 
one finds for the first terms of $z_1$:
$$z_1=q_1\left(\frac{1}{(1+q_1)^2}+q_0-4q_0q_1+3q_0q_1^2-2q_0^2q_1+\cdots \right),$$
where $q_1=\exp(2\pi \sqrt{-1}\,\tilde s_1)$.
The $\cdots$ in the above formula denote higher degree 
terms in $q_0,q_1$. 

Substitution of these series in ${\tilde s}_D=\Pi_D$ 
gives the
following form of the A-model instanton series:
\be\label{eq:instantonseries}
{\tilde s}_D=\tilde s_1^2+\tilde s_0\tilde
s_1+\frac{1}{6}+\frac{-2}{(2\pi i)^3}\frac{\partial}{\partial
  \tilde s_1}\sum_{n_i\in\mathbb{N}} N(n_i)\, \mathrm{Li}_3(q_0^{n_0}q_1^{n_1}),
\ee
where $\mathrm{Li}_n(z)=\sum_{k=1}^\infty \frac{z^k}{k^n}$.
The constant term arises from the trigamma
function $\psi_1(x)=\frac{d^2}{dx^2}\ln\,\Gamma(x)$ evaluated at
$x=1$: $\psi_1(1)=\frac{\pi^2}{6}$.
Using equation \eqref{eq:largeRZ}, the 
 polynomial part of the above equation identifies 
\[
{{\tilde s}_D} = -Z(\CO_{S_1}(K_{S_1}/2))=
 -Z(\CO_{S_1}(-\Sigma_0-2C_1))
 \]
 Up to sign this is the central charge of a D4-brane supported on 
 the compact divisor $S_1\simeq \IF_2$, with Chan-Paton line bundle 
 $\CO_{S_1}(-\Sigma_0-2C_1)$. In the notation of Section 
 \ref{twotwo} its $K$-theory class is given by 
 \be\label{eq:Kclassdualperiod}
\Lambda^1=- [\CO_{S_1}(-\Sigma_0-2C_1)]= 2\Upsilon^1.
\ee
Recall that $\ch_2(\Upsilon^1)=0$,
 hence this D4-brane has no induced D2-brane charges. 
There is however an induced fractional D0-brane charge, equal to 
\[
\int_{S_1} \left(\ch_3(\CO_{S_1}(-\Sigma_0-2C_1)) +{1\over 24} c_2(X_2)\right) = \chi(\CO_{S_1}(-\Sigma_0-2C_1)) - {1\over 24} 
\int_{S_2} c_2(X_2) = {1\over 6},
\] 
using equations \eqref{eq:adjunction}. 

%The period ${\tilde s}_D$ can be
%derived from a prepotential $\mathcal{F}$. If one realizes the
%local surface $S_1$ by decompactifying the fibre  $\tilde s_E
%\to
%\infty$ of an elliptic Calabi-Yau fibration over $S_1$,
%the period corresponding to $S_1$ is obtained from the Calabi-%Yau
%prepotential by standard special geometry formulas 
%\cite{Chiang:1999tz}.
%However note that the basis of D4-brane 
%periods obtained by local mirror symmetry 
%does not coincide with the basis of D4-brane periods employed 
%in 
%Section \ref{twotwo}, as shown in equation 
%\eqref{eq:Kclassdualperiod}. 
%For general $N\geq 2$, the double logarithmic periods of the %local mirror 
%will be associated to the $K$-theory classes 
%\[ 
%{\Lambda^i}= \sum_{j=1}^{N-1}{\sf C}_{ij} \Upsilon^j,
%\] 
%and they will satisfy 
%\[
%\langle \Upsilon_i, \Lambda^j\rangle = {\sf C}_{ij}
%\]
%as opposed to \eqref{eq:onconditionsG}. 

The central charge $Z(\Gamma,t)$ of a BPS D-brane with compact support 
will be given by:
\be
\label{eq:centralcharge}
Z(\Gamma,t)=-r {\tilde s}_D+\sum_{i=0,1} Q_{2,i} \tilde s_i-Q_0.
\ee 
in terms of the D4-, D2-, and D0-brane charges $(r,Q_{2,i},Q_0)$. 

Following \cite{geom_eng}, 
the {\bf B} model field theory limit is a scaling limit 
in a neighborhood of a special point in the compactified 
complex structure moduli space of  the family of curves \eqref{eq:locmirrorB}. For the local $\IF_2$ 
model, the special point is the 
intersection point 
\[ 
z_0=0, \qquad z_1={1\over 4}
\]
between the discriminant 
\[ 
 (1-4z_1)^2-64 z_0z_1^2=0
\]  
of the family \eqref{eq:locmirrorB} and the boundary divisor 
$z_0=0$. The scaling limit is defined by
\be\label{eq:Bmodellimit}
\bal
z_1  ={1\over 4}\bigg(1+{2\pi^2 \epsilon^2 u\over {M_0}^2}\bigg) , 
\qquad z_0   = \epsilon^4 e^{4c_0},\\
\eal 
\ee
where $M_0$ is a fiducial fixed scale and $c_0$ an
arbitrary constant as in Section \ref{twotwo}, equation 
\eqref{eq:fieldlimitA}. 
The scale $M_0$ is related to the QCD 
scale $\Lambda$ by equation \eqref{eq:azero}, which in this case 
reads 
\be\label{eq:QCDscale}
\Lambda = {2M_0\over \pi} e^{2c_0}.
\ee
Then one  can show as in \cite{geom_eng} 
that the $\epsilon \to 0$ limit of the family of 
curves \eqref{eq:locmirrorB} is the family of 
Seiberg-Witten curves of $SU(2)$ gauge theory. 

Our main goal is to show that 
 the periods ${\widetilde s}_1$, ${\tilde s}_D$ 
reproduce the weak coupling 
gauge theory central charges of the W-boson and monopole, respectively $a(u/\Lambda^2)$ and
$a_D(u/\Lambda^2)$ in the $\epsilon\to 0$ limit. 
The weak coupling region of the field theory Coulomb branch 
is given by $\frac{|u|}{\Lambda^2}\gg 1$. 
%In this regime, the leading
%terms of the Seiberg-Witten periods are given by \cite{Seiberg:%1994aj}:
%
%\begin{eqnarray}
%\label{eq:weakaaD}
%&&a_{SW}(u)=\frac{\sqrt{2u}}{2}\left(1-\frac{1}{16} %
%\left(\frac{u}{\Lambda^2}\right)^{-2}+\dots\right),
%\\
%&&a^D_{SW}(u)=\frac{\sqrt{-1}}{\pi}\sqrt{2u}\left(\ln
%\left(\frac{8u}{\Lambda^2}\right)-2+\dots \right).\nonumber
%\end{eqnarray}
%
In this regime, equations \eqref{eq:Bmodellimit}, \eqref{eq:QCDscale}
imply that 
the positive powers of $z_0$ in the 
period expansions (\ref{eq:tD}) yield subleading nonperturbative 
corrections in the $\epsilon \to 0$ limit. Therefore in order 
to reproduce the leading weak coupling terms it suffices to truncate the 
period expansions to the terms containing only powers of $z_1$. 
 Fortunately, the remaining 
series in $z_1$ can be summed up to elementary functions. 
For $\tilde
s_1$ the resulting series is  $F_1(z)$  in (\ref{eq:F1}). For 
$\Pi_D$
one finds the series 
\begin{eqnarray}
F_2(z)&=&\sum_{m=1 }^\infty
\frac{\Gamma(2m)}{\Gamma^2(m+1)}\left(2\psi(2m)-\psi(m+1)-\psi(1)
\right) z^m,
\end{eqnarray}
which is a bit harder to evaluate. We rewrite it as
\begin{eqnarray}
F_2(z)&=&\left.\frac{\partial}{\partial r}\sum_{m=1 }^\infty
\frac{\Gamma(2(m+r))}{\Gamma(m+r+1)\Gamma(m+1)\Gamma(r+1)} z^m\right|_{r=0}\nonumber.\\
&=&
\left.\frac{\partial}{\partial r}
\frac{\Gamma(r+\half)}{2\sqrt{\pi}\,\Gamma(r+1)\,z^r}\int^z 4dt\left(
  -(4t)^{r-1}+(4t)^{r-1} \frac{1}{(1-4t)^{r+1/2}}\right)\right|_{r=0}.\nonumber
\end{eqnarray}
After taking the derivatives to $r$, and performing the integral,
$F_2(z)$ can be expressed as:
\begin{eqnarray}
\label{eq:F2}
F_2(z)&=&\textstyle\half \ln(\half+\half\sqrt{1-4z})^2-\half\,
\mathrm{Li}_2(4z)\\
&&-\mathrm{Li}_2(\half-\half\sqrt{1-4z}) +2\,\mathrm{Li}_2(1-\sqrt{1-4z}).\nonumber
\end{eqnarray}
In
principle, one can also derive similar functions for the series multiplying higher powers
of $z_0$, but these expressions will not be needed in the following.

Having found these functions, we now study the behavior 
of the periods \eqref{eq:tD}
 in the field theory limit $\epsilon \to 0$. 
Since the functions $F_1(z_1)$, $F_2(z_1)$ depend on 
$\sqrt{1-4z_1}$, in addition to the 
change of variables \eqref{eq:Bmodellimit}, one has to 
introduce a branch cut starting at $z_1=1/4$ and choose 
a specific branch of the square root. We will choose the branch 
\[ 
\sqrt{1-4z_1} = -\pi \sqrt{-1} {\epsilon \over M_0} \sqrt{2u}.
\]
Then one finds 
the following small $\epsilon$ expansions:
\[
\bal
 & F_1(z_1)
%\big|_{\sqrt{1-4z_1}  =-\pi\epsilon \sqrt{-1}\sqrt{2u}/M_0} 
= \ln(2)+\sqrt{-1}{\pi \epsilon\over M_0}\sqrt{2u}+\mathcal{O}(\epsilon^2),\nonumber \\
 & F_2(z_1)
%\big|_{\sqrt{1-4z_1}=-\pi\epsilon \sqrt{-1}\sqrt{2u}/M_0} 
=\frac{1}{6}\pi^2+\ln(2)^2-{\sqrt{-1}} {\pi\epsilon 
\over M_0}
 \bigg(-2+\ln\bigg(-{2\pi^2\epsilon^2 u\over M_0^2}\bigg)
\bigg)\sqrt{2u}+\mathcal{O}(\epsilon^2)\\
%&\qquad \qquad \qquad \qquad \qquad \qquad \ \ \,
% +\mathcal{O}(\epsilon^2),\nonumber
\eal 
\]
where the second line follows from $\mathrm{Li}_2(0)=0$,
$\mathrm{Li}_2(\frac{1}{2})=\frac{1}{12}\pi^2-\frac{1}{2}\ln(2)^2$, and
$\mathrm{Li}_2(1)=\frac{1}{6}\pi^2$. 
With these expansions one obtains:
\be\label{eq:periodslimit}
\bal
\tilde s_c & =1,\nonumber\\
 \tilde s_0 & = {2\over \pi \sqrt{-1}}\ln(\epsilon) 
 +\mathcal{O}(\epsilon^0), \nonumber\\
\tilde s_1 & = {\epsilon\over M_0}\sqrt{2u}+\mathcal{O}(\epsilon^2),\\
\tilde s_D & = {\epsilon\over M_0} {\sqrt{-1}\over \pi} 
\sqrt{2u}\,\left[\ln \bigg(-{8\pi^2 u \over M_0^2}\bigg) -
2c_0 -2\right]+\mathcal{O}(\epsilon^2).\nonumber
\eal
\ee

Now recall that the ratio $M_0/\epsilon$ is the string theory 
scale $M_s$, which is sent to $\infty$ as $\epsilon \to 0$. 
Then the period expansions \eqref{eq:periodslimit}
imply that the central charges 
$M_s{\tilde s}_c, M_s{\tilde s}_0$ are divergent
in the $\epsilon \to 0$ limit, while $M_s{\tilde s}_1$ has a finite 
limit 
\[
\lim_{\epsilon \to 0} M_s{\tilde s}_1 = 
\sqrt{2u} + \cdots = 2a_1(u)
\] 
where $\cdots$ are higher order terms in the weak coupling expansion parameter $u/\Lambda^2$. 
This expression is in agreement with the Coulomb branch 
flat coordinate $a(u)$ defined in \cite{Seiberg:1994rs} 
and differs by a factor of $2$ from the normalization 
chosen in \cite{Seiberg:1994aj}. The same flat coordinate 
was denoted by $a_1$ in Section \ref{twotwo}, 
equation \eqref{eq:fieldlimitA}. 

Moreover, choosing the branch $\ln(-u) = -\pi\sqrt{-1}+\ln(u)$ for 
the logarithm, the last expression in \eqref{eq:periodslimit}
shows that $M_s{\tilde s}_D$ also has a finite $\epsilon \to 0$ 
limit,
\[
\bal 
\lim_{\epsilon \to 0} M_s {\tilde s}_D
 & = {{\sqrt {-1}}\over \pi}\sqrt{2u}
\left[ -\pi\sqrt{-1}+\ln \left({8\pi^2}{u\over M_0^2}\right) - 2c_0-2\right]+\cdots \\
%& = a^D_{SW} + \sqrt{2u} {\sqrt{-1}\over \pi}
%\left( 2c_0+2\ln(2)-\pi \sqrt{-1}\right).
\eal
\]
Rewriting this expression in terms of the flat Coulomb 
branch coordinate $a_1$ yields 
\[
\bal 
\lim_{\epsilon \to 0} M_s {\tilde s}_D
& = 2 \left[{1\over 2}+{\sqrt{-1}\over \pi} 
\left(\ln\left(2\pi {a_1\over M_0}\right) -c_0-1\right)\right]a_1+\cdots\\
\eal
\] 
By comparison with equation \eqref{eq:DfourZE} in Section 
\ref{twotwo}, it follows that 
\[ 
\lim_{\epsilon \to 0} M_s {\tilde s}_D=2
\lim_{\epsilon \to 0} Z^{pv}(\Upsilon^1) =2 a_1^D,
\]
as expected from the $K$-theory relation $\Lambda^1=
2\Upsilon^1$.
As observed in equation \eqref{eq:gaugeprep} 
the dual period $a_1^D$ is derived from a gauge theory 
prepotential with classical coupling constant 
\[ 
\tau_0 = {1\over 2} + {\sqrt{-1}\over \pi} 
(c_0+2\ln\, 2-3/2),
\]
These expressions are in agreement with 
\cite{Seiberg:1994aj} up to the classical 
$\tau_0$ dependent terms. 
 %   As an illustration, we have plotted the magnitude of the periods %for $u=\exp(-0.02i)\Lambda^2$ 
   % and $e^{4c_0}=4\times 10^{-5}$ in
%Figure \ref{fig:magnitude}. %
%
%\begin{figure}[h!]
%\centering
%\includegraphics[totalheight=6cm]{magnitudeF2} 
%\caption{Magnitude of the periods $\Pi_c$, $\Pi_i$ and
 % $\Pi_D$ along the trajectory
  %from the large volume limit to the field theory limit, for
 % $u/\Lambda^2=\exp(-0.02i)$ and $z_0=4\times 10^{-5}$.}  
%\label{fig:magnitude}
%\end{figure}
%

The BPS particles of $SU(2)$ field theory correspond to those BPS D-branes
whose central charges are finite in the $\epsilon\to 0$ limit, 
taking into account the scaling $\alpha'=\epsilon^2/\Lambda^2$, 
in Section \ref{twotwo}. 
Thus we deduce from
Eqs. (\ref{eq:centralcharge}) and (\ref{eq:periodslimit}) that only
bound states of D4-branes and D2-branes with $K$-theory charges 
\[
-r \Lambda^1+ Q_{2,1}\Upsilon_1
\]
survive in
the field theory limit, the other charges being infinitely massive.
Indeed the central charge in the field theory is
\be
Z((n_m,n_e),u)=n_m a_1^D(u)+n_ea_1(u),
\ee
and comparing with (\ref{eq:centralcharge}) gives $n_m=r$, and $n_e=2Q_{2,1}$.

\section{Large radius stability and the
weak coupling BPS spectrum}\label{sectionthree}

 The main goal of the present section is to formulate 
 a precise conjectural relation between large radius 
  supersymmetric D-brane configurations and the 
  gauge theory BPS spectrum. Since the field theory 
  scaling region is far from the large radius limit, such a 
  correspondence will necessarily involve parallel transport
  of the BPS spectrum. As explained 
  below Fig. \ref{fig:moduli} in the introduction, in this 
  process one has to take into account possible marginal 
  stability walls separating these two regions. Therefore, 
  our {\it no walls conjecture}
  will claim the existence of a suitable path connecting the 
  the large radius limit point to the field theory region 
  which avoids all possible walls of marginal stability.    
 A consequence of the no walls conjecture is 
 a complete geometric construction
 for the weak coupling BPS spectrum of the gauge theory.
 
 A strong argument for the no walls conjecture is provided 
 in Section \ref{SUtwo} for $SU(2)$ gauge theory. 
 Further evidence is presented in Sections \ref{threefive} 
 and \ref{sectionfour} by explicit computations of $BPS$ 
 degeneracies in $SU(3)$ gauge theory.

\subsection{Large radius stability}\label{threeone}
In this section it will be assumed that  geometric Bridgeland stability conditions
on $D^b(X_N)$ exist and have a large radius behavior similar to
the ones constructed in
 \cite{stabKthree,locptwo}.
More precisely, let $B+\sqrt{-1}\omega$ be a fixed complex K\"ahler class,
and $\gamma\in K^0_{c}(X_N)$ a $K$-theory class
with compact support. Suppose $\gamma$ belongs to the effective cone i.e. it is the $K$-theory class of a sheaf. Then it will be assumed that for each $\gamma$
there exists $\lambda\in \IR_{>0}$
sufficiently large such that any object $F$ of $D^b(X_N)$ with
$[F]=\gamma$ is geometrically Bridgeland (semi)stable
with respect to $B+\sqrt{-1} \lambda \omega$
if and only if it is an $(\omega, B)$ Gieseker (semi)stable
sheaf\footnote{If there were a uniform bound for all $\gamma$ 
then Bridgeland stability would coincide with $(\omega,B)$
stability. But we do not expect this to be the case.}. 
Consequently we can define a large radius BPS spectrum.

Granting the above assumptions, this subsection will be focused on
basic properties of such sheaves on the toric threefolds $X_N$.
Recall  that the large radius central charge for any sheaf $F$ with compact support on $X_N$ is
\be\label{eq:ZofF}
Z_{(\omega,B)}(F)  = - \int_{X_N} e^{-(B+ \sqrt{-1}\omega)} \ch(F)
\sqrt{{\rm Td}(X_N)}.
\ee
Note that any sheaf $F$ with compact support must be a torsion
sheaf with set theoretic support contained in the divisor $S=\sum_{i=1}^{N-1}S_i$.
Therefore
\[
Z_{(\lambda\omega,B)}(F) =
Z_{(\omega,B)}^{(2)}(F)\lambda^2 + Z_{(\omega,B)}^{(0)}(F)
+\sqrt{-1} \lambda Z_{(\omega,B)}^{(1)}(F)
\]
where
\[
\bal
Z_{(\omega,B)}^{(2)}(F) &= {1\over 2}\int_{X_N} \omega^2\ch_1(F)\\
Z_{(\omega,B)}^{(1)}(F) &= \int_{X_N} \left(\omega\ch_2(F)
-\omega B \ch_1(F)\right) \\
Z_{(\omega,B)}^{(0)}(F) & = -\int_{X_N} \left(\ch_3(F)
+{1\over 2} \ch_1(F) {\rm Td}_2(X)\right)\\
\eal
\]
If $F$ has support of dimension two, 
$\ch_1(F)$ is a nontrivial effective divisor 
class, hence 
$Z_{(\omega,B)}^{(2)}(F)\neq 0$.
In this case one can define
\be\label{eq:LRslopes}
\mu_{(\omega,B)}(F) = {Z^{(1)}_{(\omega,B)}(F)
\over Z_{(\omega,B)}^{(2)}(F)}, \qquad
\nu_{(\omega,B)}(F) = {Z^{(0)}_{(\omega,B)}(F)
\over Z_{(\omega,B)}^{(2)}(F)}.
\ee
Gieseker (semi)stability  with respect to the pair
$(\omega,B)$ is defined by
the conditions
\be\label{eq:twistedstabA}
\mu_{(\omega,B)}(F') \ (\leq) \ \mu_{(\omega,B)}(F)
\ee
for any proper nontrivial subsheaf $0\subset F' \subset F$, and
\be\label{eq:twistedstabB}
\nu_{(\omega,B)}(F') \ (\leq) \ \nu_{(\omega,B)}(F)
\ee
if the slope inequality \eqref{eq:twistedstabA} is saturated.
%These inequalities can be concisely written as
%\be\label{eq:Zineq}
%{\widetilde  Z}_{(\lambda,B)}(F') \ (\leq) \ {\widetilde Z}_{(\lambda,B)}%(F), \qquad
%\lambda >> 0,
%\ee
%where
%\[
%{\widetilde  Z}_{(\lambda,B)}(F)= Z_{(\lambda,B)}(F) /
%Z^{(2)}_{(\omega,B)}(F)
%\]
%is the normalized central charge.
For simplicity $(\omega,B)$ Gieseker stability will be called 
$(\omega,B)$-stability in the following.
One can also define twisted
$(\omega,B)$-slope stability imposing only condition
\eqref{eq:twistedstabA}.
It is easy to check that the following implications hold
\[
(\omega,B){\rm -slope\ stable} \Rightarrow (\omega,B){\rm -stable}
\Rightarrow  (\omega,B){\rm -semistable} \Rightarrow
(\omega,B){\rm -slope\ semistable}.
\]
Moreover, if the numerical invariants and $(\omega,B)$
are sufficiently generic
there are no strictly semistable objects and
the two notions of stability coincide.
Finally, note that the main properties of
$(\omega,B)$ stability conditions are analogous to
those of standard Gieseker or slope stability conditions. That is,
all the standard filtrations exist and there exist projective
or quasi-projective moduli spaces of such objects. 
%For future reference, note that
%\be\label{eq:twistedslopes}
%\bal
%\mu_{(\omega,B)}(F) & = {\int \omega \ch_2(F) + \int \omega B %%\ch_1(F) \over {1\over 2} \int \omega^2 \ch_1(F)}\\
%\nu_{(\omega,B)}(F) & = {\int \ch_3(F) + (B+{\rm Td}_2(X))%%
%\ch_2(F)
%\over {1\over 2} \int \omega^2 \ch_1(F)}.\\
%\eal
%\ee
For completeness note that the
 numerical invariants of $F$ can be written as
\be\label{eq:chernofF}
\ch_0(F)=0, \qquad
\ch_1(F)=\sum_{i=1}^{N-1}m_iS_i,
\qquad \ch_2(F)=p\Sigma_{0} + \sum_{i=1}^{N-1}n_iC_i, \qquad
\chi(F)=n,
\ee
where $m_i,n_i,p,n\in \IZ$ with $m_i\geq 0$, $i=1,\ldots, N-1$.  It is easy to check that $\ch_1(F)^2$ is always even, hence $\ch_2(F)$ is integral. Using equations 
\eqref{eq:chisheaf}, \eqref{eq:chiRR}, 
the holomorphic Euler character of a sheaf $F$ on $X^N$ with compact two dimensional support  is given by 
\[
\chi(F) =\sum_{k=0}^{2} (-1)^k {\rm dim} H^k(X_N,F).
\]

We conclude the remaining part of this subsection with some technical
results on
$(\omega,B)$-slope semistable sheaves on $X_N$.
It will be shown that any $(\omega,B)$-slope semistable  sheaf $F$ with compact support of
dimension two
 must be the extension by zero of a sheaf on the reduced
 divisor $S=\sum_{i=1}^{N-1}S_i$. Such sheaves will be
 called scheme theoretic supported on $S$. From a physical point
 of view, these are D-brane bound states supported on $S$
 where the vacuum expectation value of the Higgs field
 parameterizing normal fluctuations within $X_N$ is trivial.
 Note that in general this might not be the case since multiple
 D-branes supported on $S$ may have nontrivial nilpotent
 Higgs field expectation values \cite{DKS}.
  Moreover, the moduli stack of $(\omega,B)$-slope semistable sheaves will be shown to be smooth.

In order to prove the first claim,
note that $S$ is the zero locus of the
 section
   \[
 s=\prod_{i=1}^{N-1} x_i \in H^0(X_N, \CO_{X_N}(S)).
 \]
A sheaf $F$ is scheme theoretically supported on $S$ 
if and only if the morphism 
 \[
 \xymatrix{
 F \ar[r]^-{\phi_s} & F(S)}
 \]
 determined by multiplication by $s$ is identically zero. 
 Below we show that this must be the case for an 
 $(\omega,B)$-slope semistable 
 pure dimension two sheaf with set theoretic support on $S$. 

Suppose $\phi_s$ is not identically zero, and let $I={\rm Im}(\phi_s)\subset F(S)$, $G = {\rm Ker }(\phi_s)\subset F$.
 Obviously there is an exact sequence
 \[
 0\to G \to F \to I \to 0
 \]
 which remains exact  under multiplication by any line bundle.
 In particular there is also an exact sequence
 \[
 0\to G(-S) \to F(-S) \to I(-S) \to 0
 \]
 where $I(-S) \subset F$. Note that $\phi_s$ cannot be an isomorphism
 since $F$ and $F(S)$ have different Chern classes. Hence the inclusion
 $I(-S)\subset F$ is strict. Moreover, since $F$ is set theoretically
 supported on $S$, $G$ cannot be trivial.
  Since $\phi_s\neq 0$ by assumption, 
   $G\subset F$ is strict as well.
 Then the $(\omega,B)$-slope semistability condition yields
 the inequalities
 \[
 \mu_{(\omega,B)}(G) \ \leq \mu_{(\omega,B)}(F),
 \qquad \mu_{(\omega,B)}(I(-S)) \ \leq \mu_{(\omega,B)}(F).
 \]
 Using the above exact sequences and expressions \eqref{eq:LRslopes},
 a straightforward computation
 shows that these inequalities yield
  \[
 \int_{X_N} \omega \ch_1(I) S \geq 0.
 \]
 However, $\ch_1(I) = \sum_{i=1}^{N-1} m'_i S_i$ for some
 $m'_i\in \IZ_{\geq 0}$, $i=1,\ldots, N-1$, since $I$ must be
 pure dimension two supported on $S$. Using the intersection 
 products 
 \[
  (S_i\cdot S_j)_{X_N} = \Sigma_i \delta_{j,i+1} + \Sigma_{i-1} \delta_{j,i-1} - (\Sigma_{i-1}+\Sigma_i+2C_i)\delta_{j,i}
 \]
for $1\leq i,j\leq N-1$, one finds 
 \[
\int_{X_N} \omega \ch_1(I) S  = - m'_1\int_{\Sigma_{0}}\omega
 -m'_{N-1}\int_{\Sigma_{N-1}}\omega
 -2 \sum_{i=1}^{N-1}m_i'\int_{C_i}\omega.
  \]
 Since $\omega$ is a K\"ahler class, it follows that 
 \[ 
\int_{X_N} \omega \ch_1(I) S \leq 0 
\] 
 and equality holds if and only if all $m'_i=0$. This leads to a contradiction unless $I$ is the zero sheaf, which proves the claim.

This result implies that the moduli stacks
of $(\omega,B)$-slope semistable sheaves are smooth.
Since any such sheaf $F$ is scheme theoretically supported 
on $S$, it suffices to prove that 
${\rm Ext}^2_{S}(F,F)=0$.
This follows using the fact that Serre duality 
holds on $S$ since $S$ is a divisor in $X_N$,
and the dualizing sheaf is given by the adjunction formula
\be\label{eq:dualizing}
\omega_S \simeq \CO_S(S).
\ee
Then one has an isomorphism
\[
{\rm Ext}^2_S(F,F) \simeq {\rm Ext}_S^0(F,F(S))^\vee
\]
for any coherent $\CO_S$-module $F$. Moreover it is straightforward
to check by a slope calculation as above that 
\be\label{eq:twistslope}
\mu_{(\omega,B)}(F(S)) <
\mu_{(\omega,B)}(F).
\ee
Since $F$ is assumed $(\omega,B)$-slope semistable,
this implies that ${\rm Ext}_S^0(F,F(S))=0$, hence 
${\rm Ext}_S^2(F,F)=0$ as well. In order to prove the last claim, 
note that if $F$ is $(\omega,B)$-slope semistable then 
$F(S)$ has the same property. Suppose 
$\phi: F \to F(S)$ is a nontrivial morphism. Then 
$$\mu_{(\omega,B)}(F) \leq \mu_{(\omega,B)}({\rm Im}(\phi)) 
\leq \mu_{(\omega,B)}(F(S)),$$
contradicting inequality \eqref{eq:twistslope}.

\subsection{An example: $SU(2)$ gauge theory}\label{SUtwo}
As an example, in this section we will analyze the large radius BPS spectrum for the $N=2$ geometry and explain its relation to the 
$SU(2)$ gauge theory spectrum. An important point
 is that a priori one does not expect a one-to-one 
correspondence between large radius and field theory BPS states, because 
the field theory limit involves analytic continuation in the complex K\"ahler 
moduli space as explained in detail in Section \ref{localmirror}. Hence these two regions of the moduli space could be in principle separated by marginal 
stability walls, leading to a complicated relation between the 
two spectra. As also claimed in \cite{DGS}, it will be shown here for 
$N=2$ that 
such walls are absent for all finite mass BPS states in the field theory limit. 

It was shown in Sections \ref{twotwo}, \ref{localmirror} that  only 
$K$-theory charges of the form 
\be\label{eq:rQcharges}
\Upsilon_{r,Q} =-r {\Lambda}^1 + Q\Upsilon_1
\ee
can support finite mass $BPS$ states in the field theory limit, 
where 
\[ 
\Upsilon_1= [\CO_{C_1}(-1)],\qquad 
{\Lambda}^1 = -[\CO_{S_1}(-\Sigma_0-2C_1)],
\]
and $r,Q\in \IZ$. 
They correspond to the $W$-boson and anti-monopole charge respectively. 

As explained in Section \ref{threeone}, in the large radius limit, 
such a charge $\Upsilon_{r,Q}$ supports BPS states only if there exists 
at least one Gieseker semistable stable sheaf $F$ on $X_2$ with 
$[F]=\Upsilon_{r,Q}$. In particular, $r\geq 0$. According to Section \ref{threeone},
any such sheaf must be the extension by zero of a Gieseker semistable stable
sheaf $E$ on $S_1=\IF_2$
 with numerical invariants 
\be\label{eq:SUtwocharges}
\ch_0(E)=r,\qquad \ch_1(E) = -r\Sigma_0+(Q-2r)C_1, \qquad 
\chi(E)=0. 
\ee
For $r=0$ such a sheaf is semistable if and only if $E=\CO_{C_1}(-1)^{\oplus Q}$, $Q\geq 0$,  
where $C_1$ is a fiber of  $S_1=\IF_2$. For $Q=1$ the moduli space of 
stable sheaves is isomorphic to $\IP^1$, and the protected spin character, 
\[
\Omega(\Upsilon_{0,1};y) = y+y^{-1}.
\]
These  BPS states have the quantum numbers of a massive $W$-boson. 
 If $Q>1$, all semistable objects are isomorphic to direct sums 
 $E=\CO_{C_1}(-1)^{\oplus Q}$, which implies that there are no 
 $Q>1$ bound states. 

For $r>0$, $E$ must be a torsion free Gieseker semistable sheaf on $S_1$. For simplicity suppose the charge vector is primitive such that 
$E$ must be automatically stable. 
Then a  standard argument shows that its endomorphism ring is 
\[
{\rm Ext}^0_{S_1}(E,E) \simeq \IC.
\]
Moreover, as shown in the last paragraph of Section 
\ref{threeone}, 
\[
{\rm Ext}^2_{S_1}(E,E) = 0.
\]
Then the moduli space of stable sheaves is smooth, and 
its dimension follows from the Riemann-Roch theorem 
\[
\bal
{\rm dim}{\rm Ext}^1_{S_1}(E,E) & = 
1- \int_{S_1} \ch(E^\vee\otimes_{S_1} E) {\rm Td}(S_1) \\
& = 
1- \int_{S_1}(r^2-\ch_1(E)^2+2r\ch_2(E)){\rm Td(S_1)}.\\
\eal
\]
Equations \eqref{eq:SUtwocharges} imply 
\[ 
\int_{S_1}\ch_1(E)^2 = 2r(r-Q), \qquad 
\int_{S_1}\ch_2(E)= \chi(E) -{1\over 2} \int_{S_1}\ch_1(E)c_1(S) -
r\chi(\CO_{S_1}) = r-Q.
\]
Therefore 
\[ 
{\rm dim}{\rm Ext}^1_{S_1}(E,E) = 1-r^2
\]
for any value of $Q\in \IZ$. A nonempty moduli space is obtained 
only for $r=1$, in which case $E$ is a line bundle, 
$E\simeq \CO_{S_1}(-\Sigma_0+(Q-2)C_1)$. The moduli space 
is just a point and the protected spin character 
\[ 
\Omega(\Upsilon_{1,Q}; y)=1.
\] 
These states have the quantum numbers and degeneracies of weak coupling dyons with magnetic charge $1$. 
In conclusion the large radius BPS spectrum coincides with the weak 
coupling spectrum of $SU(2)$ gauge theory, at least for 
primitive charge vectors. More involved computations \cite{Manschot:2011ym}
show that at large radius there are no BPS states with non-primitive 
charge vector $r>1$, in agreement with 
the gauge theory BPS spectrum. 

An important conceptual point is that the one-to-one correspondence 
between large radius and gauge theory BPS states 
is not expected on general grounds. 
In principle the BPS spectrum could jump at marginal 
stability walls between large radius and the field theory scaling region. 
 Making a finiteness assumption, 
we will show below that for any charge 
$\Upsilon_{r,Q}$ with finite mass in the field theory limit,
there exists a path starting arbitrarily close to the LCS point 
$(z_0,z_1)=(0,0)$ and ending arbitrarily close 
to the center  $(z_0,z_1) =(0,1/4)$ of the field theory scaling region, 
such that the BPS degeneracy of $\Upsilon_{r,Q}$ 
is constant along $\gamma_u$.

We first construct an open
 path $\gamma_u$ in the complex structure 
moduli space
for any point $u=-|u|e^{i\theta}$ in the weak coupling region of 
the Coulomb 
branch such that $|u/M_0^2|>>1$ and $0<\theta <<1$. Such a path is determined by equations  \eqref{eq:Bmodellimit},
\[
z_1  ={1\over 4}\bigg(1+{2\pi^2 \epsilon^2 u\over {M_0}^2}\bigg) , 
\qquad z_0   = \epsilon^4 e^{4c_0},
\]
where $u,M_0,c_0$ are constant and 
\be\label{eq:trajectory}
0< \epsilon < {M_0\over \pi \sqrt{2|u|}}
\ee
is a parameter along the trajectory. Note that the upper 
end of the interval \eqref{eq:trajectory} may be made arbitrarily small by taking $|u|/M_0^2$ sufficiently large, which means 
$u$ sufficiently close to the semiclassical singular point on the 
Coulomb branch. Therefore by making a suitable choice of 
$u$ both $z_0$ and $z_1$ will be arbitrarily close to the 
LCS values $(z_1,z_0)=(0,0)$ when $\epsilon$ approaches 
the upper end of the interval. At the lower end the values of $(z_0,z_1)$ approach the center of the field theory scaling region, 
$(z_0,z_1)=(0,1/4)$. 
The absolute values and phases of the periods 
\eqref{eq:tD} are plotted below for a concrete choice of such a trajectory. In practice we will consider a closed path 
parameterized by $\xi\leq \epsilon \leq (1-\xi) 
M_0/ \pi \sqrt{2|u|}$ with $\xi>0$ a very small positive number. 
For example $\xi = 10^{-4}$. 
%\newpage
%
\begin{figure}[h!]
\centering
\includegraphics[totalheight=7cm]{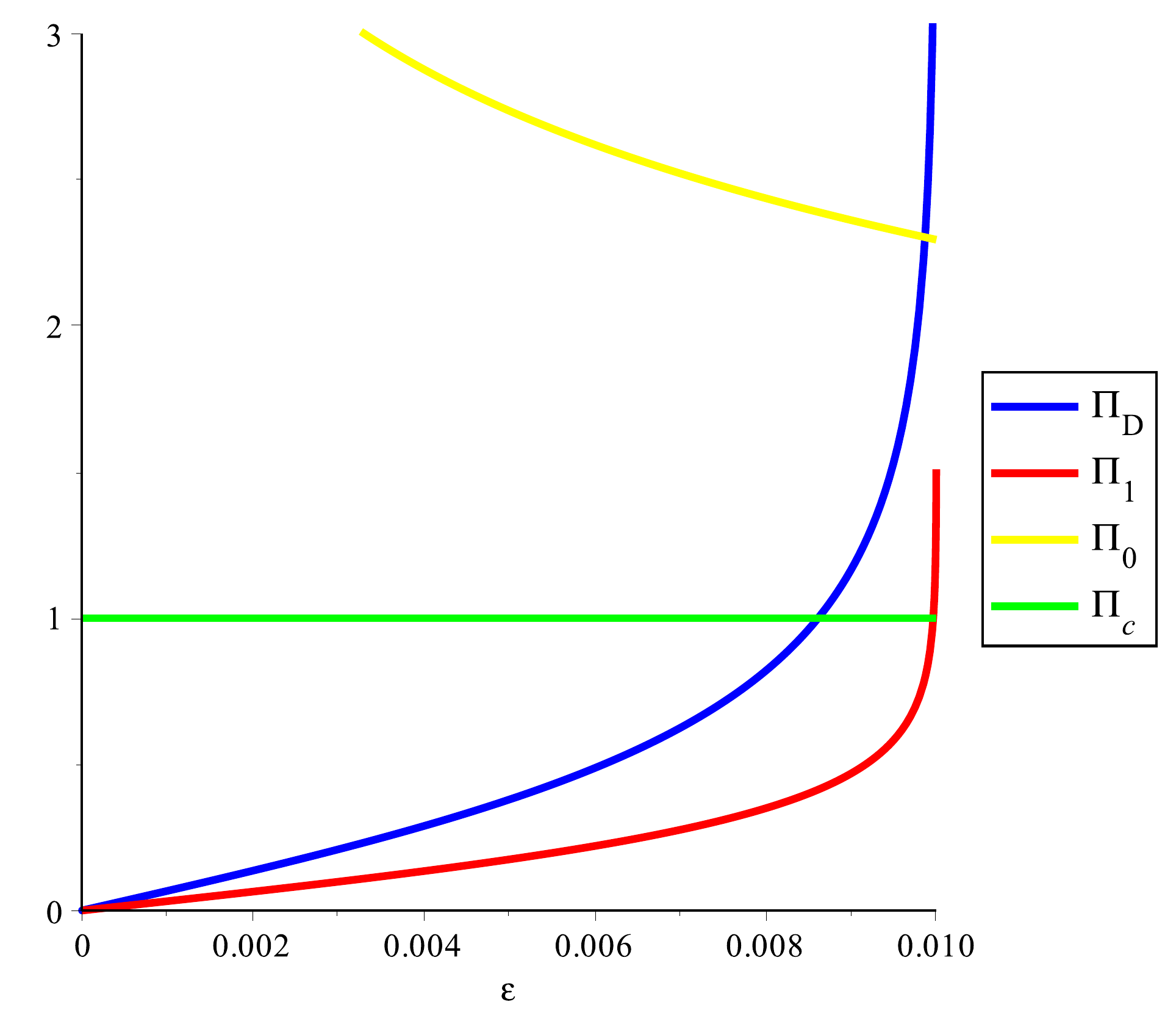} 
\caption{The magnitudes of the periods $\Pi_c$, $\Pi_i$ and
  $\Pi_D$ along the trajectory
  from the large volume limit to the field theory limit, for
  $2\pi^2\, u/M_0^2=-10^4$, $c_0=1$ and $\xi=10^{-4}$. }  
\label{fig:magnitude}
\end{figure}
\begin{figure}[h!]
\centering
\includegraphics[totalheight=7cm]{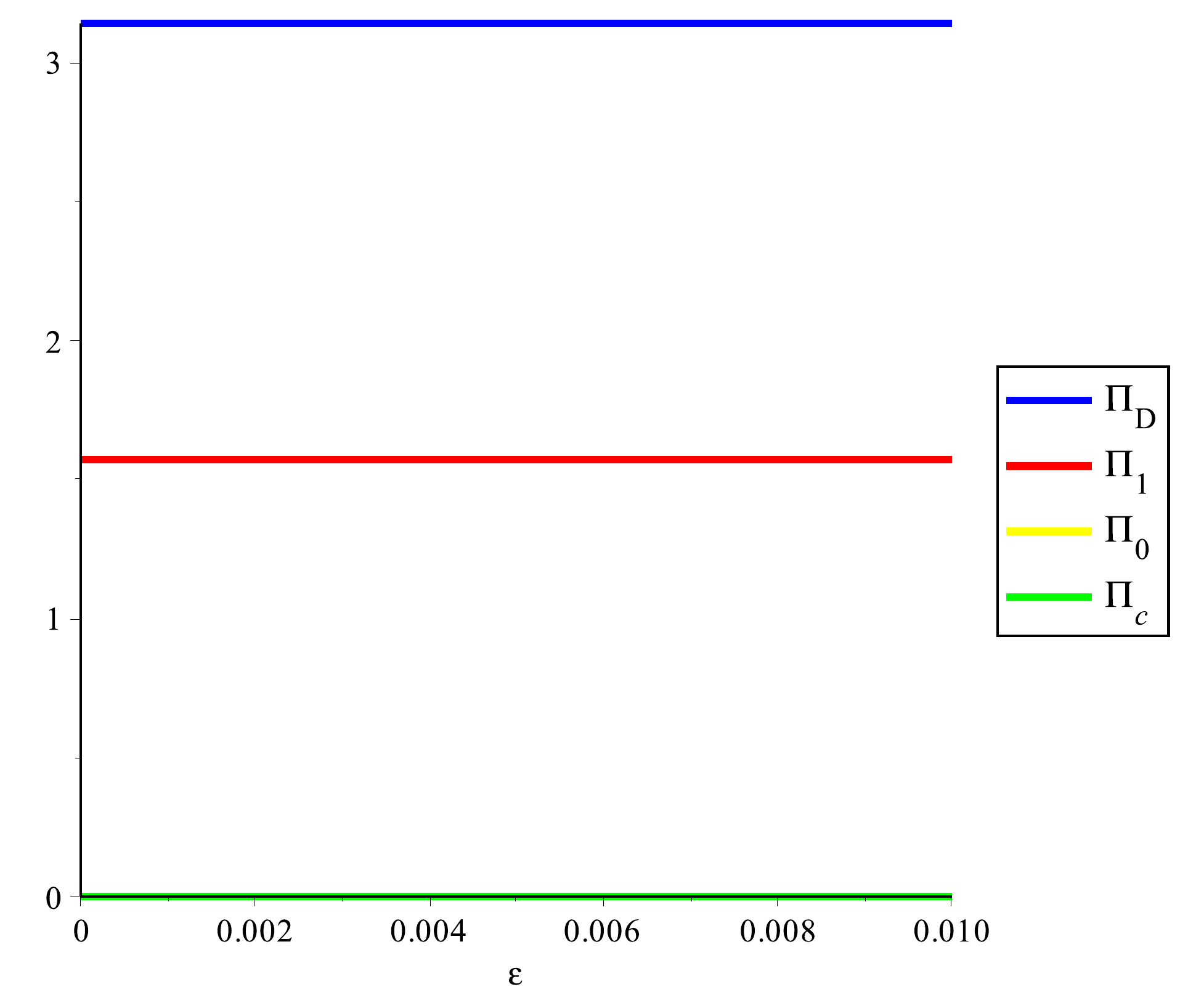} 
\caption{The arguments  of the periods $\Pi_c$, $\Pi_i$ and
  $\Pi_D$ along the trajectory
  from the large volume limit to the field theory limit, for
   $2\pi^2\, u/M_0^2=-10^4$, $c_0=1$ and $\xi=10^{-4}$.}  
\label{fig:argument}
\end{figure}

\noindent 
In particular, the phases of the 
periods $\Pi_c,\Pi_0,\Pi_1,\Pi_D$ are 
almost constant on the trajectory $\gamma_u$, and 
approximatively equal to 
\be\label{eq:phases}
0, {\pi\over 2}, {\pi\over 2}, {\pi} 
\ee
respectively. 
Moreover, numerical computations show that the 
maximum values of  
\[ 
|{\rm arg}(\Pi_D)+{\rm arg}(\Pi_c)-\pi|, \qquad 
|{\rm arg}(\Pi_0)- {\rm arg}(\Pi_1)|, \qquad |\Pi_1|/|\Pi_0|
\] 
over the trajectory can be made arbitrarily small by increasing $|u|/M_0^2$. 

Now, starting at the upper end of the interval, 
suppose the first wall of marginal stability for 
a BPS state of charge $\Upsilon_{r,Q}$ encountered 
along the trajectory $\gamma_u$
corresponds to some decomposition 
\be\label{eq:possiblewall}
\Upsilon_{r,Q} = \sum_{i=1}^n \Upsilon_{i}, \qquad 
\Upsilon_i= r_i\Lambda^1 + s_i \Upsilon_0 + Q_i \Upsilon_1
+ q_i [\CO_p],\quad 1\leq i \leq n,
\ee
$r_i,Q_i,s_i,q_i\in \IZ$ such that
the central charge $Z(\Upsilon_{r,Q})$ is aligned 
with all central charges $Z(\Upsilon_i)$.
Recall that under the current assumptions $\Upsilon_{r,Q}$
must have finite central charge in the field theory limit, hence 
either $r=0$ and $|Q|=1$ or $r=1$ and $Q\in \IZ$ arbitrary. 

Then there exist positive real numbers $\lambda_i\in \IR_{>0}$ 
 such that 
\[ 
r_i \Pi_D + s_i \Pi_0 + Q_i \Pi_1 + q_i = \lambda_i(r\Pi_D + 
Q \Pi_1)
\]
for all $1\leq i\leq n$ 
at any point on the marginal stability wall. 
Given the above behavior of the phases of 
 $\Pi_c,\Pi_0,\Pi_1,\Pi_D$ along a path $\gamma_u$, 
 by taking $|u|/M_0^2$ sufficiently large, it follows that 
 \be\label{eq:interspointA}
 s_i \Pi_0 + (Q_i-\lambda_i Q) \Pi_1 =0, 
 \qquad 
 (r_i-\lambda_i r)\Pi_D +q_i =0
 \ee
 for all $1\leq i\leq n$ at the intersection point with the wall. 
 Since the ratio $|\Pi_1|/|\Pi_0|$ can be made arbitrarily small by 
 a suitable choice of $|u| /M_0^2>>1$, it follows that 
 \be\label{eq:interspointB}
 s_i=0, \qquad Q_i = \lambda_i Q
 \ee
 for all $1\leq i \leq n$. Since $\lambda_i>0$, $n\geq 2$ and $\sum_{i=1}^n 
 Q_i =Q$, this rules out the case $r=0$, $Q=\pm 1$. 
  Therefore in order for such an intersection point to exist 
 one must have $r=1$ and $Q\in \IZ$. 
 
 Next we show that if $Q=0$ the BPS degeneracy 
 of the charge $\Upsilon_{r,Q}$ 
 does not jump across a wall of the form \eqref{eq:possiblewall}. 
 If $Q=0$, equation \eqref{eq:interspointB} implies that 
 $Q_i=0$ for all $1\leq i\leq n$. 
 The symplectic pairing of any 
 two $K$-theory classes $\Upsilon_i, \Upsilon_j$ with $s_i=s_j=0$ 
 is 
\[
\langle \Upsilon_{i}, \Upsilon_{j} \rangle= 2(r_jQ_i-r_iQ_j).
\]
Hence if $Q_i=Q_j=0$ the charges are orthogonal. 
Since this holds in this case for any pair of charges $\Upsilon_i, 
\Upsilon_j$, the Kontsevich-Soibelman wallcrossing formula 
 implies that the BPS degeneracy does not jump across such a 
 wall. This holds for refined BPS degeneracies as well. 
 
 Finally, suppose $r=1$ and $Q\neq 0$. Then equations 
 \eqref{eq:interspointA}, \eqref{eq:interspointB} 
 imply 
  \[
 (r_iQ-Q_i)\Pi_D+q_i=0
 \]
 at the intersection point for all $1\leq i\leq n$.  
  However $\Pi_D$ 
 is locally a holomorphic function, after choosing appropriate branch 
 cuts. Hence for given $(r,Q_i,q_i)$, the equations $(r_iQ-Q_i)\Pi_D+q_i=0$ can hold at most on a complex codimension one locus in the
 moduli space. Therefore there will exist 
 a smooth local deformation ${\tilde \gamma}_u$ 
 of the path $\gamma_u$ 
 in an arbitrarily small tubular neighborhood of the 
 marginal stability wall avoiding the subspace $(r_iQ-Q_i)\Pi_D+q_i=0$, $1\leq i\leq n$.  
 The above argument implies that such a perturbation cannot intersect 
 the wall. 
 
 The above argument applies to any marginal stability wall 
 of the form \eqref{eq:possiblewall}, with the 
 caveat that the required lower bound on $|u|^2/M_0^2$ 
 will depend on the wall.
Therefore, if the number of possible marginal stability walls 
of the form \eqref{eq:possiblewall} is finite, one can find a 
uniform lower bound on $|u|^2/M_0^2$ such that the argument applies simultaneously to all the walls. In this case, for 
sufficiently large $|u|/M_0^2>>1$, 
there will exist a smooth 
small deformation ${\tilde \gamma}_u$ 
of the path ${ \gamma}_u$ such that the 
BPS degeneracy of charge $\Upsilon_{r,Q}$ is constant along 
${\tilde \gamma}_u$. 

Motivated by this example we will formulate next 
an absence of walls conjecture for all values of $N\geq 2$.

\subsection{Limit weak coupling
spectrum and  absence of  walls}\label{threetwo}
Summarizing the facts, there are two distinct 
scaling limits of K\"ahler parameters, the field theory limit in Section \ref{twotwo}, and the large radius 
limit in Section \ref{threeone}. The basic idea of 
the absence of walls conjecture is that these two limits commute
in an appropriate sense, as far as the BPS spectrum is concerned. 
The goal of this section is to cast this idea in a precise mathematical 
form. 

First we define the  {\it limit weak 
coupling BPS spectrum} in gauge theory  by analogy with 
the large radius limit spectrum in string theory. 
Suppose $a=(a_i)$ as defined in Section \ref{twotwo}
is a fixed point on the Coulomb branch within the radius of convergence of the semiclassical
expansion such that 
\[ 
 \qquad {\rm Im}(a_i) >0, \qquad 1\leq i\leq N-1.
\]
In this section the units will be chosen such that $\Lambda=1$, hence 
the $a_i$ are dimensionless.
Let $\tau_0$, ${\rm Im}(\tau_0)>0$, be a 
fixed complex number in the upper half plane. 
The limit weak coupling spectrum will be defined 
as the large $\lambda$ limit of the BPS spectrum at points of the form 
\[ 
a_i(\lambda) ={\rm Re}(a_i)+ \lambda \sqrt{-1} {\rm Im}(a_i), \qquad 
\lambda\in \IR_{>0},
\]
on the Coulomb branch in a gauge theory with tree level coupling 
\[
\tau_0(\lambda)={\rm Re}(\tau_0) + \lambda \sqrt{-1} {\rm Im}(\tau_0).
\] 
In order to justify the existence of such a limit, note that the leading terms 
in the large $\lambda$ expansion of the dual periods $a_i^D$ 
derived from the prepotential \eqref{eq:gaugeprep} are 
\[ 
\bal
(a_i^D)_\lambda =  & - N{\rm Im}(\tau_0) \lambda^2 \sum_{j=1}^{N-1}{\sf C}_{ij}^{-1}
{\rm Im}(a_j)\\
& \ + N\lambda \sqrt{-1} \big[
{\rm Re}(\tau_0)   \sum_{j=1}^{N-1}{\sf C}_{ij}^{-1}
 {\rm Im}(a_j ) + {\rm Im}(\tau_0)\sum_{j=1}^{N-1}{\sf C}_{ij}^{-1}
 {\rm Re}(a_j) \big]+\cdots \\
 \eal
 \]
 The subleading terms are of order $\lambda\ln \lambda$ for the real 
part and $\ln \lambda$ for the imaginary part. The leading terms in the 
expansion of the stability parameters \eqref{eq:stabparametersA} 
are 
\be\label{eq:stabparametersB} 
\bal 
z_{i,\lambda} & 
=\big(\sum_{j=1}^{N-1}{\sf C}_{ij}(a_j^D)_\lambda  - \lambda (i+1)\sqrt{-1} {\rm Im}(a_i)\big)\\
w_{i,\lambda} & =
 \big(- \sum_{j=1}^{N-1}{\sf C}_{ij}(a_j^D)_\lambda  +
\lambda i \sqrt{-1} {\rm Im}(a_i)\big)\\
\eal 
\ee
Equations \eqref{eq:stabparametersB} determine a linear 
function on the charge lattice $\Gamma \simeq K^0(\CG)$ 
whose $\lambda$-dependence is of the form:
\[
Z_{(\tau_0,a,\lambda)}= \lambda^2 Z^{(2)}_{(\tau_0,a)} 
+ \lambda\sqrt{-1}Z^{(1)}_{(\tau_0,a)}  : \Gamma \to \IC.
\]

Recall that $\CG\subset D^b(X_N)$ is the triangulated subcategory 
spanned by the fractional branes $(P_i,Q_i)$, $1\leq i\leq N-1$.
The abelian category of $(Q,W)$-modules is the heart of a 
$t$-structure on $\CG$ and its $K$-theory is isomorphic to $\Gamma$. 
The same will hold for any quiver with potential $(Q',W')$ related to 
$(Q,W)$ by a finite sequence of mutations. Namely the $K$-theory 
of the abelian category of $(Q',W')$-modules 
 will be isomorphic to $\Gamma$. 
 Let $\Gamma_{(Q',W')}\subset \Gamma$ 
 be the cone spanned by the simple representations of $(Q',W')$ in the 
 charge lattice. 
In order to define the limit weak coupling spectrum 
we will make the following assumption:
\bigskip 

{\it For sufficiently generic $(\tau_0,a)$ there exists 
$\lambda_0>0$, 
depending on $(\tau_0,a)$, 
a quiver with potential 
$(Q,W)_{(\tau_0,a)}$, mutation equivalent to 
$(Q,W)$, and $\sigma \in \{-1,1\}$ such that
$\sigma Z^{(1)}_{(\tau_0,a,\lambda)}$ 
takes positive values on $\Gamma_{(Q,W)_{(\tau_0,a)}}\setminus \{0\}$. }
\bigskip 

Granting this assumption, for any sufficiently generic $(\tau_0,
a)$ there is a slope function 
\be\label{eq:limitslope}
\mu_{(\tau_0,a)} : \Gamma \to \IC, \qquad 
\mu_{(\tau_0,a)}(\gamma) = -
{ Z^{(2)}_{(\tau_0,a)}(\gamma)
\over Z^{(1)}_{(\tau_0,a)}(\gamma)} 
= -\sigma { Z^{(2)}_{(\tau_0,a)}(\gamma)
\over \big| Z^{(1)}_{(\tau_0,a)}(\gamma)\big|} 
\ee
Since the denominator takes positive values on the effective 
cone, this 
yields a well-defined stability condition for  $(Q',W')$-modules. 
A representation $\rho$ of the quiver with potential $(Q',W')$ 
with dimension vector $\gamma$ will be called 
$\mu_{(\tau_0,a)}$-(semi)stable if 
\[
\mu_{(\tau_0,a)}(\gamma') \ (\leq) \ \mu_{(\tau_0,a)}(\gamma)
\]
for any nontrivial proper subrepresentation $0\subset \rho'\subset \rho$ 
with dimension vector $\gamma'$. The moduli space of $\mu_{(\tau_0,a)}$-semistable representations of $(Q',W')$ with fixed charge $\gamma$ 
will be denoted by $\CM^{gauge}_{(\tau_0,a)}(\gamma)$. 
This moduli space defines the {\it limit weak coupling spectrum}.  
More precisely, the limit protected 
spin characters $\Omega_{(\tau_0,a)}^{gauge}(\gamma;y)$ are identified  
as explained in Section \ref{twofour} with refined counting invariants of 
these moduli spaces. 

As a concrete example, note that 
\[ 
\bal 
z_{i,\lambda} & =  \lambda^2 z^{(2)}_i + \lambda \sqrt{-1}  z^{(1)}_i  \\
w_{i,\lambda} & =  \lambda^2 w^{(2)}_i + \lambda \sqrt{-1}  w^{(1)}_i  \\
\eal 
\]
where 
\be\label{eq:largelambdaA}
\bal 
 z^{(2)}_i & = - N{\rm Im}(\tau_0) {\rm Im}(a_i) \\
z^{(1)}_i&=
N {\rm Re}(\tau_0) {\rm Im}(a_i) + N {\rm Im}(\tau_0) {\rm Re}(a_i) 
- (i+1) {\rm Im}(a_i)\\
w^{(2)}_i & =  N{\rm Im}(\tau_0) {\rm Im}(a_i) \\
w^{(1)}_i&=
- N {\rm Re}(\tau_0) {\rm Im}(a_i) - N {\rm Im}(\tau_0) {\rm Re}(a_i) 
+i {\rm Im}(a_i)\\
\eal 
\ee
for $1\leq i\leq N-1$. 
Next note that 
\[
z^{(1)}_{i}<0, \qquad w^{(1)}_{i} < 0,\qquad 1\leq i\leq N-1
\]
if the inequalities 
\be\label{eq:limitwedgeA} 
0< N {\rm Re}(\tau_0) {\rm Im}(a_i) + {\rm Im}(\tau_0){\rm Re}(a_i) 
-i{\rm Im}(a_i) < {\rm Im}(a_i) 
\ee
hold for all $1\leq i\leq N-1$. Therefore the main assumption formulated 
above holds in this case with $\sigma=-1$ if $a$ belongs to the 
wedge \eqref{eq:limitwedgeA}. 

The absence of walls conjecture will be formulated as an identification between the 
limit weak coupling spectrum defined above
and the limit large radius defined in Section \ref{threeone} 
spectrum in a specific $(\omega,B)$-stability chamber 
to be defined below. 

Recall that the complex K\"ahler class
was written as
\[
B + \sqrt{-1}\omega = (b_0+\sqrt{-1}t_0)H +
\sum_{i=1}^{N-1} (b_i+\sqrt{-1}t_i) D_i
\]
where $s_i=b_i+\sqrt{-1}t_i$, $i=0,\ldots, N-1$, are the
periods of $B+\sqrt{-1}\omega$ on the Mori cone generators
$\Sigma_{0}, C_i$, $i=1,\ldots, N-1$. 
The field theory limit was defined in Section
\ref{twotwo} by setting 
\be\label{eq:btepsilon}
\bal
b_0+t_0\sqrt{-1}
& ={-N\over \pi}(c_0+\ln \epsilon ) \sqrt{-1},\\
 b_i+\sqrt{-1}t_i & = \epsilon
{a_i\over M_0}\\
\eal
\ee
for $1\leq i\leq N-1$, with $0<\epsilon < e^{-c_0}$, and 
 keeping the leading terms
of the central charge \eqref{eq:ZofF} in the small $\epsilon$ expansion. 
Since $\Lambda$ has been set to $1$, 
equation \eqref{eq:azero} yields
\[
M_0 = 2\pi \left(2{e^{c_0}}\right)^{-N/(N-1)}
\]
The parameter  
 $c_0>0$ is related to the 
field theory coupling constant $\tau_0$ by the equation 
\be\label{eq:tauzero}
\tau_0= {1\over 2} + {\sqrt{-1} \over \pi} \left({c_0+N \ln 2\over N-1}
-{3\over 2}\right)
\ee
obtained from \eqref{eq:gaugeprep}. 
This has led to the conclusion that finite
mass  BPS states
in this limit must be objects of the subcategory $\CG\subset D^b(X_N)$ generated by the fractional branes $(P_i,Q_i)_{1\leq i\leq N-1}$.
%Assuming the conjectural equivalence \eqref{eq:compactequiv},

Now note that $\CG$ is identified with the subcategory
of $D^b_{cpt}(X_N)$
defined by the orthogonality conditions
\be\label{eq:orthcondA}
{\rm RHom}_{X_N}(\CO_{X_N}(aH), F) =0, \qquad a=0,1.
\ee
This follows from the expression \eqref{eq:dercatequiv} 
of the tilting functor:
\[
F \mapsto {\rm RHom}_{X_N}(T,F), \qquad 
T = \bigoplus_{i=1}^N(\CL_i\oplus \CM_i). 
\]
For each $1\leq i\leq N$, the complexes of vector 
spaces 
${\rm RHom}_{X_N}(\CL_i,F)$, ${\rm RHom}_{X_N}(\CM_i,F)$ 
are quasi-isomorphic to 
\[ 
\bigoplus_{k\in \IZ} H^k(X_N, \CL_i^{-1}\otimes_{X_N} F)[-k],\qquad 
\bigoplus_{k\in \IZ} H^k(X_N, \CM_i^{-1}\otimes_{X_N} F)[-k].
\]
Then ${\rm RHom}_{X_N}(T,F)$ is a complex $\rho_F$
of $(\CQ,\CW)$-modules such that the $k$-th term of the 
complex $\rho_F^k$ has underlying vector spaces 
\[
H^k(X_N, \CL_i^{-1}\otimes_{X_N} F), \qquad
H^k(X_N, \CM_i^{-1}\otimes_{X_N} F)
\]
respectively at the nodes $p_i,q_i$, $1\leq i\leq N$. 
Therefore, for a sheaf $F$ satisfying conditions \eqref{eq:orthcondA}, all $\rho_F^k$ have trivial 
vector spaces at the nodes $(p_N,q_N)$. This implies
that $\rho_F$ belongs to the subcategory spanned by the fractional branes $(P_i,Q_i)$, $1\leq i\leq N-1$.

Next note that any compactly supported sheaf $F$ satisfying conditions \eqref{eq:orthcondA} 
must have numerical invariants
\be\label{eq:chF}
\ch_0(F)=0, \qquad
\ch_1(F)=\sum_{i=1}^{N-1}m_iS_i,
\qquad \ch_2(F)= \sum_{i=1}^{N-1}n_iC_i, \qquad
\chi(F)=0,
\ee
 where 
 $m_i,n_i,\in \IZ$ with $m_i\geq 0$, $i=1,\ldots, N-1$. If
 $F$ has two dimensional support, $\ch_1(F)\neq 0$, hence the
 $m_i$, $i=1,\ldots, N-1$
 cannot all be zero. In comparison with \eqref{eq:chernofF}, 
 note that the invariants $p,n$ in \eqref{eq:chernofF}
 must vanish.  
 
 Finally, note that the following result holds by a standard boundedness argument which will be omitted. 
 \bigskip 
 
 {\it For fixed $(c_0,a_i)$ and a fixed effective 
 $K$-theory class $\gamma \in 
 K_0(\CG)$ there exists 
 $0<\epsilon_0< e^{-c_0}$ 
 depending on $(c_0,a_i; \gamma)$ such that 
 the moduli space of $(\omega,B)$-semistable sheaves $F$ 
 with $K$-theory class $[F]=\gamma$ is constant
 (as a scheme) and independent 
 of $\epsilon$, for $0<\epsilon< \epsilon_0$. } 
 \bigskip 
 
 Let $\CM^{string}_{(\tau_0,a)}(\gamma)$ denote the small $\epsilon$ 
 limit of the moduli 
 space of $(\omega,B)$-semistable sheaves with $K$-theory 
 class
 $\gamma=[F]$. For classes of the form $\gamma=-[F]$, with a compactly 
 supported sheaf, let $\CM^{string}_{(\tau_0,a)}(\gamma)=\CM^{string}_{(\tau_0,a)}(-\gamma)$. If $\gamma$ is not of the form 
 $\pm [F]$ with $F$ a sheaf with compact support, by convention, 
 $\CM^{string}_{(\tau_0,a)}(\gamma)$ will be empty. 
 Similarly, let $\Omega_{(\tau_0,a)}^{string}(\gamma; y)$, 
 denote the small $\epsilon$ limit of the corresponding protected spin characters.  We extend the assignment 
 \[
 \gamma \mapsto   \Omega_{(\tau_0,a)}^{string}(\gamma; y)
 \]
 to a function on the whole $K$-theory lattice $K^0(\CG)$, setting 
 \[ 
 \Omega_{(\tau_0,a)}^{string}(\gamma; y) = \left\{
 \begin{array}{ll} 
  \Omega_{(\tau_0,a)}^{string}(\gamma; y), & {\rm if}\ \gamma=[F]\ 
  {\rm for\ some\ compactly\ supported\ sheaf}\ F,\\
  & \\
   \Omega_{(\tau_0,a)}^{string}(-\gamma; y), & {\rm if}\ \gamma=-[F]\ 
  {\rm for\ some\ compactly\ supported\ sheaf}\ F,\\
  & \\
  0, & {\rm otherwise}.\\
    \end{array}
  \right.
  \]
 
 Finally, the absence of walls conjecture states that: 
 \bigskip
 
 {\it For a given charge vector $\gamma\in \Gamma$, 
 there is a one-to-one correspondence
 \be\label{eq:nowallscorresp}
 \calP_\gamma :\big\{ \CC^{gauge}(\gamma)\} 
 \longto \big\{\CC^{string}(\gamma)\big\}
 \ee
 between quiver 
 stability chambers  $\CC^{gauge}(\gamma)$ with respect to 
 $\mu_{(\tau_0,a)}$-stability and geometric 
 stability chambers $\CC^{string}(\gamma)$ 
 with respect to small $\epsilon$ 
 $(\omega,B)$-stability such that there is an isomorphism of moduli spaces 
   \be\label{eq:nowallsmoduli} 
\CM^{gauge}_{\CC^{gauge}(\gamma)}(\gamma) \simeq 
\CM^{string}_{\calP_\gamma(\CC^{gauge}(\gamma))}(\gamma)
\ee
and relation of the form 
  \be\label{eq:nowallsconjA} 
  \Omega^{gauge}_{\CC^{gauge}(\gamma)}(\gamma; y) = \Omega_{\calP_\gamma(\CC^{gauge}(\gamma))}^{string}(\gamma; y) 
  \ee
  for any chamber $\CC^{gauge}(\gamma)$.}
  \bigskip
  
  Recall that the stability chambers $\CC^{gauge}(\gamma)$ 
  are subsets of the universal cover of the Coulomb branch. 
  Similarly, the chambers $\CC^{string}(\gamma)$ are 
  subsets of the universal cover of the complex K\"ahler moduli 
  space, which is parameterized by $(\omega,B)$ in a neighborhood 
of the large radius limit.
 \bigskip

 % Note also that
 %\be\label{eq:zerochiF}
 %\chi(F(-aH)) = \chi(F) =0
 % \ee
 %for all $a\in \IZ$ by a simple application of the Riemann-Roch theorem.
%Then the main result of this section is:
%\\

As a first consistency check of this conjecture, note that a priori the 
limit moduli spaces $\CM^{string}_{(\tau_0,a)}(\gamma)$ might 
include $(\omega, B)$-semistable sheaves $F$ {\it which do not belong} 
to the subcategory $\CG$. The orthogonality conditions 
\eqref{eq:orthcondA} are equivalent to the
vanishing results
\be\label{eq:cohvanishing}
H^k(X_N,F(-aH))=0, \qquad k=0,1,2, \qquad a=0,1.
\ee
Any semistable sheaf $F$ with $K$-theory class $[F]\in K^0(\CG)$
satisfies 
\be\label{eq:vanishingchi}
0=\chi(F(-aH)) = \sum_{k=0}^2 (-1)^k {\rm dim} H^k(X_N, F(-aH)),
\ee
where the first equality follows because $K^0(\CG)$ is 
generated by the $K$-theory classes $[P_i],[Q_i]$, $1\leq 
i\leq N-1$, and $\chi(P_i)=\chi(Q_i)=0$, $1\leq i\leq N-1$. 
(See equation \eqref{eq:chPQA} above.)
However,
the dimensions of the cohomology groups can be nontrivial. 
In fact, the 
dimensions of these groups are expected to jump as $F$ moves in the 
moduli space. Hence a priori the orthogonality conditions 
\eqref{eq:orthcondA} are not guaranteed to hold throughout the moduli space even for classes $\gamma \in K^0(\CG)$. 
The following result proves that they do in fact hold at all points 
of the limit moduli spaces $\CM^{string}_{(\tau_0,a)}(\gamma)$. 
\bigskip 

{\it For  fixed parameters $a_i$,
$i=0,\ldots, N-1$ satisfying
\be\label{eq:genericparam}
 \qquad {\rm Re}(a_i)\neq 0, \qquad {\rm Im}(a_i)>0,
\qquad i=1, \ldots, N-1
\ee
and fixed
$m_i,n_i\in \IZ$, there exists $0<\epsilon_1<1$
such that for all $0<\epsilon<\epsilon_1$, any $(\omega,B)$-semistable sheaf $F$ with
numerical invariants \eqref{eq:chF} satisfies conditions
\eqref{eq:orthcondA}. In particular all such objects belong to the 
subcategory $\CG\subset D^b(X_N)$.} 
\\

In order to prove this statement, recall that 
any pure dimension
two
$(\omega,B)$-semistable sheaf $F$ with compact support must be scheme theoretically supported on $S$. This implies
\[
H^k(X_N,F(-aH))
\simeq H^k(S,F(-aH))\]
 for all $k,a$.
Therefore the required vanishing results for $k=0$ follow if
one can prove that
\[
{\rm Hom}_{S}(\CO_S(aH), F) \simeq {\rm Hom}_{X_N}(\CO_S(aH), F),
\]
with $a=0,1$, vanishes.
Moreover note that Serre duality holds on $S$ since $S\subset X_N$ is a divisor, and
the dualizing sheaf is given by the adjunction formula
\[
\omega_S \simeq \CO_S(S).
\]
Hence
\[
\bal
 H^2(S,F(-aH)) & \simeq {\rm Ext}^2_{S}(\CO_S(aH),F)\simeq
 {\rm Ext}^0_S(F, \CO_S(aH+S))^\vee \\
 & \simeq
 {\rm Ext}^0_{X_N}(F, \CO_S(aH+S))^\vee,\\
 \eal
\]
where the last isomorphism holds because both $F$ and 
$\CO_S(aH+S)$ are extensions by zero of sheaves on $S$. 
Then the vanishing results for $k=2$ also follow if one can prove that
\[
{\rm Hom}_{X_N}(F, \CO_S(aH+S))=0, \qquad a=0,1.
\]
If the vanishing results \eqref{eq:cohvanishing} hold for
$k=0,2$, then equation \eqref{eq:vanishingchi} implies that they must also
hold for $k=1$.

In conclusion it suffices to prove that
\be\label{eq:homvanishing}
{\rm Hom}_{X_N}(\CO_S(aH),F)=0,\qquad
{\rm Hom}_{X_N}(F,\CO_S(S+aH))=0, \qquad a=0,1,
\ee
for any $(\omega,B)$-semistable sheaf $F$, provided that
$\epsilon>0$ is sufficiently small.

To explain the main idea of the proof, note that given any 
nontrivial morphism $\phi: \CO_S(aH)\to F$, the 
image subsheaf ${\rm Im}(\phi) \subset F$ must satisfy 
the stability condition 
\[
\mu_{(\omega,B)}({\rm Im}(\phi)) \leq \mu_{(\omega,B)}(F). 
\]
At the same time ${\rm Im}(\phi)$ is a nontrivial 
quotient of $\CO_S(aH)$. Below we will show that for sufficiently 
small $\epsilon>0$ any 
nontrivial quotient of $\CO_S(aH)$ violates the above 
slope inequality. This implies that there cannot 
exist nontrivial morphisms $\phi: \CO_S(aH)\to F$. 
The argument for the second vanishing result in 
\eqref{eq:homvanishing} is similar. Given any nontrivial 
morphism $\psi:F \to \CO_S(aH+S)$, its image 
${\rm Im}(\psi)$ is simultaneously a quotient of 
$F$ and a subsheaf of $\CO_S(aH+S)$. In particular it 
must satisfy 
\[ 
\mu_{(\omega,B)}({\rm Im}(\psi)) \geq \mu_{(\omega,B)}(F).
\]
We will show below that this yields a contradiction for 
sufficiently small $\epsilon>0$. 

To this end, using 
equations \eqref{eq:intrelations},\eqref{eq:LRslopes}, \eqref{eq:btepsilon} and \eqref{eq:chF}, 
the leading term of the slope
$\mu_{(\omega,B)}(F)$ in the small
$\epsilon$ expansion is
\be\label{eq:Fslope}
\mu_{(\omega,B)}(F)\sim -{\sum_{i=1}^{N-1} m_i{\rm Re}(a_i)
\over \sum_{i=1}^{N-1} m_i {\rm Im}(a_i)}.
\ee
Next note that
\[
\ch_1(\CO_S(aH)) = \sum_{i=1}^{N-1}S_i,
\qquad
\ch_2(\CO_S(aH)) = \Sigma_{0} + \sum_{i=1}^{N-1}
(i+1+a)C_i
\]
\[
\ch_1(\CO_S(S+aH)) = \sum_{i=1}^{N-1}S_i,
\qquad
\ch_2(\CO_S(S+aH)) = -\Sigma_{0} - \sum_{i=1}^{N-1}
(i+1-a)C_i.
\]
Moreover
any pure dimension two quotient $\CO_S\twoheadrightarrow G$ must be the structure sheaf, $G\simeq \CO_Z$, of a
closed subscheme $Z\subseteq S$. Since $S$ is a reduced divisor defined by the equation
\[
\prod_{i=1}^{N-1} x_i =0
\]
any such subscheme will be a reduced divisor
$S_{j_1,\ldots, j_k}$ defined by an equation
of the form
\[
\prod_{j=1}^k x_{i_j}=0
\]
for some subset $\{j_1,\ldots, j_k\} \subset {1,\ldots, N-1}$.
A straightforward computation shows that
\[
\ch_1(\CO_{S_{j_1,\ldots, j_k}}) = \sum_{j=1}^k S_{i_j},
\]
and
\[
\ch_2(\CO_{S_{j_1,\ldots, j_k}}) =
p(j_1,\ldots, j_k) \Sigma_{0} + \sum_{i=1}^{N-1}
n_i(j_1,\ldots, j_k)C_i
\]
where $p(j_1,\ldots, j_k)$ is the number of connected components
of $S_{j_1,\ldots, j_k}$. 

In order to prove the last statement, note that 
\[
\ch(\CO_{S_{j_1,\ldots, j_k}}) = 1-e^{-S_{j_1,\ldots, j_k}} 
\]
since $S_{j_1,\ldots, j_k}$ is a divisor. Hence  
\[ 
\ch_2( \CO_{S_{j_1,\ldots, j_k}}) = 
-{1\over 2} \big({S_{j_1,\ldots, j_k}}\big)^2 
= -{1\over 2} \sum \big({S^c_{j_1,\ldots, j_k}}\big)^2
 \]
where the sum is over the connected components of 
${S_{j_1,\ldots, j_k}}$. Therefore it suffices to compute 
$\big({S_{j_1,\ldots, j_k}}\big)^2$ assuming 
$S_{j_1,\ldots, j_k}$ is  connected. 
This implies that $j_1,\ldots, j_k$ must be consecutive 
integers in the set $\{1,\ldots, N-1\}$ up to a permutation. 
Without loss of generality one can assume $j_{i+1}=j_{i}+1$ 
for all $1\leq i\leq k-1$. 
Then, using the linear relation  
\[ 
{S_{j_1,\ldots, j_k}}= \sum_{j=j_1}^{j_k} S_j,
\]
and the intersection products 
\[ 
(S_i\cdot S_j)_{X_N} = \left\{\begin{array}{ll}
\Sigma_i, &{\rm for}\ j=i+1 \\
\Sigma_j, &{\rm for}\ i=j+1 \\
-\Sigma_{i-1}-\Sigma_i - 2C_i, & {\rm for}\ i=j, \\
0, & {\rm otherwise} \\
\end{array}\right.
\]
one obtains 
\[ 
\bal 
\big({S_{j_1,\ldots, j_k}}\big)^2 & = 
\sum_{j=j_1}^{j_k} (S_j)^2 + 
2 \sum_{j=j_1+1}^{j_k-1} (S_j\cdot S_{j+1}) \\
& = - \sum_{j=j_1}^{j_k}(\Sigma_{j-1}+\Sigma_{j}+ 2C_j)
+ 2 \sum_{j=j_1+1}^{j_k-1} \Sigma_j\\
\eal
\]
By convention, the sum from $j_1+1$ to $j_k-1$ 
in the above equations is 0 if $j_1 +1 > j_k-1$. 
Using the relations $\Sigma_{j}=\Sigma_{j-1}+2jC_j$
recursively, it follows that 
\[
\big({S_{j_1,\ldots, j_k}}\big)^2 = -2\Sigma_0 +\cdots 
\]
where $\cdots$ is a linear combination of fiber classes $C_j$.
This proves the claim. 

The coefficients $n_i(j_1,\ldots, j_k)$
can be computed but explicit expressions are not needed in the following. The essential fact is that $p(j_1,\ldots, j_k)\geq 1$
for any nonempty subset $\{j_1, \ldots, j_k\}$.
Twisting by $\CO_{X_N}(S)$, it follows that
\[
\ch_1(\CO_{S_{j_1,\ldots, j_k}}(S)) =\sum_{j=1}^k S_{i_j},
\]
and
\[
\bal
\ch_2(\CO_{S_{j_1,\ldots, j_k}}(S)) & = \ch_2(\CO_{S_{j_1,\ldots, j_k}}) + (S\cdot S_{j_1,\ldots, j_k})_{X_N}\\
& =({\tilde p}(j_1,\ldots, j_k)-p(j_1,\ldots, j_k)) \Sigma_{0} + \sum_{i=1}^{N-1}
{\tilde n}_i(j_1,\ldots, j_k)C_i\\
\eal
\]
where ${\tilde p}_{j_1\ldots, j_k}$ is the number of connected
components of the intersection $S_{j_1,\ldots, j_k} \cap
\left(S\setminus S_{j_1,\ldots, j_k}\right)$.
Since $S$ is a linear chain of divisors, it is straightforward to check
that
\[
{\tilde p}(j_1,\ldots, j_k)-p(j_1,\ldots, j_k)\geq 0
\]
This implies that the kernel $K(j_1,\ldots j_k)={\rm Ker}(\CO_{S}(S)
\twoheadrightarrow \CO_{S_{j_1,\ldots, j_k}}(S))$ has
second Chern character
\[
\ch_2(K(j_1,\ldots, j_k)) =  q(j_1,\ldots, j_k) \Sigma_{0} -
\sum_{i=1}^{N-1}(i+1-{\tilde n}_i(j_1,\ldots, j_k))C_i
\]
with $q(j_1,\ldots, j_k) <0$.
In conclusion any quotient of $\CO_S(aH)$ must be of the form
$\CO_{S_{j_1,\ldots, j_k}}(aH)$ and the leading term of
its $(\omega,B)$-slope is
\be\label{eq:quotslope}
\mu_{(\omega,B)}(\CO_{S_{j_1,\ldots, j_k}}(aH))
\sim {M_0\over \epsilon} {p(j_1,\ldots, j_k)\over
\sum_{j=1}^{k} {\rm Im}(a_{j_k})}\qquad
p(j_1,\ldots, j_k)\geq 1.
\ee
Any nontrivial subsheaf of $\CO_S(S+aH)$ must be
of the form $K(j_1,\ldots, j_k)(aH)$ and the leading term of
its $(\omega,B)$-slope is
\be\label{eq:subslope}
\mu_{(\omega,B)}(K({j_1,\ldots, j_k})(aH))
\sim {M_0\over \epsilon} {q(j_1,\ldots, j_k)\over
\sum_{j=1}^{k} {\rm Im}(a_{j_k})}, \qquad
q(j_1,\ldots, j_k)\leq -1.
\ee
Therefore
\[
\mu_{(\omega,B)}(K({j_1,\ldots, j_k})(aH)) < \mu_{(\omega,B)}(F)
< \mu_{(\omega,B)}(\CO_{S_{j_1,\ldots, j_k}}(aH))
\]
 for sufficiently small $0<\epsilon <<1$.
 Moreover, given the numerical invariants \eqref{eq:chF},
 $F$ cannot be isomorphic to a quotient of $\CO_S(aH)$
 or to a subobject of $\CO_S(S+aH)$. 
 As explained below equation \eqref{eq:homvanishing}, 
 the above inequalities lead to a contradiction 
for sufficiently small $0<\epsilon <\epsilon_0$.  Since there are
finitely many subsets $\{j_1,\ldots, j_k\}
\subset \{1,\ldots, N-1\}$, the upper bound $\epsilon_0$
depends only on $a_i$, $m_i$, $i=1,\ldots, N-1$.

Since the presentation has been so far fairly general, it may be
helpful to study some examples in detail. The simplest example
is $N=2$, which was discussed in detail in Section 
\ref{SUtwo}. 
For $N=3$, the moduli spaces of stable sheaves in the field theory
region are very complicated for arbitrary numerical invariants
$(m_1,m_2)$. However, the case $(m_1,m_2)=(1,1)$
is tractable and will be treated next. 

\subsection{$SU(3)$ spectrum with magnetic
charges $(1,1)$}\label{threefive}
Using the results of Sections \ref{threeone} and \ref{threetwo},
any $(\omega,B)$-semistable sheaf $F$ must be
scheme theoretically supported on the reduced divisor
$S=S_1+S_2$ and have numerical invariants
\[
\ch_1(F)=m_1S_1+m_2S_2,\qquad \ch_2(F)=n_1C_1+n_2C_2,
\qquad \chi(F)=0.
\]
The $K$-theory class of such a sheaf will be denoted by 
$\gamma(m_1,m_2,n_1,n_2)\in \Gamma\simeq K^0(\CG)$.
The symplectic pairing of two $K$-theory classes 
$\gamma(m_1,m_2,n_1,n_2)$, $\gamma(m'_1,m'_2,n'_1,n'_2)$
is given by 
\be\label{eq:sympairingX} 
\bal
& \langle \gamma(m_1,m_2,n_1,n_2), \gamma(m'_1,m'_2,n'_1,n'_2)
\rangle = \\
& m_1'(n_2 -2n_1) + m_2'(n_1-2n_2) - m_1(n'_2-2n'_1) 
-m_2(n'_1-2n_2').\\
\eal
\ee

Suppose $F$ is such a sheaf with
$(m_1,m_2)\neq (0,0)$. Then there is an exact sequence
\be\label{eq:extensionA}
0\to F_1\to F \to F_2 \to 0
\ee
uniquely determined by $F$, with $F_1,F_2$ pure dimension two sheaves with scheme
theoretic support on $S_1,S_2$ respectively. Recall that $S_1\simeq \IF_2$, $S_2\simeq \IF_4$. 
Note that $F_2$ is the quotient of  $F\otimes_{X_3} \CO_{S_2}$
by its maximal dimension  subsheaf of dimension at most one, and
$F_1$ the kernel of the resulting projection $F\twoheadrightarrow F_2$. Recall that $\otimes_{X_3}$ denotes 
the tensor product of $\CO_{X_3}$-modules. 
Obviously,
\[
\ch_1(F_1)=m_1S_1,\qquad \ch_1(F_2)=m_2S_2,
\]
and
\[
\ch_2(F_1) = -p\Sigma_{1} + n_1C_1, \qquad
\ch_2(F_2) =p\Sigma_{1} + n_2C_2
\]
with $p\in \IZ$.

By analogy with \cite[Lemma 2.6]{homfly-pairs},
the adjunction formula yields an isomorphism
\be\label{eq:extgroup}
{\rm Ext}^1_{X_3}(F_2,F_1) \simeq {\rm Hom}_{S_2}(F_2,
F_1\otimes_{X_3}\CO_{S_2}(S_2)).
\ee
Therefore there is a one to one correspondence between  extension classes $e\in {\rm Ext}^1_{X_3}(F_2,F_1)$ and morphisms
$\phi_e: F_2\to F_1\otimes_{X_3}\CO_{S_2}(S_2)$.
Moreover, this correspondence is functorial. This implies
that given any subsheaf $F_2'\subseteq F_2$, the
class $e$ is in the kernel of the natural map
\[
{\rm Ext}^1_{X_3}(F_2,F_1) \to
{\rm Ext}^1_{X_3}(F'_2,F_1)
\]
if and only if $F_2'\subseteq {\rm Ker}(\phi_e)$.
In particular there is a commutative diagram
with exact rows and columns
\be\label{eq:extdiagA}
\xymatrix{
 &            & 0\ar[d] & 0\ar[d]  & \\
   &          & {\widetilde F_2} \ar[r]^-{1} \ar[d]
             &  {\widetilde F_2}\ar[d] & \\
 0\ar[r] & F_1\ar[r]\ar[d]_-{1} &F \ar[r]\ar[d] & F_2 \ar[r]\ar[d]
 & 0 \\
 0\ar[r] & F_1 \ar[r] & {\widetilde F}_1 \ar[r]\ar[d]
  & I \ar[r]\ar[d]& 0\\
  &            & 0 & 0  & \\
 }
\ee
where ${\widetilde F}_2= {\rm Ker}(\phi_e)$ and
$I={\rm Im}(\phi_e)$.  A slightly more involved argument analogous to \cite[Lemma 2.8]{homfly-pairs} shows that
${\widetilde F}_1$ must be pure of dimension two.
Since $I$ is a subsheaf of $F_1\otimes_{X_3}\CO_{S_2}(S_2)$, which is dimension one, supported on $\Sigma_{1}$,
\[
\ch_2(I) = q\Sigma_{1}
\]
for some $q\in \IZ$, $0\leq q\leq m_1$. This implies that
\[
\ch_2({\widetilde F}_1) = (q-p)\Sigma_{1}+n_1C_1,
\qquad
\ch_2({\widetilde F}_2) =(p-q)\Sigma_{1}+ n_2C_2.
\]
For future reference note that completely analogous considerations apply to the middle vertical column in
diagram \eqref{eq:extdiagA}. Namely there is an isomorphism
\be\label{eq:extgroupB}
{\rm Ext}^1_{X_3}(\wF_1,\wF_2) \simeq
{\rm Hom}_{S_1}(\wF_1,
\wF_2\otimes_{X_3}\CO_{S_1}(S_1)).
\ee
Hence the associated extension class
${\widetilde e}$ corresponds to a morphism
$\phi_{\widetilde e}: \wF_2\to
\wF_1\otimes_{X_3}\CO_{S_1}(S_1)
$ whose image will be denoted by ${\widetilde I}$.
Then a standard diagram chasing argument using the snake lemma proves that ${\widetilde I}\simeq I$.

Next recall that in the field theory limit
\be\label{eq:SUthreemoduli}
b_0+t_0\sqrt{-1}
={-3\over \pi}(c_0+\ln(\epsilon/2)) \sqrt{-1}, \qquad b_i+\sqrt{-1}t_i = \epsilon
{a_i\over M_0}, \qquad
i=1,2,
\ee
with $0<\epsilon < e^{-c_0}$, and 
\[
M_0 = 2\pi \left(2{e^{c_0}}\right)^{-3/2}
\]
since $\Lambda$ has been set to $1$. 
Moreover,
 $c_0$ is related to the 
field theory coupling constant $\tau_0$ by 
\[
\tau_0= {1\over 2} + {\sqrt{-1} \over \pi} \left({c_0+3\ln 2\over 2}
-{3\over 2}\right).
\]
Note that $t_i>0$, hence ${\rm Im}(a_i)>0$, for $i=1,2$. 
The coupling constant $\tau_0$ must belong to the upper half-plane, 
hence 
\[ 
c_0 > 3(1-\ln 2).
\]
In addition, 
we will also work in the wedge \eqref{eq:limitwedgeA} on the Coulomb branch, which specializes to 
\be\label{eq:limitwedgeB} 
\left(i-{3\over 2}\right) {{\rm Im}(a_i) \over {\rm Im}(\tau_0)}
< {{\rm Re}(a_i)} < 
\left(i-{1\over 2}\right) {{\rm Im}(a_i) \over {\rm Im}(\tau_0)},
\qquad i=1,2.
\ee
The $(\omega,B)$-slopes have a small $\epsilon$ expansion of the form 
\[
\bal
\mu_{(\omega,B)}(F_1)  \sim & -{p\over \epsilon}
{M_0\over m_1{\rm Im}(a_1)} 
- {{\rm Re}(a_1)\over {\rm Im}(a_1)},\qquad 
\mu_{(\omega,B)}({\widetilde F}_2)
\sim {(p-q)\over \epsilon} {M_0\over m_2{\rm Im}(a_2)} 
- {{\rm Re}(a_2)\over {\rm Im}(a_2)},\\
&\qquad 
 \mu_{(\omega,B)}(F)
 \sim -{m_1 {\rm Re}(a_1)+m_2{\rm Re}(a_2)\over 
 m_1 {\rm Im}(a_1)   +m_2{\rm Im}(a_2)}.\\
 \eal
 \]
 Then, taking the small $\epsilon$ limit with $(\tau_0,a_i)$ fixed 
 the  necessary
semistability conditions
\be\label{eq:neccond}
\mu_{(\omega,B)}(F_1) \leq  \mu_{(\omega,B)}(F), \qquad
\mu_{(\omega,B)}({\widetilde F}_2) \leq \mu_{(\omega,B)}(F)
\ee
yield 
\[
0\leq p \leq q.
\]
By construction we also have $q\leq m_1$ as explained above. Therefore
$p,q$ must satisfy the constraints 
\be\label{eq:pqconstr}
0\leq p \leq q \leq m_1
\ee
This is as far as we can go for arbitrary magnetic charges
$(m_1,m_2)$.

Now suppose $m_1=m_2=1$ and  $a_1,a_2$
generic, such that 
\[
{\rm Re}(a_i) \neq 0, \qquad i=1,2.
\]
In order to determine the moduli space, note that
the sheaves $F_i$ in \eqref{eq:extensionA}
must be extensions by zero of rank one
torsion free sheaves $E_i$ on the smooth divisors
$S_i$, $i=1,2$.
Any such sheaf must be a twisted ideal sheaf i.e.
$E_i = \CJ_i\otimes L_i$, where $L_i$ are
line bundles and $\CJ_i$ ideal sheaves of zero
dimensional subschemes $Z_i\subset S_i$ for $i=1,2$.
Then there are exact sequences of $\CO_{S_i}$-modules
\[
0\to E_i \to L_i \to \CO_{Z_i}\to 0
\]
for $i=1,2$.
As $\chi(F)=0$ by assumption, it follows that
\be\label{eq:zerogenusA}
\chi(L_1)+\chi(L_2)-\chi(\CO_{Z_1})-\chi(\CO_{Z_2})=0.
\ee
The Grothendieck-Riemann-Roch theorem for
the closed immersions $S_i\hookrightarrow X_3$
yields
\[
\ch_1(L_1) = -(p+1)\Sigma_{1}+n_1C_1, \qquad
\ch_1(L_2) = (p-1) \Sigma_{1} +(n_2-3)C_2.
\]
Since $S_1,S_2$ are rational, these relations uniquely determine
\[
L_1\simeq \CO_{S_1}(-(p+1)\Sigma_{1}+n_1C_1),\qquad
L_2\simeq \CO_{S_2}( (p-1) \Sigma_{1} +(n_2-3)C_2).
\]
Then the Riemann-Roch theorem yields
\[
\chi(L_1) = p^2-pn_1, \qquad \chi(L_2) = -2p^2+pn_2.
\]
Hence relation \eqref{eq:zerogenusA} is equivalent to
\[
-p^2+p(n_2-n_1) -\chi(\CO_{Z_1})-\chi(\CO_{Z_2})=0.
\]
Since $Z_1,Z_2$ are zero dimensional, this yields in particular
\be\label{eq:chargeineqA}
p(n_2-n_1-p)= \chi(\CO_{Z_1})+\chi(\CO_{Z_2})\geq 0.
\ee

Analogous considerations apply to the middle exact
column in diagram \eqref{eq:extdiagA}. The sheaves
${\widetilde F}_i$, $i=1,2$ must be extensions by zero
of rank one torsion free sheaves ${\widetilde E}_i$
on $S_i$, $i=1,2$, with
\[
\ch_1({\widetilde E}_1) = (q-p-1)\Sigma_{1}+n_1C_1,
\qquad
\ch_1({\widetilde E}_2) = (p-q-1)\Sigma_{1} + (n_2-3)C_2.
\]
Again, ${\widetilde E}_i\simeq {\widetilde L}_i\otimes_{S_i} {\widetilde J}_i$, where ${\widetilde L}_i$
are line bundles on $S_i$ and ${\widetilde J}_i$
ideal sheaves of zero dimensional subschemes
${\widetilde Z}_i\subset S_i$, $i=1,2$.
In complete analogy with \eqref{eq:chargeineqA},
the following must hold
\be\label{eq:chargeineqAB}
(q-p)(n_1-n_2-(q-p))= \chi(\CO_{{\widetilde Z}_1})+\chi(\CO_{{\widetilde Z}_2})\geq 0.
\ee

Now note that inequalities \eqref{eq:pqconstr} yield the following cases 
\begin{itemize} 
\item[$(i)$] $p=0$, $q=1$,
\item[$(ii)$] $p=1$, $q=1$,
\item[$(iii)$] $p=0$, $q=0$.
\end{itemize}
The moduli space is determined as follows in each case.

$(i)$ Since $p=0$ inequality \eqref{eq:chargeineqA} implies that
$Z_1,Z_2$ are empty. Therefore $E_i=L_i$, $i=1,2$.
The necessary conditions \eqref{eq:neccond} yield 
\[ 
{\rm Re}(a_2){\rm Im}(a_1) -
{\rm Re}(a_1){\rm Im}(a_2)<0.
\]
Since $m_1=m_2=1$ these conditions are also sufficient provided that 
the extension class $e$ is nonzero. 
This means that the extension group 
${\rm Ext}^1_{X_3}(F_2,F_1)$ must
 have dimension at least 1. The isomorphism
 \eqref{eq:extgroup} implies that
 \[
\bal
& {\rm Ext}^1_{X_3}(F_2,F_1)\simeq 
{\rm Hom}_{S_2}(L_2, L_1\otimes_{S_1} \CO_{\Sigma_{1}}(S_2))
\simeq H^0(\CO_{\IP^1}(n_1-n_2-1)).\\
\eal
\]
Therefore $n_1\geq n_2+1$.
The moduli space in this case is isomorphic
to the projective space
$\IP^{n_1-n_2-1}$ and the BPS degeneracies in this chamber are
\[
\Omega^{string}_{(\tau_0,a)}(\gamma(1,1,n_1,n_2);y) = y^{-(n_1-n_2-1)} P_y(\IP^{n_1-n_2-1}).
\]
Here $P_y(\IP^{n_1-n_2-1})$ denotes the Poincar\'e polynomial
of the projective space. The cohomology  is an irreducible $SL(2,\IC)$ representation
of highest weight $n_1-n_2-1$.
Moreover, the Hodge numbers are $h^{r,s}(\IP^{n_1-n_2-1})=\delta_{r,s}$ with $0\leq r,s\leq n_1-n_2-1$.
Therefore, according to Section \ref{twofour}, each of these states
has spin $j_{spin} = {n_1-n_2-1\over 2}$
and trivial $R$-charge.

$(ii)$ Since $p=q=1$, inequalities \eqref{eq:chargeineqAB}
 imply that ${\widetilde Z}_1, {\widetilde Z}_2$
 are trivial, hence ${\widetilde E}_i={\widetilde L}_i$,
 $i=1,2$.
 Again, necessary and sufficient stability 
 conditions are 
 \[ 
 {\rm Re}(a_2){\rm Im}(a_1) -
{\rm Re}(a_1){\rm Im}(a_2)>0,
\]
the extension class ${\widetilde e}$ being required to be nonzero.
Then 
 the extension group ${\rm Ext}^1_{X_3}(\wF_1,\wF_2)$ must
 have dimension at least 1. Using the isomorphism
 \eqref{eq:extgroupB}, 
 \[
\bal
{\rm Hom}_{S_1}({\widetilde L}_1,
{\widetilde L}_2\otimes_{S_2}\CO_{\Sigma_{1}})
\simeq H^0( \CO_{\IP^1}(n_2-n_1+2(q-p)-1)).
\eal
\]
Therefore $n_2\geq n_1+1$.
The moduli space in this case is isomorphic
to $\IP^{n_2-n_1-1}$ and the BPS degeneracies in this chamber are
\[
\Omega^{string}_{(\tau_0,a)}(\gamma(1,1,n_1,n_2);y) = y^{-(n_2-n_1-1)} P_y(\IP^{n_2-n_1-1}).
\]
Again, each of these states
has spin $j_{spin} = {n_2-n_1-1\over 2}$
and trivial $R$-charge.

$(iii)$ Since $p=q=0$, inequalities \eqref{eq:chargeineqA}, \eqref{eq:chargeineqAB}
 imply that $Z_1,Z_2$, as well as ${\widetilde Z}_1, {\widetilde Z}_2$
 are trivial. Hence  $E_i=L_i$ and ${\widetilde E}_i={\widetilde L}_i$,
 $i=1,2$. This implies that the sheaf $I$ in diagram \eqref{eq:extdiagA} 
 must be either pure dimension one or zero. Since $q=0$, $I$ has to be zero,
 which implies that the extension class $e$ is trivial. Then $E$ is isomorphic 
 to the direct sum $F_1\oplus F_2$, which cannot be stable for 
 generic K\"ahler parameters. Therefore in this case the moduli space is 
 empty for generic $a_1,a_2$.

 In conclusion the field theory region is divided in this case in two stability chambers
separated by the wall
\be\label{eq:wallA}
 {\rm Re}(a_2){\rm Im}(a_1) -
{\rm Re}(a_1){\rm Im}(a_2)=0.
\ee
The BPS spectrum with magnetic charges $(m_1,m_2)=(1,1)$ and arbitrary electric charges
$(n_1,n_2)$ is of type $(i)$ for $ {\rm Re}(a_2){\rm Im}(a_1) -
{\rm Re}(a_1){\rm Im}(a_2)<0$
and type $(ii)$ for $ {\rm Re}(a_2){\rm Im}(a_1) -
{\rm Re}(a_1){\rm Im}(a_2)>0$.

The above results will be compared with similar computations
based on algebraic stability conditions in the  Section 4. 
They can also be compared with semiclassical analysis of BPS 
states based on zeromodes of suitable Dirac-like operators 
on the moduli spaces of monopoles in $\mathbb{R}^3$. There is 
nice agreement with the results of \cite{Gauntlett:1999vc,
semiclassical}.

\section{The $SU(3)$ quiver at weak
coupling}\label{sectionfour}

The main goal of this section is to confirm the large radius
results for large radius $SU(3)$ 
BPS states with  $(m_1,m_2)=(1,1)$ 
by a direct analysis of moduli spaces of quiver representations. As a  byproduct of this approach similar results will
be derived for magnetic charges $(1,m)$, $m\in \IZ_{>0}$.

\subsection{General considerations}\label{fourone}
A representation $\rho$ of the $N=3$ truncated quiver $(Q,W)$
is a diagram of the form
\be\label{eq:quivrepA}
\xymatrix{
W_2 \ar@<.5ex>[rrr]|{c_2} \ar@<-.5ex>[rrr]|{d_2}
& & & V_2 \ar@/_1pc/[ddlll]|{r_{1}}
\ar@/^1pc/[ddlll]|{s_{1}}  \\
& &  & \\
W_1 \ar@<.5ex>[rrr]|{c_{1}} \ar@<-.5ex>[rrr]|{d_{1}}
\ar[uu]|{b_{1}}
& & & V_1 \ar[uu]|{a_{1}}  \\}
\ee
where $V_i,W_i$, $i=1,2$ are finite dimensional vector spaces. Using the notation introduced in Section \ref{twothree}, 
the dimensions of $V_i,W_i$, $i=1,2$ will be denoted by\footnote{Note that the same notation is used for the dimension 
of $V_1$ and the arrow $d_1$ in the quiver path algebra. 
The distinction will be clear from the context.}
\[
\dim(V_i) = d_i, \qquad \dim(W_i) =e_i, \qquad i=1,2.
\]
Using equation \eqref{eq:pairingBB}, 
the symplectic pairing on the 
$K$-theory lattice is 
\be\label{eq:sympairingY}
\bal  
\langle [\rho], [\rho']\rangle = & 2(d_1e_1'-d_1'e_1) + 2(d_2e_2'-d_2'e_2) + 2(e_1d_2'-e_1'd_2) \\
& + (d_2d_1'-d_1d_2')+(e_2e_1'-e_1e_2').\\
\eal
\ee

Abusing notation, the linear maps $\rho$ have been
denoted by the same symbols as the corresponding arrows
of the quiver diagram. The meaning will be clear from the context. The
potential \eqref{eq:potentialA} yields the relations
\be\label{eq:relationsA}
\bal
& r_1a_1=0, \qquad s_1a_1=0,\qquad b_1r_1=0,\qquad b_1s_1=0\\
& c_1r_1+d_1s_1=0,\qquad r_1c_2+s_1d_2=0\\
& a_1c_1-c_2b_1=0,\qquad a_1d_1-d_2b_1=0.\\
\eal
\ee

A very useful observation is that the horizontal rows
of the above quiver representation
are Kronecker modules
\[
\rho_i:\quad \xymatrix{
W_i \ar@<.5ex>[rrr]|{c_i} \ar@<-.5ex>[rrr]|{d_i}
& & & V_i ,}\qquad i=1,2.
\]
Some basic facts on such modules, their homological
algebra and Harder-Narasimhan filtrations are summarized
for completeness in Appendix \ref{appB}.
In particular equations \eqref{eq:Kexts}, \eqref{eq:Kdualexts}
imply that
the linear space of solutions $(a_1,b_1)$ to the relations
\[
 a_1c_1-c_2b_1=0,\qquad a_1d_1-d_2b_1=0
 \]
 is isomorphic to the space of morphisms ${\rm Ext}^0_{\CK}(\rho_1,\rho_2)$, while the linear space of solutions
 $(r_1,s_1)$ to the relations
 \[
  c_1r_1+d_1s_1=0,\qquad r_1c_2+s_1d_2=0
  \]
  is isomorphic to the dual
  vector space ${\rm Ext}^1_{\CK}(\rho_1,\rho_2)^\vee$,
where
 $\CK$ denotes the abelian category of Kronecker modules.
In conclusion there is a one-to-one correspondence between
representations $\rho$ of the quiver with potential \eqref{eq:quivrepA}
and data
\be\label{eq:Kroneckerform}
(\rho_1,\rho_2), \qquad (a_1,b_1)\in {\rm Ext}^0_{\CK}(\rho_1,\rho_2), \qquad
(r_1,s_1) \in {\rm Ext}^1_{\CK}(\rho_1,\rho_2)^\vee
\ee
satisfying the remaining relations
\be\label{eq:relationsB}
r_1a_1=0, \qquad s_1a_1=0,\qquad b_1r_1=0,\qquad b_1s_1=0.
\ee
For future reference
note also the
 isomorphisms of extension groups
\be\label{eq:isomext}
\bal
{\rm Ext}^1_{(Q,W)} (\rho_1,\rho_2) & \simeq
{\rm Ext}^0_\CK(\rho_1,\rho_2), \qquad 
 {\rm Ext}^1_{(Q,W)} (\rho_2,\rho_1) & \simeq
{\rm Ext}^1_\CK(\rho_1,\rho_2)^\vee,\\
\eal 
\ee
\be\label{eq:isomextB}
\bal
{\rm Ext}^k_{(Q,W)} (\rho_1,\rho_1) & \simeq
{\rm Ext}^k_\CK(\rho_1,\rho_1), \qquad 
 {\rm Ext}^k_{(Q,W)} (\rho_2,\rho_2) & \simeq
{\rm Ext}^k_\CK(\rho_2,\rho_2)^\vee,\\
\eal
\ee
where $k=0,1$. 
On the left hand side of the above equations, $\rho_1,\rho_2$ 
are representations of the $SU(3)$ quiver $(Q,W)$ with 
$V_2=W_2=0$, respectively $V_1=W_1=0$. In the 
right hand side, they are just Kronecker modules. 
The first is proven in Appendix \ref{extensions}, equations 
\eqref{eq:extcomplexC} and \eqref{eq:extcomplexD}. The proof of the second is similar, the details being left to reader.

Finally, note that in this section we will use the algebraic stability conditions constructed in Section \ref{twothree}, taking the stability parameters 
$(z_i,w_i)$ as in equation \eqref{eq:kingparameters}.

\subsection{$W$-bosons}\label{Wsection}
According to the field theory limit discussed in Section
\ref{twotwo}, at weak coupling the massive $W$-bosons 
are bound states of $D2$-branes wrapping the fibers of 
the compact divisors $S_1,S_2$. The goal of this section is 
to identify the corresponding quiver representations and 
the region in the parameter space of stability conditions 
$(\theta,\eta)=(\theta_i,\eta_i)$ where such bound states are stable. 
Recall that the parameters $(\theta,\eta)$ were 
introduced in Section \ref{twothree}, equation 
\eqref{eq:kingparameters}.

First note that the tilting functor \eqref{eq:dercatequiv} 
maps any compactly supported complex $F$ with numerical invariants 
\[
\ch_1(F) =\sum_{i=1}^{N-1}m_iS_i, \qquad 
\ch_2(F) = \sum_{i=1}^{N-1}n_i C_i, 
\qquad 
\chi(F)=0
\] 
to a complex of representations with dimension vector 
\be\label{eq:tiltdimvect}
d_i = n_i-im_i, \qquad e_i = n_i-(i+1)m_i, \qquad 1\leq i\leq N-1.
\ee 
In particular, an object with $m_i=0$ will be mapped to a complex 
of representations with $d_i=e_i=n_i$ for $1\leq i\leq N-1$. 

The massive $W$-bosons corresponding to the simple roots 
$\alpha_1,\alpha_2$ are D2-branes of the form $\CO_{C_1}(-1)$,
$\CO_{C_2}(-1)$ in the derived category which fit in 
exact triangles  
\[
P_i[-2] \to \CO_{C_i}(-1) \to Q_i[-2], \qquad i=1,2,
\]
where $P_i$, $Q_i$ are supported on the divisor $S_i$, $i=1,2$. 
Therefore the tilting functor \eqref{eq:dercatequiv}
maps $\CO_{C_i}(-1)$ to
$\rho_i[-2]$, where $\rho_i$, $i=1,2$, are $(Q,W)$-modules
of  dimension vectors $(d_i,e_i)=(1,0,1,0)$, respectively 
$(d_i,e_i)=(0,1,0,1)$.
Note that the fractional branes $Q_i$ are not to be confused with 
the Kronecker modules $Q_n$ introduced in Appendix 
\ref{appB}. 

 In addition the $W$ boson
corresponding to the positive root  $\alpha_1+\alpha_2$ is a
D2-brane wrapping a reducible vertical curve with components
$C_1,C_2$ meeting at a point. The corresponding object in the
derived category is $\CO_C(-1)$,
where $C$ is a complete intersection curve of the 
form $S\cap H$.
The tilting functor maps $\CO_C(-1)$ to
representation $\rho[-1]$, where $\rho$ is a
$(Q,W)$-module of dimension vector $(1,1,1,1)$.
Using standard results on Kronecker modules, as reviewed in
Appendix \ref{appB}, the representations $\rho_1,\rho_2$ are
$(\theta,\eta)$-stable if and only if conditions 
\be\label{eq:WchamberA}
\theta_i<\eta_i, \qquad i=1,2.
\ee
If conditions \eqref{eq:WchamberA} are satisfied $\rho_i$ are isomorphic
to Kronecker modules of the form $R_{p_i}$, $p_i\in \IP^1$,
$i=1,2$ given in Appendix \ref{appB}.
In each case, the
moduli space of stable representations
is isomorphic to $\IP^1$, and the BPS degeneracy $(-2)$, as expected.

For representations $\rho$ with dimension vector $(1,1,1,1)$ note that the
relations \eqref{eq:relationsA} imply that $r_1,s_1$ must be zero if $a_1$
or $b_1$ is nonzero, and $a_1,b_1$ are zero if $r_1$ or $s_1$ are nonzero. Therefore, using the isomorphisms \eqref{eq:isomext}, such a
representation must be either an extension
\be\label{eq:WextA}
0\to \rho'_1\to \rho \to \rho'_2\to 0
\ee
or
\be\label{eq:WextB}
0\to \rho'_2\to \rho \to \rho'_1\to 0
\ee
in the abelian category of $SU(3)$ quiver representations, 
where 
$\rho'_1,\rho'_2$ are representations 
with dimension vectors $(1,0,1,0)$, $(0,1,0,1)$ respectively. 
Note that $\rho_1',\rho_2'$ are identified at the same 
time with Kronecker modules of dimension vector $(1,1)$. 
As explained in Appendix \ref{appB},  
any such Kronecker module is either of the form $R_p$, $p\in \IP^1$,
or a direct sum of simple modules. Using relations \eqref{eq:relationsA}
and isomorphisms \eqref{eq:isomext}, it follows that $\rho$ is stable only if
$\rho_1'\simeq \rho_2'\simeq R_p$ for some $p\in \IP^1$.
Then the space of stability conditions $(\theta,\eta)$ satisfying
\eqref{eq:WchamberA} is divided into two regions as follows.

The first region consists of stability parameters satisfying
\be\label{eq:WchamberB}
\theta_i<\eta_i, \qquad i=1,2, \qquad
\eta_1+\theta_1 < \eta_2+\theta_2
\ee
In this case all nontrivial extensions of the form \eqref{eq:WextA}
are stable and all extensions of the form \eqref{eq:WextB} are unstable.
Since ${\rm Ext}^1_{\CK}(R_p,R_{p'})\simeq \delta_{p,p'}\IC$, 
one obtains a nontrivial extension in \eqref{eq:WextA} only if 
$\rho_1'\simeq \rho_2'\simeq R_p$ for some $p\in \IP^1$. 
Moreover, for each $p\in \IP^1$ there is a 
a unique nontrivial extension of the form \eqref{eq:WextA} 
up to isomorphism. Therefore the moduli space 
of stable extensions  is again isomorphic to $\IP^1$.
Using the correspondence explained in Section
\ref{twofour}, we obtain a spin ${1\over 2}$
multiplet with trivial $R$-charge, as expected. 

The second region is defined by
\be\label{eq:WchamberBB}
\theta_i<\eta_i, \qquad i=1,2, \qquad \eta_1+\theta_1 > \eta_2+\theta_2
\ee
In this case all nontrivial extensions of the form \eqref{eq:WextB}
are stable and all extensions of the form \eqref{eq:WextA} are unstable.
The
moduli space of stable extensions is again isomorphic to $\IP^1$, hence we obtain again a spin ${1\over 2}$ multiplet
with trivial $R$-charge.

In conclusion, if inequalities \eqref{eq:WchamberA} are satisfied, the BPS
spectrum contains three massive $W$ bosons with electric
charges $\alpha_1,\alpha_2,\alpha_1+\alpha_2$ and degeneracy $(-2)$.
However, the $W$ boson with electric charges $(2,2)$ is realized as a
bound state of the first two in two different ways corresponding to two regions separated by a marginal
stability wall.  The moduli space is isomorphic to $\IP^1$ on both sides
of the wall but it parameterizes different extensions of the
form \eqref{eq:WextA}, \eqref{eq:WextB}
in the two regions.

\subsection{Moduli spaces and stability chambers  for magnetic charges $(1,m)$}\label{fourtwo}

\subsubsection{Magnetic charges $(1,1)$}
This is the case studied in detail in Section \ref{threefive} at large
radius, where BPS configurations are $(\omega,B)$-stable sheaves
$F$ with Chern character
\[
\ch(F) = S_1+S_2 + n_1C_1+n_2C_2.
\]
It is straightforward to check that if
\be\label{eq:chargeineq}
n_i\geq (i+1)m_i, \qquad i=1,2,
\ee
the tilting functor
will map a sheaf $F$ to a one-term complex $\rho[-1]$,
where $\rho$ is a representation of the quiver \eqref{eq:quivrepA}
with dimension vector
\be\label{eq:tiltdimvectB}
d_i = n_i-im_i, \qquad e_i = n_i-(i+1)m_i, \qquad i=1,2.
\ee
This will be assumed in the following, as well as $m_1=m_2=1$, 
$e_i\geq 1$ for $i=1,2$.
Special cases where $e_1=0$ or $e_2=0$ can be easily treated analogously.

Note that if inequalities \eqref{eq:chargeineq} are not satisfied,
the tilting functor will map the sheaf $F$ with $(m_1,m_2)=(1,1)$
to an object of the form
$\rho[-2]$ where $\rho$ is a representation with dimension vector
\[
d_i=i-n_i, \qquad e_i =i+1-n_i.
\]
The stability analysis is similar and will be left to the reader.

Let $(\theta,\eta)=({\theta}_i,\eta_i)_{1\leq i\leq 2}$, be King stability parameters for $(Q,W)$-modules
of numerical type \eqref{eq:tiltdimvectB}. Therefore 
$(\theta_i, \eta_i)_{1\leq i\leq 2}$ satisfy the 
linear relations 
 \be\label{eq:kingparametersB}
\sum_{i=1}^2 (d_i{\theta}_i+e_i\eta_i)=0
\ee
in $\IR^4$. 
At the end of Section \ref{twothree} such stability parameters 
were denoted by 
$({\bar \theta}_i, {\bar \eta}_i)_{1\leq i\leq 2}$ 
in order to emphasize the difference with respect to more general 
stability parameters not satisfying equations \eqref{eq:kingparametersB}. In this section 
all stability parameters will be assumed to satisfy equations 
\eqref{eq:kingparametersB}, hence this notational distinction
will not be necessary.

As shown in Appendix \ref{Atwo}, if
\be\label{eq:quivchamberAB}
\eta_i>0, \qquad \theta_i < 0, \qquad
|\theta_i|< |\eta_i|, \qquad i=1,2,
\ee
the Kronecker modules $\rho_i$, $i=1,2$,
of any $(\theta,\eta)$-semistable module $\rho$
must be of the form
\be\label{eq:splitmoduleAB}
\rho_i \simeq \oplus_{j=1}^{h_i} Q_{k_{i,j}}^{\oplus r_{i,j}},\qquad  i=1,2
\ee
for some integers $h_i\geq 1, k_{i,j}\geq 0, r_{i,j}\geq 1$
such that
\[
k_{i,1}> k_{i,2}> \cdots > k_{i,h_i}\geq 0.
\]
Since $d_i-e_i=1$, $i=1,2$, for magnetic charges $(m_1,m_2)=(1,1)$,
the decomposition
\eqref{eq:splitmoduleAB} must reduce to
\[
\rho_i \simeq Q_{e_i}, \qquad i=1,2.
\]
Hence $(a_1,b_1)\in {\rm Ext}^0_\CK(Q_{e_1},Q_{e_2})$
and $(r_1,s_1)\in  {\rm Ext}^1_\CK(Q_{e_1},Q_{e_2})^\vee$. However, as explained in Appendix \ref{Aone},
below equation \eqref{eq:indecompextD}, in this case
the maps $(a_1,b_1)$, respectively $(r_1,s_1)$
must be simultaneously injective or trivial. Then relations
\eqref{eq:relationsB} imply that either $(a_1,b_1)$ or
$(r_1,s_1)$ must be trivial. Therefore,
using the isomorphisms \eqref{eq:isomext},
 one obtains two cases
\begin{itemize}
\item[$(i)$] $(r_1,s_1)=0$ and
$\rho$ is an extension of the form
\be\label{eq:plusext}
0\to \rho_2\to \rho \to \rho_1 \to 0
\ee
in the abelian category of $(Q,W)$-modules
with extension class determined by  $(a_1,b_1)$. Note that
such an extension is trivial unless $e_2\geq e_1$ i.e
$n_2\geq n_1+1$.
\item[$(ii)$] $(a_1,b_1)=0$ and
$\rho$ is an extension of the form
\be\label{eq:minusext}
0\to \rho_1\to \rho \to \rho_2 \to 0
\ee
in the abelian category of $(Q,W)$-modules
with extension class determined by  $(r_1,s_1)$. Note that
such an extension is trivial unless
$e_1\geq e_2+1$ i.e $n_1\geq n_2+1$.
\end{itemize}

Suppose the stability parameters $(\theta,\eta)$,
satisfy
\be\label{eq:wallD}
d_i\theta_i+ e_i\eta_i =0, \qquad i=1,2,
\ee
in addition to \eqref{eq:quivchamberAB}, and are otherwise
generic.  Given the linear relation \eqref{eq:kingparametersB}, 
these equations 
determine a real codimension one wall in the 
hyperplane \eqref{eq:kingparametersB}. 
Since $\theta_i<0<\eta_i$, $i=1,2$ by assumption, each module $\rho_i$ is $(\theta_i,\eta_i)$-stable of slope zero for $i=1,2$. Therefore  all extensions
of the form \eqref{eq:plusext} or \eqref{eq:minusext}
(including the trivial ones) are $(\theta,\eta)$-semistable
on the wall \eqref{eq:wallD}.

Moreover, it is also straightforward to determine the
moduli spaces of stable representations in the adjacent
chambers.
For example, let $(\theta^+, \eta^+)$ be stability parameters
 such that
\[
d_1\theta_1^+ + e_1 \eta^+_1 = - d_2\theta^+_2 - e_2 \eta^+_2 =\epsilon,
\]
with $\epsilon\in \IR_{>0}$ is a positive real number which may
be taken arbitrarily small. Then it is easy to prove that
$\rho$ is $(\theta^+,\eta^+)$-semistable if and only if
it is $(\theta^+,\eta^+)$-stable and if and only if it
fits in a nontrivial extension of the form \eqref{eq:plusext}.
Therefore in this case the moduli space
is isomorphic to $\IP^{n_2-n_1-1}$, and the protected spin character
\[
\Omega(\gamma(1,1,n_1,n_2);y)=y^{-(n_2-n_1-1)/2}P_y(\IP^{n_2-n_1-1}).
\]
The opposite chamber is defined analogously by
\[
d_1\theta_1^- + e_1 \eta^-_1 = - d_2\theta^-_2 - e_2 \eta^-_2 =-\epsilon,
\]
with $\epsilon\in \IR_{>0}$. Then $\rho$ is
$(\theta^-,\eta^-)$-semistable if and only if it is
$(\theta^-,\eta^-)$-stable
and if and only if it fits in a nontrivial extension of the form
\eqref{eq:minusext}.
In this case the moduli space
is isomorphic to $\IP^{n_1-n_2-1}$, and the 
protected spin character
\[
\Omega(\gamma(1,1,n_1,n_2);y)= y^{-(n_1-n_2-1)/2}P_y(\IP^{n_1-n_2-1}).
\]

\subsubsection{Magnetic charges $(1,m)$}
One can extend the above analysis to $m_1=1$, $m_2>1$  noting that in this case the decomposition \eqref{eq:splitmoduleAB} reduces to
 \[
\rho_1\simeq Q_{n_1-2}, \qquad
\rho_2\simeq \oplus_{j=1}^{h_2}Q_{k_{2,j}}^{\oplus r_{2,j}}
\]
where
\[
\sum_{j=1}^{h_2} r_{2,j}k_{2,j} = n_2-3m_2.
\]
Obviously, the electric charges are taken such that $n_1>2$, $n_2>3m_2$.
 If $n_1\geq n_2-3m_2+3$,
equations \eqref{eq:indecompextA} show that
\[
{\rm Hom}_{\CK}(\rho_1,\rho_2)=0.
\]
Therefore the maps $a_1,b_1$ must be trivial, and isomorphism \eqref{eq:isomext} implies that $\rho$
fits in an extension
\[
0\to \rho_1\to \rho \to \rho_2\to 0
\]
in the category of $(Q,W)$-modules.

Now consider the wall
\[
(n_1-1)\theta_1 + (n_1-2)\eta_1=0
\]
in the moduli space of King stability parameters $(\theta,\eta)$ satisfying \eqref{eq:quivchamberAB}.
Let $(\theta^+,\eta^+)$ be stability parameters satisfying
\[
(n_1-1)\theta^+_1 + (n_1-2)\eta^+_1= -
(n_2-2m_2)\theta^+_2 - (n_2-3m_2)\eta^+_2 =
\epsilon
\]
with $\epsilon >0$ sufficiently small. Obviously $\rho_1\subset
\rho$ destabilizes $\rho$ on this side of the wall. Therefore
the moduli space of $(\theta^+,\eta^+)$ representations
with dimension vector
\[
(n_1-1,n_1-2,n_2-2m_2,n_2-3m_2), \qquad
n_1\geq n_2-3m_2+3, \qquad n_2\geq 3m_2+1
\]
is empty. Moreover, note also that semistable representations $\rho$ can exist for generic values of the
stability parameters on the wall only if
\[
\rho_2\simeq Q_{k_2}^{\oplus r_2}
\]
for some $k_2,r_2\geq 1$. This implies
\[
m_2 = r_2, \qquad n_2 = (k_2+2)r_2.
\]
If this is not the case, the moduli space of semistable representations at generic points on the wall is empty,
which implies that the moduli spaces in the adjacent chambers are empty as well.
Assuming
\[
(m_2,n_2) = (r_2, (k_2+2)r_2), \qquad
k_2,r_2\geq 1, \qquad
n_1 \geq (k_2-1)r_2 +3
\]
the refined Donaldson-Thomas invariants in
the opposite chamber
\[
(n_1-1)\theta_1 + (n_1-2)\eta_1= -
(n_2-2m_2)\theta_2 - (n_2-3m_2)\eta_2 =
-\epsilon
\]
can be computed by wallcrossing.

To this end, we first list the non-vanishing DT-invariants.
The moduli space of $Q_{k_2}$ is just a point, and therefore the
refined DT-invariant is given by:
\be
\Omega(Q_{k_2},y)=1.
\ee
Moreover the isomorphisms \eqref{eq:isomext}, \eqref{eq:indecompextA}, imply that the $(Q,W)$-module 
$\rho_2\simeq Q_{k_2}$ is rigid. This implies that the 
only semi-stable objects with charge $r_2(0,1,0,1)$, 
$r_2\geq 1$ are isomorphic to the direct sums of the form 
$Q_{k_2}^{\oplus 2}$. 
Then $\Omega(Q_{k_2}^{\oplus r_2})=0$ for $r_2\geq 2$, 
and the rational
invariant $J(\gamma,y)$ defined 
in equation (\ref{eq:refmulticover}) 
is given by:
\be
J(Q_{k_2}^{\oplus r_2},y)= (-1)^{r_2-1}\frac{y-y^{-1}}{r_2\,(y^{r_2}-y^{-r_2})}
\ee
Using the refined wall-crossing formula of \cite{wallcrossing} in the form derived by \cite{Manschot:2010qz} 
in terms of rational invariants, 
one finds for the generating function of $\Omega((1,r_2,n_1,r_2n_2);y)$: 
\[
\bal 
& \sum_{r_2\geq 0} \Omega((1,r_2,n_1,r_2n_2);y)\,q^{r_2} \\
& = 1+\sum_{\sum_i \ell_i r_{2,i}=r_2>0}
\prod_{i} \frac{1}{ \ell_{i}!}
\left(\frac{y^{r_{2,i}(n_1-n_2)}-y^{r_{2,i}(n_2-n_1)}}{y-y^{-1}}\right)^{\ell_i} 
J((0,r_{2,i},0,r_{2,i}n_2);y)^{\ell_{i}}\,q^{\ell_i r_{2,i}}\\
& = \exp\left(-\sum_{r>0} 
  \frac{y^{r(n_1-n_2)}-y^{-r(n_1-n_2)}}{r(y^r-y^{-r})} (-q)^{r}\right)=
\prod_{s\geq 0} \frac{(1+y^{n_2-n_1+2s+1}q)}{(1+y^{n_1-n_2+2s+1}q)}  \\ 
& 
= \prod_{s=0}^{n_1-n_2-1} 
(1+y^{n_2-n_1+2s+1}q), \\
\eal
\]
where we assumed $|y|<1$. 
The last line in the above formula is in agreement with the 
semiprimitive wallcrossing formula of 
\cite{Denef:2007vg}. 
This corresponds to the numerical invariants:
\be
\sum \Omega((1,r_2,n_1,r_2n_2))\,q^{r_2}=(1-(-1)^{n_1-n_2}q)^{n_1-n_2},
\ee
which shows that for charges $(1,r_2,n_1,r_2n_2)$,  $r_2$ is bounded
above by $n_1-n_2$. In the ``+'' chamber, the charges
$(1,r_2,n_1,r_2n_2)$ satisfy $n_2-n_1>0$; the invariants follow
from the formulas above by interchanging $n_1$ and $n_2$.

These examples show that, in strong contrast 
to the $SU(2)$ theory, there are arbitrarily high spin BPS states and we can have higher magnetic charges.

\subsection{Comparison with large radius spectrum}\label{nowallsSUthree}
Let us recall the large radius BPS spectrum with 
charges $\gamma(1,1,n_1,n_2)$ found in Section  \ref{threefive}. 
Using the parameterization \eqref{eq:SUthreemoduli}, 
for sufficiently small $\epsilon$, the complex K\"ahler moduli
spaces is divided into two chambers separated by the wall
\be\label{eq:largeradiuswall}
{\rm Re}(a_2){\rm Im}(a_1) -
{\rm Re}(a_1){\rm Im}(a_2)=0.
\ee
which does not depend on the electric charges $(n_1,n_2)$.

In the chamber $\CC^{string}_-(\gamma(1,1,n_1,n_2))$
given by 
\be\label{eq:largerRchamberA}
{\rm Re}(a_2){\rm Im}(a_1) -
{\rm Re}(a_1){\rm Im}(a_2)<0
\ee
the BPS spectrum consists of states with charges $\gamma(1,1,n_1,n_2)$ with 
$n_2+1\leq n_1$. The corresponding supersymmetric D-brane 
configurations are nontrivial extensions 
\[
0\to L_1\to F \to L_2 \to 0 
\] 
where $L_1=\CO_{S_1}(-\Sigma_1+n_1C_1)$, 
$L_2=\CO_{S_2}(-\Sigma_1+(n_2-3)C_2)$ are line 
bundles supported on the surfaces $S_1,S_2$ respectively. 
Therefore the moduli space is isomorphic to the projective space 
$\IP^{n_1-n_2-1}$. These states 
have spins $j_{(n_1,n_2)}= {n_1-n_2-1\over 2}$ and 
protected spin characters 
\be\label{eq:largeRdegA}
\Omega_{(\tau_0,a)}^{string}(\gamma(1,1,n_1,n_2);y) = y^{-(n_1-n_2-1)}\chi_y(\IP^{n_1-n_2-1}). 
\ee

In the opposite chamber, 
$\CC^{gauge}_+(\gamma(1,1,n_1,n_2))$, 
 the spectrum consists 
 states with 
$n_1+1\leq n_2$. The supersymmetric D-brane configurations 
are now nontrivial extensions of the form 
\[
0\to L_2 \to F \to L_1\to 0,
\]
which are parameterized by $\IP^{n_2-n_1-1}$. 
  These states 
have spins $j_{(n_1,n_2)}= {n_2-n_1-1\over 2}$ and 
protected spin characters 
\[
\Omega_{(\tau_0,a)}^{string}(\gamma(1,1,n_1,n_2);y) = y^{-(n_2-n_1-1)}\chi_y(\IP^{n_2-n_1-1}). 
\]

On the other hand, the above analysis of quiver moduli spaces 
yields a wall of the form 
\[ 
(n_1-1)\theta_1 + (n_1-2)\eta_1 = (n_2-2)\theta_2 
+(n_2-3)\eta_2=0 
\]
in the moduli space of King stability parameters for each pair $(n_1,n_2)$. 
Each of these walls is the intersection of a wall of the form 
\be\label{eq:ambientwalls}
(n_1-1) z_1 + (n_1-2) w_1 =\lambda 
\big((n_2-2)z_2 + (n_2-3)w_2\big), \qquad \lambda\in \IR_{>0},
\ee
in the moduli space of Bridgeland stability parameters with 
the subspace of King stability parameters. For a physical Bridgeland 
stability condition corresponding to a point $(a_i,a_i^D)$ on the 
universal cover of the Coulomb branch, the parameters $(z_i,w_i)$ are given by \eqref{eq:stabparametersA}. These are the walls where the central charges of the Kronecker modules $\rho_1,\rho_2$ of dimension vectors 
$(n_1-2,n_1-1)$, $(n_2-3,n_2-2)$ respectively, are aligned. 

Now suppose the parameters $(\tau_0,a_i)$ satisfy conditions 
\[
{\rm Im}(\tau_0)>0, \qquad {\rm Im}(a_i)>0, \qquad i=1,2,
\]
and \eqref{eq:limitwedgeB} as in Section \ref{threefive} 
and take the large $\lambda\to \infty$ limit as defined in 
Section \ref{threetwo}. In this limit, the walls \eqref{eq:ambientwalls}
become marginal stability walls for the limit stability condition 
defined by the slope function \eqref{eq:limitslope}. That is, 
loci where 
\[ 
\mu_{(\tau_0,a)}(\rho_1)= \mu_{(\tau_0,a)}(\rho_2).
\]
Since inequalities \eqref{eq:limitwedgeB} are assumed to be 
satisfied, the main assumption in Section \ref{threetwo} 
will hold for the $SU(3)$ quiver $(Q,W)$. No mutations are 
necessary. Then, using equations \eqref{eq:largelambdaA}, 
\[
\bal 
\mu_{(\tau_0,a)}(\rho_i)^{-1}= \left({1\over 2} - {n_i\over 3}\right)
{1\over {\rm Im}(\tau_0)} + {{\rm Re}(a_i)\over {\rm Im}(a_i)}
\eal
\]
for $i=1,2$. Therefore the large $\lambda$ limit of the walls 
\eqref{eq:ambientwalls} is given by 
\be\label{eq:limitwalls}
{{\rm Re}(a_2)\over {\rm Im}(a_2)}-
{{\rm Re}(a_1)\over {\rm Im}(a_1)}=
{n_2-n_1\over 3\, {\rm Im}(\tau_0)}.
\ee
Then the quiver stability analysis in the previous section implies that 
the  $\mu_{(\tau_0,a)}$-limit semistable quiver 
representations in the chamber 
\be\label{eq:limitchamberA} 
\CC^{gauge}_-(\gamma(1,1,n_1,n_2)):\qquad 
{{\rm Re}(a_2)\over {\rm Im}(a_2)}-
{{\rm Re}(a_1)\over {\rm Im}(a_1)}<
{n_2-n_1\over 3\, {\rm Im}(\tau_0)}.
\ee
must be nontrivial extensions of the form  
\[ 
0\to \rho_1 \to \rho\to \rho_2 \to 0.
\] 
Such nontrivial extensions exist only if $n_2+1\leq n_1$, 
and the moduli space is isomorphic to $\IP^{n_1-n_2-1}$. 
The protected spin character is 
\be\label{eq:limitdegA}
\Omega^{gauge}_{(\tau_0,a)}(\gamma(1,1,n_1,n_2);y) 
= y^{-(n_1-n_2-1)}\chi_y(\IP^{n_1-n_2-1})
\ee

In the opposite chamber, 
$\CC^{gauge}_+(\gamma(1,1,n_1,n_2))$,
the semistable representations have to 
be nontrivial extensions of the form 
\[ 
0\to \rho_2 \to \rho\to \rho_1 \to 0.
\] 
which exist only if $n_1+1\leq n_2$. The moduli space is again 
isomorphic to $\IP^{n_2-n_1-1}$ and 
\[ 
\Omega^{gauge}_{(\tau_0,a)}(\gamma(1,1,n_1,n_2);y) 
= y^{-(n_2-n_1-1)}\chi_y(\IP^{n_2-n_1-1}).
\]
These results are in agreement with the absence of walls 
conjecture formulated in Section \ref{threetwo}. 
The bijection \eqref{eq:nowallscorresp} is given in this case 
by 
\[
\calP_{\gamma(1,1,n_1,n_2)} (\CC^{gauge}_{\pm}(\gamma(1,1,n_1,n_2))) = \CC^{string}_{\pm}(\gamma(1,1,n_1,n_2)). 
\]
Equations \eqref{eq:nowallsconjA} are clearly satisfied. 

 However the gauge theory limit weak coupling spectrum 
 exhibits a more refined wall structure, the walls \eqref{eq:limitwalls}
being obviously dependent on the charges
$(n_1,n_2)$, as opposed 
to \eqref{eq:largeradiuswall}. 
For very large ${\rm Im}(\tau_0)>>1$, keeping $n_1,n_2$  fixed, the walls \eqref{eq:limitwalls} approach asymptotically the large 
radius wall \eqref{eq:largeradiuswall}. To complete the picture, note that 
the Kronecker modules $\rho_1,\rho_2$ forming gauge theory bound states are related by tilting to the 
line bundles $L_1,L_2$ (up to a shift) forming D-brane bound states.

This is not a contradiction since $(\omega,B)$-stability employed in Section \ref{threefive} is not a
Bridgeland stability condition on $D^b(X_3)$. Therefore
the unique wall found there is a limit wall which does not actually exist in 
the moduli space of Bridgeland stability conditions.
Instead for each pair $(n_1,n_2)$, there is a wall where 
the string theory central charges of the line bundles $L_1,L_2$ 
are aligned. At large radius, the central charges of $L_1,L_2$ 
are given by \eqref{eq:largeRZ}. Omitting world-sheet-instanton corrections, 
one finds 
\[ 
Z_{(\omega,B)}(L_1) = t_0t_1+t_1^2-b_1^2+n_1b_1+\sqrt{-1}
\left(-t_0b_1-2t_1b_1+n_1t_1\right)
\]
\[
\bal
Z_{(\omega,B)}(L_2) = &
t_0t_2+2t_1t_2+2t_2^2-2(b_1b_2+b_2^2)+n_2b_2\\
&
+\sqrt{-1}\left(-t_0b_2-2t_1b_2-2t_2b_1-4t_2b_2+n_2t_2\right)\\
\eal
\]
In the small $\epsilon$  limit defined in Section \ref{threetwo},
equation \eqref{eq:btepsilon}, the leading terms of the central charges are 
\[
Z_{(\omega,B)}(L_i)\sim t_0t_i -t_0b_i\sqrt{-1}, \qquad 
i=1,2.
\]
Therefore all these walls approach asymptotically the limit wall
\[ 
t_1b_2=t_2b_1
\]
which is the same as \eqref{eq:largeradiuswall} 
using the parameterization \eqref{eq:SUthreemoduli}. 
  A similar situation has been encountered in a
similar context in \cite[Sect. 3]{DM-crossing} 
(see in particular Fig. 1.
in loc. cit.)

\section{Strong coupling chamber for the $SU(N)$
quiver}\label{strongsection}

According to \cite{Alim:2011kw}, \cite{spectral_networks},
there exist strong coupling chambers of the $SU(N)$ theory where
the BPS spectrum   consists of a finite set of
stable BPS states equipped with a natural two-to-one map to the
set $\Delta_N^+$ of positive roots of $SU(N)$.
The purpose of this section is to identify an analogous chamber
in the space of stability parameters $(\theta,\eta)$
of the $SU(N)$ quiver obtained by geometric engineering.
More precisely, for certain values of $(\theta, \eta)$ the
set
$\CS_{(\theta,\eta)}(Q,W)$ of all stable  quiver modules should be finite, and equipped with a natural
two-to-one map $\CS_{(\theta,\eta)}(Q,W)\to \Delta^+_N$.
As a corollary, an adjacent chamber -- called deceptive --
will also be identified
where all massive $W$-bosons are stable and the
BPS spectrum is completely determined by wallcrossing.

 Using the same conventions as in the previous section
  a representation $\rho$ of the
 the $SU(N)$ quiver will be  a diagram of the form
 \be\label{eq:SUNquiverA}
\xymatrix{
W_{N-1} \ar@<.5ex>[rrr]|{c_{N-1}} \ar@<-.5ex>[rrr]|{d_{N-1}}
& & & V_{N-1} \ar@/_1pc/[ddlll]|{r_{N-2}}
\ar@/^1pc/[ddlll]|{s_{N-2}}  \\
& &  & \\
W_{N-2} \ar@<.5ex>[rrr]|{c_{N-2}} \ar@<-.5ex>[rrr]|{d_{N-2}}
\ar[uu]|{b_{N-2}}
& & & V_{N-2}  \ar[uu]|{a_{N-2}}  \\
\vdots & & & \vdots \\
W_{i+1} \ar@<.5ex>[rrr]|{c_{i+1}} \ar@<-.5ex>[rrr]|{d_{i+1}}
& &  & V_{i+1}\ar@/_1pc/[ddlll]|{r_i}
\ar@/^1pc/[ddlll]|{s_i} \\
& & & \\
W_i \ar@<.5ex>[rrr]|{c_i} \ar@<-.5ex>[rrr]|{d_i}
\ar[uu]|{b_i}
& & &  V_i\ar[uu]|{a_i}  \\
\vdots && & \vdots \\
W_2\ar@<.5ex>[rrr]|{c_2} \ar@<-.5ex>[rrr]|{d_2}
& & & V_2\ar@/_1pc/[ddlll]|{r_1}
\ar@/^1pc/[ddlll]|{s_1} \\
& & & \\
W_1  \ar@<.5ex>[rrr]|{c_1} \ar@<-.5ex>[rrr]|{d_1} \ar[uu]|{b_1}
& & & V_1 \ar[uu]|{a_1}
\\
}
\ee
with potential
\be\label{eq:SUNquiverB}
\CW=\sum_{i=1}^{N-2} {\rm Tr}\left[r_i(a_ic_i-c_{i+1}b_i)+s_i(a_id_i-d_{i+1}b_i)
\right].
\ee
As in Section \ref{twothree}, the 
 dimension vector of $\rho$ will be denoted by $(d,e)$, where
$d=(d_1,\ldots, d_{N-1})$ are the dimensions of $(V_1,\ldots, V_{N-1})$ and $e=(e_1,\ldots,e_{N-1})$ the dimensions of
$(W_1,\ldots, W_{N-1})$.
The stability parameters $(z_i,w_i)_{1\leq i\leq N-1}$ 
will be assumed of the form \eqref{eq:kingparameters},
where $\theta=(\theta_1,\ldots, \theta_{N-1})$ are assigned to the nodes
$(V_1,\ldots, V_{N-1})$ and $\eta=(\eta_1,\ldots,\eta_{N-1})$ to
$(W_1,\ldots, W_{N-1})$. It will be assumed that $N\geq 3$ in this
section.

As a starting point,
note that if
the stability parameters satisfy
\be\label{eq:strongchamberC}
\eta_{i}<\eta_{i-1}, \qquad \theta_{i}<\theta_{i-1},
\qquad i=2,\ldots, N-1,
\ee
one can construct two 
simple, rigid, stable representations $\rho^\pm(\alpha)$
of the $SU(N)$ quiver for each positive root $\alpha\in \Delta^+_N$ as follows.

A representation with
$e_i=0$ for all $i=1,\ldots, N-1$,
reduces to a representation of
the linear quiver with $N-1$ nodes
\[
\xymatrix{
V_1 \ar[r]^-{a_1}& V_2 \ar[r]^-{a_2} & \cdots \ar[r]^-{a_{N-1}} & V_{N-1}}
\]
with stability parameters $(\theta_1, \ldots, \theta_{N-1})$.
Using gauge transformations, all linear maps can be set in canonical form,
${\bar a}_{i-1}: \IC^{d_{i-1}} \to \IC^{d_i}$ where ${\bar a}_{i-1}$ is a diagonal matrix
of the form
\[
({\bar a}_{i-1})_{kl} = \left\{\begin{array}{ll}
1,& {\rm for}\ 1\leq k=l\leq {\bar d}_{i-1},\\
& \\
0,&{\rm otherwise},\\
\end{array}\right.
\]
for some $0\leq {\bar d}_{i-1} \leq {\rm min}(d_{i-1}, d_{i})$. If ${\bar d}_{i-1}=0$, by convention
${\bar a}_{i-1}=0$.
 If the stability parameters $\theta_i$ are ordered
as in \eqref{eq:strongchamberC}, it is straightforward to prove that
that such a representation is stable if and only if
it has dimension vector of the form
\[
d_i = \left\{\begin{array}{ll}
1, & {\rm for}\ j\leq i\leq k, \\
& \\
0, & {\rm otherwise},\\ \end{array}\right.
\]
for some $j,k\in \{1,\ldots, N-1\}$, $j\leq k$.
Then there is an obvious 1-1 correspondence between
stable representations and positive roots
$\alpha \in \Delta_N^+$, sending $\rho$ to
$\alpha =\sum_{i=1}^{N-1} d_i\alpha_i$.
Similar considerations apply similarly to representations with
$d_i=0$ for all $i=1,\ldots, N-1$.

In order to complete the picture, it will be shown below that
these are the only stable representations of the $SU(N)$ quiver
if in addition the stability parameters $(\eta,\theta)$ satisfy
\be\label{eq:strongchamberA}
\eta_{i} < \theta_i < \eta_{i-1} <\theta_{i-1}
\ee
for all $2\leq i \leq N-1$. Thus equation \eqref{eq:strongchamberA}
defines a single chamber. 

The proof will be inductive. Consider first $N=3$. Then
\eqref{eq:strongchamberA} specializes to
\be\label{eq:strongchamberB}
\eta_2<\theta_2<\eta_1<\theta_1.
\ee
Let $\rho$ be a stable representation of dimension vector
$(d_1,d_2,e_1,e_2)$ such that $(d_1,d_2)$ and $(e_1,e_2)$
are simultaneously nontrivial.
 In this case  note that
 inequalities \eqref{eq:strongchamberB}
 imply
 \[
\eta_2 < \mu_{(\theta,\eta)}(\rho) < \theta_1.
\]
For any $(\theta,\eta)$-stable representation $\rho$, this implies
that $a_1$ must be injective if $V_1\neq 0$ and $b_1$ must be
surjective if $W_2\neq 0$.
In particular  if $V_1$ and $W_2$ are both nontrivial, $V_2,W_1$
must be nontrivial as well.  Moreover the relations
\[
r_1a_1=s_1a_1=0
\]
imply that $${\rm Im}(a_1)\subseteq {\rm Ker}(r_1)\cap {\rm Ker}(s_1),$$ and
\[
b_2r_2=b_2s_2=0
\]
imply that $${\rm Im}(r_2)+{\rm Im}(s_2) \subseteq {\rm Ker}(b_2).$$
This further implies that the data $\rho'=(V_1,{\rm Im}(a_1),a_1)$
determines a proper nontrivial subobject of $\rho$ of slope
\[
\mu_{(\theta,\eta)}(\rho') ={\theta_1+\theta_2\over 2}.
\]
Furthermore
$b_1:W_1\to W_2$ induces an isomorphism ${\bar b}_1:W_1/
{\rm Ker}(b_1) {\buildrel \sim \over \longto} W_2$, and 
the data
$\rho''=(W_1/{\rm Ker}(b_2), W_2, {\bar b}_1)$ determines a
nontrivial quotient of $\rho$ of slope
\[
\mu_{(\theta,\eta)}(\rho'') ={\eta_1+\eta_2\over 2}.
\]
Then inequalities \eqref{eq:strongchamberB} imply
that
\[
\mu_{(\theta,\eta)}(\rho'')<\mu_{(\theta,\eta)}(\rho')
\]
which contradicts the stability of $\rho$.

In conclusion at least one of $V_1$ or $W_2$ must be trivial.
If both are trivial, $\rho$ reduces to the Kronecker
module $(V_2,W_1,r_1,r_2)$. Since $\theta_2<\eta_1$ the
only stable modules of this form are the simple ones. This
is not allowed by the assumption that $(d_1,d_2)$ and
$(e_1,e_2)$ are simultaneously nontrivial.

Suppose  $V_1$ is trivial and $W_3$ nontrivial.
Then $V_2$ must be nontrivial since $(d_1,d_2)\neq (0,0)$ by assumption. Moreover, if $b_1:W_1\to W_2$ is not injective,
${\rm Ker}(b_1)\neq 0$,
is a subobject of $\rho$ of slope
\[
\eta_1>\mu_{(\theta,\eta)}(\rho).
\]
This  leads again to a contradiction, hence ${\rm Ker}(b_1)=0$.
Then relations
\[
b_1r_1=b_1s_1=0
\]
imply that $r_1,s_1$ are trivial. Hence
$\rho$ splits as a direct sum of subrepresentations,
contradicting stability.  The remaining case,
 $W_3$ trivial and $V_1$ nontrivial, is similar and will be left to the reader.

Next suppose the claim holds for the $SU(N-1)$ quiver for
some $N>3$. Let $\rho$ be a $(\theta,\eta)$-stable representation
of the $SU(N)$ quiver. If $(V_1,W_1)$ are simultaneously trivial
or $(V_{N-1},W_{N-1})$  are simultaneously trivial, $\rho$ is a representation of an $SU(N-1)$ quiver and the inductive step is
trivial.
Therefore in the following suppose $(d_1,e_1)\neq (0,0)$,
$(d_N,e_N)\neq (0,0)$ and $(d_1,\ldots, d_{N-1})\neq (0,\ldots, 0)$, $(e_1,\ldots, e_{N-1})\neq (0,\ldots, 0)$.
Let
\[
I_j= {\rm Im}(r_j)+{\rm Im}(s_j)\subseteq W_j, \qquad j=1,\ldots, N-2
\]
and
\[
K_j={\rm Ker}(r_j)\cap {\rm Ker}(s_j) \subseteq V_{j+1},
\qquad j=1,\ldots, N-1.
\]
Set $K_1=V_1$ and $I_{N-1}=0$. For each $j=1,\ldots, N-2$ let
${\bar a}_j=a_j|_{K_j}$ and ${\bar b}_j : W_j/I_j \to
W_{j+1}/I_{j+1}$ be the linear maps induced by $b_j:W_j\to W_{j+1}$.
Then the relations derived from the potential \eqref{eq:SUNquiverB} imply that
\[
b_j(I_j) \subseteq I_{j+1}, \qquad a_j(K_j)\subseteq K_{j+1}
\]
for $j=1,\ldots, N-1$. Therefore the data
\[
\rho'=(K_1,\ldots, K_N, {\bar a}_1, {\bar a}_2,\ldots,
{\bar a}_{N-1}|_{K_{N-1}})
\]
determines a subobject of $\rho$ with $(e_1,\ldots, e_{N-1})=
(0,\ldots, 0)$ while the data
\[
\rho''=(W_1/I_1, \ldots, W_{N-1}/I_{N-1}, {\bar b}_1, \ldots,
{\bar b}_{N-1})
\]
determines a quotient of $\rho$ with $(d_1,\ldots, d_{N-1})=
(0,\ldots, 0)$. At this point there are several cases.

$1)$ $V_1,W_{N-1}$ are both nontrivial. Let $L_1\subseteq K_1=V_1$ be a one dimensional subspace, and set $$L_j=
(a_j\circ a_{j-1} \circ \cdots \circ a_1)(L_1)\subseteq K_{j+1}$$
for
$j=1,\ldots, N-2$. Let $k\in \{1,\ldots  N-1\}$ be the smallest label such
that $L_k\neq 0$ and $L_{k+1}=0$. Then the data
\[
\lambda_k=\left(L_1,\ldots, L_k, {\bar a}_1|_{L_1}, \ldots, {\bar a}_{k-1}|{L_{k-1}}\right)
\]
is a subrepresentation of $\rho$ contained in $\rho'$, and
\[
\mu_{(\theta,\eta)}(\lambda_k) =
{ 1\over k} \sum_{i=1}^k \theta_i.
\]
Similarly, let $p_{N-1}:W_{N-1}\twoheadrightarrow Q_{N-1}$ be a one dimensional quotient of $W_{N-1}$. By
successive compositions, there is a sequence of one dimensional quotients and induced linear maps
\[
\xymatrix{
\cdots & W_j/I_j \ar[r]^-{{\bar b}_j} \ar@{>>}[d]_-{p_j} &
W_{j+1}/I_{j+1} \ar@{>>}[d]_-{p_{j+1}} &\cdots &
W_{N-1} \ar@{>>}[d]_-{p_{N-1}} \\
\cdots & Q_j  \ar[r]^{q_j} &Q_{j+1} &\cdots & Q_{N-1} \\}
\]
Let $l\in \{1,\ldots, N-2\}$ be the largest label such that
$q_l\neq 0$, but $q_{l-1}=0$.  Then the data
\[
\sigma_l = (Q_l, \cdots, Q_{N-1}, q_l, \cdots, q_{N-2})
\]
is a quotient of $\rho$ of slope
\[
\mu_{(\theta,\eta)}(\sigma_l) = {1\over N-1-l} \sum_{j=l}^{N-1}
\eta_j.
\]
Now note that inequalities \eqref{eq:strongchamberA} imply
\[
\mu_{(\theta,\eta)}(\lambda_k) \geq
{1\over N-1} \sum_{i=1}^{N-1}\theta_i
> {1\over N-1}\sum_{i=1}^{N-1} \eta_i \geq
\mu_{(\theta,\eta)}(\sigma_l)
\]
for any $k,l$, contradicting stability of $\rho$.

2) Suppose $V_1$ is trivial, but $W_{N-1}$ is nontrivial. Then the
bottom part of $\rho$ is of the form
\[
\xymatrix{ W_2\ar@<.5ex>[rrr]|{c_2} \ar@<-.5ex>[rrr]|{d_2}
& & & V_2\ar@/_1pc/[ddlll]|{r_1}
\ar@/^1pc/[ddlll]|{s_1} \\
& & & \\
W_1  \ar[uu]|{b_1}
& & &
\\
}
\]
For each $j=1,\ldots,N-1$, let $U_j={\rm Ker}(c_j)\cap {\rm Ker}(d_j)\subseteq W_j$. Note that $U_1=W_1$ since $c_1,d_1$ are
trivial. Moreover the relations determined by \eqref{eq:SUNquiverB}
show that
\[
b_j(U_j) \subset U_{j+1}, \qquad j=1, \ldots, N-1.
\]
Therefore the data
\[
(U_1,\ldots, U_{N-1}, b_1|_{U_1}, \ldots, b_{N_2}|_{U_{N-2}})
\]
is a subobject of $\rho$. Proceeding by analogy with the construction
of the subobject $\lambda_k$ above one finds a further subobject
$\gamma_k$ for some $k\in \{1,\ldots, N-1\}$
with dimension vector
\[
d_i=0, \qquad i=1,\ldots, N_1, \qquad
e_i=\left\{\begin{array}{ll}1, & {\rm for}\ 1\leq i\leq k\\
0, & {\rm otherwise}. \end{array}\right.
\]
Then
\[
\mu_{(\theta,\eta)}(\gamma_k)= {1\over k}\sum_{i=1}^k \eta_i
\geq {1\over N-1}\sum_{i=1}^{N-1} \eta_i \geq
\mu_{(\theta,\eta)}(\sigma_l)
\]
and equality holds only if $k=l=N-1$.
Again this contradicts stability.

3) The remaining case, $W_{N-1}$ trivial and $V_1$ nontrivial, is treated analogously, details being omitted.

In conclusion,  in the chamber \eqref{eq:strongchamberA}
there is indeed a finite set of $\CS_{(\theta,\eta)}(Q,W)$
of stable representations which maps two-to-one to $\Delta^+_N$.
For each positive root
\[
\alpha = \sum_{i=1}^{N-1} n_i(\alpha) \alpha_i, \qquad n_i(\alpha)\in \{0,1\}, \qquad i=1,\ldots, N-1,
\]
 there are exactly two
representations $\rho_\alpha^{\pm}$ with dimension
vectors
\[
d_i(\rho_\alpha^+) = n_i(\alpha), \qquad e_i(\rho_\alpha^+) =0,
\qquad i=1,\ldots, N-1
\]
respectively
\[
d_i(\rho_\alpha^-) = 0, \qquad e_i(\rho_\alpha^-) =n_i(\alpha)
\qquad i=1,\ldots, N-1.
\]
These states are in one-to-one correspondence to the strong coupling spectrum obtained in
\cite{Alim:2011kw}, \cite{spectral_networks} using different techniques.
In fact, one can check that the quiver used here is related to that
of \cite{Alim:2011kw} by a mutation. For brevity, this will be
explained below only for $N=3$. 

\subsection{A mutation of the $SU(3)$ quiver}
The $SU(3)$ quiver of \cite{Alim:2011kw} is of the form
\[\xymatrix{
 4 \bullet \ar[dd]|{\tilde b_1} & & & \ar@<1ex>[lll]|{\tilde c_2}
 \ar@<-1ex>[lll]|{\tilde d_2}
 \bullet  3 \\
& &  & \\
 1 \bullet \ar@<1ex>[rrr]|{\tilde d_1} \ar@<-1ex>[rrr]|{\tilde c_1}
& & & \bullet 2 \ar[uu]|{\tilde a_1} \\}
\]
with a superpotential of the form
\[
{\widetilde \CW} = {\tilde a_1}{\tilde c_1}{\tilde b_1}{\tilde c_2}
-{\tilde a_1}{\tilde d_1}{\tilde b_1}{\tilde d_2}.
\]
Let ${\tilde \gamma}_i$, $i=1,\ldots, 4$ denote the generators of the
charge lattice associated to the nodes. The strong coupling spectrum found in \cite{Alim:2011kw} consists of six states with charges
\be\label{eq:strongchargesA}
{\tilde \gamma}_i, \qquad 1\leq i\leq 4, \qquad
{\tilde \gamma}_1+{\tilde \gamma}_4, \qquad
{\tilde \gamma}_2 + {\tilde \gamma}_3.
\ee

Using the rules listed for example on page 2 of
\cite{mutations}, it is straightforward to check that the
above quiver with potential is related to the $SU(3)$ quiver
used in this paper by a mutation at  node 4. In more detail, a mutation
at node 4, reverses all arrows beginning and ending at 4, and also
adds two more arrows corresponding to the paths ${\tilde b}_1{\tilde c}_2$, ${\tilde b}_1{\tilde d}_2$. This yields the
diagram
\[
\xymatrix{
4\bullet  \ar@<.5ex>[rrr]|{c_2} \ar@<-.5ex>[rrr]|{d_2}
& & & \bullet 3\ar@/_1pc/[ddlll]|{r_{1}}
\ar@/^1pc/[ddlll]|{s_{1}}  \\
& &  & \\
1\bullet  \ar@<.5ex>[rrr]|{c_{1}} \ar@<-.5ex>[rrr]|{d_{1}}
\ar[uu]|{b_{1}}
& & & \bullet 2 \ar[uu]|{a_{1}}  \\}
\]
where $b_1,c_2,d_2$ are obtained by reversing ${\tilde b}_1,
{\tilde c}_2, {\tilde d}_2$ and  the new arrows $r_1,s_1$
correspond to the paths ${\tilde b}_1{\tilde c}_2$,
${\tilde b}_1{\tilde d}_2$. All other arrows are unchanged. The superpotential
of the new quiver is
\[
r_1(a_1c_1+c_2b_1)-s_1(a_1d_1-d_2b_1).
\]
This expression is related to the superpotential \eqref{eq:SUNquiverB}
by an automorphism of the path algebra changing $b_1$ to $-b_1$.

Let $\gamma_i$, $1\leq i\leq 4$ denote the generators of the K-theory
lattice corresponding to the nodes of the new quiver. Then the stable BPS states found in this section have charges
\be\label{eq:strongchargesB}
\gamma_i, \qquad 1\leq i\leq 4, \qquad \gamma_1+\gamma_4, \qquad \gamma_2+\gamma_3.
\ee
These are related to the generators ${\tilde \gamma}_i$, $1\leq i\leq 4$ by the linear transformations
\[
\gamma_1={\tilde \gamma_1}+{\tilde \gamma}_4, \qquad
\gamma_2={\tilde \gamma}_2, \qquad
\gamma_3 = {\tilde \gamma}_3, \qquad
\gamma_4 = -{\tilde \gamma}_4.
\]
This transformation maps the charge vectors \eqref{eq:strongchargesB}
to
\[
{\tilde \gamma}_1+{\tilde \gamma}_4, \qquad {\tilde \gamma}_2, \qquad {\tilde \gamma}_3, \qquad -{\tilde \gamma}_4, \qquad
{\tilde \gamma}_1, \qquad {\tilde \gamma}_2+{\tilde \gamma}_3.
\]
By comparison with \eqref{eq:strongchargesA}, it follows that  \eqref{eq:strongchargesB} is the strong coupling spectrum in
a different region of the Coulomb branch, where the BPS particle of
charge ${\tilde \gamma}_4$ has been replaced with its antiparticle.
Mathematically, this is expected since the two quivers related by mutations determine different t-structures on the derived category,
corresponding to different regions in the moduli space of Bridgeland
stability conditions.

\subsection{A deceptive chamber}\label{deceptivesect}
 Using the above results, the  wallcrossing formula of
\cite{wallcrossing} determines the BPS spectrum
in an adjacent  chamber where all $W$-bosons are stable.
Let
\[
\eta_i = \theta_i + \epsilon_i, \qquad\epsilon_i\in \IR, \qquad  i=1,\ldots, N-1.
\]
Let $\theta_i$, $i=1,\ldots, N-1$ be some fixed parameters
such that
\[
\theta_{N-1}< \cdots <\theta_1.
\]
Then $(\theta_i, \eta_i)$ is in the strong coupling chamber
\eqref{eq:strongchamberA} for $\epsilon_i<0$ and sufficiently small  $|\epsilon_i|<<1$, $i=1, \ldots, N-1$. The adjacent chamber
is defined by $\epsilon_i>0$ and $|\epsilon_i|<<1$, $i=1, \ldots, N_1$.
Using equation \eqref{eq:kingparameters} for the stability parameters,  central charges are of the form 
\be\label{eq:pmcentralcharges}
\bal
Z(\rho_\alpha^+) & = - r\sum_{i=1}^{N-1} n_i(\alpha) \theta_i
+\sqrt{-1} r\sum_{i=1}^{N-1} n_i(\alpha), \\
Z(\rho_\alpha^-) & = - r\sum_{i=1}^{N-1} n_i(\alpha) (\theta_i +\epsilon_i)
+\sqrt{-1} r\sum_{i=1}^{N-1} n_i(\alpha)\\
\eal
\ee
for some $r\in\IR_{>0}$. 
As $\epsilon_i$ changes from negative to positive values
the ordering of  $Z(\rho_\alpha^+), Z(\rho_\alpha^-)$ is reversed
for each $\alpha$. Moreover if $|\epsilon_i|$ are sufficiently small
the ordering of any pair $Z(\rho_\alpha^\pm), Z(\rho_\beta^\pm)$
with $\alpha\neq \beta$ is preserved.
In order to apply the wallcrossing formula of \cite{wallcrossing}
note that the symplectic pairing on the K-theory lattice of charges
is given by
\be\label{eq:pairingB}
\chi([{\rho}_1], [{\rho}_2]) = \sum_{{\sf a}}
d_{t({\sf a})}(\rho_2) d_{h({\sf a})}(\rho_1) -
d_{t({\sf a})}(\rho_1) d_{h({\sf a})}(\rho_2)
\ee
where the sum is over all arrows ${\sf a}$ of $Q$ and $h({\sf a})$,
$t({\sf a})$ denote the head and the tail of ${\sf a}$.
For the strong coupling representations this yields
\[
\chi( \rho_{\alpha}^+, \rho_{\alpha}^-) = 2
\]
Let $\gamma_\alpha^\pm$ denote the charge vectors of the 
representations $\rho_\alpha^\pm$. 
Then, applying the standard $SU(2)$ wallcrossing formula 
 \cite{wallcrossing,GMN} 
 \[
 \bal 
 K_{\gamma_\alpha^+}K_{\gamma_\alpha^-} = \big(
 K_{\gamma_\alpha^-}K_{\gamma_\alpha^++2\gamma_\alpha^-}
 K_{2\gamma_\alpha^++3\gamma_\alpha^-} \cdots \big)
 K_{\gamma_\alpha^++\gamma_\alpha^-}^{-2} 
 \big(\cdots K_{3\gamma_\alpha^++2\gamma_\alpha^-} 
 K_{2\gamma_\alpha^++\gamma_\alpha^-}K_{\gamma_\alpha^+}\big)
 \eal
 \]
 for each root,  
 the BPS spectrum in the
deceptive chamber will consist of a  tower of states
with dimension vectors
\[
(d_i,e_i)=(n_i(\alpha), n_i(\alpha) + k), \qquad k\in \IZ
\]
for each $\alpha\in \Delta_N^+$. The $k\neq 0$
 have degeneracy 1 and spin 0.
 The $k=0$ states are massive vector multiplets  with degeneracy 2 and spin $1/2$. Formally, there is a one-to-one correspondence
 between the  above states and the weak coupling states found in
 \cite{Fraser:1996pw,Chen:2011gk} by monodromy arguments.
However, a more careful analysis reveals important physical differences,
showing that the chamber studied in this section is not a weak coupling
chamber. This follows from the observation that
for a fixed $\alpha$ the central charges of the tower of states
obtained by wallcrossing
 are contained in the cone cut by the central
 charges \eqref{eq:pmcentralcharges} in the upper half-plane as shown
 in Fig. 5.

 \bigskip
 \begin{figure}[h]
  \setlength{\unitlength}{1cm}
\hspace{20pt}
\begin{picture}(2,3)
%\put(0,0){\color{red}\vector(2,1){2}}
\put(0,0){\color{red}\vector(3,1){2.2}}
\put(2.3,0.8){${}_{Z(\rho^+_{\alpha_1})}$}
\put(0,0){\color{red}\vector(3,2){2}}
\put(1.8,1.6){${}_{Z(\rho^-_{\alpha_1})}$}
\multiput(0,0)(0.2,0.1){15}{\color{red}.}
% \put(0,0){\color{blue}\vector(-3,1){2}}
\multiput(0,0)(-0.3,0.1){12}{\color{blue}.}
\put(0,0){\color{blue}\vector(-3,2){2}}
\put(-3.2,0){${}_{Z(\rho^-_{\alpha_2})}$}
\put(0,0){\color{blue}\vector(-3,0){2.2}}
\put(-3.0,1.5){${}_{Z(\rho^+_{\alpha_2})}$}
%\put(0,0){\color{green}\vector(0,1){3}}
\multiput(0,0)(0,0.2){18}{\color{green}.}
\put(0,0){\color{green}\vector(-1,3){1}}
\put(-2.5,3.0){${}_{Z(\rho^-_{\alpha_1+\alpha_2})}$}
\put(0,0){\color{green}\vector(1,3){1}}
\put(1.1,3.0){${}_{Z(\rho^+_{\alpha_1+\alpha_2})}$}
\end{picture}
\caption{Schematic representation of the BPS charge vectors in equation \eqref{eq:pmcentralcharges} for $SU(3)$ with positive roots $\alpha_1,\alpha_2, \alpha_1+\alpha_2$. }
\end{figure}
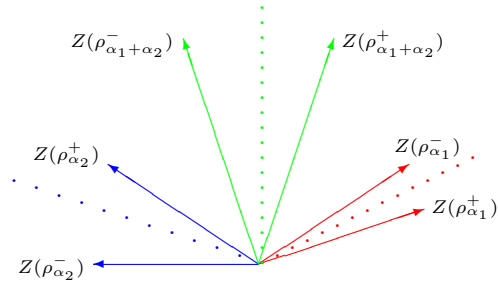
\bigskip

\noindent
For small $0<\epsilon_i<<1$, $i=1,\ldots, N-1$, the
opening angle of the cone is also very small,
\[
\Delta\phi_\alpha \sim {\sum_{i=1}^{N-1}n_i(\alpha)\epsilon_i\over
\sum_{i=1}^{N-1}n_i(\alpha)}
\]
and there is no overlap between cones associated to different
positive roots.
The massive $W$ bosons are stable but have a very small binding energy
unlike the semiclassical regime where they are 
dyon/anti-monopole
bound states, and the monopoles are very massive. In fact were
this spectrum to occur in the semiclassical regime, the opening angle
of these cones would have to be very close to $\pi$. Hence any two
cones would have to have a very big overlap since they must all be contained in a complex half-plane.

\section{Line defects and framed BPS states}\label{sectionfive}

This section is focused on a geometric construction of
magnetic line defects and the resulting mathematical model for framed BPS states. In particular it will be shown that framed
BPS states for simple magnetic line defects are modeled by
framed quiver representations. This will lead to a mathematical
derivation of the framed wallcrossing formula of \cite{framedBPS} for 
these special line defects,
as well as a recursive algorithm for unframed BPS states.
The latter completely determines the unframed BPS spectrum
at any point on the Coulomb branch in terms of noncommutative
Donaldson-Thomas invariants of framed quivers.
This section is concluded with an application of framed wallcrossing 
to the {\it absence of exotics} conjecture. 

\subsection{Geometric construction of magnetic 
line defects}\label{fiveone}
It has been already observed in Section \ref{sectiontwo} that
noncompact D4-branes wrapping  divisors $D_i$ of the form
\eqref{eq:diveqA} are natural candidates for infrared line operators
carrying magnetic charge.
In order to generalize the previous quiver construction to such
 configurations one has to specify the Chan-Paton line bundle
on the noncompact D4-brane. More precisely, the noncompact D4-brane supported on $D_i$ must be presented as an object
$N_i$
in the derived category $D^b(X_N)$. Then open string zero modes
between the D4-brane $N_i$ and the fractional branes
$(P_i,Q_i)_{1\leq i\leq N}$ are determined by the extension groups
${\rm Ext}^1(N_i,P_i)$ etc.
The low energy dynamics of a bound state of fractional
branes in the presence of such a noncompact D4-brane
will be a quantum mechanical model determined by an
enhancement $({\widetilde Q},{\widetilde W})$ of the
quiver with potential $(Q,W)$. The enhanced quiver will contain an extra
node corresponding to $N_i$ and extra arrows corresponding to the
additional open string zero modes. The effective superpotential
${\widetilde W}$ will also include new terms, ${\widetilde W} = W +\cdots$, which are determined in principle by the $A_\infty$-structure
of the derived category. While conceptually clear, an explicit form
of ${\widetilde W}$ is quite difficult to derive in practice.

A simple set of D4-branes $N_i$, $i=1,\ldots, N-1$, where this
problem is easily solved can be constructed starting with the exceptional
collection of line bundles
\[
L_i=\CO_{X_N}(D_i), \qquad M_i=\CO_{X_N}(D_i+H), \qquad 1\leq i\leq N
\]
given in equation \eqref{eq:linebundlesA}, Section \ref{twoone}. Note that the orthogonality conditions \eqref{eq:onconditions}
and the canonical exact sequences
\[
0\to \CO_{X_N}\to \CO_{X_N}(D_i) \to \CO_{D_i}(D_i) \to 0
\]
yield isomorphisms
\be\label{eq:geomlineopA}
\bal
& {\rm RHom}(\CO_{D_i}(D_i), P_j)\simeq \delta_{i,j}{\underline \IC},
\qquad  {\rm RHom}(\CO_{D_i}(D_i), Q_j)=0, \qquad 1\leq j\leq N-1\\
& {\rm RHom}(\CO_{D_i}(D_i),P_N[1]) \simeq {\underline \IC},
\qquad
{\rm RHom}(\CO_{D_i}(D_i),Q_N) \simeq 0, \\
\eal
\ee
where ${\underline \IC}$ stands for
the one term complex of vector spaces consisting of $\IC$
in degree zero.
Since $(P_i,Q_i)_{1\leq i\leq N}$ have compact support,
for each isomorphism listed in \eqref{eq:geomlineopA},
 there is a second one obtained
by Serre duality. For example the first isomorphism in
\eqref{eq:geomlineopA} yields
\[
{\rm RHom}(P_j, \CO_{D_i}(D_i)) \simeq \delta_{i,j}{\underline \IC}[-3].
\]
Now let $N_j= \CO_{D_j}(D_j)[1]$ for some fixed $1\leq j\leq N-1$.
Note that 
\[
{\rm RHom}(P_j, N_j) \simeq {\underline \IC}[-2],
\]
therefore 
\[
{\rm Ext}^2_{X_N}(P_j,N_j) \simeq \IC
\]
So by Serre duality,
\be\label{eq:star}
{\rm Ext}^1_{X_N}(N_j,P_j) \simeq \IC.
\ee
Then, using equations 
\eqref{eq:geomlineopA} and 
\eqref{eq:star}, 
the ${\rm Ext}^1$-quiver ${\widetilde \CQ}$  of the collection
of D-branes $(P_i,Q_i)_{1\leq i\leq N}$ in the presence of the extra
noncompact object $N_j$ is
\be\label{eq:framedquivA}
\xymatrix{
& & & \vdots & \ar@/_0.2pc/[dl]|{r_N} & {}\quad 
\ar@/^0.2pc/[dll]|{s_N} & \vdots &  & \\
& & & Q_N\ar@<.5ex>[rrr]|{c_N} \ar@{-}[u]|{b_N} 
\ar@<-.5ex>[rrr]|{d_N}
& & & P_N \ar@/_1pc/[ddlll]|{r_{N-1}}
\ar@/^1pc/[ddlll]|{s_{N-1}}  \ar@{-}[u]|{a_N}
\ar@/^1pc/[rdddddd]|{g_j} & &
\\
& & & & &  &  & & \\
& & & Q_{N-1} \ar@<.5ex>[rrr]|{c_{N-1}} 
\ar@<-.5ex>[rrr]|{d_{N-1}}
\ar[uu]|{b_{N-1}}
& & & P_{N-1} \ar[uu]|{a_{N-1}}  & & \\
& & & \vdots & & & \vdots & & \\
& & & Q_{j+1}
 \ar@<.5ex>[rrr]|{c_{j+1}} \ar@<-.5ex>[rrr]|{d_{j+1}}
& &  & P_{j+1} \ar@/_1pc/[ddlll]|{r_j}
\ar@/^1pc/[ddlll]|{s_j} & & \\
& & & & & & & & \\
& & & Q_j \ar@<.5ex>[rrr]|{c_j} 
\ar@<-.5ex>[rrr]|{d_j}
\ar[uu]|{b_j}
& & &  P_j \ar[uu]|{a_j}  & N_j\ar[l]|{f_j} 
& \\
& & & \vdots && & \vdots & & \\
& & & Q_2 \ar@<.5ex>[rrr]|{c_2} \ar@<-.5ex>[rrr]|{d_2}
& & & P_2 \ar@/_1pc/[ddlll]|{r_1}
\ar@/^1pc/[ddlll]|{s_1} & & \\
& & & & & & & & \\
& & & Q_1 \ar[uu]|{b_1}
\ar@<.5ex>[rrr]|{c_1} \ar@<-.5ex>[rrr]|{d_1} \ar[uu]|{b_1}
& & & P_1\ar[uu]|{a_1} 
\ar@/^0.2pc/@{-}[dl]|{s_N} \ar@/_0.2pc/@{-}[dll]|{r_N}
& & 
\\
& & &\vdots \ar[u]|{b_N} & & &\vdots \ar[u]|{a_N} & & \\
}
\ee
As explained above, a priori the
superpotential ${\widetilde \CW}$
may contain additional terms
\[
{\widetilde \CW}=\CW+\cdots
\]
corresponding to closed loops in the path algebra of
${\widetilde \CQ}$ containing the arrows
$f_j,g_j$. However, according to Sections \ref{twotwo}
and \ref{twothree}, the fractional branes $P_N,Q_N$ are
very heavy and decouple in the field theory  limit. One arrives
therefore at the natural conjecture that the framed BPS states will be configurations of the fractional
branes $(P_i,Q_i)_{1\leq i\leq N-1}$
bound to the noncompact D4-brane $N_j$.
Such bound states are quantum wavefunctions in the
quantum mechanics determined by the framed truncated
quiver with potential $({\widetilde Q}, {\widetilde W})$,
obtained by omitting $(P_N,Q_N)$ and all adjacent arrows
in the above diagram. Now an important point is that
there are no closed loops containing the framing arrow $f_j$
in the path algebra of ${\widetilde Q}$. Therefore
the truncated potential ${\widetilde W}$ must be equal
to the potential \[
W= \sum_{i=1}^{N-2} \left[r_i(a_ic_i-c_{i+1}b_i)+s_i(a_id_i-d_{i+1}b_i)
\right]
\]
of the unframed (truncated) quiver $Q$ defined in Section
\ref{twothree}.

In conclusion,
supersymmetric D-brane bound states in the presence
of a noncompact D4-brane $N_j$ must be
mathematically defined in terms of
Bridgeland stable representations
of the quiver with potential
$({\widetilde Q},W)$.
Generalizing the correspondence conjectured in Sections
\ref{twothree},\ref{twofour} for unframed BPS states,
 the framed BPS degeneracies defined in \cite{framedBPS}
 will be identified with Donaldson-Thomas invariants
 of framed quiver representations.
A more precise statement will be formulated
in Section \ref{fivethree}
after a detailed discussion of stability conditions for
framed quiver representations.

\subsection{Framed stability conditions}\label{fivetwo}
By analogy with the unframed case,  stability conditions
for framed quiver representations are determined by
the central charges assigned to each node, provided they
belong to some half-plane $\IH_\phi$ of the complex plane.
Suppose $z_i,w_i$, are  the
central charges associated to the vertices $P_i,Q_i$, $i=1,\ldots, N-1$
and $\xi$ the central charge assigned to the extra framing vertex.
The new aspect in the present case is that one has to take a
limit where the absolute value
$|\xi|$ is much larger than $|z_i|, |w_i|$ since the extra D4-brane
is noncompact.
For fixed numerical invariants, this will yield a limit stability
condition presented in detail below. Note that a similar effect of the phase 
of a noncompact D-brane was previously studied in \cite{JM}. 

Suppose ${\tilde \rho}$ is a representation of $({\widetilde Q}, W)$
with dimension vector $(d,e,1)$ where
$d=(d_i)_{1\leq i \leq N-1}$,
$e=(e_i)_{1\leq i\leq N-1}$, where the last entry corresponds to the
extra node.
The central charge of ${\tilde \rho}$ is given by
\[
Z({\tilde\rho}) =\xi + \sum_{i}^{N-1}(d_iz_i+e_iw_i)
\]
Let
\[
\mu_{(z,w,\xi)}({\tilde \rho}) = -{{\rm Re}(e^{-i\phi}Z({\tilde\rho}))\over
{\rm Im}(e^{-i\phi}Z({\tilde \rho}))}.
\]
Then ${\tilde \rho}$ is $(z,w,\xi)$-(semi)stable if
\[
\mu_{(z,w,\xi)}({\tilde \rho}') \ (\leq) \
\mu_{(z,w,\xi)}({\tilde \rho})
\]
for any proper nontrivial subrepresentation $0\subset {\tilde \rho}'\subset
{\tilde \rho}$. Note that the subrepresentation ${\tilde \rho}'$ 
is allowed to have both multiplicity $0$ and $1$ at the extra node. 

In the present case the extra framing vertex corresponds to a
noncompact D4-brane, hence the relevant stability
conditions will be limit stability conditions obtained by sending
$|\xi|\to \infty$, keeping at the same time $|z_i|, |w_i|$,
$i=1,\ldots, N-1$ finite. It is straightforward to show that with
fixed numerical invariants $(d,e,1)$, for sufficiently large
$|\xi|>>0$,
$(z,w,\xi)$-stability specializes to the conditions below,
where $\varphi \in [\phi, \phi+\pi)$
is the phase of $\xi$.
\begin{itemize}
\item[$(a)$] Any nontrivial subrepresentation $0\subset \rho'\subset
{\tilde \rho}$ with numerical invariants $(d',e',0)$ satisfies
\be\label{eq:framedstabA}
\mu_{(z,w)}(\rho') \ (\leq) \ -{\rm cot}(\varphi-\phi).
\ee
\item[$(b)$] Any nontrivial quotient ${\tilde \rho}\twoheadrightarrow
\rho''$
with numerical invariants $(d'',e'',0)$ satisfies
\be\label{eq:framedstabBC}
\mu_{(z,w)}(\rho'') \ (\geq) \ -{\rm cot}(\varphi-\phi).
\ee
\end{itemize}
For simplicity let $\delta = - {\rm cot}(\varphi-\phi)$. The above
 conditions will be referred to as framed
$(z,w,\delta)$-stability.

\subsection{Framed BPS states, Donaldson-Thomas invariants, and wallcrossing}\label{fivethree}

Donaldson-Thomas invariants and wallcrossing formulas for framed
quiver representations are obtained by 
applying the formalism \cite{wallcrossing,genDTI}
to the abelian category of $({\widetilde Q},W)$-modules.
Note that similar results for framed quiver representations were 
obtained in \cite{wall-framed}. We will present below a 
self-contained treatment because the details will be needed 
for applications in later sections. 

To fix notation, extension groups in the category of 
$({\widetilde Q},W)$-modules will be denoted by
${\rm Ext}^k_{({\widetilde Q},W)}({\tilde \rho}_1,
{\tilde \rho}_2)$ while extension groups of $(Q,W)$-modules
will be similarly denoted by
${\rm Ext}^k_{({Q},W)}(\rho_1,\rho_2)$.
A few basic facts on
extensions of quiver representations are summarized, for completeness, in Appendix \ref{extensions}.

 A necessary condition for the results of \cite{wallcrossing,genDTI} to apply to the
present case is that the pairing defined by
\be\label{eq:pairingAO}
\chi( {\widetilde \rho}_1, {\widetilde \rho}_2)
= \sum_{k=0}^1 (-1)^k\big({\rm dim}{\rm Ext}^k_{({\widetilde Q},W)}
({\widetilde \rho}_1, {\widetilde \rho}_2)  -
{\rm dim}{\rm Ext}^k_{({\widetilde Q},W)}
({\widetilde \rho}_2, {\widetilde \rho}_1)\big)
\ee
depend only on the dimension vectors of ${\widetilde \rho}_1$,
${\widetilde \rho}_2$, for any pair of objects.
More specifically, this is required for the construction of a well
defined integration map from the motivic Hall algebra
of the category of $({\widetilde Q},W)$-modules
to a Poisson algebra spanned by numerical $K$-theory classes
of objects over $\IQ$. The integration map constructed
in \cite{wallcrossing} uses motivic weight functions
while the one constructed in \cite{genDTI} uses the constructible function of \cite{micro}. However the above condition is a necessary prerequisite in both constructions.

Note that the pairing \eqref{eq:pairingAO} is not an index
because the $k$-th extension groups with $k=2,3$ might be nonzero.
If the category were a CY3-category, the above pairing would
reduce to an
index by Serre duality, in which case the required condition is obvious.
This is not the case for framed quiver modules, therefore
some work is needed to check the above condition
using the results in Appendix \ref{extensions}.

First note that any
$({\widetilde Q}, W)$-module fits in a canonical
exact sequence
\be\label{eq:canexseq}
0\to \rho \to {\tilde \rho} \to \lambda_0^{\oplus r}
\to 0
\ee
where $\lambda_0$ is the simple module supported at the
framing node and $r\geq 0$ the dimension of ${\tilde \rho}$
at the framing node. Then $\rho$ has dimension
zero at the framing node, i.e. it is a $(Q,W)$-module.
 Then using the standard long exact sequences
 \eqref{eq:longextseqA}, \eqref{eq:longextseqB},
and equations \eqref{eq:extstuffA}-\eqref{eq:extstuffC},
it follows easily that\footnote{Note that $r_1,r_2$ below
denote the dimensions of the vector spaces associated to the 
framing node. They should not be confused with the same notation 
used for arrows of the quiver $Q$.}
\be\label{eq:pairingA}
\chi({\widetilde \rho}_1, {\widetilde \rho}_2)=
\chi({\rho}_1, {\rho}_2)
 -r_1d_j(\rho_2)+r_2d_j(\rho_1).
\ee
where  $\chi({\rho}_1, {\rho}_2) $ is the 
same pairing restricted to $(Q,W)$-modules.
As explained at end of Appendix \ref{extensions}, 
$\chi({\rho}_1, {\rho}_2)$ coincides with the natural 
symplectic pairing of the K-theory classes $[\rho_1], [\rho_2]$ 
and is given by 
\be\label{eq:pairingBB}
\bal 
\chi( {\rho}_1, {\rho}_2)  =\langle [\rho_1], [\rho_2]\rangle =\sum_{{\sf a}}
(d_{t({\sf a})}(\rho_2) d_{h({\sf a})}(\rho_1) -
d_{t({\sf a})}(\rho_1) d_{h({\sf a})}(\rho_2)),
\eal
\ee
where the sum is over all arrows of $Q$.
In particular this implies that the pairing
 \eqref{eq:pairingA} is indeed determined by numerical invariants.

Since only $({\widetilde Q}, W)$-modules of multiplicity $r\leq 1$
will be considered in this paper,
 it will be more convenient to work with the subcategory $\CA_{\leq 1}({\widetilde Q},W)$
consisting of objects with $r\in \{0,1\}$.  Note that this
is not an abelian category but it is closed under extensions
of any two objects with $0\leq r_1+r_2\leq 1$.
Therefore there is no obstruction
in applying the formalism of \cite{wallcrossing} to
$\CA_{\leq 1}({\widetilde Q},W)$. Then
 for any dimension vector ${\gamma}=(d,e)$ and any stability
parameters $(z,w,\delta)$ one obtains motivic
 Donaldson-Thomas
invariants $DT^{mot}_{({\widetilde Q},W)}( \gamma, r; z,w,\delta)$, $r=0,1$.
Following the same steps as in Section \ref{twofour} one further defines Hodge type as well as refined framed Donaldson-Thomas invariants.

The relation between framed invariants and physical framed BPS degeneracies is analogous to the unframed case discussed in
Section \ref{twofour}.
For any line defect $L_\zeta$ of charge $\gamma$ and phase
$\zeta=-e^{i\varphi}$, one defines \cite{framedBPS}
a protected spin character
\be\label{eq:framedchar}
\fro(L_\zeta, u, \gamma; y) = {\rm Tr}_{\CH^{BPS}_{u,L,\zeta,\gamma} }y^{2J_{spin}}(-y)^{2J_R}
\ee
at any point $u$ on the Coulomb branch.
The map $\varrho:{\CC}_\CG^{\sf alg}\to {\rm Stab}^{\sf alg}(\CG)$
constructed in Section \ref{twothree} assigns to each
$u\in \CC_\CG$ a set of stability parameters $(z_i(u),w_i(u))_{1\leq i\leq N-1}$. Then, if $L_\zeta$ is one of the line defects
determined by the noncompact D4-brane $N_j$,  the geometric engineering conjecture states that
\be\label{eq:framedGEconj}
\fro(L_\zeta, u, \gamma; y)=DT^{ref}_{({\widetilde Q},W)}( \gamma, 1; z(u),w(u),\delta;y)
\ee
where $\delta = -{\rm cot}(\varphi-\phi)$.
The framed refined Donaldson-Thomas invariants are defined in terms of
the motivic ones in complete analogy with Section \ref{twofour}.

Note also that one has a positivity as well as absence of exotics conjecture
for framed BPS states  \cite{framedBPS}. These conjectures are
obvious generalizations of those stated in the unframed case in
Section \ref{twofour}.

Once a framed quiver $({\widetilde Q}, W)$ is fixed by making a
choice of $N_j$ for some $j=1,\ldots, N-1$ the refined
Donaldson-Thomas
invariants will be denoted by $DT^{ref}( \gamma, 1; z,w,\delta;y)$
for simplicity.

For clarification, note that one can choose to work either
with integral refined Donaldson-Thomas invariants as defined
in \cite{wallcrossing} or rational ones obtained by a conjectural
refinement of \cite{genDTI}. The two sets of invariants are
rational refined  invariants related to each other by refined
 multicover formulas \cite{wall-pairs,Manschot:2010qz}.
In the present case, the integral and rational invariants coincide
for framed
$r=1$ objects since
any charge vector $(\gamma,1)$ is primitive.
However, the rational unframed invariants are related to the integral ones by
\be\label{eq:refmulticover}
J(\gamma; z,w; y) := \sum_{k\geq 1, \gamma=k\gamma'}
{1\over k[k]_{(-y)}} {DT^{ref}}(\gamma'; z,w; -(-y)^k)
\ee
  where for any $n\in \IZ$,
  \[
  [n]_y = {y^n-y^{-n}\over y-y^{-1}}.
  \]

  As a test of the identification \eqref{eq:framedGEconj},
  we will now show that the wallcrossing formulas
  derived from the mathematical formalism \cite{wallcrossing},
  \cite{genDTI} coincide with the physical wallcrossing formula for
  framed BPS degeneracies derived in \cite{framedBPS}.

For fixed charge vector $\gamma$ and fixed stability
parameters $(z,w)$, strictly $(z,w,\delta)$-semistable
representations can exist only for finitely many values of
$\delta$, called critical values of type $(\gamma; z,w)$.
Physically, these are the framed BPS walls found in
\cite{framedBPS}, where there exists a charge vector
$\gamma'=(d'_i,e'_i)_{1\leq i\leq N-1}$, such that $\gamma''=\gamma-\gamma'$ has non-negative
entries and $$\zeta^{-1}\sum_{i=1}^{N-1} (z_id_i'+w_ie_i')\in
\IR_{<0}.$$
 Note that $\delta_0(\gamma)=
\mu_{(z,w)}(\gamma)$ is always a critical value of type
$(\gamma; z,w)$ with $\gamma''=0$. 
Moreover,  this is the only critical value with this property;  for
all other critical values $\gamma''\neq 0$.

  Suppose $\delta_c\in \IR$
is a critical value  for given $\gamma$, $(z,w)$.
For each $\gamma=(d,e)$ let $d_j(\gamma)$
be the component of $\gamma$ at the node $j$ which
receives the framing map $f$. Then set
\[
\langle \langle \gamma, \gamma'\rangle \rangle
= (-1)^{\chi(\gamma, \gamma')+d_j(\gamma)} (\chi(\gamma, \gamma')+d_j(\gamma)).
\]
Moreover, for any ordered
sequence $(\gamma_s)_{1\leq s\leq l}$ of dimension vectors, with $l\geq 2$, set
\[
\bal
C_y((\gamma_s))=
\prod_{s=1}^{l-1} [\langle \langle\gamma_s, \sum_{v=s+1}^l \gamma_v\rangle\rangle]_y.
\eal
\]
Since $(z,w)$ are fixed, let also $\mu(\gamma) = \mu_{(z,w)}(d,e)$.

Consider two stability parameters $\delta_-<\delta_c<\delta_+$ sufficiently close to $\delta_c$.
Then the wallcrossing formula for framed BPS degeneracies
reads
\be\label{eq:wallcrossingA}
\bal
& {DT^{ref}}(\gamma,1; z,w,\delta_-; y) (\gamma) - {DT^{ref}}(\gamma,1; z,w,\delta_+; y) =\\
&
\sum_{l\geq 2} {1\over (l-1)!}
\sum_{\substack {\gamma_1+\cdots+\gamma_l=\gamma
\\
\gamma_s\neq (0,0), \ 1\leq s\leq l-1\\
\mu(\gamma_s) = \delta_c,\ 1\leq s\leq l-1}}
C_{(-y)}((\gamma_s))
DT^{ref}(\gamma_l,1; z,w,\delta_+; y)
\prod_{s=1}^{l-1}J(\gamma_s; z,w;y).\\
\eal
\ee
In this form, the wallcrossing formula follows
from explicit Hall algebra computations
by analogy with \cite{chamberII,ranktwo}. 
We will explain below that it agrees with the refined wallcrossing formula
 of \cite{wallcrossing}, as well
 as the physical wallcrossing formula of \cite{framedBPS}.

Note that for generic
$(z,w)$ all dimension vectors $\gamma_j$, $j=1,\ldots, l-1$ in the right hand side of \eqref{eq:wallcrossingA} are parallel. Therefore they are all multiples, $\gamma_j=q_j\alpha$,
$q_j\in \IZ_{>0}$, of a dimension vector
$\alpha$. Moreover there exists a
second dimension vector $\beta$ such that  $\gamma_l
= \beta + q_l \alpha$ for some $q_l\in \IZ_{\geq 0}$. 
Then $\gamma = \beta+q\alpha$ with $q=\sum_{j=1}^l q_j$.

The refined wallcrossing formula of \cite{wallcrossing}
is formulated in terms of an
associative algebra over $\IQ$ generated by
${\hat e}_{q}$, ${\hat f}_{q}$, $q\in \IZ$,
satisfying
\be\label{eq:multable}
\bal
& {\hat e}_{q}{\hat e}_{q'} = {\hat e}_{q'}{\hat e}_{q}=
{\hat e}_{q+q'}\\
& {\hat e}_q{\hat f}_p = y^{qd_j(\alpha)}{\hat f}_{q+p}\\
& {\hat f}_p{\hat e}_q = y^{-qd_j(\alpha)} {\hat f}_{q+p}
\\
& {\hat f}_p{\hat f}_q = {\hat f}_q {\hat f}_p =0\\
\eal
\ee
The last of the above equations may look puzzling, but it
reflects the choice of working in the truncated subcategory $\CA_{\leq 1}({\widetilde Q},W)$ made in the previous section. In particular, no extensions of rank $r\geq 2$
occur in the
Hall algebra identities underlying the
wallcrossing formula \eqref{eq:wallcrossingA}
(see in particular \cite[Lemma 2.4 ]{chamberII}.)
Consider the Laurent expansion of the DT invariants 
\[
DT^{ref}(\beta+q\alpha; z,w,\delta_\pm;y)=\sum_{n\in \IZ}
\Omega_n^{\pm}(q)y^n\]
and
\[
DT^{ref}(\beta+q\alpha,1 ; z,w,\delta_\pm;y)=\sum_{n\in \IZ}
\fro_n^{\pm}(q) y^n.
\]
The geometric engineering conjecture \eqref{eq:framedGEconj}
identifies the coefficients $\Omega_n^\pm(q)$, $\fro_n^\pm(q)$
with gauge theory BPS degeneracies of given spin.
Let
\[
{\bf E}(x) = \prod_{i=0}^{+\infty} (1+y^{2i+1}x)^{-1}
\]
be the quantum dilogarithm.
For each $q\in \IZ$ let
\[
S_q= \prod_{n\in \IZ} {\bf E}((-y)^n{\hat e}_q)^{(-1)^n
{\Omega}_n(q)},
\qquad
U_q^\pm =  \prod_{n\in \IZ}
{\bf E}((-y)^n{\hat f}_q)^{(-1)^n
\fro_n^\pm(q)}
\]
Then the refined wallcrossing formula of \cite{wallcrossing,DGS} for the wall
$\delta=\delta_c$ on the $\delta$-axis is
\be\label{eq:wallcrossingB}
\prod_p U_p^{+} \prod_{q} S_q = \prod_q S_q \prod_p U_p^-.
\ee
Using the last equation in \eqref{eq:multable},
\[
\bal
 U_q^{\pm} & =1 -\sum_{n\in \IZ} \sum_{i=0}^\infty
\fro^\pm_n(q)
y^{2i+1+n} {\hat f}_q = 1+{1\over y-y^{-1}}
\sum_{n\in \IZ} y^n\fro_n^\pm(q) {\hat f}_q \\
& = 1+{1\over y-y^{-1}} DT(\beta+q\alpha, 1;z,w,\delta_\pm;y) {\hat f}_q\\
\eal
\]
By substitution in \eqref{eq:wallcrossingB} and
using again \eqref{eq:multable} it follows that the
wallcrossing formula becomes
\be\label{eq:wallcrossingC}
F^+\prod_{q}S_q = (\prod_{q} S_q) F^-
\ee
where
\[
F^+ = \sum_{q}DT(\beta+q\alpha,1;z,w,\delta_\pm;y) {\hat f}_q.
\]
In the view of equation \eqref{eq:framedGEconj},
this is the same as the wallcrossing formula
\cite[Eq. (3.43)]{framedBPS}.

For comparison with \eqref{eq:wallcrossingA} note that
the latter may be rewritten as
\be\label{eq:wallcrossingAB}
\bal
& DT^{ref}(\beta+q\alpha,1; z,w,\delta_-; y) - DT^{ref}(\beta+q\alpha,1; z,w,\delta_+; y) =\\
&
\sum_{l\geq 2} {1\over (l-1)!}
\sum_{\substack {q_1+\cdots+q_l=q
\\
q_s\neq 0, \ 1\leq s\leq l\\ }}
DT^{ref}(\beta+q_l\alpha,1; z,w,\delta_+; y)\\
& \qquad \qquad \qquad \qquad 
\prod_{s=1}^{l-1}[\sum_{v=s+1}^l-1
(q_s-q_v)\langle\langle \alpha, \beta \rangle \rangle]_{(-y)}
J(q_s\alpha; z,w;y).\\
\eal
\ee
Then
note that
\[
\bal
{\rm ln}\, S_q & = \sum_{n\in \IZ} (-1)^{n}\Omega_n(q)
\sum_{i=0}^\infty \sum_{k=1}^\infty
{(-1)^{(n+1)k}\over k} \big(y^{2i+1+n}{\hat e}_q\big)^k \\
& = -\sum_{k=1}^{+\infty} {1\over k((-y)^k-(-y)^{-k})}
\sum_{n\in \IZ}[-(-y^k)]^n\Omega_n(q) {\hat e}_{kq}\\
& = {1\over y-y^{-1}} \sum_{k=1}^{+\infty} {1\over k[k]_{(-y)}}
DT^{ref}(q\alpha;z,w;-(-y)^k){\hat e}_{kq}
\eal
\]
which explains the refined multicover formula
\eqref{eq:refmulticover}. Equation \eqref{eq:wallcrossingAB}
follows by collecting the terms in \eqref{eq:wallcrossingB}
as in \cite[Sect. 4]{chamberII}, \cite[Sect 4.]{ranktwo}.

\subsection{A recursion formula for unframed BPS
states}\label{recursionsect}
As an application of the mathematical formalism developed so far,
a recursion formula will be derived next for unframed BPS invariants
at any values of the stability parameters $(z,w)$. This will follow from
the chamber structure of the framed BPS spectrum on the
$\delta$-line, keeping the parameters $(z,w)$ fixed.
Note that any framed representation ${\tilde \rho}$ of numerical
type $(d,e,1)$ has a
canonical subrepresentation $\rho$
 of type $(d,e,0)$ obtained by simply
removing the framing data. According to condition $(a)$ above, if
$(d,e)\neq (0,0)$,
$\rho$ destabilizes ${\tilde \rho}$ if
\be\label{eq:framedchamberA}
\delta < \mu_{(z,w)}(\rho)\equiv \mu_{(z,w)}(d,e).
\ee
Therefore in the chamber \eqref{eq:framedchamberA}, the only semistable framed representation is the simple
module associated to the framing node.
Therefore
\[
DT^{ref}(d,e,1; z,w,\delta)=0
\]
for all $(d,e)\neq (0,0)$, for any $\delta$ satisfying inequality \eqref{eq:framedchamberA}. In this chamber the
only nontrivial invariant is
\[
DT^{ref}(0,0,1;z,w,\delta) =1,
\]
which is in fact independent of the stability parameters.

At the same time condition $(b)$ at the end of Section 
\ref{fivetwo} 
rules out any
quotient ${\tilde \rho}\twoheadrightarrow \rho''$ for
\[
\delta >>0.
\]
This is equivalent to the statement that the quiver module ${\tilde\rho}$
is generated by the framing vector as a module over the path
algebra i.e. it is a cyclic module.
Therefore for any charge vector $\gamma$ and 
any parameters $(z,w)$ there is an asymptotic chamber $\delta>>0$
where $(z,w,\delta)$-stability
is equivalent to cyclicity. This usually called the noncommutative Donaldson-Thomas (NCDT) 
chamber in the quiver literature \cite{szendroi-noncomm}.

An important feature of the NCDT chamber is that it usually leads to 
explicit combinatorial formulas for unrefined framed 
DT invariants by virtual 
localization \cite{GP,symm}. 
Typically one uses an algebraic torus action on the 
moduli space of stable objects induced by a scaling action on the 
linear maps of the quiver representations 
which preserves the potential $W$. The virtual localization formula then 
expresses the unrefined DT invariants as a sum of local contributions 
associated to the fixed loci of the torus action on the moduli space. 
For generic values of $\delta$, the classification of stable representations 
fixed by the torus action up to isomorphism is very difficult, and the fixed 
loci are often higher dimensional. In contrast, for a sufficiently 
generic torus action 
$$\IC^\times \times 
\CM_{cyclic}(\gamma,1)\to \CM_{cyclic}(\gamma,1)$$
on the moduli space of framed cyclic modules, the 
 fixed loci are isolated and have a relatively simple classification in terms of collections of colored partitions 
\cite{youngA, youngB,szendroi-noncomm, NCDT-tilings}. Then the fixed point 
theorems of \cite{GP, symm} yield an expression of the form 
\be\label{eq:virtloc}
DT(\gamma, 1; z,w, +\infty) = \sum_{{\tilde \rho} \in 
\CM_{cyclic}(\gamma,1)^{\bf \IC^\times}} 
(-1)^{w({\tilde \rho})} 
\ee
for the unrefined NCDT invariants, where $w({\tilde \rho})$ is the dimension of the Zariski tangent space $T_{\tilde \rho}\CM_{cyclic}(\gamma,1)$.
An explicit torus action with isolated fixed loci is 
given for framed $SU(N)$ quivers in Appendix \ref{appC}. 
Since the classification of the fixed points is still fairly involved 
for general $N$, explicit computations using formula 
\eqref{eq:virtloc} are carried out only for $N=3$. 
In this case we also prove that there are only finitely many cyclic 
modules up to isomorphism, i.e the asymptotic framed BPS spectrum is 
finite. 

Motivic NCDT invariants can also be explicitly computed in certain examples 
using different techniques  
 \cite{szendroi-noncomm, NCDT-tilings, motivic-conifold, motivic-crepant, motivic-dim-red}. A computation based on \cite{motivic-dim-red} 
 is outlined for $SU(N)$ quivers  in Section 
 \ref{exotics}.
 
In conclusion, the framed BPS invariants are
explicitly
computable both for $\delta<<0$ and $\delta>>0$.
Then the unframed ones can be recursively determined
summing the contributions of all intermediate walls, according to
equation \eqref{eq:wallcrossingA}.
In close analogy with \cite{wall-pairs}, this yields the following formula
\be\label{eq:recformA}
\bal
& (-1)^{d_j(\gamma)}d_j(\gamma)J(\gamma;z,w;y)+ \sum_{l\geq 2} {1\over l!}
\sum_{\substack {\gamma_1+\cdots+\gamma_l=\gamma
\\
\gamma_s\neq (0,0), \ 1\leq s\leq l\\
\mu(\gamma_s) = \mu(\gamma),\ 1\leq s\leq l}}
 {C}_{(-y)}((\gamma_s))
 \prod_{s=1}^{l}J(\gamma_s;z,w;y)  +\\
& DT^{ref}(\gamma,1; z,w,+\infty; y)  +
\sum_{l\geq 2} \sum_{\substack {\gamma_1+\cdots+\gamma_l=\gamma
\\
\gamma_s\neq (0,0), \ 1\leq s\leq l\\
 \mu(\gamma)< \mu(\gamma_1)\leq \mu(\gamma_2)\leq \cdots \leq
 \mu(\gamma_{l-1})}}\\
 &
\qquad \qquad \qquad
\bigg[{C_{(-y)}((\gamma_s))\over
s(\gamma_1,\ldots, \gamma_{l-1})}
 DT^{ref}(\gamma_l,1; z,w,+\infty; y) \prod_{s=1}^{l-1}J(\gamma_s;z,w;y)\bigg] =0,
 \\
\eal
\ee
where $s(\gamma_1,\ldots, \gamma_{l-1})$ is the order of the
subgroup $\CS(\gamma_1,\ldots, \gamma_{l-1})\subset \CS_{l-1}$
of the permutation group of $(l-1)$ letters
consisting of permutations $\sigma$ preserving the slope ordering,
i.e.
\[
\mu(\gamma_{\sigma(1)} )\leq \mu(\gamma_{\sigma(2)})
\leq \cdots \mu(\gamma_{\sigma(l-1)}).
\]
More explicitly, for any sequence $(\gamma_1,\ldots,\gamma_{l-1})$
satisfying the slope inequalities
\[
\mu(\gamma_{1} )\leq \mu(\gamma_{2})
\leq \cdots \leq \mu(\gamma_{l-1})
\]
there exists a unique partition $l-1 = \sum_{i=1}^k l_i$ with
$l_i\geq 1$, such that
\[
\mu(\gamma_1)=\cdots = \mu(\gamma_{l_1}) <
\mu(\gamma_{l_1+1}) = \cdots = \mu(\gamma_{l_1+l_2})
< \cdots < \mu(\gamma_{l-l_k}) =\cdots = \mu(\gamma_{l-1})
\]
Then $\CS(\gamma_1,\ldots, \gamma_{l-1})\simeq
\times_{i=1}^k \CS_{l_i}$ and
\[
s(\gamma_1,\ldots, \gamma_{l-1}) = \prod_{i=1}^k l_i!.
\]

For any charge vector $\gamma=(d_i,e_i)_{1\leq i\leq N-1}$
define the height $|\gamma|= \sum_{i=1}^{N-1}(d_i+e_i)$.
Then, the above
formula determines all $J(\gamma;z,w;y)$ recursively
in the height $|\gamma|$ provided that
 the asymptotic
 invariants $DT^{ref}(\gamma,1; z,w,+\infty; y) $ are known.

Note that numerical Donaldson-Thomas invariants and
wallcrossing formulas are obtained taking the
limit $y\to (-1)$ in the above formulas. Abusing notation,
the numerical invariants will be denoted by the same symbols,
the distinction residing in the absence of the argument $y$.
Note that the $y\to (-1)$ limit of $C_{(-y)}((\gamma_s))$ is
\[
\bal
C((\gamma_s))=
\prod_{s=1}^{l-1} \langle \langle\gamma_s, \sum_{v=s+1}^l \gamma_v\rangle\rangle.
\eal
\]

For illustration, the numerical version of the recursion
formula \eqref{eq:recformA} will be tested below
in the strong coupling chamber found in Section
\ref{strongsection}.  The stability parameters $(z,w)$
are of the form
\[
z_i = -\theta_i+ \sqrt{-1}, \qquad
w_i = -\eta_i + \sqrt{-1}, \qquad i=1,2,
\]
where $\theta_i, \eta_i\in \IR$, $i=1,2$, satisfy
\[
\eta_2< \theta_2 < \eta_1  < \theta_1.
\]
Consider for example the dimension vector 
 $\gamma=(1,1,0,0)$ and let $j=2$.
There is only one decomposition of $\gamma$ up to permutations
\[
\gamma= (1,0,0,0)+(0,1,0,0).
\]
The $(\theta,\eta)$-slopes,
\[
\mu(1,0,0,0)=\theta_1, \qquad \mu(0,1,0,0)=\theta_2,
\qquad \mu(1,1,0,0)={\theta_1+\theta_2\over 2}
\]
are ordered as follows
\[
\mu(1,0,0,0)>\mu(1,1,0,0) > \mu(0,1,0,0).
\]
Given the summation conditions in \eqref{eq:recformA},
it follows that the sum over $l\geq 2$ in the first row is trivial,
while the sum over $l\geq 2$ in the second row reduces to a single term,
\[
C((\gamma_1,\gamma_2))
DT(\gamma_2,1;z,w,+\infty)
J(\gamma_1;z,w)
\]
where
\[
\gamma_1=(1,0,0,0), \qquad \gamma_2=(0,1,0,0).
\]
By direct substitution,
\[
d_2(\gamma)=1, \qquad
\chi(\gamma_1,\gamma_2)=-1, \qquad d_2(\gamma_1)=0,
\]
hence
\[
(-1)^{d_2(\gamma)}d_2(\gamma)=-1,
\qquad
C((\gamma_1,\gamma_2))= 1.
\]
Therefore \eqref{eq:recformA} reduces to
\[
-J(\gamma;z,w) + DT(\gamma,1;z,w,+\infty) + DT(\gamma_2,1;z,w,+\infty)
 J(\gamma_1;z,w) =0.
\]
According to Appendix \ref{appC}, $DT(\gamma,1;z,w,+\infty)=0$ and $DT(\gamma_2,1;z,w,+\infty)=1$.
 Moreover
 $J(\gamma;z,w)=J(\gamma_1;z,w)=1$ at strong coupling, hence the $y\to (-1)$ limit of formula \eqref{eq:recformA}
holds in this case.

 As an alternative to the recursion formula, note that the spectrum of
 framed BPS states at $\delta>>0$ can be also related
 to the spectrum at $\delta<<0$ applying directly
 the wall crossing
 formula of \cite{wallcrossing}. Again, using the
 $SU(3)$ strong coupling chamber found in Section 
 \ref{strongsection} as an example,
 recall that there are 6 unframed BPS states
 with dimension vectors 
  \[
 \bal
 & \gamma_1=(1,0,0,0),\qquad
 \gamma_2=(0,1,0,0), \qquad
 \gamma_3=(0,0,1,0),\\
 &
 \gamma_4=(0,0,0,1),\qquad
 \gamma_5=(1,1,0,0), \qquad
 \gamma_6=(0,0,1,1).\\
 \eal
 \]
In addition in the chamber $\delta<<0$ there is only one
framed BPS states with dimension vector $\gamma=0$ corresponding to the simple module supported at the framing
node. Consider the Lie algebra over $\IQ$ generated
by $\{e_\gamma, f_\gamma\}$ satisfying
\[
[e_\gamma, e_{\gamma'}] =(-1)^{\chi(\gamma,\gamma')}
\chi(\gamma,\gamma')e_{\gamma+\gamma'}, \qquad
[e_\gamma, f_{\gamma'}] =
(-1)^{d_j(\gamma)+\chi(\gamma,\gamma')}(d_j(\gamma)
+\chi(\gamma, \gamma'))f_{\gamma+\gamma'}.
\]
 Define $U_i$, $i=1,\ldots, 6$ by
$$
U_i = \text{exp} (\sum \frac{{e}_{m\gamma_i}}{m^2}),
$$
Then the framed spectrum in the chamber $\delta>>0$ is determined by
$$
(U_4 U_2 U_6 U_5 U_3 U_1)^{-1} \text{exp}
({f}_{0}) U_4 U_2 U_6 U_5 U_3 U_1 \ .
$$
Note that $U_2$ and $U_6$ commute and $U_3$ and $U_5$ commute. After the algebraic manipulation we obtain the 7 invariants listed in
Appendix \ref{appC}, equation \eqref{eq:FVtwo}.

\subsection{Absence of exotics I}\label{exotics}

This section will be concluded with an application of the
above results to the absence of exotics conjecture formulated
in Section \ref{twofour}. As explained there, 
assuming a Lefschetz type construction 
for the $SL(2,\IC)_{spin}$ action on the 
space of BPS states, absence of exotics
translates into the vanishing of off-diagonal virtual Hodge
numbers ${\mathfrak h}^{r,s}(\gamma;z,w)$
of the moduli space of $(z,w)$-stable representations
with charge $\gamma$.
Equivalently, the Hodge type Donaldson-Thomas invariants
$DT(\gamma; z,w; x,y)$ in equation \eqref{eq:hodgeDT}
are required to be Laurent polynomials in $(xy)^{1/2}$.
Laurent polynomials in $x^{1/2}, y^{1/2}$ satisfying this 
condition will be called {\it rational} in the following. 
There is of course an entirely analogous
statement for framed BPS states, where the virtual
Hodge numbers depend on the extra stability parameter $\delta$.
%A stronger statement, which implies absence of exotics, is that
%the motivic framed and unframed Donaldson-Thomas invariants
%are only functions of $\IL^{1/2}$. For ease of exposition,
%motivic invariants satisfying this property will be called rational
%in the following.
Below it will be shown that  the required 
vanishing results follow by wallcrossing
from the chamber structure of framed BPS states studied in the
previous subsection.

Using the formalism of \cite{wallcrossing},  the
wallcrossing formula \eqref{eq:wallcrossingA}
admits a natural motivic
version written in terms of motivic 
DT invariants 
$DT^{mot}(\gamma_l, 1; z,w, \delta_\pm))$, 
$J^{mot}(\gamma_s; z,w)$. The coefficients 
$C_{(-y)}((\gamma_s))$ must be accordingly 
replaced by their motivic versions 
$C_{(-\IL^{1/2})}((\gamma_s))$.
Taking virtual
Hodge polynomials, the motivic wallcrossing formula yields
a polynomial wallcrossing formula for Hodge type Donaldson-Thomas
invariants:
\be\label{eq:motwallcrossing}
\bal
& {DT}(\gamma,1; z,w,\delta_-;x,y) (\gamma) -
{DT}(\gamma,1; z,w,\delta_+;x,y) =\\
&
\sum_{l\geq 2} {1\over (l-1)!}
\sum_{\substack {\gamma_1+\cdots+\gamma_l=\gamma
\\
\gamma_s\neq (0,0), \ 1\leq s\leq l-1\\
\mu(\gamma_s) = \delta_c,\ 1\leq s\leq l-1}}
C_{(-(xy)^{1/2})}((\gamma_s))
DT(\gamma_l,1; z,w,\delta_+;x,y)
\prod_{s=1}^{l-1}J(\gamma_s; z,w;x,y),\\
\eal
\ee
where the $J(\gamma_s; z,w;x,y)$ are the images of 
the motivic invariants $J^{mot}(\gamma_s; z,w)$ 
via the virtual Hodge polynomial map. 
Analogous considerations hold of course for the recursion formula  \eqref{eq:recformA} which is an iteration of the wallcrossing formula.
Using these formulas, absence of exotics for framed and
unframed invariants reduces to absence of exotics for the framed asymptotic ones. The latter will then
be proven shortly
using the results of \cite{motivic-dim-red}. Therefore, in short,
rationality of both framed and unframed invariants is established,
granting the motivic wallcrossing formula of \cite{wallcrossing}
for $SU(N)$ quivers.

The proof of absence of exotics for asymptotic framed invariants
will be based on the main result of \cite{motivic-dim-red},
where they are expressed
in terms of Chow motives of certain affine
varieties. In order to apply the results of \cite{motivic-dim-red}
one first has to check that the
 potential
 \[
 W = \sum_{i=1}^{N-2}[r_i(a_ic_i-c_{i+1}b_i)+s_i(a_id_i-d_{i+1}b_i)]
 \]
of the $SU(N)$ quiver has a linear factor
 according to \cite[Def. 2.1]{motivic-dim-red}.
First note that any potential $W'$ which differs from $W$ by cyclic
permutations in each term is equivalent to $W$ since they define the
same relations in the path algebra. Therefore $W$ is equivalent to
\[
W' = \sum_{i=1}^{N-2}(a_ic_ir_i-b_ir_ic_{i+1}+a_id_is_i-b_is_id_{i+1})
\]
Then note that $W'$ has a factorization of the form $W'=LR$
in the path algebra of the quiver without relations, where
 \[
 L=\sum_{i=1}^{N-2}(a_i+b_i), \qquad
 R = \sum_{i=1}^{N-2}(c_ir_i-c_{i+1}b_i+ a_id_i-d_{i+1}b_i).
 \]
 Since the product is defined by concatenation of paths it is straightforward to check that all terms in the expansion of $LR$
not belonging to $W'$ are trivial. For example
\[
a_ic_{i+1}b_i=0
\]
since the tail of $a_i$ does not coincide with the head of $c_{i+1}$.
Moreover, any two distinct nodes of the quiver are connected by at most one of the arrows $a_i, b_i$, $1\leq i\leq N_2$ in $L$.
These are precisely
the conditions required by \cite[Def 2.1]{motivic-dim-red}.

Then
\cite[Thm. 7.1]{motivic-dim-red} provides an explicit expression for the
 motivic Donaldson-Thomas invariants
 $DT^{mot}(\gamma, 1; z,w,+\infty)$ in terms
 of Chow motives of general linear groups
$GL(n,\IC)$, $n\geq 1$ and Chow motives of ``reduced quiver varieties",
which are constructed as follows.  For  a fixed dimension vector
$\gamma=(d_i,e_i)_{1\leq i\leq N-1}$ let
\[
\CV(\gamma) =
\bigoplus_{{\sf a}\in
\{a_j,b_j,c_j,d_j,r_j,s_j\}} {\rm Hom}(\IC^{d_{t({\sf a})}},
\IC^{d_{h({\sf a})}})
\]
be the linear space of all quiver representations.
The potential $W$ determines a gauge invariant polynomial function
$W_\gamma$ on $\CV(\gamma)$. The ``reduced
quiver variety"
$\CR(\gamma)$ is defined as the zero locus of the F-term equations
\[
a_i=b_i=0, \qquad
\partial_{a_i}W_\gamma = 0, \qquad 
\partial_{b_i}W_\gamma=0
\]
in $\CV(\gamma)$. In the present case, one obtains the quadratic equations
\be\label{eq:quadrel}
c_ir_i+d_is_i=0, \qquad r_ic_{i+1}+s_id_{i+1}=0.
\ee
According to equation
\eqref{eq:GLmotive} in Appendix \ref{motives},
\[
[GL(n,\IC)]= \prod_{k=1}^n (\IL^k-1),
\]
hence its virtual Hodge polynomial is $\prod_{k=1}^n((xy)^k-1)$.
Therefore, thanks to the above result of 
\cite{motivic-dim-red},
 in order to prove absence of exotics in the asymptotic
chamber it suffices to prove that the virtual Hodge polynomials with
compact support of the reduced varieties $\CR(\gamma)$ are polynomial functions on $(xy)$.
This will be done below using the compatibility of the virtual Hodge
polynomial with motivic decompositions.

Note that $\CR(\gamma)$
admits a natural projection  $\pi:\CR(\gamma)\to \CB(\gamma)$ to the
linear space
 $\CB(\gamma) =\bigoplus_{{\sf a}\in \{c_j,d_j\}}
{\rm Hom}(\IC^{d_{t({\sf a})}}, \IC^{d_{h({\sf a})}})$
given by
\[
\pi(c_i,d_i,r_i,s_i) \mapsto (c_i,d_i).
\]
For each $i=1,\ldots, N-1$ the pair of linear maps
$(c_i,d_i)$
determine a Kronecker module $\kappa_i$
of dimension vector $(e_i,d_i)$.
Therefore $\CB(\gamma)$ is a direct product
\[
\CB(\gamma) = \times_{i=1}^{N-1} \CV(e_i,d_i)
\]
where $\CV(e_i,d_i)$ denotes the linear space of all Kronecker modules
of dimension vector $(d_i,e_i)$. 
Using equation \eqref{eq:Kdualexts}, the space of 
 solution $(r_i,s_i)$ to equations 
\eqref{eq:quadrel} is isomorphic to the dual 
extension group of Kronecker modules ${\rm Ext}^1(\kappa_i, 
\kappa_{i+1})^\vee$. Therefore the fiber of $\pi$ over a point
$(\kappa_i)\in \CB(\gamma)$ is isomorphic to the linear space
\[
\bigoplus_{i=1}^{N-2}{\rm Ext}^1_\CK(\kappa_i,\kappa_{i+1})^\vee,
\]
where $\CK$ denotes the abelian category of Kronecker modules.

If the dimension of the fiber $\pi^{-1}(\kappa_1,\ldots, \kappa_{N-1})$
were constant, $\CR(\gamma)$ would be isomorphic to a product of linear spaces, which is obviously rational. This is in fact not the case;
the dimensions of the fiber jumps as the point $(\kappa_1,\ldots, \kappa_{N_1})$ moves in $\CB(\gamma)$.
However, suppose there is a finite stratification of $\CB$ with locally
closed  strata ${\CS_\alpha}$ such that the fiber of $\pi$ has constant dimension $p_\alpha$ over the stratum $\CS_\alpha$.
Then the following relation holds in the ring of motives
\[
[\CR(\gamma)] =\sum_{\alpha}\IL^{p_\alpha} [\CS_\alpha].
\]
This yields a similar relation,
\[
P_{(x,y)}(\CR(\gamma)) = \sum_{\alpha}(xy)^{p_\alpha}
P_{(x,y)}(\CS_\alpha)
\]
for virtual Hodge polynomials with compact support.
Therefore in order to prove the claim
it suffices to construct a stratification
$\CS_\alpha$ such that
each polynomial  $P_{(x,y)}(\CS_\alpha)$ is only a function
of $(xy)$.
It will be shown next that the natural stratification of
$\CB(\gamma)$ by gauge group orbits satisfies this condition.

Let $\CV(e,d)\simeq {\rm Hom}(\IC^e,\IC^d)^{\oplus 2}$ be the linear
space of all Kronecker modules of dimension vector $(e,d)$. Suppose
$\CS$ is an orbit of the natural $GL(e,\IC)\times GL(d,\IC)$ action on
$\CV(e,d)$ and let $G_\CS\subset GL(e,\IC)\times GL(d,\IC)$ be its stabilizer. Given any Kronecker module $\kappa\in \CS$
corresponding to a
point in $\CS$, the stabilizer $G_\CS$ is isomorphic to the group
of invertible elements in the endomorphism algebra ${\rm End}_\CK(\kappa)$. Recall that $\CK$ denotes the abelian category of
Kronecker modules. According to \cite{J-motivic}
(see below Def. 2.1), this implies that $G_\CS$ is special, which means
that any principal $G_\CS$-bundle over a complex variety is locally
trivial in the Zariski topology. In particular this holds for the
natural principal $G_\CS$-bundle
\[
\xymatrix{
G_\CS \ar[r] & GL(e,\IC)\times GL(d,\IC) \ar[d] \\
& \CS,\\}
\]
which yields a relation of the form
\[
[GL(e,\IC)][GL(d,\IC)] = [G_\CS][\CS]
\]
in the ring of motives. Taking virtual Hodge polynomials with
compact support one further obtains
\[
P_{(x,y)}(\CS)P_{(x,y)}(G_\CS) =
\prod_{k=1}^d((xy)^k-1)\prod_{l=1}^e((xy)^l-1).
\]
Note that the right hand side of this identity is a product of irreducible
factors $(xy-\zeta)$, with $\zeta$ a root of unity. Since the polynomial
ring is a unique factorization domain, it follows that the same must hold
for both factors in the left hand side. Therefore $P_{(x,y)}(\CS)$
is a polynomial function of $(xy)$ as claimed above.

 In order to conclude this section,
one can ask the question whether absence of exotics
may hold for the BPS spectrum of any toric Calabi-Yau threefold.
We expect this to be the case for general BPS states on
toric Calabi-Yau threefolds, based on similar arguments.
 Using dimer technology
\cite{dimer_special,tiltingII} any toric  Calabi-Yau
threefold $X$ has an exceptional collection of line bundles
which identifies the derived category $D^b(X)$
with the derived category of a quiver with potential
$(Q,,W)$. There is moreover a region in the K\"ahler
moduli space where one can construct Bridgeland
stability conditions where the heart of the underlying
$t$-structure is the abelian category of
$(Q,W)$-modules.
In this region, BPS states will be mathematically
modeled by supersymmetric
quantum states of moduli spaces of stable
quiver representations.
Moreover, explicit formulas for
motivic Donaldson-Thomas invariants
of moduli spaces of
framed cyclic representations have been obtained in
\cite{NCDT-tilings}, and they depend only on
$\IL^{1/2}$. Therefore one can employ a similar strategy,
defining $\delta$-stability conditions for framed representations, and studying motivic wallcrossing.
This provides a framework
for a mathematical study of absence of exotics
for dimer models. The details will be left for future work.

A much harder problem is absence of exotics in geometric
regions of the K\"ahler moduli space \cite{locptwo} where
there are no stability conditions
with algebraic t-structures. In those cases, one has to
employ perverse t-structures in the construction of
stability conditions, and the
role of framed quiver representations is played by
large radius stable pair invariants.
Explicit motivic formulas for such invariants
are known only in cases
where $X$ has no compact divisors \cite{motivic-conifold,motivic-crepant}.
If $X$ has compact divisors, no explicit large radius  motivic computations
have been carried up to date. However absence of exotics is
expected based on the refined vertex formalism \cite{IKV}.
Moreover, the cohomology of smooth moduli spaces of semi-stable sheaves on rational surfaces is
known to be of Hodge type $(p,p)$ \cite{modulicohA,modulicohB}, which suggests that also for toric Calabi-Yau's with compact 
divisors no exotic BPS states arise.
For completeness note that  absence of exotics is
known to fail \cite{Andriyash:2010yf, wall-pairs} on non-toric Calabi-Yau threefolds.

\section{BPS states and cohomological Hall 
algebras}\label{sectionsix}

This section explains
the relation between BPS states and
the mathematical formalism of  cohomological  Hall algebras
 \cite{COHA}. Although cohomological and
motivic Donaldson-Thomas invariants are known to be equivalent
\cite{COHA}, the cohomological construction
provides more insight into the
 geometric construction of the $SL(2,\IC)_{spin}$-action
on the space of BPS states \cite{berkeley}. Moreover, it also
offers a new perspective on the absence of exotics,
which is now related to a conjectural Atiyah-Bott fixed
point theorem for the cohomology groups defined
in \cite{COHA}.

\subsection{Cohomological Hall algebras}
The algebra of BPS states was first constructed in \cite{BPSalg}
in terms of scattering amplitudes for
D-brane bound states. In the semiclassical approximation, the algebraic
structure is encoded in the overlap of three quantum BPS wave functions
on an appropriate correspondence variety. This formulation can be
made very explicit for quiver quantum mechanics.
 More recently, a rigorous mathematical formalism
for BPS states has been proposed in \cite{COHA} for quivers
with potential. A detailed comparison between the physical definition
of \cite{BPSalg} and the formalism of \cite{COHA}
has not been carried out so far in the literature.
Leaving this for future work,  the construction of \cite{COHA}
will be briefly summarized below.

%Mathematically, the spaces of BPS states for a quiver
%with potential have been constructed in \cite{COHA}.
The most general definition of the corresponding algebraic structure is given in the framework of Cohomological Hall algebra (COHA for short) in \cite{COHA}. In the loc. cit. the authors defined the category of Exponential Mixed Hodge Structures (EMHS for short) as a tensor category which encodes ``exponential periods'', i.e. integrals of the type $\int_Cexp(W)\alpha$, where $C$ is a cycle in an algebraic variety $X$, $W:X\to {\bf C}$ is a regular function (or even a formal series) and $\alpha$ is a top degree form on $C$. There are different ``cohomology theories'' which give ``realizations'' of EMHS. Every such theory is a tensor functor ${\bf H}$ from the category $EMHS$ to the category of graded vector spaces. Similarly to the conventional theory of motives there are several standard realizations:

a) Betti realization which is given by the cohomology of pairs $H^{\bullet}(X,W^{-1}(t))$, where $t$ is a negative number with a very large absolute value (it is also called ``rapid decay cohomology'');

b) De Rham realization which is given by the cohomology $H^{\bullet}(X, d+dW\wedge\,\bullet)$ of the twisted de Rham complex (or, better, the hypercohomology of the corresponding complex in the Zariski topology on $X$);

c) critical realization which is given by the cohomology of  $X$ with the coefficients in the sheaf of vanishing cycles of $W$.

It is convenient to combine all those versions of COHA into the following definition  proposed in \cite{COHA}.

\begin{defi}\label{BPScoh} {\bf Cohomological Hall algebra} of $(Q,W)$ (in realization $\bf H$) is an associative twisted graded  algebra in the target tensor category $\mathcal C$ of the 
cohomology functor ${\bf H}$ 
defined by the formula

$$\mathcal{H}:=\oplus_\gamma \mathcal{H}_\gamma\,,\,\,\,\,\mathcal{H}_\gamma:={\bf H}^\bullet(M_\gamma/G_\gamma,W_\gamma):=
{\bf H}^\bullet(M_\gamma^{univ},W_\gamma^{univ}),$$
where $I$ is the set of vertices of the quiver, and $\gamma=(\gamma^i)_{i\in I} \in {\bf Z}_{\ge 0}^I$ is the dimension vector.

\end{defi}
In the above definition  $M_{\gamma}$ is  the space of representations of $Q$ of dimension $\gamma=(\gamma^i)_{i\in I}\in {\bf Z}_{\ge 0}^I$, and    $M_\gamma^{univ}=EG_{\gamma}\times_{BG_{\gamma}}M_{\gamma}$ is be the standard universal space used in the  definition of equivariant cohomology.
The group $G_{\gamma}=\prod_{i\in I}GL(\gamma^i,{\bf C})$ acts by changing basis at each vertex of the quiver and $M_\gamma/G_\gamma$ denotes the corresponding stack.
It was proved in the loc. cit. that COHA is an associative algebra.

It is important to explain the relationship of COHA to the space of BPS states. By definition
COHA does not depend on the central charge $Z$, hence it does not depend on stability parameters. It was conjectured in loc.cit.  that after a choice of the central charge, the algebra $\mathcal{H}$ ``looks like'' the universal enveloping algebra of a Lie algebra $g$ which is a direct sum (as a vector space) of the ``fixed slope`` Lie algebras. More precisely $g=\oplus_lg_l$ where the summation is taken over the set of rays $l$ in the upper-half plane, and each algebra $g_l$  depends on the moduli spaces of semistable objects with central charges in $l$. Therefore a choice of stability conditions should determine a space of ``good`` generators of COHA. This space of generators is not known aside from a of few simple examples. Furthermore, it is not known whether there is a space $V$ of generators of $\mathcal{H}$ which carries a Lie algebra structure (hence we cannot identify the Lie algebra $g$ with such a set). It is expected (and proved in the case of symmetric quivers) that there is a space of (graded) generators of the form $V=V^{prim}\otimes {\bf C}[x]$, where $deg\,x=2$ and $V$ is vector space graded by ${\bf Z}_{\ge 0}^I\times {\bf Z}$. One should think of this grading as a charge-cohomological degree grading.
From the point of view of the above discussion one can think that $V^{prim}=\oplus_{(\gamma,k)}V_{\gamma,k}$  corresponds to $\mathcal{H}_{BPS}(\gamma)$. This space of generators plays the 
role of the space of single-particle BPS states. Full COHA ``looks like``  the algebra of multiparticle states.
% (BPS-algebra in terminology of Harvey and Moore). 
In practice the space of stability parameters $u$ has the size of one-half of the full space of stability conditions. After a choice of such subspace of ``physical'' stability conditions
one defines the DT-invariants (which correspond of course to BPS invariants) as $\Omega(\gamma,u)=\sum_k y^k dim V_{\gamma,k}$. The summands in this sum can be interpreted as refined BPS invariants.
Furthermore, the above considerations can be performed directly in the category EMHS. In this case $V^{prim}$ is an object of EMHS. As in the usual Hodge theory which is mathematically described by the category of Mixed Hodge Structures (MHS for short), objects of EMHS carry (exponential version of) Hodge and weight filtrations defined in \cite{COHA}. Then the motivic and numerical DT-invariants are defined by taking Serre polynomial (a.k.a. virtual Poincar\'e polynomial) and Euler characteristic of the corresponding objects. This definition agrees with the above conjectural description via the set $V$ of generators of COHA. Notice that the motivic version of COHA carries the action of the motivic Galois group (the group of automorphisms of the corresponding cohomology functor ${\bf H})$. Therefore, instead of considering the trace of the element $-1$ which gives the Euler characteristic, one can consider the trace of another element of the motivic Galois group (or the character function as a function on the Galois group). It might lead to some new interesting invariants.

\subsection{Framing and $SL(2,\IC)_{spin}$ action}
According to  Section \ref{twofour}, physics arguments predict 
an
 $SU(2)_{spin}$ action on the space of BPS states.
If one is mainly interested in algebraic aspects, forgetting the Hilbert
space structure, the
natural induced
$SL(2,\IC)_{spin}$ action may be considered instead.
As explained in Section \ref{twofour}, the $SL(2,\IC)_{spin}$ action
is identified with the Lefschetz action on the cohomology of the
D-brane moduli
space \cite{Witten:1996qb}, which is well defined when
the latter is smooth.  Therefore, as a physical test of the formalism
proposed in \cite{COHA}, the cohomology groups constructed there
should carry a natural Lefschetz type action.
This construction was  developed by Kontsevich and Soibelman in \cite{berkeley}, and is briefly recalled below.

Let $\mathcal{C}$ be a triangulated $A_{\infty}$-category over ${\bf C}$. We also fix a stability condition $\tau$, (and hence 
a slope function $\mu$), 
and a functor $F$ from $\mathcal{C}$ to the category of complexes. For a fixed slope $\mu=\theta$ we denote by $\mathcal{C}^{ss}_{\theta}$ the abelian category of $\tau$-semistable objects having the slope $\theta$. We will impose the following assumption: {\it $F$ maps  $\mathcal{C}^{ss}_{\theta}$ to the complexes concentrated in non-negative degrees}.

\begin{defi}\label{framedobjA} Framed object is a pair $(E,f)$ where $E\in Ob(\mathcal{C}^{ss}_{\theta})$ and $f\in H^0(F(E))$.

\end{defi}

The above definition can be given in the case of abelian categories as well.
In the case of the quiver with potential $(Q,W)$ we can define the above structure by adding an extra vertex $i_0$ and $d_i$ arrows from $i_0$ to the vertex $i\in I$. Then the functor $F$ maps a representation $E=(E_i)_{i\in I}$ to $F(E)=\oplus_iHom({\bf C}^{d_i},E_i)$.

There is an obvious notion of morphism of framed objects and hence of isomorphic framed objects.
\begin{defi}\label{framedstabA}
 We call framed object stable if there is no exact triangle $E^{\prime}\to E\to E^{\prime\prime}$ such that both $E^{\prime},E^{\prime\prime}\in Ob(\mathcal{C}^{ss}_{\theta})$ and such that there is
$f^{\prime}\in H^0(F(E^{\prime}))$ which is mapped to $f\in H^0(F(E))$.

\end{defi}

\begin{prop}\label{automprop} If $(E,f)$ is a stable framed object then $Aut(E,f)=\{1\}$.

\end{prop}

Therefore one can speak about a scheme (not a stack) of stable framed objects. There are versions of this notion which involve polystable objects (i.e. sums of stables with the same slope). Conjecturally for a wide class of triangulated categories the moduli space of stable framed objects ${\mathcal M}^{sfr}$ is a projective variety.
In particular this should be true in the case of quivers with potential. The following conjecture was formulated in the loc. cit.

\begin{conj}\label{purityconj} Suppose that  ${\mathcal C}$ is a $3CY$ category. Then there is a formal manifold $\widehat{\mathcal M}^{sfr}$ and a formal function $W\in \widehat{\mathcal O}({\mathcal M}^{sfr})$ such that:

a) ${\mathcal M}^{sfr}$ is the set of critical points of $W$.

b) For every $i\ge 0$ the cohomology group $H^i({\mathcal M}^{sfr},\phi_W)$ with the coefficients in the sheaf of vanishing cycles $\phi_W({\bf Z}_{{\mathcal M}^{sfr}})$ carries a pure Hodge structure of weight $0$ as well as the Lefshetz decomposition.

\end{conj}

Assuming the Conjecture we arrive to the following:

\begin{coro}\label{puritycor} The series $A^{sfr}:=\sum_{\gamma \in {\bf Z}_{\ge 0}^I}[H^{\bullet}({\mathcal M}^{sfr},\phi_W)]\widehat{e}_{\gamma}$ enjoys the wall-crossing formulas. Here the symbol 
$[H^{\bullet}({\mathcal M}^{sfr},\phi_W)]$ denotes an 
element of the $K_0$ ring of an appropriate 
subcategory of EMHS, while 
$\widehat{e}_{\gamma}$ are generators of the  quantum torus
over this $K_0$ ring, 
$\widehat{e}_{\gamma_1}\widehat{e}_{\gamma_2}=
{\IL}^{\chi(\gamma_1,\gamma_2)/2}\widehat{e}_{\gamma_1+\gamma_2}$ and $\chi(\gamma_1,\gamma_2)$ is the Euler-Ringel form.
\end{coro}

In particular a mutation of the quiver with potential gives rise to a conjugation of $A^{sfr}$ by the quantum dilogarithm. Applying Serre polynomial we obtain the series with coefficients which are characters of finite-dimensional $SL(2,\IC)$-representations.
Specialization of the Serre polynomial to ${\mathbb L}^{1/2}=-1$ is therefore a non-negative integer number.

The geometric construction of framed BPS
states  in Section \ref{sectionfive}
suggests in fact a generalization of the above conjectures to weak
stability conditions, as explained below.

Suppose $\CA_\tau$ is the heart of the underlying t-structure
of $\tau$ and suppose $F$ maps $\CA_\tau$
 to the complexes concentrated in non-negative degrees.
 Then a framed object will be defined as a pair
  $(E,f)$ where $E\in Ob(\CA_\tau)$ and $f\in H^0(F(E))$.
Again, there is an obvious notion of morphism of framed objects and hence of isomorphic framed objects.

The stability condition for framed objects is also generalized
as follows to a condition depending on an extra parameter
$\xi\in \IC$.
In addition to a t-structure the stability condition $\tau$ also
contains a compatible stability function
 $Z_\tau:K(\CC)\to \IC$, where $K(\CC)$ is an
appropriate quotient of the Grothendieck group of $\CC$.
The compatibility condition requires
$Z_\tau(E)$ to take values in a fixed half-plane
$\IH_\phi\subset \IC$ for any object $E$ of $\CA_\tau$.
Let $\xi\in \IH_\phi$ be an arbitrary complex parameter.
For any object $E$ of $\CA_\tau$, define
\[
Z_{(\tau,\xi)}(E) = Z_\tau(E) + h^0(F(E))\xi
\]
and
\[
\mu_{(\tau,\xi)}(E) = -{{\rm Re}(e^{-i\phi}Z_{(\tau,\xi)}(E))\over
{\rm Im}(e^{-i\phi}Z_{(\tau,\xi)}(E))},
\]
where $h^k(F(E))={\rm dim}H^k(F(E))$, $k\geq 0$.
Moreover for any morphism $d:E_1\to E_2$ in $\CA_\tau$ let
$F^k(d):H^k(E_1)\to H^k(E_2)$, $k\geq 0$ be the
induced linear maps on cohomology.

\begin{defi}\label{framedstabBB}
A framed object $(E,f)$, $E\in Ob(\CA_\tau)$ is called $(\tau, \xi)$-(semi)stable if the following conditions hold.

 $(a)$ Any framed object $(E',f')$ where $0\subset E'\subset E$
 is a nontrivial proper subobject in $\CA_\tau$ such that
 $F^0(f')=f$ satisfies
 \[
\mu_{(\tau,\xi)}(E') \ (\leq)\ \mu_{(\tau,\xi)}(E).
\]

$(b)$ Any framed object $(E'',f'')$ where
$E\twoheadrightarrow E''\neq 0$ is a quotient of $E$ in
$\CA_\tau$, not isomorphic to $E$, such that $F^0(f)=f''$
 satisfies
 \[
\mu_{(\tau,\xi)}(E) \ (\leq)\ \mu_{(\tau,\xi)}(E'').
\]
\end{defi}

If appropriate boundedness results hold for fixed numerical
invariants, making $|\xi|$  very large yields the following notion of
(weak) limit stability condition.

\begin{defi}\label{framedstabB}
A framed object $(E,f)$, $E\in Ob(\CA_\tau)$ is called limit
$(\tau, \delta)$-(semi)stable if the following conditions hold.

 $(a)$ Any framed object $(E',f')$ where  $0\subset E'\subset E$
 is a nontrivial proper subobject in $\CA_\tau$
 such that  $F^0(f')=0$ satisfies
 \[
\mu_{\tau}(E') \ (\leq)\ \delta
\]

$(b)$ Any framed object $(E'',f'')$ where
$E\twoheadrightarrow E''\neq 0$ is a quotient of $E$ in
$\CA_\tau$, not isomorphic to $E$, such that $F^0(f)=0$
 satisfies
 \[
\mu_{\tau}(E'') \ (\geq)\ \delta.
\]
\end{defi}

\noindent
Here $\delta = -{\rm cot}(\varphi-\phi)$, where $\varphi\in
[\phi, \phi+\pi)$ is the phase of $\xi$.

Then note that
Proposition \ref{automprop} holds for both $(\tau,\xi)$-stable
objects and $(\tau,\delta)$-limit stable objects.
The case of framed BPS states
studied in Section \ref{sectionfive}
suggests that Conjecture \ref{purityconj} and Corollary \ref{puritycor}
should also hold in this more general framework,
perhaps with appropriate amendments.

\subsection{Absence of exotics II}\label{exoticsB}
In addition to the $SL(2,\IC)$ action, the space of BPS states is also
expected to carry an action of the $R$-symmetry group, which
is $SU(2)_R$ in the field theory limit. Again, if one is mainly interested
in algebraic aspects, this extends naturally to an action
of the  complex $R$-symmetry
group $SL(2,\IC)_R$. More generally the space of BPS
states in the low energy IIA effective theory of a
toric threefold $X$ is expected to carry only an action
of a Cartan subgroup, $\IC^\times_R \subset SL(2,\IC)_R$.
As explained in Section \ref{twofour}, for 
 a smooth projective D-brane moduli space $\CM$,
the space of BPS states is identified with the cohomology 
of $\CM$ and 
$\IC_R^\times$ acts with weight $p-q$ on the Dolbeault cohomology group
$H^{p,q}(\CM)$. Then it is natural to conjecture that
$\IC^\times_R$ acts analogously on the cohomology groups
constructed in \cite{COHA,berkeley}.
Hence absence of exotics is equivalent to
the statement that the Hodge numbers of the cohomology groups in Definition (\ref{BPScoh}) are trivial unless $p=q$.
Given the equivalence between motivic and cohomological Donaldson-Thomas
invariants \cite[Prop. 14, Sect. 7]{COHA}, it suffices to to prove that the former
depend on $\IL^{1/2}$ only, as discussed in detail in Section
\ref{exotics}.  However, it is worth noting
absence of exotics would follow in the presence of a suitable
torus action provided one can prove an
Atiyah-Bott fixed point theorem in the cohomological
formalism of \cite{COHA,berkeley}.

Recall that on a smooth projective variety $\CM$
equipped with an algebraic torus action $\IC^\times \times \CM\to\CM$,
the Atiyah-Bott theorem
yields a direct sum decomposition
\be\label{eq:fixedcoh}
H^{p,q}(\CM) \simeq \bigoplus_{\Xi} H^{p-n_\Xi,q-n_\Xi}(\Xi)
\ee
of Dolbeault cohomology groups.
Here $\Xi\subset \CM$ are the connected
components of the $\IC^\times$-fixed locus, which are
smooth compact subvarieties of $\CM$, and the index
$n_\Xi$ is defined as follows.
The normal bundle $N_{\Xi/\CM}$ has a natural 
${\IC^\times}$-equivariant
structure. Since ${{\IC^\times}}$ leaves $\Xi$ pointwise fixed
the normal bundle decomposes as a direct sum over
irreducible representations of ${{\IC^\times}}$
\[
N_{\Xi/\CM}\simeq \bigoplus_{k\neq 0} N_{\Xi/\CM}^{k}
\]
where $t\in{{\IC^\times}}$ acts on each direct summand $N_{\Xi/\CM}^{k}$
by scaling by $t^k$. By definition, the index $n_\Xi$ is the
rank of
\[
N_{\Xi/\CM}^- = \bigoplus_{k< 0} N_{\Xi/\CM}^{k}.
\]
In particular if the fixed loci are isolated points, all Hodge
numbers $h^{p,q}(\CM)$ vanish unless $p=q$.

A similar result for the cohomology groups defined
in (\ref{BPScoh}) would reduce absence of exotics to
the existence of a torus action with isolated fixed points.
More precisely, one would need a torus action  ${{\IC^\times}} \times
 M_\gamma\to M_\gamma$ on the affine space of representations
 of type $\gamma$, which leaves the superpotential $W_\gamma$ invariant and commutes
with the gauge group action.  If a result analogous to
\eqref{eq:fixedcoh} holds, absence of exotics follows immediately
if the ${\IC^\times}$-fixed locus in the moduli space is a finite set of isolated points.
According to Appendix \ref{appC}, this is the case for moduli spaces
of cyclic framed representations of the $SU(N)$ quiver.

%\rem\
%1. There should be a clearer characterization
%of which kind of string compactifications should
%lead to no exotics.
%2. Related to this it would be good to mention
%clear counterexamples where we know there are
%exotics.
%3. I think the discussion of Harvey-Moore of the
%algebra of BPS states can be applied fairly directly
%to quiver quantum mechanics. It leads to something
%similar to the story in terms of the ``cohomological
%Hall algebra.''  Perhaps it would be good to include
%it to make the identification more plausible from
%a physical perspective.

\appendix

\section{Exceptional collections and quivers for $X_N$}\label{appA}

The main goal of this section is to show that
\eqref{eq:linebundlesA} is a full strong exceptional collection of line
bundles
on $X_N$, and the objects \eqref{eq:fractbranesA} are the dual
fractional branes.

As a first step, it will be helpful to review the analogous constructions
for canonical resolutions of two dimensional
quotient $A_N$ singularities. These are noncompact toric surfaces
$Y_N$ determined by the fan in Fig. 6.
\bigskip
 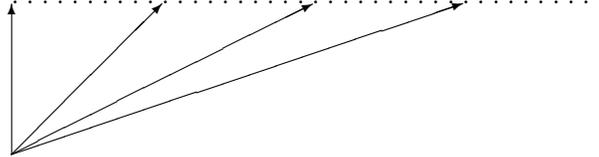
\begin{figure}[h]
  \setlength{\unitlength}{1mm}
\hspace{-50pt}
\begin{picture}(60,25)
\put(0,0){\vector(0,1){20}}
\put(0,0){\vector(1,1){20}}
\put(0,0){\vector(2,1){40}}
\put(0,0){\vector(3,1){60}}
\multiput(0,20)(2,0){40}{.}
\end{picture}
\caption{
The toric fan for the resolution of the $\IC^2/\IZ_3$ singularity. For arbitrary $N\geq 1$ there are $N-1$ inner rays determined by the
$N-1$ equidistant inner points on the horizontal line.}
\end{figure}
\bigskip

\noindent 
The toric data for $Y_N$ is encoded in the following 
charge matrix 
\be\label{eq:tordataB} 
\begin{array}{llllllll}
 & x_0 & x_1 & x_2 & x_3 & \ldots & x_{N-1} & x_{N} \\
% &  &  &  &  &  &  &  &  &  \\
\IC^\times_{(1)}& 1 & -2 & 1 & 0 & \ldots & 0 & 0 \\
\IC^\times_{(2)} & 0 & 1 & -2 & 1 & \ldots & 0 & 0 \\
&  &  &  &  &  &  &    \\
 &    &     &    &    & \vdots &   &       \\
 \IC^\times_{(N-1)}&  0&0  & 0 &0   &\ldots   &  -2 & 1  \\
 \end{array}
 \ee
The rays of the fan correspond to $N$ curves 
curves $C_0,\ldots, C_{N}$ on $Y_N$ 
defined by the equations $x_i=0$, $0\leq i\leq N$ 
respectively. The curves $C_1,\ldots, C_{N-1}$ 
corresponding to the inner rays are the exceptional 
compact cycles of the resolution, and have intersection matrix
\[
C_i\cdot C_j =
\left\{\begin{array}{ll}
0, & \mathrm{if}\ |i-j|\geq 2\\
& \\
1, & \mathrm{if}\ |i-j|=1\\
& \\
-2, & \mathrm{if}\ i=j.\\
\end{array}\right.
\]
The curves $C_0,C_N$ corresponding to
the outer rays are isomorphic to the complex line. 

The results of \cite{KV} imply that the following line bundles
\[
L_i = \CO_{Y_N}(E_i), \qquad 1\leq i\leq N
\]
form a full strong exceptional collection on $Y_N$, 
where 
\[
E_i= \sum_{j=0}^{N-i} jC_{i+j}, \qquad 1\leq i\leq N.
\]
Note that the $E_i$ are effective divisors determined by 
\[ 
x_{i+1}x_{i+2}^2 \cdots x_N^{N-i} =0. 
\] 
The dual fractional branes are the objects
\[
F_N=\CO_{C_1+\cdots+C_N},\qquad
 F_i=\CO_{C_i}(-1)[1],\qquad  i=1,\ldots, N-1.\]
Straightforward computations confirm that
\[
{\rm RHom}_{Y_N}(L_i,F_j) =\delta_{i,j}{\underline \IC}.
\]
Moreover, according to 
\cite{tilting,rep_assoc_coh,morita_derived}, the
 endomorphism algebra ${\rm End}_{Y_N}(L,L)$,
with $L=\oplus_{i=1}^{N}L_i$, 
is isomorphic to the path algebra of the affine $A_{N-1}$-quiver,
\medskip

\[
\xymatrix{
L_1 \ar@/^1pc/[r]^-{f_{2,1}}
\ar@/^5pc/[rrrrr]^-{g_{N,1}}  & L_2\ar@/^1pc/[r]^-{f_{3,2}}
\ar@/^1pc/[l]^-{g_{1,2}} & L_3 \ar@/^1pc/[l]^-{g_{2,3}} & \cdots &  L_{N-1} \ar@/^1pc/[r]^-{f_{N,N-1}} & L_{N}
\ar@/^1pc/[l]^-{g_{N-1,N}}
\ar@/^5pc/[lllll]^-{f_{1,N}}
\\}
\]
where
\[
f_{i,i-1}= \prod_{j=0}^{i-1}x_j, \quad g_{i-1,i} = \prod_{j=i}^{N}
x_, \quad 2\leq i \leq N, \quad
f_{1,N} = x_0, \quad g_{N,1}= \prod_{j=1}^N x_j.
\]
For simplicity, set
$f_{1,0}=f_{1,N}$ and $g_{0,1}=g_{N,1}$.
Then note that the quadratic relations
\[
f_{i,i-1} g_{i-1,i} - g_{i,i+1} f_{i+1,i} =0
\]
are obviously satisfied for all $i=1,\ldots, N$.

The toric threefolds $X_N$ are smooth toric fibrations over $\IP^1$ with fibers isomorphic to $Y_N$. Exceptional
collections for compact toric fibrations over projective spaces have been
constructed in \cite{derived_toric_fibrations}.
Proceeding by
analogy with \cite{derived_toric_fibrations}, one
obtains the collection \eqref{eq:linebundlesA}. However since the fibers
$Y_N$ are noncompact one has to check directly that
\eqref{eq:linebundlesA} is a full strong exceptional collection. The first property, namely that the line bundles \eqref{eq:linebundlesA} generate
$D^b(X_N)$ is entirely analogous to \cite{derived_toric_fibrations}.
The vanishing results
\[
{\rm Ext}^m_{X_N}(L_i, L_k(aH))=H^m(X_N, L_i^{-1}\otimes_{X_N}L_k(aH)) =0,
\]
for any $i,k=0,\ldots , N-1$ and any $a=0,\pm 1$, require
more work. 

Recall first that for any line bundle $L$ on $X_N$ there is a Leray spectral
 sequence
\[
H^p(\IP^1, R^q\pi_*L) \Rightarrow H^{p+q}(X_N, L)
\]
where the direct images $R^q\pi_*L$ are quasi-coherent sheaves
on $\IP^1$. Moreover, since $\CO_{X_N}(H)=\pi^*\CO_{\IP^1}(1)$, we have 
$R^q\pi_*L(aH) \simeq R^q\pi_*L \otimes_{\IP^1} 
\CO_{\IP^1}(a)$
for any $a\in \IZ$.

By construction, the line bundles $\{L_i\}_{0\leq i\leq N-1}$ restrict to an exceptional collection on each fiber. Furthermore, recall  that
for any morphism $f: {\sf X} \to {\sf Y}$ of
complex algebraic varieties where ${\sf Y}={\mathrm {Spec}}({\sf R})$ is affine,
and for any quasi-coherent ${\CO}_{\sf X}$-module $F$,
the direct image $R^qf_*(F)$ is isomorphic to the quasi-coherent
sheaf determined by the
${\sf R}$-module $H^q(F)$. Then, using the standard affine open cover
of $\IP^1$, and the fact that the restriction of the
collection $\{L_i\}_{0\leq i\leq N-1}$ is an exceptional collection on each
fiber, it follows that
\[
R^q\pi_*(L_i^{-1}\otimes L_k) =0
\]
for all $0\leq i,k\leq N-1$ and all $q>0$. This yields an isomorphism
\[
H^m(X_N, L_i^{-1}\otimes L_k(aH))\simeq H^m(\IP^1,
R^0\pi_*( L_i^{-1}\otimes L_k) \otimes_{\IP^1}\CO_{\IP^1}(a)).
\]
In order to prove vanishing for all $m\geq 1$, let $U_1$, $U_2$
be the affine open subsets $y_1\neq 0$, respectively $y_2\neq 0$
in $\IP^1$. Then note that $\pi^{-1}(U_s)$, $s=1,2$,
 is a toric variety
determined by the data
\be\label{eq:tordataE}
\begin{array}{ccccccccc}
 & u_s & x_1 & x_2 & x_3 & \ldots & x_{N-1} & x_{N} & z_s  \\
& 1 & -2 & 1 & 0 & \ldots & 0 & 0 & 0  \\
 & 0 & 1 & -2 & 1 & \ldots & 0 & 0 & 0  \\
 &    &     &    &    & \vdots &   &    &     \\
 & 0 & 0 & 0 &  0 & \ldots &  -2 & 1 &   0 \\
 \end{array}
\ee
where
\[u_s=x_0y_s^2, \qquad s=1,2,\qquad  z_1=y_1^{-1}y_2, \qquad
z_2=y_1y_2^{-1}.\]
 The transition functions on the overlap $\pi^{-1}(U_1\cap U_2)$
 are
\be\label{eq:transfct}
z_2=z_1^{-1}, \qquad u_2 =z_1^2u_1,
\ee
$(x_1,\ldots, x_{N})$ being obviously unchanged. The 
varieties $\pi^{-1}(U_s)$, $s=1,2$ are isomorphic to $Y_N\times \IC$.
%Let $Q_j$, $j=1,\ldots,N-1$ denote the
%charge vectors encoding the $(\IC^\times)^{N-1}$ action
%in \eqref{eq:tordataE}.

Next let 
$L=\CO_{X_N}(\sum_{i=1}^{N-1}m_iD_i +aH)$ on $X_N$ for some arbitrary
$m_i \in \IZ$, $1\leq i\leq N-1$, and $a\in \{-1,0,1\}$.
The spaces of local sections $\Gamma(U_s, R\pi_*L)$, $s=1,2$,
are spanned by monomials of the form
\[
u_s^{k_s}z_s^{l_s} \prod_{i=1}^N
x_i^{n_{s,i}},
\qquad k_s,l_s,n_{s,i} \in \IZ_{\geq 0}, \qquad s=1,2, \qquad
1\leq i \leq N
\]
which have the same scaling behavior as the monomial 
$\prod_{i=1}^{N-1}\prod_{j=1}^{N-i} x_{i+j}^{jm_i }$
 under the torus action 
\eqref{eq:tordataE}. 
%
%satisfying the constraints
%\[
%\langle Q_j, (k_s,l_s,n_{s,i})_{1\leq i\leq N})\rangle = m_j, 
%\qquad
%s=1,2, \qquad 1\leq j\leq  N-1.
%\]
Using the transition functions \eqref{eq:transfct},
\[
u_2^{k_2}z_2^{l_2} \prod_{i=1}^N
x_i^{n_{2,i}}\bigg|_{U_1\cap U_2} =
u_1^{k_2} z_1^{2k_2-l_2-a} \prod_{i=1}^N
x_i^{n_{2,i}}\bigg|_{U_1\cap U_2}.
\]
Since $l_2\geq 0$,
the exponent $2k_2-l_2-a$ takes all values
in $\IZ \cap (-\infty, \ 2k_2-a]$ for fixed $k_2$. Since
$k_2\geq 0$ as well, and $a\in \{-1,0,1\}$,
this implies that the $\check{\rm C}$ech differential has trivial
cokernel, hence $H^m(\IP^1,R^0\pi_*L)=0$, $m\geq 1$.
This proves the required vanishing results.

Note also that that the endomorphism algebra
${\rm End}(T)$, where $T$ is the direct sum of all line bundles
$L_i,M_i$, $1\leq i\leq N$,  is generated by the toric monomials
\be\label{eq:endalgB}
\xymatrix{
& & & \vdots & \ar@/_0.2pc/[dl]|{\gamma_{N,1}} & {}\quad 
\ar@/^0.2pc/[dll]|{\gamma_{N,2}} & \vdots &  & \\
& & & L_1 \ar@<.5ex>[rrr]|{\phi_{1,1}} 
\ar@<-.5ex>[rrr]|{\phi_{1,2}}
\ar@{-}[u]|{\psi_N} 
& & & M_1\ar@/_1pc/[ddlll]|{\gamma_{1,1}}
\ar@/^1pc/[ddlll]|{\gamma_{2,1}}  \ar@{-}[u]|{\lambda_N} 
& &
\\
& & & & &  &  & & \\
& & & L_2 \ar@<.5ex>[rrr]|{\phi_{2,1}} \ar@<-.5ex>[rrr]|{\phi_{2,2}}
\ar[uu]|{\psi_1}
& & & M_2 \ar[uu]|{\lambda_1}  & & \\
& & & \vdots & & & \vdots & & \\
& & & L_i \ar@<.5ex>[rrr]|{\phi_{i,1}} \ar@<-.5ex>[rrr]|{\phi_{i,2}}
& &  & M_i \ar@/_1pc/[ddlll]|{\gamma_{i,1}}
\ar@/^1pc/[ddlll]|{\gamma_{i,2}} & & \\
& & & & & & & & \\
& & & L_{i+1} \ar@<.5ex>[rrr]|{\phi_{i+1,1}} \ar@<-.5ex>[rrr]|{\phi_{i+1,2}}
\ar[uu]|{\psi_i}
& & &  M_{i+1}\ar[uu]|{\lambda_i}  & & \\
& & & \vdots && & \vdots & & \\
& & & L_{N-1} \ar@<.5ex>[rrr]|{\phi_{N-1,1}} \ar@<-.5ex>[rrr]|{\phi_{N-1,2}}
& & & M_{N-1}\ar@/_1pc/[ddlll]|{\gamma_{N-1,1}}
\ar@/^1pc/[ddlll]|{\gamma_{N-1,2}} & & \\
& & & & & & & & \\
& & & L_{N} \ar@<.5ex>[rrr]|{\phi_{N,1}} \ar@<-.5ex>[rrr]|{\phi_{N,2}} \ar[uu]|{\psi_{N-1}}
& & & M_{N}\ar[uu]|{\lambda_{N-1}}
\ar@/^0.2pc/@{-}[dl]|{\gamma_{N,2}} 
\ar@/_0.2pc/@{-}[dll]|{\gamma_{N,1}}
& & 
\\
& & &\vdots \ar[u]|{\psi_N} & & &\vdots \ar[u]|{\lambda_N} & & \\
}
\ee
where
\[
\phi_{i,1}=y_1, \qquad \phi_{i,2}=y_2, \qquad 1\leq i\leq N
\]
\[
\psi_i=\prod_{j=i+1}^N x_j, \qquad \lambda_i = \prod_{j=i+1}^N
x_j, \qquad 1\leq i\leq N-1
\]
\[
\psi_N = \prod_{j=1}^N x_j, \qquad \lambda_N = \prod_{j=1}^N
x_j
\]
\[
\gamma_{i,1} = y_1\prod_{j=0}^{i} x_j, \qquad
\gamma_{i,2}=-y_2\prod_{j=0}^{i} x_j, \qquad 
1\leq i\leq  N-1.
\]
\[ 
\gamma_{N,1} = x_0y_1, \qquad 
\gamma_{N,2} = - x_0y_2.
\]
The above generators satisfy the quadratic relations
\be\label{eq:Endrelations}
\bal
& \phi_{i,1} \psi_i = \lambda_i \phi_{i+1,1}, \qquad
\phi_{i,2} \psi_i = \lambda_i \phi_{i+1,2}, \qquad
1\leq i \leq N-1\\
& \phi_{i+1,1}\gamma_{i,2} + \phi_{i+1,2}\gamma_{i,1} =0, \qquad
1\leq i\leq N-1\\
& \phi_{1,1} \gamma_{N,2} + \phi_{1,2}\gamma_{N,1}=0,\\
& \psi_i\gamma_{i,1} = \gamma_{i-1,1}\lambda_{i-1}, \qquad
\psi_i\gamma_{i,2} = \gamma_{i-1,2}\lambda_{i-1},
\qquad 1\leq i \leq N,\\
\eal
\ee
where by convention
$\gamma_{0,1}=\gamma_{N,1}$,
$\gamma_{0,2}=\gamma_{N,2}$ and $\lambda_0=\Lambda_N$.

The next task is to
check that the objects \eqref{eq:fractbranesA}
indeed satisfy the orthogonality conditions
\eqref{eq:onconditions}.
This claim follows by straightforward
although somewhat tedious computations.
A simple computation using equations 
\eqref{eq:linebundlesB} and 
the intersection products 
\[ 
(S_i\cdot S_j)_{X_N} = \left\{\begin{array}{ll}
\Sigma_i, &{\rm for}\ j=i+1 \\
\Sigma_j, &{\rm for}\ i=j+1 \\
-\Sigma_{i-1}-\Sigma_i - 2C_i, & {\rm for}\ i=j \\
0, & {\rm otherwise} \\
\end{array}\right.
\]
of divisors on $X_N$ yields 
\be\label{eq:restrA}
L_i^{-1}\big|_{S_j} \simeq \left\{\begin{array}{ll}
\CO_{S_j}, & {\rm for} \ i\geq j+1,\\
\CO_{S_j}(-\Sigma_{j}), & {\rm for} \ i=j,\\
\CO_{S_j}(-2C_j), & {\rm for} \ i\leq j-1,\\
\end{array}\right.
%\qquad
%M_i^{-1}\big|_{S_j}\simeq L_i^{-1}\big|_{S_j}\otimes_{S_j}
%\CO_{S_j}(-C_j),
\ee
where $i=1,\ldots,N-1$.

Now let
$\IF_m=\IP(\CO_{\IP^1}\oplus \CO_{\IP^1}(m))$
be a Hirzebruch surface of any degree $m\in \IZ$,
$\Sigma$ a section such that $\Sigma^2 =m$,
and $C$ the fiber class. Then note the following isomorphisms
\be\label{eq:cohA}
H^k(\IF_m, \CO_{\IF_m}(aC))\simeq H^k(\IP^1, \CO_{\IP^1}(a)),
\ee
\be\label{eq:cohB}
H^k(\IF_m, \CO_{\IF_m}(-\Sigma+aC))=0,
\ee
and
\be\label{eq:cohC}
H^k(\IF_m, \CO_{\IF_m}(-2\Sigma+aC))\simeq H^{k-1}(\IP^1,
\CO_{\IP^1}(a-m))
\ee
for all $k,a\in \IZ$.
Equations \eqref{eq:cohA},\eqref{eq:cohB} follow easily
from the Leray spectral sequence for the canonical projection
$\pi:\IF_m\to \IP^1$.
Equation \eqref{eq:cohC} follows from the long exact sequence
associated to the exact sequence
\[
0\to \CO_{\IF_m}(-2\Sigma+aC) \to
\CO_{\IF_m}(-\Sigma+aC) \to \CO_\Sigma(-\Sigma
+ aC)\to 0
\]
using equation \eqref{eq:cohB}.

Equations \eqref{eq:restrA} -
\eqref{eq:cohC} imply the orthogonality conditions
\eqref{eq:onconditions}
for $i=1,\ldots,N$, $j=1,\ldots, N-1$.
For example 
\[
\bal 
{\rm RHom}^k(L_i, P_i) & \simeq {H}^{k+1}(X_N, L_i^{-1}\otimes 
F_i) \\
& \simeq H^{k+1}(S_i, \CO_{S_i}(-\Sigma_{i-1}-\Sigma_i)) \\
& \simeq H^{k+1}(S_i, \CO_{S_i}(-2\Sigma_{i-1} - 2iC_i)) \quad 
{\rm since}\quad \Sigma_i = \Sigma_{i-1}+2iC_i\\
& \simeq H^{k}(\IP^1,\CO_{\IP}^1)\simeq \IC \delta_{k,0}\\
\eal
\]
where the next to last isomorphism follows from equation 
\eqref{eq:cohC} with $a=-2i$ and $m = (\Sigma_{i-1})^2_{S_i} 
=-2i =a$. 

The remaining cases, $i=1,\ldots, N$
and $j=N$, require an inductive argument. For concreteness let
$i=1$, the other cases being completely analogous.
The inductive step is based on the observation that for any two
effective divisors $D,D'$ in $X_N$ there is an exact sequence
of $\CO_{X_N}$-modules
\[
0\to \CO_{D'}(-D) \to \CO_{D+D'} \to \CO_D\to 0.
\]
Applying this to the decomposition $S= S_1+\sum_{j=2}^{N-1}S_j$,
and using relations \eqref{eq:restrA}, one obtains an exact sequence
\[
0\to \CO_{\sum_{j=2}^{N-1}S_j}(-S_1-2H) \to
\CO_S(-D_1) \to \CO_{S_1}(-\Sigma_{1})\to
0
\]
The vanishing results \eqref{eq:cohA} imply that all
 cohomology groups of $\CO_{S_1}(-\Sigma_{1})$ are trivial.
 Therefore the associated long exact sequence breaks into isomorphisms
 \[
 H^k(X_N,\CO_{\sum_{j=2}^{N-1}S_j}(-S_1-2H)) \simeq
 H^k(X_N,\CO_S(-D_1) )
 \]
 for all $k\geq 0$. Repeating the above argument, there is an
 exact sequence
 \[
 0\to \CO_{\sum_{j=3}^{N-1}S_j}(-S_2-2H)\to
\CO_{\sum_{j=2}^{N-1}S_j}(-S_1-2H)\to \CO_{S_2}(-\Sigma_{1}-2H)
\to 0
\]
since $\CO_{X_N}(S_1)|_{S_j}\simeq \CO_{S_j}$ for all $j\geq 3$.
Again the vanishing results \eqref{eq:cohA} and the associated
long exact sequence yield isomorphisms
\[
H^k(X_N, \CO_{\sum_{j=3}^{N-1}S_j}(-S_2-2H)) \simeq
H^k(X_N, \CO_{\sum_{j=2}^{N-1}S_j}(-S_1-2H))
\]
for all $k\geq 0$. Proceeding inductively, the required vanishing results for cohomology groups 
will be reduced to \eqref{eq:cohA} in finitely many steps.

\section{Motives for pedestrians}\label{motives}

The basic construction of motives of complex algebraic varieties
will be briefly reviewed here for completeness, following,
for example \cite{J-motivic,motivic-Hall}.
The
ring of motives $K_0({\sf Var}/\IC)$ is a quotient of the
${\IQ}$-vector space generated by all isomorphism
classes of algebraic varieties ${\sf X}$ over $\IC$
by the equivalence relation
\[
[{\sf X}]\sim [{\sf Y}]+[{\sf X}\setminus {\sf Y}]
\]
for any closed subvariety ${\sf Y}\subset {\sf X}$.
The ring structure is determined by the direct product
${\sf X}\times{\sf Y}$ which is compatible with equivalence
relation $\sim$, hence descends to $\IQ$-linear associative product
$[{\sf X}][{\sf Y}]= [{\sf X}\times{\sf Y}]$.
 The equivalence class of a variety ${\sf X}$ in this ring
 will be called the Chow motive of ${\sf X}$.
The Chow motive of the complex line is denoted by $\IL$ and called the Tate motive for historical reasons.

Note that one can construct in complete analogy
a ring of motives of complex schemes of finite type. An important result for Donaldson-Thomas
invariants is that the ring of motives of schemes of finite type
is in fact isomorphic to the ring of motives of varieties,
\[
K_0({\sf Sch}/\IC) \simeq K_0({\sf Var}/\IC).
\]
%Moreover, the ring of motives of stacks is a localization
%of $K_0({\sf Var}/\IC)$,
%\[
%K_0({\sf St}/\IC) \simeq K_0({\sf Var}/\IC)[[GL(n,\IC)]^{-1},
%n\in \IZ],
%\]
%including the formal inverses of the equivalence classes of
%the general linear groups. This reflects the presence of
%higher rank stabilizers.

The
Chow motive of a scheme ${\sf X}$ encodes more refined
information than the topology of the topological space 
${\sf X}$,
but is obviously coarser than the algebraic scheme structure.
In order to understand this in more detail, note that for
${\sf X}\subset \IP^k$ a projective or quasi-projective scheme, the Chow
motive $[{\sf X}]$ is equal to the Chow motive of the
reduced scheme
${\sf X}^{red}$, obtained by taking a quotient of the
structure sheaf $\CO_{\sf X}$ by its nilpotent ideal.
For example the Chow motive
of any  multiple line $x^n=0$ in $\IC^2$ is $\IL$ for any
$n\geq 1$. 

For concrete computations it is worth noting that if
${\sf X}\to {\sf Y}$ is a smooth morphism of schemes
such that all
fibers are isomorphic to a scheme ${\sf Z}$, then
$[{\sf X}]=[{\sf Z}][{\sf Y}]$. This result yields for example
the following identity \cite[Lemma 2.6]{motivic-Hall}
\be\label{eq:GLmotive}
[GL(n,\IC)] = \prod_{k=1}^n(\IL^k-1),
\ee
where $GL(n,\IC)$ is the underlying algebraic variety of the
general linear group.  This follows from the fact that there is a
smooth map $f_v :GL(n,\IC)\to \IC^n\setminus \{0\}$, sending
$g\in GL(n,\IC)$ to $g(v)$, where $v\in \IC^n\setminus \{0\}$
is a fixed nonzero vector. This map is surjective, and the fiber
over any point $v'\in \IC^n\setminus \{0\}$ is the stabilizer
of $v'$ in $GL(n,\IC)$, which is isomorphic to $GL(n-1,\IC)$.
Therefore the general result stated above yields
\[
[GL(n,\IC)]= [GL(n-1,\IC)][\IC^{n}\setminus \{0\}]
 = (\IL^n-1)[GL(n-1,\IC)].
\]
This implies \eqref{eq:GLmotive} by recursion.

Any geometric invariant of schemes or stacks which depends only
on their Chow motive is called a motivic invariant. A good example
example is the Hodge polynomial with compact support.
For a smooth compact variety ${\sf X}$, this is just the usual Hodge polynomial
with compact support,
\[
P_{(x,y)}({\sf X}) = \sum_{p,q} x^py^q h_c^{p,q}({\sf X})
\]
where $h^{p,q}_c({\sf X})$ are the Hodge numbers of ${\sf X}$
for compactly supported cohomology. For singular varieties a suitable
generalization must be defined using Deligne's theory of mixed Hodge
structures. See for example \cite[Ex. 4.3]{J-motivic} for a brief
summary in a similar context and for further references.
According to loc. cit. the Hodge polynomial with compact
support determines a ring morphism
\be\label{eq:Hodgemap}
P:K_0({\sf Var}/\IC) \longto \IQ(x,y).
\ee

The construction of \cite{wallcrossing} yields motivic Donaldson-Thomas invariants for moduli spaces of Bridgeland stable objects
in triangulated CY3-categories equipped with a cyclic
$A_\infty$-structure and a choice of ``orientation data".
For the purpose of the present paper the discussion can be
confined to derived categories of quivers with potential
$(Q,W)$.
Suppose $\tau$ is a Bridgeland stability condition
on $D^b(Q,W)$ and $\gamma$ is a fixed dimension vector
such that the (coarse)
moduli space of stable objects $\CM^s_\tau(\gamma)$
is a projective or quasi-projective scheme.
Then the construction of \cite{wallcrossing} assigns to
$\CM^s_\tau(\gamma)$ an element
$DT^{mot}_\tau(\gamma)$ in an
extension
$K_0({\sf Var}/\IC)[\IL^{1/2},\IL^{-1/2}]$ of the ring of motives.
It is crucial to note that, $DT^{mot}_\tau(\gamma)$
is not identical to the Chow motive $[\CM^s_\tau(\gamma)]$
of the moduli space. By construction, $DT^{mot}_\tau(\gamma)$ depends in an essential manner on the
homotopy class of the cyclic $A_\infty$ structure on the
derived category, while $[\CM^s_\tau(\gamma)]$ does not.
The motivic Donaldson-Thomas invariant $DT^{mot}_\tau(\gamma)$ is also called
the {\it virtual motive} of the moduli space in order to avoid
any confusion with the Chow motive.

The motivic Donaldson-Thomas invariants 
for moduli spaces of stable $(Q,W)$-representations
admit a very explicit presentation in terms of
Chow motives due to  \cite{motivic-Hilbert}.
The basic observation is that the moduli space
$\CM^s_\tau(\gamma)$ of $\tau$-stable representations
is in this case isomorphic to the critical locus of a polynomial
function $W_\gamma$ on a smooth quasi-projective variety.
The ambient variety is
the moduli space $\CN^s_\tau(\gamma)$ of
$\tau$-stable representations of the quiver
$Q$ with no relations and the polynomial function $W_\gamma:
\CN^s_\tau(\gamma)\to \IC$ is naturally determined
by $W$. Suppose $W$ is chosen such that the critical locus 
of $W_\gamma$ is contained in the fiber at zero, $W_\gamma^{-1}(0)$.
In addition, one requires a torus action
on $\CN^s_\tau(\gamma)$ preserving
$W_\gamma$ and satisfying some mild technical conditions.
Then, the results of \cite{motivic-Hilbert},
imply the following formula
\be\label{eq:virmotive}
DT^{mot}_\tau(\gamma) = -\IL^{w(\gamma)}
([W_\gamma^{-1}(\lambda)]- [W_\gamma^{-1}(0)])
\ee
for any $\lambda \neq 0$,
where the terms on the right hand side are Chow motives.
The exponent $w(\gamma)$ is a half-integral weight
depending on $\gamma$ whose exact expression will
not be needed here.

Once the virtual motivic invariants are constructed,
one can easily obtain various polynomial Donaldson-Thomas
invariants applying the Hodge polynomial map \eqref{eq:Hodgemap}.
More precisely, \eqref{eq:Hodgemap} can be extended to a
ring morphism
\[
{\mathfrak P}:K_0({\sf Var}/\IC)[\IL^{1/2},\IL^{-1}]
\to \IQ(x^{1/2},y^{1/2})
\]
sending $\IL^{1/2}$ to $(xy)^{1/2}$. Then the virtual
Hodge numbers ${\mathfrak h}^{r,s}(\gamma,\tau)\in \IZ$,
$r,s\in {1\over 2}\IZ$ are
defined by
\be\label{eq:virthodge}
{\mathfrak P}(DT^{mot}_\tau(\gamma)) = \sum_{r,s\in {1\over 2}\IZ}
{\mathfrak h}^{r,s}(\gamma,\tau) x^ry^s.
\ee
Equation \eqref{eq:virmotive} implies that
$r-s\in \IZ$ for all nonzero ${\mathfrak h}^{r,s}(\gamma,\tau)$. If the moduli space $\CM^s_\tau(\gamma)$ is smooth and projective,
of dimension $m$, the virtual Hodge numbers are related to the
usual ones by
\[
{\mathfrak h}^{r,s}(\gamma,\tau)= h^{r+m/2,s+m/2}(\CM^s_\tau(\gamma)).
\]
However, in general the numbers
${\mathfrak h}^{r,s}(\gamma,\tau)$ will be different
from the ones
obtained by applying the map \eqref{eq:Hodgemap} to the Chow
motive $[\CM^s_\tau(\gamma)]$. 

\section{Kronecker modules}\label{appB}

Kronecker modules are finite dimensional representations
of a quiver with two nodes, two arrows and no relations
as below
\[
\xymatrix{
 \bullet \ar@<.5ex>[rr] \ar@<-.5ex>[rr] & & \bullet\ \\}
\]
They form an abelian category $\CK$
of homological dimension one. The extension groups
${\rm Ext}^j_\CK(\kappa_1,\kappa_2)$, $j=0,1$,
of any two Kronecker modules
$\kappa_i=(W_i,V_i,f_i,g_i:W_i\to V_i)$, $i=1,2$,
are the cohomology groups of the two term complex
\be\label{eq:Kexts}
{\rm Hom}(V_1,V_2)\oplus {\rm Hom}(W_1,W_2)
{\buildrel \delta \over \longto}
{\rm Hom}(W_1,V_2)^{\oplus 2}
\ee
where the first term is in degree 0, and 
\[
\delta(\alpha, \beta) = \big(f_2\beta-\alpha f_1,
g_2\beta-\alpha g_1\big).
\]
In particular note that the dual vector space ${\rm Ext}^1_\CK(\kappa_1,\kappa_2)^\vee$
is the kernel of the map
\be\label{eq:Kdualexts}
{\rm Hom}(V_2,W_1)^{\oplus 2}
{\buildrel \delta^\vee \over \longto}
{\rm Hom}(V_2,V_1)\oplus {\rm Hom}(W_2,W_1),
\ee
\[
\delta^\vee(\gamma,\eta) = (
-f_1\gamma-g_1\eta,\gamma f_2 + \eta g_2).
\]

\subsection{Harder-Narasimhan filtrations}\label{Aone}
Now suppose $(\psi,\phi)$ are stability parameters for Kronecker modules assigned to the nodes as follows
\[
\xymatrix{
\psi\ \bullet \ar@<.5ex>[rr] \ar@<-.5ex>[rr] & & \bullet\ \phi\\}
\]
For any nontrivial
 Kronecker module $\kappa=(W,V, f,g:W\to V)$ let
\[
\mu_{(\psi,\phi)}(\kappa) = {\phi\dim(V)+\psi\dim(W)\over
\dim(V)+\dim(W)}.
\]
As usual, $\kappa$ is called $(\psi,\phi)$-(semi)stable if
\[
\mu_{(\psi,\phi)}(\kappa') \ (\leq)\
\mu_{(\psi,\phi)}(\kappa)
\]
for any nontrivial proper submodule $\kappa'\subset \kappa$.
If $\psi<\phi$ the only $(\psi,\phi)$-stable Kronecker modules
are the two simple ones determined by the two nodes.
 If $\psi>\phi$, the $(\psi,\phi)$-stable Kronecker modules
form three groups up to isomorphism, as follows
\begin{itemize}
\item[(a)] $Q_n$, $n\geq 0$ are indecomposable Kronecker modules
of dimension vector $(n,n+1)$ of the form
\[
\xymatrix{
H^0(\CO_{\IP^1}(n-1)) \ar@<.5ex>[rrr]|{z_1} \ar@<-.5ex>[rrr]|{z_2}
& & & H^0(\CO_{\IP^1}(n))},
\]
the linear maps being multiplication by the homogeneous coordinates
$[z_1,z_2]$.
\item[(b)] $R_{p}$
are  indecomposable Kronecker modules of the form
\[
\xymatrix{
\IC \ar@<.5ex>[rrr]|{z_1} \ar@<-.5ex>[rrr]|{z_2}
& & & \IC },
\]
where $p=[z_1,z_2]$ is a point on $\IP^1$.
\item[(c)] $J_n$, $n\geq 0$ are the indecomposable Kronecker modules
of dimension vector $(n+1,n)$ obtained by dualizing $Q_n$.
 \end{itemize}
Moreover, a straightforward computation shows that
\be\label{eq:slopeorder}
\bal
\mu_{(\psi,\phi)}(Q_n)  < \mu_{(\psi,\phi)}(Q_{n'}) <
\mu_{(\psi,\phi)}(R_{p}) < \mu_{(\psi,\phi)}(J_m)
< \mu_{(\psi,\phi)}(J_{m'})
\eal
\ee
for all $n< n'$, $m>m'$, as long as $\psi>\phi$.

The above list of stable Kronecker modules is closely related to the
 classification of indecomposable modules in
\cite[Ch. VIII, Thm. 7.5]{artinrep}.
According to loc. cit. there are
three groups of indecomposable modules, $\{Q_n\}_{n\geq 0}$,
$\{J_n\}_{n\geq 0}$ and a third group $\{R_{p; j}\}$ labelled by a point
$p\in \IP^1$ and a positive integer $j\in \IZ_{\geq 1}$, such
that $R_{p, 1}=R_{p}$.
The explicit form of the modules $R_{p,j}$ with $j>1$
will not be needed in the following.
Moreover, \cite[Ch. VIII, Thm. 7.5]{artinrep} also computes
all
nontrivial extension groups
of indecomposable  Kronecker modules, obtaining
\be\label{eq:indecompextA}
\bal
{\rm Ext}^0_\CK(Q_n,Q_{n'}) & \simeq \IC^{n'-n+1}, \qquad  n'\geq n,\\
{\rm Ext}^1_\CK(Q_n,Q_{n'}) & \simeq \IC^{n-n'-1},\qquad n\geq n'+1,
\\
\eal
\ee
\be\label{eq:indecompextB}
\bal
{\rm Ext}^0_\CK(J_n,J_{n'}) & \simeq \IC^{n-n'+1}, \qquad
n\geq n',\\
{\rm Ext}^1_\CK(J_n,J_{n'}) & \simeq \IC^{n'-n-1},\qquad
n'\geq n+1,
\\
\eal
\ee
\be\label{eq:indecompC}
\bal
& {\rm Ext}^0_\CK(R_{p},R_{q}) \simeq \delta_{p,q}
\IC,\\
&
{\rm Ext}^1_\CK(R_{p},R_{q}) \simeq \delta_{p,q}
\IC\\
\eal
\ee

 \be\label{eq:indecompextD}
\bal
{\rm Ext}^0_\CK(Q_n,R_{p}) & \simeq \IC, \qquad
{\rm Ext}^0_\CK(R_{p},J_n) & \simeq \IC, \\
{\rm Ext}^1_\CK(R_{p},Q_n) & \simeq \IC, \qquad
{\rm Ext}^1_\CK(J_n, R_{p}) & \simeq \IC. \\
\eal
\ee
All extension groups not listed above are trivial. Moreover, note that
the space of morphisms ${\rm Ext}^0_\CK(Q_n,Q_{n'})$, $n'\geq n$,
consists of maps of the form
\[
\xymatrix{
H^0(\CO_{\IP^1}(n-1))
\ar@<.5ex>[rrr]|{z_1} \ar@<-.5ex>[rrr]|{z_2} \ar[d]^-{f}
& & & H^0(\CO_{\IP^1}(n))\ar[d]^-{f} \\
H^0(\CO_{\IP^1}(n'-1))
\ar@<.5ex>[rrr]|{z_1} \ar@<-.5ex>[rrr]|{z_2}
& & & H^0(\CO_{\IP^1}(n')) \\}
\]
where $f$ denotes multiplication by a degree $n'-n$ homogeneous polynomial
in $z_1,z_2$. In particular any such map is either  injective or zero.
Similarly the dual vector space
${\rm Ext}^1_\CK(Q_n,Q_n')^\vee$
consists of maps of the form
\[
\xymatrix{
H^0(\CO_{\IP^1}(n'-1))
\ar@<.5ex>[rrr]|{z_1} \ar@<-.5ex>[rrr]|{z_2}
& & & H^0(\CO_{\IP^1}(n'))   \ar@<-.5ex>[ddlll]_-{\eta}
\ar@<+.5ex>[ddlll]^-{\gamma}  \\
& & & \\
H^0(\CO_{\IP^1}(n-1))
\ar@<.5ex>[rrr]|{z_1} \ar@<-.5ex>[rrr]|{z_2}
& & & H^0(\CO_{\IP^1}(n)) \\}
\]
where $(\gamma,\eta)$ are degree $n-n'-1$ homogeneous polynomials of
$z_1,z_2$ satisfying
\[
z_1\gamma+z_2\eta=0.
\]
Again, $(\gamma, \eta)$ are either both injective or zero.

Given an arbitrary Kronecker module $\kappa$, let
\be\label{eq:filtrA}
0=HN_0(\kappa)\subset HN_1(\kappa)\subset \cdots \subset HN_k(\kappa) =\kappa, \qquad k\geq 1,
\ee
be its Harder-Narasimhan filtration
with respect to $(\psi,\phi)$-stability, where $\psi>\phi$.
Then the  slope inequalities \eqref{eq:slopeorder}
and equations \eqref{eq:indecompextA}-\eqref{eq:indecompextD}
imply that there
exists $1\leq k_1\leq k$ and  integers
\[
0\leq n_1< n_2<\cdots <  n_{k_1-1},\qquad
 n_{k_1+1}> \cdots> n_k\geq 0,
 \]
\[
s_1,\ldots, s_{k_1-1}, s_{k_1+1}, \ldots, s_k>0
\]
such that
\be\label{eq:succquot}
HN_{l}(\kappa)/HN_{l-1}(\kappa) \simeq \left\{\begin{array}{ll}
J_{n_l}^{\oplus s_l} & {\rm for}\ 1\leq l\leq k_1-1 \\
& \\
R & {\rm for}\ l=k_1\\
& \\
Q_{n_l}^{\oplus s_l} & {\rm for}\ k_1+1\leq l \leq k.
\end{array}\right.
\ee
 Here $R$ is a $(\psi,\phi)$-semistable module with
 slope $(\psi+\phi)/2$ which admits a Jordan-H\"older
 filtration such that all successive quotients are isomorphic to some
 $R_{p}$. In particular \eqref{eq:filtrA} admits a subfiltration
 \be\label{eq:filtrB}
 0={ \kappa}_0 \subset {\kappa}_1 \subset
{ \kappa}_2\subset   { \kappa}_3=\kappa
 \ee
 such that $\kappa_1/\kappa_0=J(\kappa)$ is a successive extension of
 stable modules of type $(c)$, $\kappa_2/\kappa_1=R(\kappa)$ a successive extension of stable modules of type $(b)$, and $\kappa_3/\kappa_2=Q(\kappa)$
 a successive extension of stable modules of type $(a)$.
 Moreover, the slope inequalities \eqref{eq:slopeorder}
 and equations \eqref{eq:indecompextA}-\eqref{eq:indecompextD}
 imply that
 \be\label{eq:succquotB}
 J(\kappa) \simeq \oplus_{l=1}^{k_1-1} J_{n_l}^{\oplus s_l},
 \qquad
 Q(\kappa) \simeq \oplus_{l=k_1+1}^k Q_{n_l}^{\oplus s_l}.
 \ee

 \subsection{Application to representations of the
 $SU(3)$ quiver}\label{Atwo}
 This section consists of some results on Kronecker 
 modules used in Sections \ref{fourone}, 
 \ref{Wsection}, \ref{fourtwo}. 
Let $\rho$ be a representation of the quiver with
potential \eqref{eq:quivrepA}. As observed in Section
\ref{sectionfive}, the horizontal rows of $\rho$ determine
two
Kronecker modules $\rho_1,\rho_2$.
Each of them has a filtration
\be\label{eq:filtrC}
0=\rho_{i,0} \subset \rho_{i,1} \subset \rho_{i,2}\subset \rho_{i,3}
=\rho_i, \qquad i=1,2,
\ee
of the form \eqref{eq:filtrB}.
Let $\{V_{i,j}\}$, $\{W_{i,j}\}$,
$i=1,2$, $j=0,\ldots, 3$ be the
induced filtrations on the underlying vector spaces.
Then
equations \eqref{eq:indecompextA}-\eqref{eq:indecompextD} imply
via straightforward exact sequences
\[
a_1(V_{1,j}) \subseteq V_{2,j},
\qquad
b_1(W_{1,j}) \subseteq W_{2,j},\qquad r_1(V_{2,j}) \subseteq
W_{1,j}, \qquad  s_1(V_{2,j}) \subseteq
W_{1,j}
\]
for $j=1,\ldots,3$.
This shows implies that $\rho$ has a filtration 
of the form 
\[ 
0=K^0(\rho) \subset K^1(\rho) \subset K^2(\rho)\subset K^3(\rho)=\rho
\]
in the abelian category of $(Q,W)$-modules 
such that the two Kronecker modules determined by the horizontal 
maps of each quotient $K^{j+1}(\rho)/K^j(\rho)$, $0\leq j\leq 2$ are 
isomorphic to 
$\rho_{1,j+1}/\rho_{1,j}$, $\rho_{2,j+1}/\rho_{2,j}$ respectively. 
In particular, there is a quotient 
$\rho\twoheadrightarrow \rho''$ with 
$\rho''=K^3(\rho)/K^2(\rho)$, such that the underlying Kronecker 
modules of $\rho''$ are 
modules $(Q(\rho_1), Q(\rho_2))$ and the linear maps
$(a_1'',b_1'', r_1'', s_1'')$ are  induced by $(a_1,b_1,r_1,s_1)$.

Now suppose $\rho$ is a $(Q,W)$-module of dimension vector
$(d_i,e_i)_{1\leq i\leq 2}$ such that
\[
d_i-e_i=m_i,
\]
with $m_i\geq 0$, $i=1,2$.
The next goal is to show that there exist King
stability parameters
$(\theta_i,\eta_i)_{1\leq i\leq 2}$,
\[
\sum_{i=1}^2 (d_i\theta_i+e_i\eta_i)=0,
\]
such that
the quotient $\rho\twoheadrightarrow \rho''$ destabilizes
$\rho$ unless $\rho=\rho''$.
Let $(d''_i,e_i'')$, $i=1,2$ denote the dimension vector of
$\rho''$ and suppose the projection $\rho\twoheadrightarrow \rho''$
is not an isomorphism. This implies
\[
\sum_{i=1}^2((d_i-d''_i)+(e_i-e_i'')) >0.
\]
Note that by construction
\be\label{eq:builtineqA}
d_i''-e_i'' =m_i''\geq d_i-e_i =m_i, \qquad i=1,2,
\ee
and at least one of these inequalities is strict under the current assumptions.
Suppose
\be\label{eq:quivchamberA}
\eta_i>0, \qquad \theta_i < 0, \qquad
|\theta_i|< |\eta_i|, \qquad i=1,2.
\ee
Then, using inequality \eqref{eq:builtineqA},
it follows that
\[
\bal
e_i''|\eta_i|-d_i''|\theta_i| & \leq (e_i-d_i)|\eta_i| + d_i''(|\eta_i|-|\theta_i|).
\eal
\]
The right hand side of this inequality is
\[
(e_i-d_i)|\eta_i| + d_i''(|\eta_i|-|\theta_i|)= e_i|\eta_i|-d_i|\theta_i|
+ (d_i-d_i'')(|\theta_i|-|\eta_i|).
\]
Therefore inequalities \eqref{eq:quivchamberA} imply
\[
e_i''\eta_i+d_i''\theta_i \leq e_i\eta_i +d_i\theta_i
\]
for $i=1,2$. Since at least one of inequalities \eqref{eq:builtineqA}
is strict, this implies
\[
\sum_{i=1}^2 (e_i''\eta_i+d_i''\theta_i) <
\sum_{i=1}^2 (e_i\eta_i+d_i\theta_i) =0.
\]
Therefore indeed the quotient $\rho\twoheadrightarrow \rho''$
destabilizes $\rho$ under the current assumptions. In conclusion,
if inequalities \eqref{eq:quivchamberA} are satisfied, for
any $(\theta,\eta)$-semistable representation $\rho$,
the filtrations \eqref{eq:filtrC}
collapse to
\[
0\subset \rho_{i,3} =\rho_i.
\]
Therefore the Harder-Narasimhan filtration of each
Kronecker module $\rho_i$, $i=1,2$  with respect to $(\psi,\phi)$
stability reduces to
\be\label{eq:filtrD}
0=\rho_i^0\subset \rho_i^1\subset \cdots \rho_i^{h_i} =\rho_i,\qquad h_i\geq 1, \qquad i=1,2
\ee
such that the successive quotients are
\[
\rho_i^j/\rho_i^{j-1} \simeq Q_{k_{i,j}}^{\oplus r_{i,j}}
\]
for some $k_{i,j}\in \IZ_{\geq 0}$, $r_{i,j}\in \IZ_{\geq 1}$,
$i=1,2$, $j=1,\ldots, h_i$,
satisfying
\[
k_{i,1}> k_{i,2}> \cdots > k_{i,h_i}.
\]
Moreover equations \eqref{eq:indecompextA} imply that the filtrations
\eqref{eq:filtrD} must be split, that is
\be\label{eq:splitmoduleA}
\rho_i \simeq \oplus_{i=1}^{h_i} Q_{k_{i,j}}^{\oplus r_{i,j}},\qquad  i=1,2
\ee
This yields strong constraints on the structure of representations
$\rho$ with underlying Kronecker modules $\rho_1,\rho_2$
as above.

\section{Background material on extensions}\label{extensions}

The purpose of this section is to summarize some
background material on extension groups in abelian categories
of quiver modules following for example
\cite[Ch. 2.3, Ch. 2.4]{hom_alg}.
Here ${\sf Q}$ will denote a quiver with finitely many nodes and arrows and ${\sf R}$ an ideal of relations in the path algebra.
The vertices of ${\sf Q}$ will be denoted by $\nu$, the arrows
by ${\sf a}$ and the relations by ${\sf r}$. The latter are
linear combinations of paths with integral coefficients such 
that all these paths have the same starting and ending 
points. Therefore 
one can naturally define the tail $t({\sf r})$ and head
$h({\sf r})$.

A $({\sf Q}, {\sf R})$-module is a module for the path algebra 
of the quiver ${\sf Q}$ with relations ${\sf R}$. 
These modules form an abelian category. 
Extension groups in the abelian category of $({\sf Q},{\sf R})$-modules
will be denoted by ${\rm Ext}^k_{({\sf Q},{\sf R})}(\rho_1,\rho_2)$, $k\in \IZ$. They are defined in terms of projective resolutions
as follows.
A$({\sf Q},R)$-module $\Pi$ is projective if
for any surjective morphism ${ \rho}\twoheadrightarrow
{\rho}''$ and any morphism $\phi'':\Pi\to
 \rho''$ there exists a morphism $\phi:\Pi \to
 {\rho}$ such that the following diagram commutes
\[
\xymatrix{
\Pi \ar[dr]^-{\phi''} \ar[d]_-{\phi} & \\
{\rho} \ar@{>>}[r] & {\rho}'' .\\}
\]
It is a basic fact that to any node $\nu$ of the quiver
diagram ${\sf Q}$ one can assign a projective
module $\Pi_\nu$, which is the module consisting
of all paths starting at $\nu$. Moreover, any finite dimensional
representation ${\rho}$ has a projective resolution
\be\label{eq:projresA}
\cdots \Pi^{-1} {\buildrel d^{-1}\over \longto}
\Pi^{0} \to {\rho} \to 0,
\ee
where each term $\Pi^k$ is a direct sum of modules
of the form $\Pi_\nu$ and the differentials
are defined in terms of natural concatenation of paths.

For illustration,
 suppose $\rho_\nu$ is the simple module
 assigned to the node $\nu$.
 In this case
 \be\label{eq:projresB}
\bal
\Pi^0  \simeq \Pi_\nu,\qquad
\Pi^{-1} \simeq \bigoplus_{{\sf a},\, t({\sf a})=\nu}
\Pi_{h({\sf a})},\qquad
\Pi^{-2} \simeq \bigoplus_{{\sf r},\, t({\sf r}) = \nu}
\Pi_{h({\sf r})}
\eal
\ee
and the differentials $d^{-2}, d^{-1}$ are defined
by natural concatenation of paths. Given a collection of 
paths $(p_{h(\sf a)})_{t({\sf a})=\nu}\in \Pi^{-1}$,
\[ 
d^{-1}((p_{h({\sf a})})_{t({\sf a})=\nu}) = \sum_{{\sf a},\, t({\sf a})=\nu}
p_{h(\sf a)}{\sf a}.
\]
There is a similar expression for $d^{-2}$ derived by linearizing 
the relations in the path algebra. 
The higher terms $\Pi^k$, $k\leq -2$, are determined by
the higher syzygies of the ideal of relations, i.e.
relations on relations etc. For a systematic approach see 
\cite{projresolutionsI,projresolutionsII,NSpaper} 
and references therein.

Returning to the general case, the extension groups
${\rm Ext}^k_{({\sf Q}, {\sf R})}({\rho},
{ \rho}')$
are the cohomology groups of the complex of vector
spaces
\be\label{eq:extcomplexA}
0\to {\rm Hom}_{({\sf Q},{\sf R})}(\Pi^0, {\rho}')
{\buildrel \circ d^{-1}\over \longto}
{\rm Hom}_{({\sf Q},{\sf R})}(\Pi^{-1}, {\rho}')
{\buildrel \circ d^{-2}\over \longto} \cdots
\ee
Using projective resolutions, one can prove that the extension 
groups ${\rm Ext}^k_{({\sf Q},{\sf R})}(\rho,\rho')$, $k=0,1$ are isomorphic to the first two cohomology groups of the complex 
\be\label{eq:extcomplexB}
\bal 
0 \to \bigoplus_\nu {\rm Hom}(V_\nu(\rho), V_\nu(\rho')) 
{\buildrel \delta_0\over \longto} 
\bigoplus_{\sf a} {\rm Hom}(V_{t({\sf a})}(\rho), V_{h({\sf a})}(\rho')) 
{\buildrel \delta_1\over \longto} 
\bigoplus_{\sf r} {\rm Hom}(V_{t({\sf r})}(\rho), V_{h({\sf r})}(\rho'))
\eal
\ee
where $V_\nu(\rho)$ is the vector space assigned to the node 
$\nu$ 
in the representation $\rho$. 
The differential $\delta_0$ is given by 
\[ 
\delta_0(\alpha_\nu) = (\rho'({\sf a}) \circ \alpha_{t({\sf a})} - 
\alpha_{h({\sf a})} \circ \rho({\sf a}))
\]
where $\rho({\sf a}):V_{t({\sf a})}\to V_{h({\sf a})}$ is the linear map 
assigned to the arrow ${\sf a}$ in the representation $\rho$. 
There is a a similar expression for $\delta_1$ obtained by linearizing 
the relations. As an application, will give here a proof of the first
isomorphism in  \eqref{eq:isomext}, as well as 
equations \eqref{eq:pairingA},  \eqref{eq:pairingBB}.

 Recall that representations of the 
$SU(3)$ quiver are of the form 
\[
\xymatrix{
W_2 \ar@<.5ex>[rrr]|{c_2} \ar@<-.5ex>[rrr]|{d_2}
& & & V_2 \ar@/_1pc/[ddlll]|{r_{1}}
\ar@/^1pc/[ddlll]|{s_{1}}  \\
& &  & \\
W_1 \ar@<.5ex>[rrr]|{c_{1}} \ar@<-.5ex>[rrr]|{d_{1}}
\ar[uu]|{b_{1}}
& & & V_1 \ar[uu]|{a_{1}}  \\}
\]
with relations
\be\label{eq:SUthreerelations}
\bal
& r_1a_1=0, \qquad s_1a_1=0,\qquad b_1r_1=0,\qquad b_1s_1=0\\
& c_1r_1+d_1s_1=0,\qquad r_1c_2+s_1d_2=0\\
& a_1c_1-c_2b_1=0,\qquad a_1d_1-d_2b_1=0.\\
\eal
\ee
Consider two representations of $(Q,W)$ of the form 
\[
\rho_i:\quad \xymatrix{
W_i \ar@<.5ex>[rrr]|{c_i} \ar@<-.5ex>[rrr]|{d_i}
& & & V_i ,}\qquad i=1,2
\]
where only the horizontal maps $(c_1,d_1)$, respectively 
$(c_2,d_2)$ 
are nontrivial. Specializing the complex \eqref{eq:extcomplexB} 
to the pair $(\rho_1,\rho_2)$ yields
\be\label{eq:extcomplexC}
0\to {\rm Hom}(V_1,V_2) \oplus {\rm Hom}(W_1,W_2) 
{\buildrel \delta_1\over \longto } {\rm Hom}(W_1,V_2)^{\oplus 2}.
\ee
Note that the degree $0$ term in \eqref{eq:extcomplexB} 
is trivial in this case since $\rho_1,\rho_2$ are supported at 
disjoint sets of nodes. The differential $\delta_1$ is obtained
by linearizing the relations \eqref{eq:SUthreerelations}, 
\be\label{eq:diffone}
\delta_1(\beta,\alpha) = (c_2\circ \beta-\alpha\circ c_1, 
d_2\circ \beta - \alpha \circ d_1),
\ee 
where $\alpha = {\dot a}_1$, $\beta = {\dot b}_1$. 
The extension group ${\rm Ext}^1_{(Q,W)}(\rho_1,\rho_2)$ is isomorphic 
to ${\rm Ker}(\delta_1)$.

On the other hand, regarding $(\rho_1,\rho_2)$ as Kronecker modules, 
the complex \eqref{eq:Kexts} takes the form 
\be\label{eq:extcomplexD}
0\to {\rm Hom}(V_1,V_2) \oplus {\rm Hom}(W_1,W_2) 
{\buildrel \delta_0 \over \longto } {\rm Hom}(W_1,V_2)^{\oplus 2}
\to 0
\ee
with 
\[
\delta_0(\beta,\alpha) = (c_2\circ \beta-\alpha\circ c_1, 
d_2\circ \beta - \alpha \circ d_1). 
\]
Comparing \eqref{eq:diffone} and 
\eqref{eq:extcomplexD}, we deduce
 ${\rm Ext}^0_{\CK}(\rho_1,\rho_2)\simeq 
{\rm Ker}(\delta_0)$. It follows that there is an 
isomorphism ${\rm Ext}^1_{(Q,W)}(\rho_1,\rho_2)\simeq 
{\rm Ext}^0_{\CK}(\rho_1,\rho_2)$. 
Similarly, using ${\rm Ext}^1_{\CK}(\rho_1,\rho_2)\simeq 
{\rm Coker}(\delta_0)$ one can establish the second 
isomorphism in \eqref{eq:isomext}. 

We now prove equation \eqref{eq:pairingBB}. 
Given an exact sequence
\[
0\to \rho_1\to \rho_2\to \rho_3 \to 0
\]
there are two long exact sequences,
\be\label{eq:longextseqA}
\cdots\to  {\rm Ext}^k_{({\sf Q},{\sf R})}(\rho_3,\rho)
\to {\rm Ext}^k_{({\sf Q},R)}(\rho_2,\rho) \to
{\rm Ext}^k_{({\sf Q},R)}(\rho_1,\rho)\to
{\rm Ext}^{k+1}_{({\sf Q},R)}(\rho_3,\rho)\cdots \to
\ee
respectively
\be\label{eq:longextseqB}
\cdots\to {\rm Ext}^k_{({\sf Q},{\sf R})}(\rho,\rho_1)
\to {\rm Ext}^k_{({\sf Q},{\sf R})}(\rho,\rho_2) \to
{\rm Ext}^k_{({\sf Q},{\sf R})}(\rho,\rho_3)\to
{\rm Ext}^{k+1}_{({\sf Q},{\sf R})}(\rho,\rho_1)\to
\cdots
\ee
for any $({\sf Q},{\sf R})$-module $\rho$.

Now consider in more detail the case where $({\sf Q},{\sf R})$ is
the framed (truncated) quiver with potential
$({\widetilde Q}, W)$ obtained in Section
\ref{fiveone}. Let  $\lambda_0$
the simple module supported at the framing node.
Then the following hold.

$(P.1)$ Suppose $\rho$ is a representation
of $({\widetilde Q}, W)$ with dimension vector 0
at the framing node.
Then a projective resolution of $\rho$ as a
$({\widetilde Q}, W)$-module is identical with the
projective resolution of $\rho$ as an unframed
$(Q,W)$-module.
This follows from the fact that ${\widetilde Q}$ has only one
extra arrow $f_j$ compared to $Q$, which joins the framing node
to the fixed node $j$. So any path starting at any node of $Q$ will never contain $f_j$. In particular this implies that
\be\label{eq:extstuffA}
{\rm Ext}^k_{({\widetilde Q}, W)}(\rho_1,\rho_2)
\simeq {\rm Ext}^k_{({ Q}, W)}(\rho_1,\rho_2)
\ee
for any $(Q,W)$-modules $\rho_1,\rho_2$ and all $k\in
\IZ$, and also
\be\label{eq:extstuffB}
{\rm Ext}^k_{({\widetilde Q}, W)}(\rho,\lambda_0)
=0
\ee
for any $(Q,W)$-module $\rho$ and all $k\in
\IZ$.

$(P.2)$ The simple module $\lambda_0$ has a two term
projective resolution
\[
0\to \Pi_j \to \Pi_0 \to \lambda_0 \to 0
\]
since there are no relations containing $f_j$. Here
$\Pi_0$ is the projective module starting at the framing node.
Therefore, for any $(Q,W)$-module $\rho$,
\be\label{eq:extstuffC}
{\rm Ext}^k_{({\widetilde Q}, W)}(\lambda_0,\rho)
\simeq V_j \delta_{k,1}
\ee
where $V_j$ is the vector space of $\rho$ at the node $j$
which receives the framing arrow $f_j$.
As stated in the main text, the long exact sequences 
\eqref{eq:longextseqA}, \eqref{eq:longextseqB} 
and equations \eqref{eq:extstuffA}, \eqref{eq:extstuffB}, 
\eqref{eq:extstuffC}, 
imply equation \eqref{eq:pairingA}.

Next we prove equation \eqref{eq:pairingBB}. 
Given two finite dimensional
 $(Q,W)$-modules $\rho_1,\rho_2$ one can check using projective resolutions that the pairing
\[
\chi({\rho}_1, {\rho}_2)
= \sum_{k=0}^1 (-1)^k\big({\rm dim}{\rm Ext}^k_{({Q},W)}
({\rho}_1, {\rho}_2)  -
{\rm dim}{\rm Ext}^k_{({Q},W)}
({ \rho}_2, { \rho}_1)\big)
\]
is given by equation \eqref{eq:pairingBB}. Since this is a fairly
long computation, details will be omitted. For the skeptical
reader, note that there is a different derivation
of equation \eqref{eq:pairingBB} based on the equivalence
of derived categories $D^b(\CQ,\CW)\simeq
D^b(X_N)$ explained in detail in Section
\ref{twoone}. This equivalence assigns compactly supported
objects $E_1,E_2$ in $D^b(X_N)$ to $\rho_1,\rho_2$,
with $K$-theory classes
\[
[E_l] = \sum_{i=1}^{N-1} (d_i(\rho_l)[P_i] + e_i(\rho_l)[Q_i]), \qquad l=1,2,
\]
where $P_i,Q_i$ are the fractional branes defined in Section
\ref{twoone}.
Then, using Serre duality, the above pairing becomes
\[
\bal
\chi(\rho_1,\rho_2) & = \sum_{k=0}^3
(-1)^k {\rm dim}{\rm Ext}^k_{X_N}(E_1,E_2)\\
& = \int_{X_N}( \ch_1(E_2)\ch_2(E_1) - \ch_1(E_1)
\ch_2(E_2)).\\
\eal
\]
The second identity follows from the Riemann-Roch theorem
since $E_1,E_2$ must have at most two dimensional support.
Then a straightforward intersection theory computation
confirms \eqref{eq:pairingBB}.

\section{Classifications of fixed points}\label{appC}
In this section we classify torus fixed points in moduli 
spaces  of cyclic modules of the 
$N=3$ gauge theory framed quiver.
The unframed quiver diagram is in
\eqref{eq:quivrepA}, with the potential $W$ given by
$$
W=r_1(a_1c_1- c_2b_1)+ s_1(a_1d_1-d_2b_1) \ .
$$

This gives the following 8 relations:
$$
r_1 : a_1c_1-c_2b_1=0 \ , \ \ s_1: a_1d_1-d_2b_1=0 \ , \ \ a_1: c_1r_1+d_1s_1=0 \ , \ \ b_1: r_1c_2+s_1d_2=0 \ .
$$
$$
c_1 : r_1a_1=0 \ , \ \ c_2: b_1r_1=0 \ , \ \ d_1: s_1a_1=0 \ , \ \ d_2: b_1 s_1 = 0 \ .
$$
\medskip

\noindent
{\bf Path algebra.} First we consider the case in which the framing map takes values in $V_2$ and construct the path algebra.
Denote by $e$ the framing vector and by $\mathcal{H}_n$ the set of vectors obtained by acting with $n$ arrows
on $e$. We have the decomposition of the path algebra $\IC Q / d W = \oplus\mathcal{H}_n$,
where
\be
\bal
& \mathcal{H}_0 = \{e\} \ , \\
& \mathcal{H}_1 = \{r_1e, \  s_1e\} \ , \\
& \mathcal{H}_2 = \{ d_1r_1e, \ c_1r_1e=d_1s_1e, \ c_1s_1e\} \ , \\
& \mathcal{H}_3 = \{ a_1 d_1r_1e = d_2 b_1r_1e = 0, \ a_1 c_1 r_1 e = c_2 b_1 r_1 e = 0, \ a_1c_1s_1e
= c_2b_1s_1e=0\} \ , \\
% & \mathcal{H}_4 = \{ 0 \} \ (\because r_1 a_1 = s_1 a_1 = 0).
\eal
\ee
%{\color{blue} Wu-yen, aren't all monomials in $\CH_3$ zero
%because of relations? For example 
%\[ 
%(a_1d_1)r_1 = d_2(b_1r_1)=0.
%\] 
%Does this change anything?}

The path algebra is finite dimensional
and therefore there are only finitely many fixed modules of the path algebra.

Now suppose the framing vector $e$ is in $V_1$. The decomposition of the path algebra is given by
\be
\bal
& \mathcal{H}_0 = \{e\} \ , \\
& \mathcal{H}_1 = \{a_1e\} \ , \\
& \mathcal{H}_2 = \{ 0 \} \ (\because r_1 a_1 = s_1 a_1 = 0) .
\eal
\ee

\medskip

\noindent
{\bf $\IC^{\times}$ torus action.}
Let $T_F = (\IC^{\times})^{8}$ be the flavor torus
acting on the $i$-th arrow by a scaling factor $\lambda_i$, where $i$ is in the
set of all arrows $S_a$.
Therefore the $T_F$ action acts on the whole path algebra and has a subtorus
$T_{F,dW} \subset T_F$ which leaves invariant the relations $dW=0$.
$T_{F,dW}$ is isomorphic to $(\IC^{\times})^4$ and is given by

\be
\bal
T_{F,dW} = \{\lambda_i \in \IC^{\times}, \ i \in S_a \ |  & \ \
\lambda_{a_1} \lambda_{c_1} = \lambda_{c_2} \lambda_{b_1}, \ \
\lambda_{a_1} \lambda_{d_1} =\lambda_{d_2}  \lambda_{b_1}, \\
& \ \ \lambda_{c_1} \lambda_{r_1} = \lambda_{d_1} \lambda_{s_1}, \ \
\lambda_{r_1} \lambda_{c_2} =\lambda_{s_1}  \lambda_{d_2}  \} \simeq (\IC^{\times})^4 .
\eal
\ee

However the subtorus $T_{F,dW}$ might contain part induced by gauge group action,
which we have to mod out.

Let $(\mu_1, \mu_2, \tilde{\mu}_1, \tilde{\mu}_2) \in (\IC^{\times})^4$
be the diagonal torus in the gauge group
$GL(V_1) \times GL(V_2) \times GL(W_1) \times GL(W_2)$. This $(\IC^{\times})^4$ action
will have an induced torus action $T_{\text{sub}}$ on the arrows, which can be seen
to be isomorphic to $(\IC^{\times})^3$.
We can always use this action to make
$\lambda_{a_1}=1, \ \lambda_{b_1}=1, \ \lambda_{r_1}=1$. Note that this is not a
unique choice. So we define our torus action $T_Q$ to be

\be
\bal
T_Q \equiv T_{F,dW} / T_{\text{sub}} = \{  & (\lambda_{c_1}, \lambda_{d_1}, \lambda_{c_2}, \lambda_{d_2},
\lambda_{s_1}) \in (\IC^{\times})^5 \ | \\
& \lambda_{c_1} = \lambda_{c_2}, \ \lambda_{d_1} =\lambda_{d_2} , \ \lambda_{c_1} = \lambda_{d_1} \lambda_{s_1} \} \simeq (\IC^{\times})^2 .
\eal
\ee

\medskip

\noindent
{\bf Classifications of $T_Q$-fixed points.} Recall the fact that there exists a one-to-one
correspondence between the framed $T_Q$-fixed $\IC Q / d W$-module and
the $T_Q$-fixed annihilator $I$ of the framed vector $e$. The annihilator $I$ is a left
ideal in the path algebra $\IC Q / d W$.

In the current example the path algebra is finite dimensional, so we could perform
the analysis explicitly. First we list the weights of the path algebra elements in terms of
$(\lambda_{d_1},\lambda_{s_1}) = (\lambda_1, \lambda_2)$.
In the framed $V_2$ case we have
\be
\bal
& \{ w(e)=1 \} \ , \\
& \{ w(r_1e)=1 , \  w(s_1e) = \lambda_2 \} \ , \\
& \{ w(d_1r_1e) = \lambda_1 , \ w(c_1r_1e)= \lambda_1\lambda_2, \
w(c_1s_1e) = \lambda_1\lambda_2^2 \} \ . 
%& \{ w(a_1 d_1r_1e)=\lambda_1, \ w(a_1 c_1 r_1 e)=\lambda_1\lambda_2 ,
%\ w(a_1c_1s_1e) =\lambda_1\lambda_2^2 \} \ .
\eal
\ee

The $T_Q$-fixed left ideal must be generated by linear combinations of
the elements of the same weight. For example an element $r \in I$ of weight
$\lambda_1\lambda_2$ should be of the form $r= \xi \ ( c_1r_1e)$.
Therefore from the list we conclude the $T_Q$-fixed annihilator $I$ of framing 
vector $e$ is generated by monomials of the path algebra and the class $[I]$ is an 
isolated point in the moduli space of cyclic representations.

\medskip

\noindent
{\bf Weights of the fixed points.} The deformation complex of the quiver
is 4-term complex,

$$
0 \to \mathcal{T}_1 \stackrel{\delta_1}{\to} \mathcal{T}_2 \stackrel{\delta_2}{\to} \mathcal{T}_3 \stackrel{\delta_3}{\to} \mathcal{T}_4 \to 0 ,
$$ where $\mathcal{T}_1 = \text{End}(V_1) \oplus \text{End}(V_2) \oplus\text{End}(W_1)
\oplus \text{End}(W_2)$, $\mathcal{T}_2$ is the space of all arrows including the framing, $\delta_1$
the linearized gauge transformation and $\delta_2$ the linearized relations of $\partial W =0$.

The complex is self-dual and therefore the weight of a fixed point $p$ with dimension vector $(d_1, d_2, e_1, e_2,1)$
is $(-1)^{\text{dim}T_p}$, where $T_p$ is the tangent space at the fixed point $p$. If the framing map takes values in $V_i$, 
$i=1,2$, we have 
$$
\bal
\text{dim}T_p &= d_1d_2+e_1e_2+2d_1e_1+2d_2e_2+2d_2e_1+d_i -d_1^2-d_2^2-e_1^2-e_2^2 \\
&= d_1^2+d_2^2+d_i + d_1d_2+e_1^2+e_2^2+e_1e_2. (\text{mod} 2) \ ,
\eal
$$ 

The framed numerical DT invariants for cyclic modules are
given by \cite{symm}
\[
DT(\gamma, 1; z,w,+\infty) = \sum_{p}(-1)^{{\rm dim}T_p}.
\]
\noindent
{\bf Invariants $F_{\infty}(\gamma)$.} Here we list all the nonvanishing invariants
$DT(d_1,d_2,e_1,e_2,1; z,w, +\infty)$. Since they are independent of $(z,w)$, it is convenient to simplify the notation omitting these arguments.

\begin{itemize}
\item Framed $V_1$ case: 
\be\label{eq:FVone}
DT((1,0,0,0),1;+\infty)=1,\qquad DT((1,1,0,0),1;+\infty)=1.
\ee
\item Framed $V_2$ case: 
\be\label{eq:FVtwo}
\bal
& DT((0,1,0,0),1;+\infty)=1,\qquad  \ \ \,
DT((0,1,1,0),1;+\infty)=-2,\\
& DT((1,1,1,0),1;+\infty)=-2,\qquad
DT((0,1,2,0),1;+\infty)=1,\\
& DT((1,1,2,0),1;+\infty)=3,\qquad \ \ \,
DT((2,1,2,0),1;+\infty)=3,\\
& DT((3,1,2,0),1;+\infty)=1.\\
\eal
\ee
\end{itemize}

\medskip

\noindent
{\bf $T_Q$-fixed loci for $N\geq4$.} We now show that the $T_Q$-fixed loci are isolated points for $N \geq 4$.
The superpotential for the truncated framed quiver is
$$
W = \sum_{i=1}^{N-2} [ r_i(a_i c_i - c_{i+1} b_i) + s_i (a_i d_i - d_{i+1} b_i) ] \ .
$$

This gives the following $(6N-10)$ relations.
\be\label{eq:Nrelations}
\bal
& a_i c_i -c_{i+1}b_i =0 \ , \ a_i d_i -d_{i+1} b_i =0 \ , \ \forall \ i=1, \cdots, N-2 \ , \\
& c_i r_i + d_i s_i =0 \ , \ r_i c_{i+1} +s_i d_{i+1} =0 \ , \ \forall \ i=1, \cdots, N-2 \ , \\
& r_1 a_1 =0 \ , \ s_1 a_1 = 0 \ , \ b_{N-2} r_{N-2} =0 \ , \ b_{N-2} s_{N-2} =0 \ , \\
& b_i r_i + r_{i+1}a_{i+1} = 0 \ , \ b_i s_i + s_{i+1} a_{i+1} =0 \ , \ \forall \ i=1, \cdots, N-3 \ .
\eal
\ee

Suppose the framing node is in $V_{k+1}$. Consider the subset $\mathcal{P}_{k+1} \subset \IC Q / dW$,
generated by 3 elements $a_{k} c_{k} r_{k}=a_{k} d_{k} s_{k}$ , $a_{k} d_{k} r_{k}$ and
$a_{k} c_{k} s_{k}$. They are three triangle paths starting from $V_k$. Using
\eqref{eq:Nrelations} repeatedly we can show
$$
\bal
& ( a_{k} c_{k} r_{k} ) (a_{k} d_{k} r_{k}) =(a_{k} d_{k} r_{k}) (a_{k} c_{k} r_{k}) \ , \\
& ( a_{k} c_{k} r_{k} ) (a_{k} c_{k} s_{k}) =(a_{k} c_{k} s_{k}) (a_{k} c_{k} r_{k}) \ , \\
& ( a_{k} c_{k} s_{k} ) (a_{k} d_{k} r_{k}) = (a_{k} c_{k} r_{k})^2 \ .
\eal
$$

For examples, up to minus signs, we have the following:
$$
\bal
& (a_{k} c_{k} r_{k}) (a_{k} d_{k} r_{k}) = a_{k} c_{k} (b_{k-1} r_{k-1}) d_{k} r_{k} =
a_{k} (a_{k-1} c_{k-1}) r_{k-1} d_{k} r_{k} = \\
& a_{k} a_{k-1} (d_{k-1} s_{k-1}) d_{k} r_{k} =  a_{k}( d_{k} b_{k-1}) s_{k-1} d_{k} r_{k} =
a_{k} d_{k} b_{k-1}( r_{k-1} c_{k}) r_{k}= \\
& a_{k} d_{k} ( r_{k} a_{k} ) c_{k} r_{k} = (a_{k} d_{k}  r_{k})( a_{k} c_{k} r_{k}) \ . \\
& \ \  \\
& ( a_{k} c_{k} s_{k} ) (a_{k} d_{k} r_{k}) = a_{k} c_{k} (b_{k-1}  s_{k-1}) d_{k} r_{k}
=a_{k} c_{k} b_{k-1}  (r_{k-1} c_{k}) r_{k} = \\
& a_{k} c_{k} (r_{k}  a_{k}) c_{k} r_{k} = (a_{k} c_{k} r_{k})^2 \ .
\eal
$$

Therefore we conclude $\mathcal{P}_{k+1}$ can be represented
as
$$
\mathcal{P}_{k+1} =\{ (a_{k} d_{k} r_{k})^{n_1} (a_{k} c_{k} r_{k})^{n_2} , (a_{k} c_{k} s_{k})^{n_1}(a_{k} c_{k} r_{k})^{n_2} \ |
\ n_1, n_2 \in \IZ_{\geq 0} \}
$$

Now we want to express a general element in the path algebra $\IC Q / dW$.
Define $\overline{\mathcal{P}}_{k+1} \subset \IC Q / dW$ to be the set of paths starting
from $V_{k+1}$. Obviously we have $\mathcal{P}_{k+1} \subset \overline{\mathcal{P}}_{k+1}$.

We have the following properties for the paths.
\begin{itemize}
\item {\bf P1.} By using the relation $c_k r_k a_k = c_k b_{k-1} r_{k-1} = a_{k-1} c_{k-1} r_{k-1}$ and the relations of the same type, we can transform the triangles starting from $V_i$ into the
forms of $acr$, $adr$ and $acs$. A similar statement holds for the triangles starting from $W_i$.
\item {\bf P2.} Triangles of type $\{ acr,adr,acs\}$ and the hooks of type $\{cr,dr,cs\}$ commute.
For example, $(c_k r_k)(a_k c_k r_k) =(a_{k-1} c_{k-1} r_{k-1})(c_k r_k)$.
\item {\bf P3.} Triangles starting from $V_i$ commute with $a_i$. Triangles starting from $W_i$ commute with $b_i$. For example, $(c_{k+1}r_{k+1}a_{k+1})a_k = a_k (a_{k-1} c_{k-1} r_{k-1})$.
\item {\bf P4.} Composition rules for hooks.
$(c_{k-1} r_{k-1}) (c_{k} s_{k}) = (c_{k-1} s_{k-1}) (c_{k} r_{k})$,
$(c_{k-1} r_{k-1}) (d_{k} r_{k}) = (d_{k-1} r_{k-1}) (c_{k} r_{k})$,
$(c_{k-1} s_{k-1}) (d_{k} r_{k}) = (c_{k-1} r_{k-1}) (c_{k} r_{k})$.
\end{itemize}

\medskip

Using {\bf P1}, {\bf P2}, {\bf P3}, {\bf P4} and \eqref{eq:Nrelations} to group together all the triangles
for a given path, one can show that any monomial elements in
$\overline{\mathcal{P}}_{k+1}$ can be arranged into an element in $\mathcal{P}_{k+1}$, followed
by $r_k$, $s_k$, $c_k r_k$, $c_k s_k$, $d_k r_k$, $\cdots$ (\textit{ie.} certain power of hooks),
$b_k r_k$, $b_k s_k$, $b_{k+1}b_{k} r_{k}$,
$b_{k+1}b_k s_k$, $\cdots$ (\textit{ie.} a sequence of $b_i$ times $r$ or $s$), or a sequence of $a_i$. Namely we have
\be
\bal
\overline{\mathcal{P}}_{k+1} = & \{ p\ , \ r_k p\ , \ s_k p \ , \ c_k r_k p \ , \ c_k s_k p\ , \ d_k r_k p \ , \\
& \ c_{k-1} r_{k-1}  c_{k} r_{k} p \ , \ c_{k-1} r_{k-1}  c_{k} s_{k} p\ ,
\ c_{k-1} r_{k-1}  d_{k} r_{k} p \ , \ \cdots  \\
& \ b_{k+1}b_{k} r_{k} p \ , \ b_{k+1}b_k s_k p \ , \cdots \ , \ a_k p \ , a_{k+1} a_k p \ , \cdots \ | \ p \in \mathcal{P}_{k+1} \ \}
\eal
\ee

We associate to each arrow a $\IC^{\times}$ action and obtain the flavor
torus $T_F = (\IC^{\times})^{6N-10}$. And the subtorus $T_{F,dW} \subset
T_F$ fixing the relations is again isomorphic to $\IC^4$.

We now analyze the torus action in $T_{F,dW}$ which can be induced by
the gauge group action $GL(V_1) \times \cdots \times GL(V_{N-1}) \times
GL(W_1) \times \cdots \times GL(W_{N-1})$. The induced action
is isomorphic to $(\IC^{\times})^{2N-3}$, which we use to fix the following
torus weight,
\be\label{eq:gaugefix}
\lambda_{a_i} = \lambda_{b_i} = \lambda_{r_1} =1 \ , \ \forall \  i =1, \cdots, N-2 \ .
\ee

Using \eqref{eq:gaugefix} to scale \eqref{eq:Nrelations} we find all the torus weight
of $c_i$, $d_i$, $r_i$, and $s_i$ should be the same.

$$
\bal
&\lambda_{c_1} =\lambda_{c_2} = \cdots = \lambda_{c_{N-1}} \equiv \lambda_{c} \ , \\
&\lambda_{d_1} =\lambda_{d_2} = \cdots = \lambda_{d_{N-1}} \equiv \lambda_{d} \ , \\
&\lambda_{r_1} =\lambda_{r_2} = \cdots = \lambda_{r_{N-2}} = 1 \ , \\
&\lambda_{s_1} =\lambda_{s_2} = \cdots = \lambda_{s_{N-2}} \equiv \lambda_{s} \ . \\
\eal
$$

So we have our final torus action $T_Q$ as,

\be
T_Q = \{  (\lambda_{c}, \lambda_{d}, \lambda_{s}) \in (\IC^{\times})^3 \ | \ \lambda_{c} = \lambda_{d} \lambda_{s} \} \simeq (\IC^{\times})^2 .
\ee

Similarly define $(\lambda_{d},\lambda_{s}) \equiv (\lambda_1, \lambda_2)$.
Torus weights of the elements in $\overline{\mathcal{P}}_{k+1}$ are given by
\be\label{eq:weightlist}
\bal
\{ \ & w(p)\ , \ w(r_k p) = w(p) \ , \ w(s_k p) =  \lambda_2 w(p) \ , \\
\ & w(c_k r_k p) = \lambda_1\lambda_2 w(p) \ , \ w(c_k s_k p) =\lambda_1\lambda_2^2 w(p) \ , \
w(d_k r_k p) =\lambda_1 w(p) \ , \\
& w(c_{k-1} r_{k-1}  c_{k} r_{k} p) = \lambda_1^2\lambda_2^2 w(p) \ , \
w(c_{k-1} r_{k-1}  c_{k} s_{k} p) = \lambda_1^2\lambda_2^3 w(p)\ , \\
& \ w(c_{k-1} r_{k-1}  d_{k} r_{k} p)= \lambda_1^2\lambda_2 w(p)\ , \ \cdots \\
& \ w(b_{k+1}b_{k} r_{k} p) = w(p) \ , \ w(b_{k+1}b_k s_k p)=\lambda_2 w(p)  \ , \cdots \\
& \ w(a_k p)=w(p) \ , w(a_{k+1} a_k p) = w(p) \ , \cdots \ | \ p \in \mathcal{P}_{k+1} \ \} \ ,
\eal
\ee and the weight of $p \in \mathcal{P}_{k+1}$ is
$$
w(p) = \left\{\begin{array}{ll}
\lambda_1^{n_1+n_2} \lambda_2^{n_2} &\mbox{if $p=(a_{k} d_{k} r_{k})^{n_1} (a_{k} c_{k} r_{k})^{n_2}$} \\
\lambda_1^{ n_1 + n_2} \lambda_2^{2n_1+n_2} &\mbox{if $p= (a_{k} c_{k} s_{k})^{n_1}(a_{k} c_{k} r_{k})^{n_2}$}
\end{array} \right. .
$$

The $T_Q$-fixed annihilator $I$ is generated by linear combinations of the path monomials of the same weights.
Given a torus weight $\lambda_1^{\alpha_1}\lambda_2^{\alpha_2}$ we can solve for finitely many monomial paths $p_i \in \mathcal{P}_{k+1}$ from \eqref{eq:weightlist} . The elements in the $I$ with weight $\lambda_1^{\alpha_1}\lambda_2^{\alpha_2}$
are most generally written as a finite sum $\sum_i \xi_i p_i$. If $\xi_j$ is not vanishing, $p_j$ should be included
as one of the monomial generators of the $T_Q$-fixed annihilator, since each $p_j$ is a linear map
from framing vector $e$ to a different vector space. We can exhaust all the monomial generators of $I$ this
way. This illustrates that torus fixed $I$ is generated by monomials and corresponds to an isolated point in the
moduli space of representations.

\bibliography{newrefB.bib}
 \bibliographystyle{abbrv}
\end{document}